# SCALING IN THE UNIVERSE

Stefano Borgani

*INFN, Sezione di Perugia, c/o Dip. di Fisica dell'Università,
via A. Pascoli, I-06100 Perugia, Italy
SISSA–ISAS, Via Beirut 2–4, I-34014 Trieste, Italy*

**Abstract**

The aim of this review article is to give a comprehensive description of the scaling properties detected for the distribution of cosmic structures, like galaxies and galaxy clusters. I will also discuss the more popular theoretical models, which have been proposed to account for the huge body of observational data. Due to the great variety of statistical methods, developed in the last twenty years to statistically describe the large-scale structure of the Universe, I will mainly concentrate on those methods which reveal remarkable regularities and scaling in the structure of the Universe. Although in most cases I prefer not to enter into the technical aspects of how implementing such methods, more details will be furnishes about the description of galaxy clustering in terms of fractal concepts. Statistical methods based on fractal analyses have been recently employed in cosmological context. Despite recent claims for a Universe, which behaves like a fractal at arbitrarily large scales, I will show that the fractal language can be usefully employed to disprove this picture. The emerging scenario is that of a Universe, which behaves like a self-similar structure at small scales, where fractality is dynamically generated by non-linear gravitational clustering, while preserving large-scale homogeneity. Nevertheless, even at scales $\gtrsim 10h^{-1}$ Mpc, where gravity still acts linearly, the distribution of galaxy clusters shows remarkable scale-invariant features, which could give precise hints about the initial conditions for the evolution of the large-scale structure of the Universe.

# Contents

# 1 Introduction

The study of the large–scale structure (LSS) of the Universe represents one of the most exciting research fields in cosmology. In the last twenty years or so the collection of a huge body of observational data has greatly contributed to improve our knowledge of "cosmography", so as to adequately test theoretical models about the origin and evolution of the Universe. The currently accepted view is that the structures observed today represent the result of gravitational evolution, starting from nearly homogeneous initial conditions at the outset of matter–radiation decoupling, with fluctuations of the energy density of the order $\delta \sim 10^{-5}$, which have subsequently grown by gravitational instability (see refs.[295, 314] for recent textbooks about the current status of cosmology and LSS studies). This picture is considered extremely plausible for a number of reasons, and recently received further support from the first detection of temperature anisotropies in the Cosmic Microwave Background (CMB), as provided by the COBE satellite [419].

The idea that the Universe should be uniform led to the formulation of the *Cosmological Principle*, on which most of the current cosmogonic pictures are based. In one of its versions, the Cosmological Principle states that the Universe is homogeneous and isotropic in its spatial part. Under this assumption about the symmetry of the space–like hypersurfaces, a system of coordinates can always be found in which the line element is written as

$$ds^2 \;=\; c^2\,dt^2 - a^2(t)\left[\frac{dr^2}{1-kr^2} + r^2\left(d\vartheta^2 + \sin^2\vartheta\,d\varphi^2\right)\right] \tag{1}$$

(see, *e.g.*, ref.[413]). With a suitable definition of the units of $r$, in the above expression the curvature constant $k$ can be considered to have only three possible values; $k = 0$ for a spatially flat Universe, $k = +1$ for a closed (positive curvature) Universe and $k = -1$ for an open (negative curvature) Universe. The quantity $a(t)$ represents the cosmic expansion factor. It gives the rate at which two points at fixed comoving coordinates $(r_1, \vartheta_1, \varphi_1)$ and $(r_2, \vartheta_2, \varphi_2)$ increase their mutual physical distance as $a(t)$ increases. Its time dependence can be worked out by solving Einstein's equations for the Friedmann–Lemaitre–Robertson–Walker (FLRW) metric of eq.(1). If the matter content of the Universe is that of a perfect fluid, such equations reduce to the system of two equations

$$\left(\frac{\dot{a}}{a}\right)^2 \;\equiv\; H^2 \;=\; \frac{8\pi G}{3}\rho + \frac{\Lambda}{3} - \frac{k}{a^2} \tag{2}$$

$$-\frac{\ddot{a}}{a} \;=\; \frac{4\pi G}{3}\left(\rho + 3p\right), \tag{3}$$

which are usually called Friedmann's equations. In eq.(2) I have also included the cosmological constant term $\Lambda$, which is assumed to be negligible in most cases. From a heuristic point of view, such equations can be seen as the equivalent of the energy conservation principle and of the second law of dynamics in classical (non relativistic) mechanics. Following the expression of the FLRW metric, two points at distance $d = a(t)r$ ($r$ is the fixed comoving distance) will move apart with velocity $v = \dot{a}r = Hd$. Determinations of the Hubble constant at present time, $H_o$, by using redshift–independent methods to measure galaxy distances give

$$H = 100\,h\,\mathrm{km\,s^{-1}\,Mpc^{-1}} \quad , \quad 0.4 \lesssim h \lesssim 1.$$

Note that this large uncertainty is not due to measurement errors, instead it comes from discrepancies between different distance estimators (see, e.g., ref.[353] and refs.[402, 398] for different points of view about the measurement of $H_o$; see also [167] for a recent review on this subject).

Based on eq.(3), it is possible to define the *critical* density $\rho_c = 3H_o^2/8\pi G = 1.9 \times 10^{-29}\,h^{-2}\,\mathrm{g\,cm^{-3}}$, such that present density values $\rho$ above, below or equal to $\rho_c$ refer to closed, open or flat geometries, respectively. Measurements of the cosmic mean density are usually expressed through the density parameter $\Omega \equiv \rho/\rho_c$. Current limits on its present value, $\Omega_o$, are

$$0.2 < \Omega_o \lesssim 1,$$

with values indistinguishable from the "closure" limit, $\Omega_o = 1$, being usually preferred on the ground of the standard inflationary scenario.

Once we specify the equation of state, which gives the relation between the density $\rho$ and the pressure $p$, the system of equations (2) and (3) can be solved for $a(t)$. Some particularly interesting cases are:

$$\Omega = 1,\ p = 0\ (\text{matter dominated}) \quad \Rightarrow \quad a(t) \propto t^{2/3}$$
$$\Omega = 1,\ p = \rho/3\ (\text{radiation dominated}) \quad \Rightarrow \quad a(t) \propto t^{1/2}$$
$$\Omega = 0\ (\text{free expansion}) \quad \Rightarrow \quad a(t) \propto t$$
$$p = -\rho\ (\text{vacuum dominated}) \quad \Rightarrow \quad a(t) \propto \exp(Ht).$$

Note that the $\Omega = 1$ cases approximate the expansion in non–flat geometries at sufficiently early times, when the curvature term in eq.(3) becomes negligible. Vice versa, the $\Omega = 0$ case represents the asymptotic expansion of an open Universe, when a very large value of $a(t)$ makes the density term so small that it gives negligible deceleration ($\ddot{a} \simeq 0$).

One of the fundamental consequences of the Cosmological Principle is the prediction that the Universe has undergone in the past a hot phase, during which the cosmic temperature took a much higher value than that, $T_o \simeq 2.7\,K$, which is today observed for the CMB. The resulting cosmological framework of the hot Big Bang in a spatially homogeneous and isotropic Universe is so widely accepted that it received the denomination of Standard Model (not to be confused with the Standard Model for electroweak interactions !). Indications point in favour of this model and the most striking and direct supports can be summarized as follows.

i) The observed proportionality between the recession velocity of galaxies and their distance (Hubble law), which is a natural consequence of assuming the FLRW metric of eq.(1).

ii) The detection and the high degree of isotropy of the CMB radiation, which is the evidence of a primordial hot stage of the Universe, characterized by a high degree of isotropy.

iii) The observed light element abundances, which match remarkably well the predictions of primordial nucleosynthesis, that is an unavoidable step in the evolution of the hot Universe.

Although the assumption of a homogeneous and isotropic Universe is correct at an early stage of the Universe or today at sufficiently large scales, nevertheless it is manifestly violated at scales below the typical correlation length of density fluctuations ($\sim 10h^{-1}$ Mpc), where the structure of the Universe is much more complex. However, this does not represent a problem for the Cosmological Principle, which, instead, would be in trouble if we were observing non negligible anisotropies at scales comparable to the horizon size ($d_H \sim (cH_o)^{-1} = 3000h^{-1}$ Mpc).

Observations of the Universe on scales similar to the typical galaxy dimension, $\sim 10$ kpc, reveal large inhomogeneities and the current view is that below such scales non–gravitational forces are dynamically dominant. On the other hand, scales $R \gg 10$ kpc are considered relevant to the LSS. The main difference between small (galaxy) scales and large scales lies essentially in the dynamics giving rise to structure formation. Indeed, the galaxy mass is determined by the capacity of the baryonic content to cool down during gravitational collapse, as the density increases. A quantitative analysis shows that, for masses $M \gtrsim 10^{12} M_\odot$, the heat produced during the initial collapse prevents a further compression [342]. Apart from the details of the heat production and dissipation, it is clear that, while the efficiency of the dissipation in a proto–object of dimension $R$ is proportional to $R^2$ (*i.e.*, to the extension of its surface), the heat production is proportional to $R^3$ (*i.e.*, to the mass of the object). Then, it is easy to understand that a characteristic scale $R^*$ must exist, such that above $R^*$ the rate of heat production is greater than the dissipation rate, which prevents the gravitational collapse from proceeding. The precise value of $R^*$ depends on the geometry of the collapse process, on environmental effects, and on the dissipation characteristics of the collapsing material. Detailed analyses give $R^*$ values that are very similar to the typical scale of normal galaxies. The study of dissipative processes, which determine the internal structure and dynamics of galaxies, are then essential to understand the origin of galaxies. However, such analyses are very difficult and the *genesis* and evolution of structures below the galaxy scales is still a widely debated issue.

On the contrary, on scales much larger than the galaxy ones, it is possible to study the formation and evolution of cosmic structure only on the basis of the gravitational interaction. Its evolution follows initially a linear pattern, while later, when the fluctuation amplitude increases sufficiently, it undergoes non–linear phases. For this reason, the large scale dynamics is not so easy to understand. However, on such scales the problem is much better determined and one's hope is to solve adequately the dynamical picture.

On such scales the essential observation is that galaxies have a spatial distribution with highly non–random characteristics. They show a strong tendency to group together forming clusters, while clusters themselves are clumped into "superclusters" on even larger scales. The resulting hierarchical appearance of the galaxy distribution suggests the presence of a sort of scale–invariance, which is also supported by several quantitative statistical analyses. The classical example is represented by the 2–point correlation function, which is observed to decline with a power–law shape, having the same slope for both galaxies and clusters, although at different scales and with different amplitudes. This scaling property is one of the most relevant characteristics that must be accounted for by any galaxy formation model. The hierarchical arrangement of the clustering is even more remarkable if we consider that it extends from the small scales, where gravitational dynamics is in the non–linear regime, up to large scales where linearity still holds. Therefore, a detailed statistical representation of the clustering displayed by the distribution of galaxies and galaxy systems is fundamental in order to compare the present Universe with the predictions of theoretical models for structure formation.

Instead of using positions of luminous objects, an investigation of the large–scale matter distribution in the Universe can be efficiently realized also by observing the effects of the background gravitational field on galaxy peculiar motions. A direct estimate of the radial peculiar velocity of a galaxy at distance $d$ is obtained by subtracting the Hubble velocity, $H_o d$, from the observed recessional velocity, once a redshift–independent estimate of $d$ is available. This kind of distance measurements are usually based on relations between intrinsic structural parameters of galaxies, such as the Tully–Fisher relation for spirals [390] (which relates the absolute luminosity and the observed rotation velocity), the Faber–Jackson relation for ellipticals [151] (which relates the absolute luminosity and the internal velocity dispersion) and the $D$–$\sigma$ relation for ellipticals [262] (which relates a suitably defined apparent diameter $D$ to the line–of–sight velocity dispersion $\sigma$). An exciting development in this field is represented by the recent completion of large galaxy redshift surveys and the availability of a considerable amount of redshift–independent distance estimates (see ref.[79] for a review). As a consequence, a lot of theoretical work has been devoted to find methods for extracting the large–scale three–dimensional velocity and mass density fields from measurements of radial peculiar velocities. At large scales peculiar motions are related to the gravitational potential field by quasi linear dynamical equations. In this regime, it makes sense to address the problem of reconstructing the matter distribution from the observed galaxy motions. Since the linearity of the gravitational clustering at large scales should have preserved the initial shape of the primordial fluctuation spectrum, the reconstruction procedure could furnish precise indications about the initial conditions. Several attempts in this direction have been already pursued (see, *e.g.*, refs.[39, 40, 121]) with promising results, despite the rather limited and sparse amount of available data. A decisive step forward in this direction should be however possible in the next years, with the availability of a more complete sampling of the velocity field traced by galaxy motions.

A further very efficient way to probe the nature of primordial fluctuations is represented by the investigation of the temperature fluctuations in the CMB. Such fluctuations are expected to be originated at the recombination time (corresponding to a redshift $z_{rec} \sim 1000$), when protons and free electrons couple together to form neutral hydrogen atoms. After that epoch, matter and radiation decouple and the Universe becomes transparent to the electromagnetic radiation. For this reason, inhomogeneities in the CMB should reflect the matter fluctuations just before decoupling. In past years, many efforts have been devoted to detect such anisotropies, with the result of continuously pushing down the lower limits for their amplitude. Only quite recently, the COBE satellite succeeded in detecting a significant signal for CMB temperature fluctuations at the angular scale $\vartheta \simeq 10°$. Although a detection at such a large angle (corresponding to a physical scale largely exceeding the horizon size at recombination) does not definitely fix the nature of the primordial fluctuations, nevertheless it provides fundamental support to the idea that the presently observed structures have grown from very small initial perturbations in the Friedmann background. Hopefully, in a short time more refined measurements at smaller angular scales should be able to further restrict the number of allowed initial condition models.

In this article I will mainly concentrate on the statistical analysis of the distribution of galaxies and galaxy systems, also comparing observational data to both numerical and analytical approaches to gravitational clustering. Although information coming from the study of peculiar motions and CBM temperature anisotropies are mentioned in order to constrain theoretical models for the initial perturbation spectrum, I will essentially deal with the study of the LSS in the "configuration space".

This work has been essentially motivated by the need to explain the nature and the origin of the hierarchical arrangement of the observed galaxy distribution. Particularly surprising is the coexistence of the observed complexity of the large–scale texture with the regularity of its statistical properties. The scaling displayed by the detected clustering, from the small scales ($\sim 10\,h^{-1}$ kpc) of galaxy halos, to those ($0.1 \lesssim r \lesssim 10 h^{-1}$ Mpc) of galaxy clustering, up to the large scales ($10 \lesssim r \lesssim 100 h^{-1}$ Mpc) for the clustering of rich galaxy systems, calls for dynamical mechanisms, which should be able to generate them over such a wide scale range. In order to make a close investigation of such scaling properties and of their possible dynamical origin, I review results obtained from the application of a series of statistical tests. Furthermore, I will concentrate on correlation and fractal analyses of observational data and N–body simulations of non–linear gravitational clustering, both reviewing already published material and presenting new results.

More in detail, this Article is organized as follows.

In Section 2 I will give a "by eye" description of the large–scale galaxy distribution, as obtainable from the available data sets. The presentation of a series of plots of galaxy and cluster samples will contribute to have a more precise idea about the great variety of the large–scale structure. I will also briefly discuss the essential characteristics of currently employed catalogues.

After this qualitative presentation of the large-scale clustering, I introduce in Section 3 the more important statistical methods, which are employed to quantify the properties of the galaxy distribution. I also review the results of the application of these methods to the analysis of extended data sets. Other than the classical approach based on correlation functions, I also present the power–spectrum analysis, the topological description of galaxy clustering based on the genus characteristics, and the study of mass and luminosity functions for cosmic structures. Technical descriptions about the implementation of such methods are beyond the scope of this article, while emphasis will be given to the discussion of the statistical information that they provide.

Section 4 is devoted to a detailed presentation of the concept of fractal structure and to a discussion of its relevance in the general framework of statistical mechanics and, more in particular, to the study of the galaxy distribution. Starting from the observed power–law shape of correlation functions, it has been argued that the galaxy distribution has well defined scale–invariant properties. It is clear that using fractal concepts to characterize the large–scale clustering does not imply that it is represented by a self–similar structure extending up to arbitrarily large scales, as some authors suggested (see, *e.g.*, ref.[99]). Vice versa, methods of fractal analysis are well suited to characterize the presence of scaling and the width of the scale–range where it takes place. After the introduction of the concept of fractal (self–similar) structure, I will show that it is characterized by the so–called *fractal dimension*. I introduce several definitions of fractal dimension, as originally considered in the study of non–linear systems and deterministic chaos. The *multifractal* spectrum extends the concept of fractal dimension by providing a hierarchy of scaling indices, which give a complete description of the scaling properties of a self–similar structure. In order to evaluate the fractal dimension for a point distribution, I will describe a list of algorithms, which are based on different definitions of dimension and rely on different approximations to its "true" value. Finally, I review the results obtained up to now from the fractal interpretation of the large–scale structure of the Universe. After that, I will discuss in some details the existing relations between correlation functions and fractal dimensions.

Section 5 is devoted to the discussion of the dynamics of structure formation in the framework of the gravitational instability picture. After writing the equations for the evolution of density inhomogeneities, I will describe their solution in the simple linear regime and also describe some approximations to treat the non–linear stage of the gravitational clustering. I will also discuss the origin of primordial fluctuations, in relation with the matter content of the Universe. Both baryonic and non–baryonic models, based on hot, cold and mixed dark matter, are described, and the effect of taking non–Gaussian initial conditions is also considered. In the framework of *biassed* models of galaxy formation, I will describe the motivations which support the idea that galaxies and galaxy systems are more clustered than the underlying dark matter.

In order to provide a comprehensive treatment of gravitational clustering in the strongly non–liner regime, I resort in Section 6 to the analysis of numerical N–body simulations. The purpose of this analysis is to verify the existing connections between scale-invariant clustering and non–linear gravitational dynamics. After an introductory description of cosmological N–body codes, I will describe the results of correlation and fractal analysis, starting from different initial fluctuation spectra. The analysis of high–order correlation functions shows that these closely follow the hierarchical expression, as predicted by many models of non–linear clustering. As an important outcome of fractal analysis, self–similarity is shown to be always associated with non–linear gravitational dynamics. The application of several fractal dimension estimators points toward a multifractality of the small–scale clustering, with a dimension $D \simeq 1$ which always characterizes the distribution inside the overdense, virialized structures. This represents a robust outcome of our analysis, quite independent of the initial conditions and supports the idea that a $D \simeq 1$ fractal dimension is a sort of attracting solution of the non–linear gravitational dynamics.

In order to check whether this scaling behaviour can be extended to larger scales, I describe in Section 7 the scaling analysis for the distribution of galaxy clusters. The application of the fractal analysis to both angular and redshift samples of galaxy clusters shows that a remarkable scale–invariance characterizes their non–linear clustering. The resulting multifractal spectrum has a characteristic dimension $D \simeq 1$–1.4 corresponding to the more dense parts of the distribution. The breaking of the scale–invariance occurs at the scale $\sim 40h^{-1}$ Mpc. At larger scales, the cluster distribution becomes essentially homogeneous, thus disproving the picture of a purely fractal Universe, extending up to arbitrary large scales. I finally compare observational data with results coming from the fractal analysis of simulated cluster samples. This represents a necessary step in order to check whether the detected self–similarity is rooted into initial conditions for structure formation.

In Section 8 I will summarize and discuss the emerging picture about the scaling properties of the LSS of the Universe.

# 2 Observations of large scale structure

In this section I describe the visual appearance of the large–scale structure of the Universe, as emerging from the available data sets. First of all, I give a qualitative description of the global texture of the galaxy distribution, showing how recent improvements of data sets enlarged our view of the large–scale clustering. After that, I describe the more important catalogues of galaxies and galaxy clusters, that are used to trace the large–scale structure of the Universe. The rather complex picture emerging for the galaxy clustering cannot be satisfactorily accounted for just by a crude photograph of the galaxy distribution, instead it requires a quantitative statistical analysis, that will be described in much more details in following sections.

## 2.1 An "eye ball" description of galaxy clustering

Starting from the first investigations of the galaxy distribution in the sky, it has been recognized that galaxies in our neighborhood are distributed in a very inhomogeneous way, while only considering sufficiently large patches of the Universe the homogeneity expected on the ground of the Cosmological Principle seems to be attained. A first pioneering attempt to realize a systematic survey of galaxies was realized by Shapley & Ames already in 1932 [371]. They included in their catalogue all the galaxies brighter than the 13th photographic magnitude. The first visual impression, which this catalogue provided, was that of a Universe in which galaxies are not randomly distributed on the sky, but tend to be clumped to form large structures. After that, more deep and detailed surveys confirmed and strengthened this kind of picture. The Zwicky catalogue [428], that reach the apparent magnitude $m = 15.5$, showed the presence of clusters of galaxies containing up to 1000 objects and more, superclusters and filaments, with characteristic sizes of several tens of Mpc. In Figure $1a, b$ the distribution of galaxies on portions of the sky surveyed by the Zwicky sample is shown. The Coma cluster is the dense concentration at $\alpha = 13^h$ and $\delta = 28°$, while the Virgo cluster is centered at $\alpha = 12.5^h$ and $\delta = 12°\!.5$. A decrease in the galaxy density west of $9^h$ and east of $16^h$ is due to galaxy obscuration. In the southern hemisphere the most relevant structure is the Pisces–Perseus filament, which crosses the sky in the declination interval $30°$-$40°$. Again, the apparent absence of galaxies at $21^h$ and $3^h$ is due to galactic absorption.

A great step forward in the description of the large–scale distribution of galaxies has been subsequently realized with the compilation of the Lick map by Shane & Wirtanen [368]. This sample, that goes even deeper than the Zwicky compilation, is complete down the magnitude $m = 18$ and includes roughly one million galaxies, for which the count in cells of $10 \times 10$ arcmin are given, instead of the position of each object. Despite the great depth reached by the Lick map, that would be expected to wash out many details of the galaxy clustering, a great variety of structures still appears. Although the Lick map surely provided for a long time a fundamental basis for many studies of large–scale structure, in recent years our knowledge of the galaxy distribution on the sky has reached a much greater depth, thanks to recently compiled angular samples, such as the APM sample [263] (see Figure 2) and the Edinburgh/Durham Southern Galaxy Catalogue [208].

Although such enormous angular samples contain a great amount of informations about the large–scale distribution of galaxies, nevertheless the availability of complete redshift surveys led to a dramatic change in our view of the geometry of the galaxy clustering pattern. It is however clear that measuring redshifts and, thus, distances of galaxies, is a much harder task than only measuring angular positions on the sky. For this reason, three-dimensional galaxy samples include nowadays only a small fraction of all the galaxies listed in the angular surveys, and continuous efforts are devoted to enlarge and improve our three-dimensional view of the Universe. Because of the intrinsic difficulty of having redshift samples with the same depth and sky coverage as angular ones, a possible strategy is to restrict the attention only on particularly interesting regions of the sky.

Kirshner et al. [238], in their redshift survey of the Böotes region, revealed the presence of a big underdense structure, where the density of bright galaxies is $\lesssim 20\%$ of the mean, and which has a diameter of $\sim 60\,h^{-1}$ Mpc. Another example of a redshift survey devoted to the investigation of a single structure is that realized by Haynes and Giovanelli [203], in order to properly study the spatial structure of the Perseus–Pisces region. In this redshift survey, the authors have shown that many of the galaxies in the region lie in a narrow redshift range around $\sim 50\,h^{-1}$ Mpc. Therefore, as well as in the angular projection, the Perseus–Pisces filaments turns out to be a thin structure in the redshift direction too.

Instead of investigating the details of the galaxy distribution in a specific region, the CfA survey [125, 216] is an attempt to map the general galaxy distribution, rather than to explore a particular feature on the sky and it should represent the three-dimensional version of the Zwicky map. Partial results of such an investigation are shown in Figures $3b, c, d$, where the wedge diagrams of declination slices of the sky are plotted. One of the most relevant features is again the presence of several large regions that appear to be almost devoid of galaxies, while galaxies seem to be distributed in elongated structures. De Lapparent et al. [125] suggested that "the apparent filament is a cut through boundaries of several bubble-like structures". The pronounced radial "finger" along the line of sight in Figures $3b, c$ is the Coma cluster. If we could map the actual position of galaxies rather than their redshifts, this feature would be approximately spherically symmetric, with a radius $\sim 1\,h^{-1}$ Mpc. The elongation in redshift space occurs because of peculiar velocities of galaxies inside the cluster, that affect the determination of the distance, if simply based on the Hubble relation. The slice of Figure $4b$ could suggest that filamentary structures in this region are cuts through two-dimensional sheets, not one dimensional filaments. Indeed in contrast to Figure $4a$, that samples the apparent filaments of Figure $1a$, there are no detected filaments on the sky in the region covered by Figure $1b$. Thus, being difficult to expect that the intersection of a slice with a three-dimensional network of filaments give rise in turn to a network of filaments, a sheet-like distribution of galaxies around the voids could better account for the data.

Despite a number of observations have been devoted to enlarge the sampled volume of the Universe, there is up to now no striking evidence for an upper limit to the scale of structure discernible in the redshift analysis. Indeed, the existence of coherent structures involving scales comparable with the size of the sample itself is apparent. The most relevant of such structures is the so–called *Great Wall* revealed by the slices of the CfA sample [171]. The apparent extension of the *Great Wall* in both right ascension and declination is only limited by the extension of the survey. The detected spatial extent in these two dimensions is $\sim 60\,h^{-1}$ Mpc $\times\,170\,h^{-1}$ Mpc. The typical thickness, approximately along the redshift direction, is $\lesssim 5\,h^{-1}$ Mpc. The density contrast between the wall and the mean of the survey is $\Delta\rho/\rho \simeq 5$. The 360° view of Figure 4 indicates the geometrical relation between the Perseus–Pisces chain and the Great Wall.

Quite differently from the optical samples, the recent availability of galaxy catalogues selected in the infrared band by IRAS (Infrared Astronomical Satellite) opens the possibility of having extended nearly all–sky redshift surveys, due to the limited absorbtion of the Galactic plane at that frequencies. One of these, the QDOT (Queen Mary and Westfield–Durham–Oxford–Toronto) redshift survey [359] has been realized by measuring the redshifts of 1–on–6 randomly selected IRAS galaxies and reaches a depth of $\sim 140h^{-1}$ Mpc (see Figure 5). Although infrared selected galaxies are found to avoid rich clusters with respect to optical ones, so that they are more uniformly distributed, nevertheless the availability of such a redshift sample permits one to identify low–density, but very large, structures [359].

Due to the already mentioned problems in measuring a large number of redshifts, an alternative way to have hints about the spatial distribution of galaxies is to realize the so–called pencil–beam surveys: instead of considering a quite large patch of the sky, with a rather limited depth, a different strategy is to realize very deep, but very narrow, surveys so to include a not exceedingly large number of galaxies. Adopting this kind of approach, significant results about the galaxy redshift distribution have been obtained by Broadhurst et al. [74], that combined data coming from four distinct surveys at the north and south Galactic poles to produce a well sampled distribution of galaxies by redshift on a linear scale extending to $2{,}000h^{-1}$ Mpc. By analysing their pencil–beam survey, they found a remarkably regular redshift distribution with most galaxies lying in discrete peaks, having a remarkable periodicity over a scale of $\sim 128\,h^{-1}$ Mpc. A comparison with the CfA galaxy distribution shows that the first peak occurs just in correspondence of the Great Wall. The detection of such peaks in the galaxy distribution, at least in one direction, lead some authors to model the large scale structure of the Universe by means of suitable cellular structures, such as provided by the Voronoi tessellation [95, 218, 400]. In this picture, the cross section of a pencil–beam with the three–dimensional cellular pattern could generate one–dimensional redshift distributions which strongly depend on the direction of the beam itself. By studying the power–spectrum of a three–dimensional Gaussian random field, Kaiser & Peacock [234] claimed that the periodicity at the scale of $128h^{-1}$ Mpc does not necessarily imply the presence of an excess of power at this scale in the three-dimensional structure. It is however clear that, before definitely assess the presence of a preferred scale in the large-scale galaxy distribution, more independent pencil beams in different directions are needed.

Other than galaxies, olso galaxy clusters are usefully employed to trace the LSS, especially at large ($\gtrsim 50h^{-1}$ Mpc) scales (see, e.g., the review by Bahcall [19]). The enhanced clustering of rich galaxy systems with respect to the clustering of galaxies [22] makes it possible to reveal structures, which are otherwise hardly detectable on the ground of the observed galaxy distribution. To this purpose, many attempts have been devoted to compile homogeneous samples of galaxy clusters. The most famous of these samples is probably the Abell catalogue [3], selected already in 1958 from the Palomar Survey Plates. It includes visually selected clusters lying north of declination $\delta = -27°$. More recently, the extension of this sample to the southern sky led to the compilation of the ACO cluster sample [5], that used the same selection criteria as Abell's, so to give a consistent whole sky coverage of the cluster distribution (see Figure 6). Based on the these samples, many features have been observed in the distribution of rich galaxy clusters, indicating the presence of relevant structures up to scales of $\sim 300h^{-1}$ Mpc. For instance, Tully [393] detected the presence of structures of this size lying in the plane of the Local Supercluster. Another relevant structure is the Shapley concentration [370], that is an overdensity in the cluster distribution lying $\sim 20°$ away from the direction of the cosmic microwave background (CMB) dipole ($l \simeq 256°$, $b \simeq 48°$; see ref. [376]) and at a distance of $\sim 140h^{-1}$ Mpc. Several investigations have shown the dynamical relevance of this cluster concentration to determine our motion as inferred from the CMB dipole [360, 337] and the optical dipole of the cluster distribution [325, 361]. Thanks to the recent availability of sufficiently large redshift data for galaxy clusters [332], more and more efforts are nowadays produced to give a detailed description of the clustering and the geometry of the cluster distribution [391, 327, 81].

From the above description of the large-scale texture of the Universe, the great complexity and the variety of observed structures is apparent. Galaxies, instead of being uniformly distributed, are arranged to form filaments of some tens of Mpcs and rich clusters, while leaving nearly devoid regions of sizes up to $\sim 50h^{-1}$ Mpc. In turn, clusters are themselves non-trivially clustered, but give rise to structures of even higher hierarchy, the superclusters. The appearance of this complexity makes it clear that any comparison with dynamical models for the origin and evolution of the LSS should pass through two fundamental steps. Firstly, compiling homogeneous samples of galaxies and galaxy clusters as extended as possible, both in two- and, even better, in three-dimensions. Secondly, realizing detailed statistical analyses of observational data, that were able to quantify both the global geometrical properties and the details of the clustering.

## 2.2 Galaxy samples

The statistical investigation of the distribution of galaxies has been initially realized by means of angular homogeneous samples, in which angular positions of the objects on the sky are reported. An angular sample contains all the galaxies that satisfy a given selection criterion. Since galaxy distances are not known a priori, selection criteria are based on "apparent" properties of these objects.

A first criterion is based on the apparent luminosity. Samples, whose compilation is based on it, include all the galaxies in a given region of the sky, which have observed luminosities exceeding a fixed value. For historical reasons, apparent luminosities are expressed in logarithmic units, by means of the apparent magnitude $m = \mathbf{M} + 5\log d + 25$. Here $d$ represents the distance in Mpc of the object from the observer, while $\mathbf{M}$ is the *absolute* magnitude (*i.e.*, the apparent magnitude of the same object if placed at the fixed distance $d = 10\,pc$). Once a limiting apparent magnitude is chosen, an important related quantity that characterizes a galaxy sample is the *depth* of the survey. Galaxies selected by apparent magnitude are found to have a fairly definite absolute magnitude, $\mathbf{M}^*$, with rather limited deviations around this value. Thus, $\mathbf{M}^*$ and $m$ define a characteristic distance

$$D = 10^{0.2(m-\mathbf{M}^*)-5}\,\mathrm{Mpc}, \qquad (4)$$

which represents the effective depth of the sample.

With the recent availability of satellites to realize complete surveys of the galaxy distribution, the possibility to select object in bands that are different from the optical one has also been opened, as traditionally done by ground–based telescopes. However, since galaxies of different morphology are characterized by having different luminosity at different frequencies, it turns out that catalogues compiled in different bands do not sample with the same efficiency galaxy populations having different morphology. The classical example is represented by the IRAS survey (see below), which select objects according to their observed flux in the infrared band. Since early–type galaxies have on average a greater infrared emissivity, this sample preferentially selects spirals instead of ellipticals.

A further problem arising in the comparison of different samples is due to the fact that galaxies are selected not always according to their apparent magnitude, but also according to other intrinsic properties, such as the angular diameter (diameter limited samples). Accordingly, a different definition of depth of the sample is given, reflecting the relation between distance and apparent size. Note that using angular sizes instead of apparent magnitudes causes some bias in the completeness of the sample. For instance, spiral and irregular galaxies are known to have a lower surface brightness, with respect to ellipticals and spheroidals [175, 107]. Therefore, they are preferentially included in diameter–limited samples rather than in magnitude–limited samples.

For the above reasons, particular care must be payed when one is comparing the results of statistical analyses made from galaxy catalogues, which are compiled by using different selection criteria. In the following, I give a list of the most important angular galaxy samples.

i) **The Zwicky sample** [428], that is based on the Palomar Observatory Sky Survey (POSS). This sample contains the angular positions of 3753 galaxies, having apparent magnitudes $m \leq 15.5$ and coordinates with declination $\delta \geq 0$ and galactic latitude $b^{II} \geq +40°$ (see Figure 1).

ii) **The Lick sample** [368], which gives count–in–cells for galaxies with apparent magnitudes $m \leq 18$. Each galaxy belongs to an elementary cell of $10' \times 10'$. In turn, these cells are grouped in sets of $36 \times 36$ to form maps, that have an extension of $6° \times 6°$. The centers of each map are separated one from each other by $5°$ in declination, from $\delta = -20°$ up to $\delta = +90°$. The separation in right ascension is such that each map is overlapped to adjacent one at most for $1°$.

iii) **The Uppsala General Catalogue (UGC)** [288], that is based on the POSS plates and contains all the galaxies in the northern hemisphere ($\delta \geq -2°20'$), having apparent diameter $\geq 1\rlap{.}'0$. In addition, it also includes all the galaxies of the Zwicky sample brighter than $m = 14.5$, even if their diameter is smaller than $1\rlap{.}'0$. Informations are also listed about major and minor photometric axes, morphology, colour index and radial velocity, when available.

iv) **The Jagellonian field** [349], that includes more than 10,000 galaxies comprised in a small angular region of $6° \times 6°$. This sample turns out to be almost 8 times deeper than the Zwicky sample in such a way that, even though the angular extension of the latter is $\sim 10$ times larger, the *spatial* dimension of the two sampled regions are almost the same.

v) **The ESO/Uppsala Catalogue** [248], which is based on the *ESO Quick Blue Atlas*. This sample is considered complete for those galaxies of the southern hemisphere having declination $\delta < -17\rlap{.}°5$ and major diameters greater than $1'$. Additional informations, such as photometric axes, position angles, morphological types, other than colour indices and radial velocities are also listed, when available.

vi) **The APM Galaxy Survey** [263], that has been realized by using the Automate Plate Measuring (APM) machine for an automatic scan of $5\rlap{.}°8 \times 5\rlap{.}°8$ for each of the 185 plates of the UK Schmidt J Survey. The plates cover an area of 4,300 square degrees in the region $\delta < -20°$ and $b \lesssim 40°$. It includes around 2 millions galaxies brighter than $m = 20.5$ and is considered 95% complete in the range $17 < m < 20.5$. Adjacent plate centers are separated by $5°$, so that the resulting overlap can be used to check for the presence of systematic errors (see Figure 2).

**vii) The Edinburgh/Durham Southern Galaxy Catalogue** [208], that has been realized by scanning with the Edinburgh plate measuring machine (COSMOS) 60 plates of the UK Schmidt J Survey. The plates are located around the south galactic pole. The sample reaches a limiting magnitude $m \simeq 20$ and contains $\sim 10^6$ galaxies. The scanning of $5°.3 \times 5°.3$ ensures some overlapping between adjacent plates to correct for systematics. The sample is considered to be a 95% complete, with $< 10\%$ stellar contamination.

Other than the above angular samples, there exist several three–dimensional samples, that, in addition to the angular coordinates, give also the redshift for each object. Although the redshift data can be translated into distance data according to the Hubble law, nevertheless galaxy peculiar motions affect to some extent the three–dimensional picture of the galaxy clustering, producing elongation of the clustering along the line–of–sight. In general, redshift samples use angular catalogues as reference databases from which to select galaxies. Suitable objective criteria are applied to choose those galaxies for which to measure the redshift. In the following I describe the most relevant galaxy redshift samples.

**i) The CfA1 (Center for Astrophysics) survey** [214], that select galaxies from the Zwicky and UGC samples. This survey includes all the 2400 galaxies having magnitudes $m \leq 14.5$ and angular positions characterized by ($\delta > 0°$, $b^{II} > +40°$) and by ($\delta \geq -2°.5$, $b^{II} < -30°$). For each galaxy, equatorial coordinates, heliocentric velocity and apparent magnitude are given.

**ii) The Southern Sky Redshift Survey (SSRS)** [107], that includes galaxies selected from the ESO/Uppsala catalogue. The survey lists 1657 galaxies, selected from the ESO Catalogue [248], in an area of 1.75 steradians, with declination south of $-17°.5$ and galactic latitude below $-30°$. The sample is diameter–limited with all galaxies having $\log D > 0.1$, where $D$ is a "face–on" diameter, in arcminutes, which is suitably related to galaxy morphological type and angular dimension. According to the angular sample, which is used as the source, the SSRS also provides morphological types and diameters. The depth of the sample is estimated to be $120h^{-1}$ Mpc.

**iii) The Arecibo survey** [170], that has been realized with the aim of studying the spatial galaxy distribution inside the Perseus–Pisces supercluster. It includes about 4700 galaxies with equatorial coordinates ($22^h \leq \alpha \leq 4^h$, $0° \leq \delta \leq 45°$). The completeness magnitude is $m = 15.5$ and diameter $1'.0$. Morphology and diameter informations are also provided.

**iv) The CfA2 survey** that is slowly emerging and that, when completed, should represent the extension of the CfA1 survey to the apparent magnitude $m = 15.5$. The already published data are organized in declination slices, that are $6°$ thick. A first slice, that contains 1057 objects in the declination range $26°.5 \leq \delta \leq 32°.5$ (Figure 4b), was completed in 1986 [125], while data have been subsequently published [216]. Two other complete slices [171] contain 1443 galaxies in the declination range $32°\!.5 \leq \delta \leq 44°\!.5$ (Figure 3c). Nowadays, the survey should be complete in eleven slices, six in the north and 5 in the south. The CfA2 North covers the portion of the sky $8^h < \alpha < 17^h$, $20°\!.5 < \delta < 44°\!.5$ and $8°\!.5 < \delta < 14°\!.5$, and includes 5248 galaxies. The CfA South covers $20^h < \alpha < 4^h$, $6° < \delta < 42°$ and includes 3045 galaxies [409].

v) **The QDOT IRAS redshift survey** [359], which contains 2163 randomly selected IRAS galaxies with measured redshifts, with a 1–to–6 sampling rate. All the included objects have a flux greater than 0.6 $Jy$ and the sample covers nearly all the sky at galactic latitudes $|b^{\text{II}}| > 10°$. The overall completeness after excluding galactic sources is estimated to be 98%.

vi) **The Strauss et al. IRAS redshift survey** [378], which also selects galaxies from the IRAS Point Source Catalogue. The sample includes 2649 galaxies flux limited at 60 $\mu m$ and covers 11 steradians of the sky, thus providing excellent coverage. The sampling rate is here 1–to–1, but at the expenses of a reduced depth.

vii) **The Stromlo–APM redshift survey** [255], that is presented to be essentially complete to the limiting magnitude $m = 17.15$ and contains 1769 galaxies randomly selected at a rate of 1 in 20 from the APM catalogue. The extension of this survey is $\sim 30$ times that of the CfA1 survey, so that it is particularly suitable for the determination of the mean galaxy density or to the study of large–scale features.

## 2.3 Cluster samples

Galaxies do not represent the only class of cosmic structures that can be used to investigate the large scale structure of the Universe, rather they tend in turn to group together to form structures on larger scales, such as groups or clusters. In turn, such structures can be considered as single objects and, then, the statistics of their distribution can be analysed in full analogy with the case of galaxies. The advantage of using clusters mainly resides in the fact that they trace the structures up to the very large scales of some hundreds of Mpcs.

In general, different cluster samples use different selection criteria for identifying clusters, so that different it can be also the respective distribution. In general, cluster samples can be divided in two main categories: those in which clusters are selected by the visual inspection of the galaxy distribution and those in which they are identified by means of an objective computer algorithm. It is clear that samples of the first type are likely to suffer by human biases, which are quite difficult to be accounted for. On the other hand, even in the case of automatically selected cluster catalogues, some biases are however present. As an example, if they are selected from an angular galaxy distribution, serious contaminations due to projection effects can be present in any case. It is clear that the best one can hope to do is by using

three–dimensional galaxy samples from which to select galaxy systems. However, as we have already noted, at present only relatively small galaxy redshift samples are available, so that the cluster catalogues selected from these are also far from being extended and complete samples.

The most important angular samples of clusters are described as follows.

i) **The Abell Catalogue** [3] includes a total of 2712 clusters that are the richest, densest clusters selected by visual inspection from the POSS plates (see Figure 6). Out of these rich clusters, 1682 constitute the Abell's complete statistical sample and are distributed over 4.26 steradians. The Abell selection criteria can be summarized as follows: ($a$) a cluster must contain at least 50 members in the magnitude range $m_3$ to $m_3+2$, where $m_3$ is the magnitude of the third brightest galaxy; ($b$) all these members should be contained within a circle of radius $1.5h^{-1}$ Mpc around the center of the cluster; ($c$) the cluster redshift $z$ must be in the range $0.02 \lesssim z \lesssim 0.20$; and ($d$) the cluster must lie north of declination $-27°$. The 1682 clusters in the sample are divided in 6 distance classes, $D$. Each distance class contains clusters lying in a magnitude interval of width $\Delta m = 0.7$, starting from $m = 13.3$; 104 clusters belong to the first 4 distance classes (nearest clusters), while the remaining belong to the subsample $D = 5+6$. Also a richness class is assigned to each cluster, with $R = 0$ through 5, that is related to the number of members belonging to each cluster. The corresponding ranges of member counts are 30–49, 50–79, 80–129, 130–199, 200–299, and above 300.

ii) **The Zwicky Catalogue** [428] contains 9700 clusters visible to the limit of the Palomar plates ($m \simeq 20$). The criteria for including clusters in the sample are less restrictive than Abell's; ($a$) The cluster must contain at least 50 galaxies in the magnitude range $m_1$ to $m_1 + 3$, where $m_1$ is the magnitude of the brightest galaxy; ($b$) these galaxies must lie within the isopleth, where the projected density of galaxies is about twice that of the neighboring field; ($c$) no limits on the redshift are specified, but structures such as the Virgo cluster (which cover very large areas) are not included; and ($d$) the clusters must lie north of declination $-3°$ and within well specified areas. Cluster richness is defined as the number of galaxies, corrected for the mean field count, that are located within the isopleth of twice the field density. In general, Zwicky clusters differ in size from Abell's, the former being mostly larger, lower density systems.

iii) **The Schectman Catalogue** [363] identifies 646 clusters of galaxies, based on the Lick counts and using an automated procedure. The clusters are located at galactic latitudes $|b^{\mathrm{II}}| \geq 40°$ and declinations $\delta > -22°\!.5$. The selection is based on local density maxima of the galaxy distribution above a given threshold value. A selected threshold of five galaxies per $10 \times 10$ arcmin cell was used; this threshold is considerably higher than the tail of the background distribution of galaxy counts, which has a median of 1.3 galaxies per bin. Such a threshold of five galaxies succeeds in detecting 70% of Abell's $D \leq 4$ clusters and 10% of the $D = 5$ clusters. The Schectman procedure selects clusters that are considerably poorer than the Abell $R \geq 1$ clusters.

iv) **The ACO Catalogue** [5] is the extension to the southern hemisphere of the Abell sample. It contains 1635 clusters of richness class $R \geq 0$ and includes clusters in the $-27° < \delta < -17°$ overlap region with the Abell clusters. Together with the Abell Catalogue, it constitutes an all–sky sample of 4073 rich Abell clusters (see Figure 6), nominally complete to a redshift $z = 0.2$ for clusters with populations of 30 or more galaxies in the magnitude range $m_3$ to $m_3 + 2.0$.

v) **The Plionis, Barrow & Frenk (PBF) samples** [323], which select clusters from the Lick map, by using overdensity criteria and represents an extension of the Schectman sample. After smoothing the cell count over a 30 arcmin scale, clusters are identified from those cells, whose count exceeds $\kappa$ times the average count. Four different cluster samples are generated, corresponding to the overdensity factors $\kappa = 1.8, 2.5, 3$ and $3.6$ (see also ref.[324]).

vi) **The Edinburgh/Durham Southern Cluster Catalogue (EDSCG)** [261], which select clusters with an overdensity criterion from the Edinburgh/Durham galaxy catalogue. The sample contains 737 clusters and covers an area of $80° \times 20°$, centered at the south galactic pole. The EDSCG is constructed using an automated peak–finding algorithm and is complete to $m_{10} = 18.75$. Cluster selection criteria have been suitably chosen so to reduce projection contamination effects.

As in the case of the galaxy samples, the detection of redshifts for clusters included in angular catalogues permits the compilation of three–dimensional surveys. The most relevant of these samples are described as follows.

i) **The Hoessel, Gunn & Thuan (HGT) sample** [210] includes the redshifts of all the Abell clusters with distance class $D \leq 4$ and richness class $R \geq 1$, which are located at galactic latitude $|b^{\mathrm{II}}| \geq 30°$. A total amount of 104 clusters belong to this sample.

ii) **The Struble & Rood (SR) catalogue** [379], that contains all the 588 Abell clusters (including richness class $R = 0$) with measured redshifts.

iii) **The Geller & Huchra (GH) deep redshift survey** [170], that consists of the 145 Abell clusters with $R \geq 0$, $D \leq 6$ and with redshift $z \lesssim 0.2$, in the area $10^h < \alpha < 15^h$, $58° < \delta < 78°$.

iv) **The Ramella, Geller & Huchra (RGH) group sample** [338], that identifies groups of galaxies in the first two complete strips of the CfA redshift survey [216]. The group catalogue is produced by applying an algorithm which searches for "friends–of–friends" in redshift space [215]. The catalogue contains 128 groups with at least three members and 56 with at least five members.

v) **The Southern Hemisphere Group (SHG) sample** [108] contains 87 groups with 3 or more members and with a surrounding density contrast greater than 20, identified from the SSRS galaxy catalogue. The groups are located southwards of declination $-17°\!.5$, below galactic latitude $-30°$ and have mean radial velocities less than 800 km s$^{-1}$.

vi) **The Postman, Huchra & Geller (PHG) Cluster sample** [332], which is a complete sample of 351 Abell clusters with tenth–ranked galaxy magnitude $m_{10} \leq 16.5$. The survey includes all the clusters lying north of $\delta = -27°30'$, being 15 objects selected from the ACO sample. Today, it represents the largest available complete redshift survey of rich clusters.

vii) **The APM Cluster Redshift survey** [109], that contains about 200 clusters with measured redshift $z \lesssim 0.1$, that have been selected from the APM galaxy sample, using an overdensity criterion. The survey covers an area of $\sim 4300$ square degrees of the southern sky. The resulting spatial density is $2.4 \times 10^{-5}\,(h^{-1}\,\mathrm{Mpc})^{-3}$, four times that of $R \geq 1$ Abell clusters and twice than that of $R \geq 0$ clusters, thus indicating their lower richness.

viii) **The Edinburgh/Milano Cluster Redshift survey** [287], which contains clusters selected from EDCG. Selection criteria require: at least 22 member galaxies inside a radius $r = 1h^{-1}$ Mpc with magnitude between $m_3$ and $m_3 + 2$, tenth–ranked galaxy magnitude $m_{10} \leq 18.75$, and equatorial coordinates in the range ($\alpha \geq 21^h53^m$, $\alpha \leq 3^h35^m$; $-42°12' \leq \delta \leq -22°53'$). The resulting number of selected clusters is 97. On average, 10 galaxy redshifts are measured in the direction of each cluster core.

In Section 7 I will discuss the scaling analysis of the PBF and PHG cluster samples. At that point, a more detailed description of these samples will be given.

# 3 Statistical measures of the Universe

In this section I present a review of several statistical measures, which are used to investigate the large scale distribution of cosmic structures. After introducing the basic statistical formalism, I will describe the results provided by such methods. Firstly, I introduce the approach based on the analysis of correlation functions, that from an historical point of view represents the first serious attempt to quantitatively investigate the statistics of the galaxy distribution. Based on the path–integral approach, as developed in the framework of statistical mechanics and quantum field theory, I formally describe the correlation properties for a generic density field. A review of the main results of correlation analysis in clustering studies will be finally given. After that, I discuss the power spectrum analysis that, although strictly related to correlation functions, recently provided extremely significant results and represents a sort of complementary approach to the correlation one. As a further characterization of the galaxy distribution, I also review some measures of the topology and show how efficient they are to characterize the geometry of the galaxy clustering pattern. A further statistical measure concerns the luminosity distribution of galaxies and galaxy systems. Its relevance in connection with the primordial spectrum of density fluctuations is outlined and observational results about luminosity and mass functions are discussed.

The methods based on fractal analysis, which represents an important part of this article, will be discussed in the Section 4.

## 3.1 Correlation functions

The classical correlation analysis of the galaxy distribution, as pioneered by Totsuji & Kihara [388] and extended during the seventies by Peebles and coworkers [315, 306, 189], was based on the determination of the angular 2–point correlation function, $w(\vartheta)$, from projected galaxy samples. Its definition is related to the joint probability

$$\delta^{(2)}P \; = \; n^2 \, \delta\Omega_1 \, \delta\Omega_2 \, [1 + w(\vartheta_{12})] \qquad (5)$$

of finding two objects in the solid angles $\delta\Omega_1$ and $\delta\Omega_2$, respectively, at angular separation $\vartheta_{12}$ [308]. In eq.(5) the factorization of the $n^2$ term ($n$ being the angular mean galaxy density) makes $w(\vartheta)$ a dimensionless quantity and the total probability turns out to be normalized to the square of the total number of object in the distribution. According to its definition, the value of the correlation function is a measure of the non–random behaviour of the distribution. In particular, object positions are said to be correlated if $w(\vartheta) > 0$, anticorrelated if $-1 \leq w(\vartheta) < 0$, while a Poissonian distribution is characterized by $w(\vartheta) = 0$ at any angular separation.

In a similar way, the availability of redshift samples made it possible to describe the clustering pattern in terms of the spatial 2–point function. In analogy with eq.(5), it is defined through the joint probability

$$\delta^{(2)}P \;=\; n^2 \, \delta V_1 \, \delta V_2 \, [1 + \xi(r_{12})] \tag{6}$$

of finding an object in the volume element $\delta V_1$ and another one in $\delta V_2$, at separation $r_{12}$. In this case too, the spatial 2–point function $\xi(r_{12})$ is a measure of the departure from a Poissonian statistics and, for an isotropic clustering, depends only on the modulus of the separation vector $\mathbf{r}_{12}$.

The concept of correlation functions can be extended to higher orders, by considering the joint probabilities between more than two points. In the following I will formally introduce the concept of correlations for a generic density field.

### 3.1.1 Density field and correlation functions

Let us consider a generic density field that can either represent the matter density field or the galaxy distribution. In the latter case, instead of a continuous distribution, this field is described by a point–like process,

$$\rho(\mathbf{x}) \;=\; \sum_i \delta_{\mathrm{D}}^{(3)}(\mathbf{x} - \mathbf{x}_i)\,, \tag{7}$$

where $\delta_{\mathrm{D}}^{(3)}$ is the Dirac delta–function in three–dimensions, and the typical galaxy dimension (few tens of kpc) is considered negligible with respect to their mean separation ($\simeq 5h^{-1}$ Mpc). Relative fluctuations are described by

$$\delta(\mathbf{x}) \;=\; \frac{\rho(\mathbf{x}) - \bar{\rho}}{\bar{\rho}}\,, \tag{8}$$

being $\bar{\rho}$ the average density. By definition, it is $\langle \delta(\mathbf{x}) \rangle = 0$, while the requirement of a positively defined $\rho(\mathbf{x})$ leads to $\delta(\mathbf{x}) > -1$. In the following, $\delta(\mathbf{x})$ is assumed to be described by a random function, so that the Universe can be considered as a particular realization taken from an ensemble (functional space) $\mathcal{F}$ containing all the $\delta(\mathbf{x})$ fields satisfying the above two requirements.

In order to describe the statistics of the $\delta(\mathbf{x})$ field, let $\mathcal{P}[\delta(\mathbf{x})]$ be the probability that the density fluctuations are described by a given $\delta(\mathbf{x}) \in \mathcal{F}$. With the assumption of statistical homogeneity, the probability functional $\mathcal{P}[\delta(\mathbf{x})]$ turns out to be independent of the position $\mathbf{x}$, while, due to the requirement of isotropic clustering, the joint distribution of $\delta(\mathbf{x}_1)$ and $\delta(\mathbf{x}_2)$ depends only on the the separation $r_{12} = |\mathbf{x}_1 - \mathbf{x}_2|$. By definition, the probability distribution in the functional space must be be normalized so that the total probability is unity:

$$\int_{\mathcal{F}} \mathcal{D}[\delta(\mathbf{x})] \, \mathcal{P}[\delta(\mathbf{x})] \;=\; 1\,. \tag{9}$$

Here $\mathcal{D}[\delta(\mathbf{x})]$ represents a suitable measure introduced in $\mathcal{F}$ in order to define the functional integral.

A complete characterization of the statistics of the density distribution can be given in terms of the $n$–point correlation functions

$$\mu_n(\mathbf{x}_1,\ldots,\mathbf{x}_n) = \langle \delta(\mathbf{x}_1)\ldots\delta(\mathbf{x}_n)\rangle. \tag{10}$$

The notation $\langle \cdot \rangle$ indicates the average over the $\mathcal{F}$ space. Under the assumption of *ergodicity* of our system, the averages taken over the (physical) configuration space is completely equivalent to the expectations taken over an ensemble of universes, i.e. over the functional space $\mathcal{F}$. From now on I will indifferently use the symbol $\langle \cdot \rangle$ to indicate both kind of average.

The key relevance of correlation functions in statistical mechanics lies in the fact that their knowledge is required in order to uniquely specify the statistics of the distribution. In fact, let us consider the *partition functional*

$$\mathcal{Z}[J(\mathbf{x})] \equiv \int \mathcal{D}[\delta(\mathbf{x})]\,\mathcal{P}[\delta(\mathbf{x})]\,e^{i\int d\mathbf{X}\,\delta(\mathbf{X})J(\mathbf{X})} = \langle e^{i\int d\mathbf{X}\,\delta(\mathbf{X})J(\mathbf{X})}\rangle, \tag{11}$$

where $J(\mathbf{x})$ is a generic function, that plays the role of an external source perturbing the underlying statistics. According to the definitions (11) of $\mathcal{Z}[J(\mathbf{x})]$ and (6) of $n$–point correlation function, it turns out that

$$\mu_n(\mathbf{x}_1,\ldots,\mathbf{x}_n) = i^{-n} \left.\frac{\delta^n \mathcal{Z}[J]}{\delta J(\mathbf{x}_1)\ldots\delta J(\mathbf{x}_n)}\right|_{J=0}, \tag{12}$$

and the McLaurin functional series of the partition function reads

$$\mathcal{Z}[J] = 1 + \sum_{n=2}^{\infty} \frac{i^n}{n!}\int d\mathbf{x}_1\ldots\int d\mathbf{x}_n\,\mu_n(\mathbf{x}_1,...,\mathbf{x}_n)J(\mathbf{x}_1)...J(\mathbf{x}_n). \tag{13}$$

Thus, $\mathcal{Z}[J]$ is defined as the generating functional of the correlation functions, in the sense that such functions are defined as the coefficients of the McLaurin expansion of $\mathcal{Z}[J]$ [note that in eq. (13) the sum runs from $n=2$ because of the vanishing of $\langle \delta(\mathbf{x})\rangle$].

It is also convenient to introduce the *connected* or *irreducible* correlation functions, $\kappa_n(\mathbf{x}_1,\ldots,\mathbf{x}_n)$, through their generating functional

$$\mathcal{W}[J(\mathbf{x})] \equiv \ln \mathcal{Z}[J(\mathbf{x})], \tag{14}$$

so that

$$\kappa_n(\mathbf{x}_1,\ldots,\mathbf{x}_n) = i^{-n}\left.\frac{\delta^n \mathcal{W}[J]}{\delta J(\mathbf{x}_1)\ldots\delta J(\mathbf{x}_n)}\right|_{J=0}. \tag{15}$$

Similar definitions of correlation functions have been originally introduced in the statistical study of liquids [344] and are completely analogous to the Green's functions usually considered in quantum field theory [340, 329]. It is clear that a unique characterization of the statistics, i.e. the knowledge of the partition functions, requires that correlation functions of any order are known.

For $n = 2$, it is easy to show that the definition (10) of correlation functions is completely equivalent to that provided by eq. (6). In fact, the 2–point joint probability of having the density values $\rho(\mathbf{x}_1)$ in the position $\mathbf{x}_1$ and $\rho(\mathbf{x}_2)$ in $\mathbf{x}_2$ is

$$\langle \rho(\mathbf{x}_1) \rho(\mathbf{x}_2) \rangle = \bar{\rho}^2 [1 + \mu_{2,12}], \tag{16}$$

which coincides with eq.(6), once we take $\xi(r_{12}) = \mu_2(r_{12})$. According to its definition, it is also easy to verify that the 2-point correlation function must satisfy the integral constraint

$$\int_0^\infty dr\, \xi(r) = 0.$$

In order to study the structure of the 3–point correlation function, let us suppose that the point $\mathbf{x}_3$ is sufficiently far away from $\mathbf{x}_1$ and $\mathbf{x}_2$, so that the event probability in $\mathbf{x}_3$ does not depend on that in the other two points. If this is the case, the 3–point joint probability is

$$\langle \rho_1 \rho_2 \rho_3 \rangle = \langle \rho_1 \rho_2 \rangle \times \bar{\rho}, \tag{17}$$

where the meaning of the indices is obvious. Hence, requiring symmetry for the exchange of $\mathbf{x}_3$ with $\mathbf{x}_1$ and with $\mathbf{x}_2$, the 3–point probability can be cast in the form

$$\langle \rho_1 \rho_2 \rho_3 \rangle = \bar{\rho}^3 [1 + \xi_{12} + \xi_{23} + \xi_{13} + \zeta_{123}]. \tag{18}$$

Here, $\zeta \equiv \kappa_3$ is the term that correlates the three points all together and must vanish when one of these points is removed:

$$\zeta(\mathbf{x}_i, \mathbf{x}_j, \mathbf{x}_l \to \infty) = 0 \qquad i \neq j \neq l \quad ; \quad i, j, l = 1, 2, 3. \tag{19}$$

A graphic representation of eq. (18) is

$$\langle \rho_1 \rho_2 \rho_3 \rangle = \bar{\rho}^3 \left[ 1 + \quad + \quad + \quad + \quad \right], \tag{20}$$

where each leg represents a $\xi$ term, while the triangle corresponds to the $\zeta$ contribution.

On the basis of similar considerations, the 4–point joint probability is written as

$$\begin{aligned}\langle \rho_1 \rho_2 \rho_3 \rho_4 \rangle &= \bar{\rho}^4 \{1 + [\xi_{12} + ... + 6\ terms] \\ &+ [\zeta_{123} + ... + 4\ \text{terms}] + \mu_{4,1234}\}. \end{aligned} \tag{21}$$

Here the 4–point correlation function

$$\mu_{4,1234} = \xi_{12}\xi_{34} + \xi_{23}\xi_{14} + \xi_{13}\xi_{24} + \eta_{1234} \tag{22}$$

represents the term connecting the four points and gives a vanishing contribution when at least one point is moved to infinite separation from the others. In more details, the $\mu_4$ term contains three terms connecting two pairs separately, while $\eta \equiv \kappa_4$ is the usual notation to indicate the connected 4–point function, which accounts for the amount of correlation due to the simultaneous presence of the four points. Graphically, eq. (22) takes the form

$$\mu_{4,1234} = \quad + \quad + \quad + \quad , \qquad (23)$$

where the square represents the $\eta$ term.

More generally, correlations of generic order $n$ can be introduced through the $n$–point joint probability,

$$\langle \rho(\mathbf{x}_1) \ldots \rho(\mathbf{x}_n) \rangle = \bar{\rho}^n \left[ 1 + (\text{terms of order} < n) + \mu_n(\mathbf{x}_1, ..., \mathbf{x}_n) \right], \qquad (24)$$

in such a way that they give null contribution when any subset of $\{\mathbf{x}_1, \ldots, \mathbf{x}_n\}$ is removed to infinity. In turn, an important theorem of combinatorial analysis (see, e.g., ref.[329]) shows that, removing from the $\mu_n$ function all the disconnected contributions, the remaining connected part is just the $\kappa_n$ function defined by eq.(15). The general proof of this theorem is rather tricky and will not be reported here. It is however not difficult to see that, expressing the derivatives of the $\mathcal{W}[J]$ partition function in terms of that of $\mathcal{Z}[J]$, we get at the first correlation orders

$$\mu_2 = \kappa_2, \qquad \mu_3 = \kappa_3, \qquad \mu_4 = 3\kappa_2^2 + \kappa_4, \qquad \mu_5 = 10\kappa_2\kappa_3 + \kappa_5,$$

$$\mu_6 = 15\kappa_2^3 + 10\kappa_3^2 + 15\kappa_2\kappa_4 + \kappa_6. \qquad (25)$$

From eq.(24), it follows that the $n$–point functions measure by how much the distribution differs from a completely random (Poissonian) process. In fact, for a Poissonian distribution the probability of some events in any subset of $\{\mathbf{x}_1, ..., \mathbf{x}_n\}$ does not affect the probability in the other points. Accordingly,

$$\langle \rho(\mathbf{x}_1) \ldots \rho(\mathbf{x}_n) \rangle = \bar{\rho}^n. \qquad (26)$$

and correlations of any order vanish.

As already observed, the relevance of correlation functions lies in the fact that their expressions, deducible from observational data at least at the lowest order, determine the statistics of the large scale clustering. Vice versa, once a theoretical model is assumed for the probability distribution, the explicit form of $\mu_n$ and $\kappa_n$ are fixed.

As a first example, a suitable expression for the $n$–point probability is provided by the so–called Kirkwood model [237]

$$\langle \rho_1 \ldots \rho_n \rangle = \bar{\rho}^n \prod_{\substack{i \neq j}}^{\binom{N}{2}} [1 + \xi_{ij}], \qquad (27)$$

that has been originally introduced in the study of rarefied gases.

Another popular expression in cosmological context is the hierarchical pattern

$$\kappa_n(\mathbf{x}_1,\ldots,\mathbf{x}_n) = \sum_{n-trees\ a}^{t_n} Q_{n,a} \sum_{labelings} \prod_{edges}^{(n-1)} \kappa_{2,ij}, \qquad (28)$$

which expresses the $n$–point connected function in terms of products of $(n-1)$ 2–point functions [161]. In eq.(28), distinct "trees" designated by $a$ have in general different coefficients $Q_a$, while the complete sequence of these coefficients uniquely specify the details of the hierarchical model. Configurations that differ only in interchange of labels $1,...,n$ all have the same amplitude coefficients, and $ij$ is a single index which identifies links. The number of trees $t_n$ with $n$ vertices is fixed by a theorem of combinatorial analysis [345], while the total number of labeled trees is $T_n = n^{n-2}$. Thus, eq. (28) has $t_n$ amplitude coefficients ($a = 1,...,t_n$) and $T_n$ total terms. As I will show in the following of this section, the hierarchical model also povides a rather good fit to the (low–order) correlation functions of both galaxies and clusters. More details about the statistics implied by hierarchical correlations will be discussed in §4.4.2.

### 3.1.2 Correlations of a Gaussian field

A particularly interesting and simple case is that in which the density fluctuations are approximated by a random Gaussian process. The important role of Gaussian perturbations in cosmological context lies in the fact that, according to the classical inflationary scenario, they are expected to be originated from quantum fluctuations of a free scalar field at the outset of the inflationary expansion (see, e.g., ref.[28] and references therein). Even without resorting to inflation, the Central Limit Theorem guarantees that Gaussianity is the consequence of a large variety of random processes, which makes it a sort of natural choice.

The Gaussian probability distribution in the functional space $\mathcal{F}$ takes the form

$$\mathcal{P}[\delta(\mathbf{x})] = (\det K)^{1/2} \exp\left\{-\frac{1}{2} \int d\mathbf{x} \int d\mathbf{x}'\, \delta(\mathbf{x}) K(\mathbf{x},\mathbf{x}') \delta(\mathbf{x}')\right\}. \qquad (29)$$

Here $K(\mathbf{x},\mathbf{x}')$ in an invertible operator acting on $\mathcal{F}$ and symmetric with respect to the variables $\mathbf{x},\mathbf{x}'$. From eq. (29), it follows that this operator determines the variance of the distribution and, more generally, the correlation properties of the fluctuation field. The above expression of the probability distribution is such as to satisfy the normalization requirement (9). Expressing the 2–point correlation function as the second derivative of the partition functional $\mathcal{Z}[J]$, evaluated for $J(x) = 0$, it is straightforward to see that the operator $K$ determines the 2–point function according to

$$\xi(x_{12}) = \int \frac{d^3k}{(2\pi)^3} \frac{e^{i\mathbf{k}\cdot(\mathbf{x}_1-\mathbf{x}_2)}}{\hat{K}(k)}. \qquad (30)$$

Here $\hat{K}(k)$ is the representation of $K(\mathbf{x},\mathbf{x}')$ in momentum space, where it acts as a multiplicative operator. In order to prove eq. (30), let us observe that the partition functional $\mathcal{Z}[J]$ relative to the Gaussian distribution functional (29) is

$$\begin{aligned}\mathcal{Z}[J] &= (\det K)^{1/2}\, e^{-\frac{1}{2}\int d\mathbf{x}\int d\mathbf{x}'\, J(\mathbf{x})K^{-1}J(\mathbf{x}')} \int \mathcal{D}[\delta(\mathbf{x})]\, e^{-\frac{1}{2}\int d\mathbf{x}\int d\mathbf{x}'\, \delta(\mathbf{x})K\delta(\mathbf{x}')} \\ &= e^{-\frac{1}{2}\int d\mathbf{x}\int d\mathbf{x}'\, J(\mathbf{x})K^{-1}J(\mathbf{x}')}\,.\end{aligned} \qquad (31)$$

Twice differentiating the above expression with respect to $J(\mathbf{x})$, the 2–point function reads

$$\xi(x_{12}) = K^{-1}\delta_{\mathrm{D}}(\mathbf{x}_1 - \mathbf{x}_2) \qquad (32)$$

and eq.(30) follows after Fourier transforming $\xi(x_{12})$.

According to the definition (14) of $\mathcal{W}[J]$, the generator of the connected correlation functions reads

$$\mathcal{W}[J] = -\frac{1}{2}\int d\mathbf{x}\int d\mathbf{x}'\, J(\mathbf{x})K^{-1}J(\mathbf{x}')\,, \qquad (33)$$

so that the corresponding connected correlation functions are

$$\kappa_n(x_1,...,x_n) = 0 \qquad \text{if} \quad n \neq 2\,. \qquad (34)$$

Therefore, the fundamental property of a Gaussian density field is that its statistics is completely determined by 2–point correlations.

Although Gaussianity of density fluctuations seems to be the natural outcome of the primeval evolution of the Universe, nevertheless the observed distribution of cosmic structures displays clear non–Gaussian behaviour, as the detection of non–vanishing higher–order correlations shows (see below). However, even starting with an initial Gaussian density field, there are at least two valid motivations to understand the development of subsequent non–Gaussian statistics for the galaxy distribution. Firstly, gravitational clustering is known to generate higher–order correlation even in the mildly non–linear regime [308, 161], while the strongly non–linear evolution, as described by numerical N–body simulations, shows a remarkable spatial intermittency of small–scale structures, which represents the signature of non–Gaussian statistics. Secondly, non–Gaussianity is also expected in the framework of "biassed" models of galaxy formation [230, 27], in which the observed cosmic structures are identified with those peaks of the underlying Gaussian matter field, that exceeds a critical density value. In this case, analytical argument [330, 222] shows that non–Gaussianity arises as a threshold effect superimposed on a Gaussian background.

### 3.1.3 Galaxy correlations

Starting from the first attempts of Totsuji & Kihara [388], the correlation analysis of the galaxy distribution became a widely employed approach to investigate clustering and is nowadays considered as the "classical" study of the LSS of the Universe. During the 70's, the availability of extended angular galaxy samples made possible the realization of accurate correlation analyses, mainly pursued by Peebles and his collaborators. Although based on different angular samples, such as the Zwicky catalogue [428], the Jagellonian field [349] and the Lick map [368], all these analyses converge to indicate that the angular 2–point function is well represented by the power law

$$w_g(\vartheta) = A_g \vartheta^{1-\gamma}, \tag{35}$$

with $\gamma = 1.77 \pm 0.04$ and amplitude $A_g \propto \mathcal{D}^{-1}$ decreasing with the depth $\mathcal{D}$ of the sample, with a break from the power law behaviour at large angular separations.

More recently, Maddox et al. [263] found from the analysis of the APM sample that the angular 2–point correlation function has the same slope as eq.(35), in the range of validity of the power law. The break they found from the power law occurs roughly at the same physical separation as found by Groth & Peebles [189] from the analysis of the Lick map, but with a much more gently decline from a power law on larger scales, thus implying an excess of power at scales $\gtrsim 20h^{-1}$ Mpc (see Figure 7). The authors argued that this discrepancy with respect to the results of Groth & Peebles is probably due to the removal of clustering from the sample they used, when correcting for the presence of large scale gradients. A similar result has also been found by Collins et al. [103] from the analysis of the Edinburgh/Durham galaxy catalogue.

In order to extract information about the spatial properties of the galaxy distribution from eq. (35), a method is needed for deprojecting angular data, so to obtain the spatial 2–point correlation function, $\xi(r)$. A suitable deprojection method is provided by the Limber equation [251]. This method, that holds under the hypothesis of absence of any correlation between position and luminosity of objects, permits to express the angular function $w(\vartheta)$ in terms of the spatial function $\xi(r)$, according to the formula

$$w(\vartheta) = \frac{\int_0^\infty r^4 \phi^2(r/\mathcal{D}) \, dr \int_{-\infty}^{+\infty} d\rho \, \xi(\sqrt{\rho^2 + r^2\vartheta^2})}{[\int r^2 \phi(r/\mathcal{D}) \, dr]^2}. \tag{36}$$

Here, $\phi(r/\mathcal{D})$ is the radial selection function, defined as the probability for an object at distance $r$ to be included in a sample of depth $\mathcal{D}$. Its detailed shape depends on the luminosity distribution of galaxies. The inversion of the Limber equation, in which we are interested, is also possible in several cases [152]. For instance, taking the power law model (35) for $w(\vartheta)$, the spatial 2–point function turns out to be

$$\xi_g(r) = \left(\frac{r_{o,g}}{r}\right)^\gamma, \tag{37}$$

with the same value of $\gamma$ as in eq.(35) and clustering length, $r_{o,g}$, depending on the amplitude $A_g$ of the angular function and on the depth $\mathcal{D}$ according to the scaling relation

$$A_g \propto \left(\frac{r_{o,g}}{\mathcal{D}}\right)^\gamma. \tag{38}$$

All the investigations of the spatial correlations by means of angular data indicate $r_{o,g} \simeq 5h^{-1}$ Mpc, with some scatter around this value, in the range of separations $0.1 < r < 10h^{-1}$ Mpc [315, 306, 189], but with a break of the power law at larger separations. Vice versa, no deviations from a pure power law have been found down to the smallest scales sampled by the galaxy distribution. In fact, there is evidence [182, 246] that support the validity of eq.(37) down to $r \sim 3\,h^{-1}$ kpc.

A direct test of the reliability of eq.(37) has been realized with the availability of sufficiently large and complete redshift surveys. By using the CfA1 sample, Davis & Huchra [114] deduced the galaxy spatial number density, while Davis & Peebles [116] obtained the expression (37), for the 2–point function, with $r_{o,g} = 5.4 \pm 0.3\,h^{-1}$ Mpc, in fairly good agreement with angular results in the same range of physical separations (see Figure 7). An analysis of the CfA1 sample for the region centered on the Virgo cluster led Einasto et al. [147] to find a shoulder in the 2–point function at $r = 4 - 5h^{-1}$ Mpc. This discrepancy with respect to the Davis & Peebles' result indicates that, in some cases, the galaxy distribution could have local features, which affect a safe determination of the clustering parameters and disappear when a larger (fair) sample is considered. A statistical investigation of the first CfA2 slice led to quite large uncertainties in the determination of galaxy number density and 2–point correlation function. De Lapparent et al. [126] found an indetermination of $\sim 25\%$ in the galaxy number density. In this analysis the 2–point function was found to have a slope $\gamma \simeq 1.6$ and a correlation length $r_{o,g} \simeq 7.5\,h^{-1}$ Mpc, in the $3 - 14\,h^{-1}$ Mpc scale range. Because of the large uncertainty in the mean density, the ranges in the slope and amplitude are respectively $1.3 - 1.9$ and $5 - 12\,h^{-1}$ Mpc. On scales larger than $20\,h^{-1}$ Mpc, the correlation function is not well determined. Although consistent within the errors with the more stable indications coming from angular data, such results show how crucial it is to deal with a *fair* galaxy redshift sample. In fact, according to the definition (6) of 2–point function, it is clear that its unambiguous estimate relies on the possibility of uniquely define an average object number density. As shown by Figure 3, spatial samples trace structures having characteristic sizes of the same order of the sample size, so that it is not clear whether the galaxy distribution is statistically homogeneous within the sample boundaries. This problem can be even more important when single slices, such as those of the CfA2 survey are considered, in which case the sampled scales are widely different in different directions.

Instead of considering the galaxy distribution as a whole, several attempts have been also devoted to the investigation of the clustering properties of galaxies having different morphology. The relevance of this kind of analysis lies in the fact that a dependence of the clustering properties on galaxy morphology should be related to the physical processes that gave rise to galaxy formation (see Section 5, below). It has been recognized for a long time that small compact groups tend to contain more elliptical galaxies than do looser groups [154]. The central regions of rich clusters appear to be dominated by elliptical and lenticular galaxies and contain few normal spirals; irregular, less dense clusters have a composition similar to that of the field and contain many spirals [290]. By analysing the Uppsala Catalogue, Davis & Geller [113] determined the angular 2–point correlation function for the distribution of galaxies of various morphological types. They found that elliptical galaxy clustering is characterized by a power law with a slope ($\gamma_{\rm E} \simeq 2.1$) steeper than that for spiral clustering ($\gamma_{\rm S} \simeq 1.69$), while the lenticular slope has an intermediate value ($\gamma_{\rm L} \simeq 1.71$). Dressler [132] studied the galaxy populations in 55 rich clusters. He found that a well defined relationship exists between local density and morphology, which confirms an increasing elliptical and S0 population and a corresponding decrease in spirals with increasing density inside the cluster. Still by using the Uppsala Catalogue, Giovanelli et al. [175] revealed evidence for a continuous morphological segregation in a wide range of galaxy densities in the Pisces–Perseus supercluster. Furthermore, significant differences in the slope of the angular 2–point function for different morphological types are found, in agreement with previous results. Further analyses of of both angular and spatial surveys [66, 224, 280] confirmed that elliptical galaxies tend to be more clustered and to have a steeper correlation function than spirals.

Although the analysis of the 2–point function surely provides remarkable hints about the large scale galaxy distribution, nevertheless it does not represents a full statistical description. Further pieces of information are obtainable from the investigation of higher–order correlations. At the third order, the joint probability

$$\delta^{(3)}P \;=\; n^3 \, \delta\Omega_1 \, \delta\Omega_2 \, \delta\Omega_3 \left[ 1 + w_{12} + w_{13} + w_{23} + z_{123} \right] \tag{39}$$

defines the angular 3–point correlation function $z_{123}$, which depends on the shape of the triangle defined by the three points. It is clear that, as the correlation order increases, the statistical analysis becomes more and more complicated. In fact, while the estimate of the 2–point function requires the knowledge of the number of galaxy pairs at a given separation, computing the 3–point function amounts to counting the triplets of a given shape, with a subsequent increase of the noise, as well as of the required computational time. The analysis of the 3–point function in the Zwicky, Lick and Jagellonian samples [316, 189, 306] indicates that data are well fitted by the hierarchical model

$$z_{123} \;=\; Q \left[ w_{12}w_{23} + w_{12}w_{13} + w_{13}w_{23} \right] \tag{40}$$

with $Q = 1.3 \pm 0.2$ (see, however, ref.[52] for a for a different result about the Zwicky sample).

Going to even higher correlation orders causes a significant increase in the sampling noise. Some attempts in this direction have been however pursued by several authors. In their estimate of the 4–point function for the Lick map, Fry & Peebles [166] have shown that, even within the uncertainties, its expression is consistent with the hierarchical model of eq.(28). Sharp et al. [373] analyzed the Zwicky catalogue and found a marginal signal for the 4–point function, while an attempt to estimate the 5–point function gave results that are completely lost in the noise. More recently, Szapudi et al. [385] and Szalay et al. [273] analyzed the Lick and IRAS samples and claimed that signals of correlation are detected up to the eighth order, which are consistent with the hierarchical model. However, it is not clear how correlations of such a high order could provide statistical information that were both simple–to–handle and easy to compare with theoretical models.

### 3.1.4 Clustering and dynamical equilibrium

The presence of a regular correlation pattern in the galaxy distribution calls for the presence of an underlying dynamics, which were able to generate the power law shape of $\xi(r)$ over a scale range three orders of magnitude wide (from $\sim 10\,h^{-1}$ kpc to $\sim 10 h^{-1}$ Mpc). At the scales relevant for galaxy clustering we would expect that gravity makes the job. On the other hand, if we require that galaxy clustering pattern be stable, then precise relations should exist between galaxy peculiar motions and matter distribution. Peebles [312] applied this argument to the relative motions between galaxy pairs at the Mpc scale. If $v(r)$ is the r.m.s. velocity dispersion within pairs at separation $r$, then the expected relative mass excess associated to those pairs is $\delta M(r) \propto v^2(r)/r$. Since the relative fluctuation in the galaxy number count at the scale $r$, $\delta N(r)$, is associated to the 2–point function according to $\delta N(r) \propto \xi(r)$, if galaxies traces the mass we should expect that $v^2(r) \propto r^2 \xi(r) \propto r^{2-\gamma}$. Davis & Peebles [116] addressed this problem, by studying the pairwise velocity dispersion for galaxies in the CfA1 sample. They found that the velocity dispersion is a slowly increasing function of separation, $v(r) \propto r^\alpha$, with $\alpha = 0.13 \pm 0.3$. According to the above scaling relation, this indicates $\gamma \simeq 1.76$ for the slope of $\xi(r)$, thus in remarkable agreement with that found from the correlation analysis. Confirmation of this result came from similar investigations [35, 198], which have shown that the pairwise velocity dispersion is always a weak function of the separation.

Based on such results, one may ask whether it is possible to extend at even smaller scales ($\lesssim 10 h^{-1}$ kpc) the dynamical investigation of clustering. In order to have the relevant dynamical informations at such very small scales, the observed rotation curves of spiral galaxies can be usefully employed. Their rather flat shape observed in most cases indicates that $v(r) \sim const$, thus suggesting the extension at even smaller scales of the clustering pattern traced at the Mpc scale by the galaxy distribution.

However, if the Universe is dominated by dark matter on the larger scales, then, on the scale of galaxies, continuity arguments invite us to consider only their dark mass components, rather than their overall mass distribution, in order to meaningfully compare the clustering properties of matter at small and large scales. To this purpose, it is necessary to resort to a suitable mass–decomposition method, in order to separate the dynamical contributions of dark (halo) and luminous (disk) matter from the overall rotation curve. Here, I will briefly describe the attempts that we pursued [51, 58, 352] to probe the dark matter distribution inside galaxy halos, by following the mass–decomposition method originally devised by Persic & Salucci [317, 318]. By applying this prescription to a suitable sample of spiral galaxy rotation curves [319], we worked out the average dark halo matter density at the optical radius, $\bar{\rho}_{h,opt}$. In Figure 8 I plot this quantity as a function of the optical disk radius, $R$, for the spiral sample, which exhibits a remarkable power law dependence.

In order to investigate the correlation properties of the halo DM, let $M_{\rm R} = V_{\rm R}\, \bar{\rho}_{h,opt}$ be the matter contained inside the spherical volume $V_{\rm R} = 4\pi R^3/3$ encompassed by the optical radius of a given galaxy. According to the definition (6) of the 2–point correlation function, its first–order moment reads

$$\langle M \rangle_{\rm R} = \tilde{M}_{\rm R} + \bar{\rho} \, 4\pi \int_0^R r^2 dr \, \xi_{gb}(r) \,. \tag{41}$$

Here, $\bar{\rho} = 1.9 \times 10^{-29} \, h^{-2} \, \Omega_o$ g cm$^{-3}$ is the mean matter density and $\tilde{M}_{\rm R} = \bar{\rho} \, V_{\rm R}$ is the expected mass contained in a randomly placed sphere of radius $R$. Moreover, $\xi_{gb}(r)$ is the *galaxy–background* cross–correlation function, that arises since we are measuring the mass contained around the galaxy center, instead of around a randomly selected point. Taking $\xi_{gb}(r) = (r_o/r)^\gamma$, according to the scaling suggested by Figure 8, and for $R \ll r_o$, from eq. (41) we obtain

$$\frac{\langle M \rangle_{\rm R}}{\tilde{M}_{\rm R}} = K_1 \, \xi_{gb}(R) \,, \tag{42}$$

with $K_1 = 3/(3 - \gamma)$. The estimate of $\xi_{gb}(R)$ from the considered spiral sample gives $r_o = (3.2 \pm 0.3)\Omega_o^{-1/\gamma} h^{-1}$ Mpc and $\gamma = 1.71 \pm 0.05$. This result is rather interesting since the resulting correlation function for the distribution of DM inside galaxy halos follows at small scales of a few tens of kpc the same power law shape as that of galaxies at the scales of some Mpcs, the amplitude of the clustering depending on the value of the density parameter $\Omega_o$.

By comparing the detected $r_{o,gb}$ with that, $r_{o,gg} \simeq 5 h^{-1}$ Mpc, coming from the analysis of the galaxy distribution at scales ($\sim 10 \, h^{-1}$ kpc, see refs.[189, 182, 246]) comparable to that sampled by rotation curves, and requiring continuity of the clustering between the DM and the galaxy distributions, the resulting density parameter is

$$\Omega_o = 0.3 \pm 0.1 \tag{43}$$

(see ref.[352]). Note that this result agrees with estimates of the density parameter based on the virial analysis of galaxy pair velocity dispersions as described in refs. [116, 35, 198]. On the other hand, if we allow galaxies to be more clustered than the underlying matter distribution, according to the prescription of biassing (see Section 5 and ref.[123] for a review), then matter fluctuations and object number count fluctuations are related by $\delta N/N = b \, (\delta M/M)$, $b > 1$ being the biassing parameter. In this case, assuming a flat Universe ($\Omega_o = 1$) amounts to require that galaxies are more strongly clustered than matter by a factor $b \simeq 2$.

### 3.1.5 Cluster correlations

Many attempts have been devoted in recent years in order to trace the LSS on the basis of the observed the cluster distribution [19]. Indications that clusters are not randomly arranged on the sky was found already by Abell [3, 4]. A further evidence of super–clustering in the Abell survey was also detected by Bogart & Wagoner [46], Hauser & Peebles [202], and Rood [346] by means of nearest–neighbor distributions and angular correlation functions. Already from these preliminary investigations, it clearly appeared that rich galaxy systems display a clustering that is remarkably stronger than that of galaxies and developing on much larger scales. Thanks to this characteristic, it has been immediately recognized that clusters are very efficient tracers of the structure of the Universe at very large scales, where gravitational interaction is still in the linear regime and preserves memory of initial conditions.

Bahcall & Soneira [22] and, independently, Klypin & Kopylov [240] determined the rich ($R \geq 1$) Abell cluster 2–point correlation function from the HGT redshift sample. They found strong correlations in both the $D \leq 4$ redshift sample and in the larger and deeper $D = 5 + 6$ sample. The spatial correlation function was found to fit the power law expression

$$\xi_c(r) = \left(\frac{r_{o,c}}{r}\right)^\gamma, \tag{44}$$

with $r_{o,c} \simeq 25h^{-1}$ Mpc and $\gamma \simeq 1.8$, in the distance range $5 \lesssim r \lesssim 150h^{-1}$ Mpc. Thus, the rich–cluster autocorrelation function exhibits the same slope as the galaxy function, but with a remarkably larger correlation length. They also noted that cluster correlations are elongated along the line–of–sight direction, an effect that they ascribed to cluster peculiar motions. Accounting for these effects, an estimate of peculiar velocities between clusters gives $\sim 2000$ km s$^{-1}$. Similar conclusions about cluster correlation have been also reached by Postman et al. [331] and by Batuski et al. [33] from the analysis of the Zwicky and ACO samples, respectively.

Rather different results have been however claimed by Sutherland [381], who argued that redshift distorsions at separations $\gtrsim 50h^{-1}$ Mpc are too large to be completely accounted for by random peculiar motions. Instead, he suggested that such anisotropies in the redshift space correlation function are to be ascribed to spurious line–of–sight clustering; if the richness of Abell clusters is apparently enhanced by a significant amount because of foreground and background galaxies, spurious line–of–sight correlation is produced in a richness–limited sample. By analysing the SR spatial sample of Abell clusters and after correcting for the anisotropies in the redshift space correlations, Sutherland found that the power law (44) for the rich cluster 2–point function is still satisfied, but with a reduced correlation length, $r_{o,c} \simeq 14h^{-1}$ Mpc. A numerical simulation of richness contamination [122] allowed the construction of a "decontaminated" sample of $D \leq 4$, $R \geq 1$ clusters, with the result that the correlation amplitude is reduced by a factor $\sim 2$, in agreement with the Sutherland's claim. A similar conclusion has been also reached by Sutherland & Efstathiou [382]. In their analysis of the deep GH survey, they found further evidence of line–of–sight contamination and a resulting value of the clustering length $r_{o,c} \simeq 13h^{-1}$ Mpc. In addition, effects of projection contamination on the angular cluster correlation function has also been found by Olivier et al. [291] in the analysis of the Abell and ACO catalogues. After removing these effects by means of a suitable model for the galaxy distribution around clusters, the correlation strength at small angular separation is reduced by a factor 2–3, in agreement with the results obtained from spatial samples. By using an ensemble of simulated cluster catalogues, Jing et al. [225] checked whether the redshift correlation claimed by Sutherland is an effect of richness contamination or of real clustering. They found that the redshift correlation is a quite common feature in free-of-contamination simulated samples and concluded that the original Bahcall & Soneira estimate should not be affected by such effects. Based on the recently compiled samples of clusters selected from the APM and Edinburgh/Durham galaxy surveys, several authors concluded in favour of a lower correlation length, in the range $r_{o,c} \simeq 13$–$16h^{-1}$ Mpc [109, 23, 261]. Despite the agreement of the results, completely opposite conclusions have been however reached by different groups. Dalton et al. [109] claim for a better reliability of these cluster samples and take the result as a probe of a smaller clustering amplitude. Efstathiou et al. [136] found an anisotropy in the clustering of the PGH sample, which they interpreted as due to artificial clustering. After correcting for this effect, they found consistence with the smaller correlation amplitude found for the APM cluster sample. Vice versa, Bahcall & West [23] ascribe the smaller correlation length as due to the richness–clustering dependence and conclude that the result is perfectly consistent with the Bahcall & Soneira result for the richer Abell clusters. Further support for a higher value of the cluster correlation length comes also from the analyses of the distribution of cD [414] and X-ray selected clusters [245], which should be unaffected by projection contamination and point toward $r_{o,c} \simeq 20h^{-1}$ Mpc.

All these determinations of the 2–point correlation function for galaxy systems also suggest a dependence of the clustering strength on the richness, richer systems being more strongly clustered. In particular, an increase in the clustering is observed for richer systems. In their analysis of the spatial correlation function for Abell clusters, Bahcall & Soneira [22] discussed the increase of the correlation amplitude with cluster richness. They classified individual galaxies as $N = 1$ systems (where $N$ is Abell's criterion for richness classification) and suggested that galaxies have a correlation function amplitude that follows the richness–clustering relation holding for clusters. From a physical perspective, it would however be misleading to treat individual galaxies in the same way as galaxy systems, the processes governing galaxy formation probably being different from those relevant for groups and clusters. Indeed, Szalay & Schramm [384] pointed out that galaxy clustering may be intrinsically different from cluster clustering. They discussed a possible *universal* correlation function characterized by a slope $\gamma = 1.8$ and by a dimensionless amplitude

$$\beta(L) = \xi(L) = \left(\frac{r_o}{L}\right)^\gamma. \tag{45}$$

Here $L = n^{-1/3}$ ($n$ is the mean spatial density of objects) represents the average value of the separation. They derive $\beta \simeq 0.35$ for Schectman clusters and for Abell $R \geq 1$ and $R \geq 2$ clusters, while $\beta \simeq 1.1$ for galaxies. This result suggests that galaxies are relatively more strongly correlated than clusters (see Figure 9$b$). Bahcall & Burgett [20] extended this analysis to include also the correlation function of superclusters. By using the dimensionless correlation amplitude (45), they found that superclusters have $\beta \simeq 0.3$, in agreement with the value for clusters (see Figure 9$a$). Plionis & Borgani [324] analyzed the PBF cluster samples. These samples are particularly suitable to investigate the clustering–richness relation, since they are just selected by following precise richness criteria. In this analysis, we found that the slope of the resulting angular correlation is always consistent with $\gamma \simeq 2$, but with amplitudes which remarkably depend on the cluster richness (see Figure 10).

As in the case of the galaxy distribution, some attempts have also been devoted to the investigation of higher order correlations for galaxy clusters. It is however clear that, since cluster samples contain a much smaller number of objects than galaxy samples, the analysis of higher–order functions becomes particularly difficult and care must be taken about the statistical significance of any result (see, e.g., ref.[110]). Jing & Zhang [228] analyzed the Abell catalogue and found that the hierarchical expression of eq.(40) reproduces quite well the cluster 3–point function, with a value $Q \simeq 0.7 \pm 0.2$ for the hierarchical coefficient. This result has been confirmed by Toth et al. [387], who considered the northern Abell, the southern ACO and the Schectman angular samples or rich clusters. They also found consistency with the hierarchical expression, as in the galaxy case, with $Q \simeq 1.0 \pm 0.1$ for Abell clusters and a systematically lower value $Q = 0.64 \pm 0.04$ for the Schectman groups. This smaller value could well be interpreted on the light of the lower average richness, which characterizes Schectman groups with respect to Abell clusters. An investigation of the spatial 3–point function has been recently performed by Jing & Valdarnini [226], that considered a sample of 227 Abell clusters with known distances. They found that a hierarchical expression with $Q \simeq 0.7$ give a reasonable fit to the data, even within the rather large statistical uncertainties. A similar analysis based on the spatial distribution of Abell clusters, as well as on a synthetic cluster catalog extracted from large N–body simulations, led Gott et al. [177] to conclude that the 3–point function of rich clusters agrees with the hierarchical expression. We analyzed the richness dependence of the 3–point function for the PBF cluster sample [61] and found that the hierarchical model is always consistent with data, although within quite large uncertainties. The value of the coefficient $Q$ is found to be an increasing function of the richness, thus consistent with the expectation that richer cluster, selected as higher peaks of the galaxy distribution, are characterized by a more pronounced non–Gaussian statistics.

Other than analyzing the correlation properties of galaxies and clusters separately, it is also possible to investigate the cross–correlations between their relative positions. This kind of investigation is relevant in order to study the galaxy distribution inside cluster halos. In order to introduce the angular *cross–correlation* function $w_{cg}(\vartheta)$, let us consider the joint probability

$$\delta^{(2)}P = n_c n_g \delta\Omega_1 \delta\Omega_2 [1 + w_{cg}(\vartheta_{12})] \tag{46}$$

of having a cluster in the solid angle element $\delta\Omega_1$ and a galaxy in $\delta\Omega_2$. High values of $w_{cg}$ indicate a strong concentration of galaxies around cluster centers, while its shape is determined by the variation with the distance of the galaxy density out of the cluster.

The first joint statistical investigation of the distribution of galaxies and clusters was performed by Seldner & Peebles [366]. In this analysis, the distribution of galaxies was that of the Lick Catalog, while the Abell Catalog was used for the cluster positions. A good fit to the data was provided by the expression

$$w_{cg}(\vartheta) = A\vartheta^{-\rho} + B\vartheta^{-\sigma}, \qquad (47)$$

with $\rho \simeq 1.4$ and $\sigma \simeq 0.2$. The first term, which is dominant at small scales, is essentially due to the galaxy density around each cluster, while the second term, that dominates for larger angular separations, takes into account the contribution of the clustering between clusters.

A further investigation of the cross–correlation properties for the distributions of galaxies and clusters has been realized by Lilje & Efstathiou [250], which considered the same catalogues for both galaxies and clusters. They used redshifts for Abell clusters to compute a cross–correlation function $w_{cg}(\sigma)$, where $\sigma \equiv v\vartheta/H_o$ is the projected separation between a cluster with recession speed $v$ and a galaxy at angular distance $\vartheta$ from the cluster center. Their results show that on scales $r < 20h^{-1}$ Mpc the shape of the spatial cross-correlation function is well described by a unique power law,

$$\xi_{cg}(r) \simeq \left(\frac{8.8h^{-1} \text{ Mpc}}{r}\right)^{2.2}. \qquad (48)$$

This shape of the galaxy–cluster cross–correlation is consistent with the cluster density profile as reconstructed from X–ray data about the temperature profile of the intracluster gas (see, e.g., ref.[150]). This analysis indicates that, at least for the few considered clusters, the average density inside the radius $r$ scales as $\rho(r) \propto r^{-\gamma}$, with $\gamma$ slightly exceeding two, thus indicating that at scales $\lesssim 1h^{-1}$ Mpc from the cluster centre galaxies are fairly good tracers of the dark matter distribution.

As a concluding remark of this overview about the correlation analysis of the LSS of the Universe, I would like to stress the relevance of the detected scaling properties in the distribution of cosmic structures, as revealed by the power law shape of the correlation functions. This fact is even more remarkable considering that, although their amplitudes turn out to increase when passing from the halo dark matter to galaxies and to galaxy aggregates of increasing richness, their slopes remain surprisingly unchanged. This suggests a sort of self–similarity for the LSS, extending from few kpc scales, traced by spiral rotation curves, to scales of some tens of Mpc, where rich galaxy clusters still display a non–negligible clustering.

## 3.2  The power spectrum analysis

It is often useful to analyze the statistics of the galaxy distribution in Fourier space, instead of in configuration space, as done by correlation functions. To this purpose, let us introduce the Fourier transform of the fluctuation field $\delta(\mathbf{x})$,

$$\tilde{\delta}(\mathbf{k}) = \frac{\bar{\rho}a^3}{M} \int d^3\mathbf{x}\, \delta(\mathbf{x})\, e^{i\mathbf{k}\cdot\mathbf{x}}, \qquad (49)$$

where $M = \bar{\rho}a^3 \int d^3x$ is the total mass and $a(t)$ the cosmic expansion factor. The modulus of the wavevector $\mathbf{k}$ is related to the comoving wavelength of the fluctuation mode according to $k = 2\pi a(t)/\lambda$. By inverting eq.(49), one has

$$\delta(\mathbf{x}) = \frac{M}{\bar{\rho}a^3} \int \frac{d^3\mathbf{k}}{(2\pi)^3} \tilde{\delta}(\mathbf{k}) e^{-i\mathbf{k}\cdot\mathbf{x}}, \qquad (50)$$

so that the two representations $\delta(\mathbf{x})$ and $\tilde{\delta}(\mathbf{k})$ of the fluctuation field contain the same amount of information. Accordingly, the correlation in Fourier space is related to that in $\mathbf{x}$–space by means of the relation

$$\begin{aligned}\langle \tilde{\delta}(\mathbf{k}_1)\tilde{\delta}(\mathbf{k}_2)\rangle &= \left(\frac{\bar{\rho}a^3}{M}\right)^2 \int \mathcal{D}[\delta]\,\mathcal{P}[\delta] \int d^3k_1\,d^3k_2\,\delta(\mathbf{x}_1)\,\delta(\mathbf{x}_1+\mathbf{x}_{12})\,e^{i[\mathbf{k}_1\cdot\mathbf{x}_1+\mathbf{k}_2\cdot(\mathbf{x}_1+\mathbf{x}_{12})]}\\ &= (2\pi)^3 \delta_{\mathrm{D}}(\mathbf{k}_1+\mathbf{k}_2) \left(\frac{\bar{\rho}a^3}{M}\right)^2 \int d^3x_{12}\,\xi(x_{12})\,e^{i\mathbf{k}_2\cdot\mathbf{x}_{12}}.\end{aligned} \qquad (51)$$

In the above expression the presence of the Dirac delta function $\delta_{\mathrm{D}}(\mathbf{k}_1+\mathbf{k}_2)$ is analogous to the momentum–conserving term appearing in the Fourier representation of the Green's function in quantum field theory. Since $(2\pi)^3\delta_{\mathrm{D}}(\mathbf{0}) = \int d^3x$ and factoring out this term, we get the expression of the power spectrum

$$P(k) \equiv \langle |\tilde{\delta}(\mathbf{k})|^2\rangle = \frac{\bar{\rho}a^3}{M} \int d^3x_{12}\,\xi(x_{12})\,e^{i\mathbf{k}_2\cdot\mathbf{x}_{12}}. \qquad (52)$$

The relevance of the power spectrum in characterizing the large scale clustering lies in the fact that the fluctuation spectrum is the fundamental observational quantity provided by any theoretical model about the origin of primordial fluctuations. In Section 5 I will discuss in more detail how the primordial power spectrum is originated at the outset of the recombination epoch and how it depends on the matter content of the Universe. In the simple case of a Gaussian fluctuation field, the power spectrum is the only quantity that is needed to completely describe the statistics. If this is the case, the Central Limit Theorem ensures that the Fourier transform of the fluctuation field can be written as $\tilde{\delta}(\mathbf{k}) = \sqrt{P(k)}\exp i\varphi_{\mathbf{k}}$, with phases $\varphi_{\mathbf{k}}$ randomly distributed in the interval $(0, 2\pi]$.

A particularly simple model for the power spectrum is given by the power law shape

$$P(k) = Ak^n, \qquad (53)$$

with $A$ the amplitude of the spectrum and $n > -3$ in order to allow for the convergence of the integral of $P(k)$ at large wavelength. The value $n = 1$ for the spectral index corresponds to the scale–free Harrison–Zel'dovich spectrum [201, 426], that describes the fluctuations generated in the framework of the canonical inflationary scenario [196, 252, 253]. Inverting eq. (52) for $\xi(r)$,

we get

$$\begin{aligned}\xi(r) &= \frac{M}{\bar{\rho}a^3}\frac{A}{(2\pi)^3}\int d^3k\, k^n e^{i\mathbf{k}\cdot\mathbf{r}}\\ &= \frac{M}{\bar{\rho}a^3}\frac{A}{2\pi^2}\frac{\Gamma(n+3)}{n+2}\sin\left[\frac{(n+2)\pi}{2}\right]r^{-(n+3)}\,.\end{aligned} \quad (54)$$

Thus, the detected power law shape for the 2–point function, $\xi(r)\propto r^{-1.8}$, turns into a constant logarithmic slope of the power spectrum, with spectral index $n=-1.2$, at least at scales $r \lesssim 10\,h^{-1}$ Mpc. Although this value of the spectral index is decidedly far from that, $n=1$, predicted by the inflationary paradigm, we should bear in mind that it refers to a scale–range where the primordial inflationary spectrum is likely to be distorted not only by the fluctuation evolution through the equivalence and recombination epoch, but also by the departure of gravitational clustering from the linear regime.

As for the amplitude $A$, one's hope is that it should be fixed by a theoretical model predicting the primordial fluctuation spectrum. However, there are at present no compelling theory about that, and is common to consider $A$ as a free parameter to be fixed on the ground of observational data. A first normalization, that is often used, refers to the variance of the galaxy number counts inside volumes of a given size. The variance $\sigma_{\rm R}^2$ of the fluctuation field $\delta(\mathbf{x})$ at a given scale $R$ is defined as

$$\sigma_{\rm R}^2 = \frac{M}{\bar{\rho}a^3}\frac{1}{(2\pi)^3}\int d^3k\, P(k)\,|W_{\rm R}(\mathbf{k})|^2\,. \quad (55)$$

In the above equation, the scale $R$ enters through the *window* function $W_{\rm R}(\mathbf{k})$, which provides an ultraviolet cut–off on the spectrum, suppressing the modes with wavelength $\lambda \lesssim R$. Its detailed shape defines the profile of the sampling volume. For a window given by a top–hat (sharp) sphere of radius $R$, it is

$$W_{\rm R}(\mathbf{k}) = \frac{3\,(\sin kR - kR\,\cos kR)}{(kR)^3}\,, \quad (56)$$

while a sphere with Gaussian profile has

$$W_{\rm R}(\mathbf{k}) = \exp\left(-\frac{k^2 R^2}{2}\right)\,. \quad (57)$$

and a cube–hat of side $R$ has the Fourier representation

$$W_{\rm R}(\mathbf{k}) = \frac{\sin(k_x R/2)\,\sin(k_y R/2)\,\sin(k_z R/2)}{(k_x R/2)\,(k_y R/2)\,(k_z R/2)}\,. \quad (58)$$

Observational results indicates a unity variance for the galaxy counts inside a sphere having radius $R=8\,h^{-1}$ Mpc [116]. Thus, if matter density fluctuations, $\delta\rho/\bar{\rho}$ are related to galaxy count fluctuations $\delta n/\bar{n}$ according to

$$\frac{\delta n}{\bar{n}} = b_g\frac{\delta\rho}{\bar{\rho}}\,, \quad (59)$$

then the normalization condition $\sigma(R = 8\,h^{-1}{\rm Mpc}) = b_g^{-1}$ determines the power spectrum amplitude $A$. In eq.(59), the parameter $b_g$ is the so-called galaxy *biassing* factor, that accounts for a possible difference in the clustering of galaxies with respect to the underlying matter distribution and, in the linear–bias approximation, is independent of the scale. Usually, values $b_g > 1$ are considered, according to the suggestion of the *biassed* model (*e.g.*, see refs. [27, 123]) that galaxies are preferentially located in correspondence of density peaks, with a subsequent amplification of their clustering (see Section 5 below, for a more detailed discussion about biassed galaxy formation).

An alternative way to normalize the spectrum is provided by the $J_3$ integral,

$$
\begin{aligned}
J_3(R) &= \int_0^R dr\, r^2\, \xi(r) \\
&= \frac{M}{\bar{\rho}a^3} \frac{1}{2\pi^2} R^3 \int_0^\infty dk\, k^2\, P(k) \left[ \frac{\sin(kR)}{(kR)^3} - \frac{\cos(kR)}{(kR)^2} \right],
\end{aligned} \quad (60)
$$

which represents the total amount of correlation within the scale $R$. The knowledge of $J_3(R)$ at a given scale from observational data [116] provides a further way to fix the spectrum amplitude. This procedure is particularly useful since, as observed by Groth & Peebles [188], the $J_3$ integral evolves according to linear theory if $R$ is chosen so that $\xi(R) \ll 1$, even for a strongly non–linear clustering at scales $r \ll R$. This enable us to linearly evaluate $J_3$ from the primordial power spectrum and compare it with estimates from observations. From the correlation analysis of the CfA1 sample, Davis & Peebles [116] found $J_3(10h^{-1}\,{\rm Mpc}) \simeq 270\,h^{-3}\,{\rm Mpc}^3$. Since $\xi(r) = 1$ at $r \simeq 5h^{-1}$ Mpc, one would like to push the estimate of $J_3$ to even larger scales, where no departure from linear evolution is expected. It is however an unfortunate fact of live that correlation analysis of available redshift samples does not provide reliable answers at scales much larger than $10h^{-1}$ Mpc. A final warning that we should bear in mind concerns the fact that any estimate of $\sigma_{\rm R}$ or $J_3(R)$ is realized in redshift space, while local peculiar velocities are expected to distort the line–of–sight clustering in the real space (see ref.[399] for a discussion about this point).

The most efficient way to normalize the power spectrum amplitude is surely represented by the measurement of temperature fluctuations in the CMB. The detection of a statistically significant $\Delta T/T$ provided by the COBE team at large ($\simeq 10°$) angles represents nowadays a unique possibility to normalize $P(k)$ at very large scales. At the scales probed by COBE, the only contribution to temperature fluctuations comes from potential fluctuations (Sachs–Wolfe effect [368]). In order to characterize the temperature fluctuations patter on the sky, let us expand it in spherical harmonics according to

$$
\frac{\Delta T(\hat{\mathbf{x}})}{T} = \sum_{l=2}^{\infty} \sum_{m=-l}^{+l} a_{lm}\, Ylm(\vartheta, \varphi)\,, \quad (61)
$$

Here $\vartheta$ and $\varphi$ are the angles on the sky and monopole term and the dipole term (entirely ascribed to the motion of the observer) have been already removed. Assuming that the Universe is spatially flat, with vanishing cosmological constant, the coefficients of the expansion are related to the power spectrum according to

$$C_l \equiv \langle |a_{lm}|^2 \rangle = \frac{1}{2\pi}\left(\frac{H_o}{c}\right)^4 \int dk \, \frac{P(k)}{k^2} j_l^2(kx). \qquad (62)$$

(see, e.g., ref.[311]). Here, $j_l$ is the Bessel function of order $l$. According to the above equation, once one detects a multipole component, $C_l$, in the temperature fluctuation pattern, it is possible to fix the power spectrum amplitude. For this reason, the COBE result, which provided the quadrupole moment $Q = (5C_2/4\pi)^{1/2} T_o = 17 \pm 5\,\mu K$, allows to determine the large scale amplitude of $P(k)$ with a 30% uncertainty [419].

Although it seems at first sight that the amount of information provided by the analysis of the 2–point function is completely equivalent to that provided by the power spectrum analysis, nevertheless these two approaches are to be considered as complementary. In fact, while the usual correlation analysis is more sensitive to detect clustering at small scales, where the 2–point function exceeds unity, the power spectrum approach is more suitable for investigating the low wavenumber, i.e. the large scale, regime, where memory of initial conditions is still preserved. Due to such advantages, several efforts have been recently devoted to trace the power spectrum for the observed distributions of cosmic structures. Baumgart & Fry [34], using the observed distribution of CfA and Perseus–Pisces surveys, detected the power spectrum up to scales $\sim 100 h^{-1}$ Mpc. Efstathiou et al. [142] and Saunders et al. [359] analysed the power spectrum of the QDOT galaxies in terms of the variance of the counting inside cubical cells and Gaussian spheres, respectively. Testing scales of some tens of Mpcs, they found that such data are at variance with respect to the predictions of the standard Cold Dark Matter scenario. Peacock & Nicholson [303] analysed the distribution of radio galaxies, thus reaching very large scales. Peacock [302] found an expression for $P(k)$ to fit the data on the angular 2–point function of APM galaxies. Extrapolating this power spectrum to larger scales, he observed that the same $P(k)$ provides a quite good fit also to IRAS, CfA and radio–galaxies, apart from suitable rescalings in the amplitude (see Figure 11). The resulting fluctuation spectrum exhibits a break toward homogeneity at wavelength $\lambda \gtrsim 200 h^{-1}$ Mpc, result which is also confirmed by the power spectrum traced by rich clusters [304]. On these scales the effective spectral index is consistent with $n \simeq 1$, thus in agreement predictions of standard inflation. If one tries to account for these data by means of a CDM model, the amplitude of $P(k)$ requires a high normalization at large scales, whose linear extrapolation at small scales generates an excess of clustering. A similar result has also been found by Einasto et al. [146] from the analysis of a redshift sample of ACO clusters and by Jing & Valdarnini [227] from the analysis of the Strauss et al. [377] redshift survey of IRAS galaxies, combined with the PGH cluster sample. With the availability of extended galaxy redshift survey, namely the SSRS and CfA2 catalogues, Park et al. [298] and Vogeley et al. [409], analysed the resulting power spectrum from $\sim 10 h^{-1}$ Mpc scales, where the effective spectral index $n \simeq -1$ agrees with the results of correlation analysis, up to $\sim 100h^{-1}$ Mpc, where the amplitude of $P(k)$ makes the standard CDM model inconsistent with the data.

It is clear that the measurement of the power spectrum traced by cosmic structures at large (50–$100h^{-1}$ Mpc) scales could become of crucial relevance, in the light of the possibility of detecting in the next few years anisotropies of the CMB temperature at small or intermediate angular scales ($\vartheta \lesssim 1°$). In fact, the detection of $\Delta T/T$ at different scales would allow us to recover the primordial shape of $P(k)$, so that its comparison with the power spectrum traced by cosmic structures could furnish precise hints about the formation and the evolution of the LSS.

## 3.3 Topology of the LSS

The description of the variety of structures in the galaxy distribution, like filaments, voids, clusters, extending over a broad range of scales, requires a global description of the geometry of the LSS. Although correlation analysis provides rather useful information about clustering strength and scaling properties, nevertheless it says only a little about the "shape" of the galaxy distribution. On the other hand, alternative approaches, such as that based on the analysis of the topology of the large scale clustering, have been proved to be useful in characterizing its geometry and "connectivity".

A detailed description of topological concepts in a formal mathematical language is out of the scope of this article. Here I only briefly introduce the measures of topology introduced in cosmological context and what we learn from their application. In this context, the concept of "genus" has been introduced to describe the topology of isodensity surfaces, drawn from a density field. As an example, in Figure 12 the isodensity contour for a Gaussian density field is plotted, for both the regions above and below the mean density. The genus $G$ of a surface can be introduced as

$$G = (\text{number of holes}) - (\text{number of isolated regions}) + 1. \qquad (63)$$

In this way, we note that a single sphere has genus $G = 0$, a distribution made of $N$ disjoint spheres has $G = -(N-1)$, while $G = 1$ for a torus. More in general, the genus of a surface corresponds to the number of "handles" it has, or, equivalently, to the number of cuts that can be realized on that surface without disconnecting it into separate parts. A more formal definition of genus can be given by means of the Gauss–Bonnet theorem (see, e.g., ref.[286]), which relates the curvature of the surface to the number of holes. According to this theorem, for any compact two–dimensional surface the genus $G$ is related to the curvature $C$ according to

$$C = \int K\,dA = 4\pi(1 - G). \qquad (64)$$

Here $K$ represents the local Gaussian curvature of the surface that, at each point, is defined as the reciprocal of the product of the two principal curvature radii, $K = (a_1 a_2)^{-1}$. Since $K$ has the dimension of length$^{-2}$, the curvature $C$ and, thus, the genus are dimensionless quantities. For a sphere of radius $r$ it is $K = r^{-2}$, so that $C = 4\pi$ and $G = 0$, as previously argued. Strictly speaking, while the genus of a surface gives the number of its "handles", eq.(63) defines a related quantity, that is the Euler–Poincaré (EP) characteristic [286]. In a sense, we can say that, while the genus deals with the properties of a surface, the EP characteristics describe the properties of the excursion set, i.e. of the part of the density field exceeding a density threshold value. Based on the Gauss–Bonnet theorem, it can be proved that genus and EP characteristics are completely equivalent in the three-dimensional case.

In topology analysis it is useful to study the dependence of the genus of isodensity surfaces on the value of the density thresholds. If a high density value is selected, only few very dense and isolated regions will be above the threshold and the genus is negative. For a very low threshold, only few isolated voids are identified and, again, the corresponding genus is negative. For thresholds around the median density value we expect in general that the isodensity surfaces have a multiply connected structure, with a resulting positive genus. These general considerations can be verified on a more quantitative ground for models having an analytically evaluable genus. The simplest case occurs for a Gaussian random field [8, 27, 200], which, in three dimensions, has a threshold–dependent genus per unit volume

$$g(\nu) = \frac{1}{(2\pi)^2} \left( \frac{\langle k^2 \rangle}{3} \right)^{3/2} (1 - \nu^2) e^{-\nu^2/2}. \tag{65}$$

The density threshold is set so as to select only fluctuations exceeding $\nu$ times the r.m.s. value $\sigma$. Therefore, $g(\nu)$ describes the topology of the isodensity surfaces, where the fluctuations take the value $\delta = \nu\sigma$. Moreover,

$$\langle k^2 \rangle = \frac{\int P(k) k^2 d^3k}{\int P(k) d^3k} \tag{66}$$

is the second order spectral moment, whose definition implies that $g(\nu)$ depends on the shape of the power spectrum, but not on its amplitude. Following eq.(65), several interesting features of the $g(\nu)$ curve appear. First of all, as expected for a Gaussian field, which has the same structure in the overdense and underdense regions, $g(\nu)$ is an even function of $\nu$, with its maximum at $\nu = 0$. This is characteristic of the so-called "sponge–like" topology. For $|\nu| < 1$ it is $g(\nu) > 0$, due to the multiple connectivity of the isodensity surfaces, while $g(\nu) < 0$ for $|\nu| > 1$, due to the predominance of isolated clusters or voids. Different topologies are however expected when non–Gaussian fields are considered. Coles [93] proposed analytical expressions for the genus characteristic of a series of non–Gaussian fields, obtainable as local transformations of a Gaussian process.

In the case of a distribution realized by superimposing dense clusters on a smooth background, isolated structures start dominating also at rather low density values and the $g(\nu)$ curve peaks at negative $\nu$'s. Vice versa, at large and positive $\nu$ values the distribution is that of isolated regions and $g(\nu)$ becomes more negative than expected for a Gaussian field. Quite significantly, this case is usually referred to as "meatball" topology. The opposite case occurs when the distribution is dominated by big voids, with objects arranged in sheets surrounding the voids. The resulting topology is usually called "cellular" or "Swiss–cheese" and the corresponding $g(\nu)$ curve peaks at positive $\nu$'s.

It is clear that topological measures can also be usefully employed when dealing with two-dimensional density fields. However, in this case some ambiguities arise, for example in distinguishing whether an underdense area is due to a tunnel or to a spherical void in three-dimensions. In addition, the interpretation of the genus in terms of the number of handles of an isodensity surface can not be applied in two dimensions. In this case, the topology measure is represented by the EP characteristics, which is defined as the difference between the number of isolated high–density regions and the number of isolated low–density regions. The EP characteristics per unit area at the overdensity level $\nu$ for a Gaussian random field is

$$g(\nu) \;=\; \frac{1}{(2\pi)^{3/2}} \frac{\langle k^2 \rangle}{2} \nu \, e^{-\nu^2/2} \,, \tag{67}$$

so that $g(\nu)$ is an odd function of $\nu$ and $g(0) = 0$.

In order to quantify the genus of the observed large scale clustering, the first step is to extract a continuous density field starting from the discrete object distribution. This can be done by collecting the points in cells and then by smoothing the resulting cell count with a suitable window function. It is clear that in order to keep Poissonian shot–noise from dominating the geometry of the smoothed field, the smoothing radius should be chosen not to be much smaller than the typical correlation length. Since the amplitude of the genus curve turns out to depend on the profile of the power spectrum through the second–order spectral moment, repeating genus measures for different smoothing radii gives information about the shape of $P(k)$. Moreover, since different profiles of $g(\nu)$ are expected for Gaussian and non–Gaussian fluctuations, topology analysis could be also suited to properly test the random–phase prediction of the inflationary paradigm, at least at the large scales, where negligible phase correlations are introduced by gravitational evolution.

Application of the genus statistics to the study of LSS has been employed in recent years (see ref.[275] for a review), both analysing the evolution of N–body simulations and observational data sets. Gott et al. [183] realized a detailed genus analysis for galaxy and cluster redshift samples. They found that, at scales fax xexceeding the correlation length, no departure from a sponge–like topology is detected, thus supporting the random–phase hypothesis. Vice versa, at higher resolutions, evidences of some meatball shift appear in the $g(\nu)$ curve, due to the effect of non–linear gravitational evolution acting at small scales (see Figure 13). A comparison with N–body results shows that a CDM model with Gaussian initial conditions provides an overall quite good fit, apart from a slightly smaller meatball shift [277]. Vice versa, hot dark matter (HDM) models, which develop a cellular topology, seem to be in trouble.

Although the analysis of contour genus in two dimensions was originally proposed to charac-terize the geometry of the CMB temperature fluctuations [93, 180], it has been recently applied to the clustering of cosmic structures. Coles & Plionis [98] evaluated $g(\nu)$ for the galaxy distribution of the Lick map at different angular resolutions. They confirmed the meatball shift at small scales, while Gaussianity is rapidly recovered at larger angles. It is however clear that, at the depth of the Lick map, projection effects are likely to "gaussianize" the clustering geometry. A comparison of these results with simulated Lick maps, as obtained from CDM N–body simulations with non–Gaussian initial conditions [97] showed that two–dimensional topology is rather efficient to distinguish between different models. Gott et al. [178] and Park et al. [336] analysed the two–dimensional genus for slices of three–dimensional galaxy surveys, further confirming the presence of meatball shift at small angular scales. Quite differently, Moore et al. [282] analysed the QDOT IRAS redshift survey and concluded that the galaxy distribution is consistent with a Gaussian random field, with sponge–like topology, up to scales $\sim 200h^{-1}$ Mpc. Using the genus amplitude to test the power spectrum shape in the range $(10\text{-}50)h^{-1}$ Mpc, they found a roughly constant spectral index $n \simeq -1$ thus consistent with the results of correlation analysis. Plionis et al. [326] analysed projected distributions of Abell and ACO clusters and compared them with synthetic catalogues. They found that the genus for the Abell distribution is consistent with that of high peaks of a Gaussian random field, while ACO clusters show a slight excess of meatball shift.

As a final comment, it is worth recalling that the statistical information provided by the topology analysis mostly concerns the geometry of the LSS, rather than the clustering strength, like correlation functions do. In fact, once a given threshold $\nu$ is fixed, the genus measure is sensitive neither to the excess of "matter" in overdense regions nor to the *deficit* in the underdense regions. This is the reason for the independence of $g(\nu)$ of the amplitude of the power spectrum $P(k)$ and, thus, of the clustering strength. In this sense, measuring topology represents a useful and complementary approach to the correlation analysis.

## 3.4 Mass and luminosity of cosmic structures

Other than studying the spatial distribution of galaxies and galaxy systems, a further important test for any theory of evolution and formation of these structures is provided by their luminosity distribution. If we were able to determine the existing relation between mass and luminosity for a given class of objects, we could in principle deduce their mass spectrum, in order to have hints about that statistics of the density fluctuations. Here I review the main observational data about mass and luminosity distribution of galaxies and galaxy systems. Then, I introduce the concept of mass function and discuss how it is related to global statistical properties of the matter distribution.

### 3.4.1 The galaxy luminosity function

An essential statistical tool for the investigation of the luminosity distributions of galaxies is the luminosity function, which is defined as the comoving number density of galaxies with luminosity between $L$ and $L + \delta L$. Accordingly, the luminosity function can be introduced through the probability,

$$\delta P = \Phi(L)\,\delta L\,\delta V, \tag{68}$$

of finding an object with luminosity between $L$ and $L+\delta L$ in the volume element $\delta V$. Following the definition of luminosity function $\Phi(L)$, the number density of galaxies is expressed as $n_g = \int_0^\infty \Phi(L)\,dL$.

A first attempt to find an analytical fitting expression to the observed galaxy luminosity function is due to Schechter [362]. He used the galaxy sample by de Vaucoleurs & de Vaucoleurs [403] and determined both the general luminosity function and the luminosity function only for galaxies contained inside clusters. He found that the latter differs from the former only by a multiplicative factor. A good fit to the data was obtained with a luminosity function of the type

$$\Phi(L)\,dL \sim \left(\frac{L}{L^*}\right)^\alpha e^{-L/L^*} d\!\left(\frac{L}{L^*}\right), \tag{69}$$

with $\alpha = -5/4$ and $L^*$ corresponding to an absolute magnitude $M^* = -21.4$ (taking for the Hubble parameter $h = 0.5$). After Schechter's investigation of the galaxy luminosity function, many attempts have been devoted to provide fits to galaxy data using Schechter–like expressions. All such investigations converge to indicate that the power law plus exponential tail always provides a good fit, although with some differences in the deduced values of the parameters (see ref.[153] for a review about observational aspects of the galaxy luminosity function). By analysing several magnitude–limited redshift surveys, Efstathiou et al. [139] found that the field galaxy luminosity function is well described by a Schechter function, with $\alpha = -1.07 \pm 0.05$, $M^* = -19.68 \pm 0.10$ for $H_o = 100\,\mathrm{km\,s^{-1}\,Mpc^{-1}}$. Recently, de Lapparent et al. [127] calculated the luminosity function for two complete slices of the extension of the CfA redshift survey. They found that the resulting shape can still be approximated by a Schechter function with $M^* = -19.2 \pm 0.1$ and $\alpha = -1.1 \pm 0.2$. Large scale inhomogeneities in the sample (comparable with the size of the sampled volume) introduce fluctuations in the derived amplitude of $\Phi(L)$.

The concept of luminosity function of galaxies can be generalized to describe the luminosity distribution of galaxy systems. A first attempt in this direction has been pursued by Gott & Turner [181], who introduced the *group multiplicity function*, $\Phi_{gr}(L)$, defined as the luminosity function for galaxy groups. These authors estimated $\Phi_{gr}(L)$ for groups selected from the Zwicky catalogue by using overdensity criteria. A further investigation of the multiplicity function has been carried out by Bahcall [18], that found a universal multiplicity function holding for both rich Abell clusters and Turner & Gott's groups. In her analysis, the author firstly determined the expression for $\Phi(L)$ that represents the best fit to the data on rich Abell clusters,

$$\Phi(L) = \Phi^* \left(\frac{L}{L_o}\right)^{-2} e^{-L/L_o} \, (10^{12} L_\odot)^{-1} \,, \tag{70}$$

with $\Phi^* = 5.2 \times 10^{-7} \, \mathrm{Mpc}^{-3}$ and $L_o = 0.8 \times 10^{13} L_\odot$. This fit has been extended to include the groups of the Turner & Gott sample, with the same shape as in eq.(70), but with parameters $\Phi^* = 1.6 \times 10^{-7} \, \mathrm{Mpc}^{-3}$ and $L_o = 1.6 \times 10^{13} L_\odot$. Let us observe that the shape of the multiplicity function, as deducible from data samples, is very sensitive to the different techniques and definitions that are used for identifying galaxy systems on different scales. This show the necessity of obtaining homogeneous samples in which groups of galaxies are defined in an objective and scale–independent way.

### 3.4.2 The mass function

In order to describe the mass spectrum of cosmic structures, let us introduce the concept of *mass function*, $n(M)$, that is defined as the number density of objects having mass between $M$ and $M + dM$. Although we expect that the shape of the mass function should depend on the statistics of the underlying matter distribution, a precise link with the shape of the primordial spectrum implies the knowledge of the mechanisms of fluctuation evolution and structure formation. Due to the relevance of a theoretical deduction of the mass function, many attempts have been devoted in this direction in order to account for the observed luminosity distribution of galaxies and galaxy systems. Despite the variety of models that have been proposed in the literature, here I will mainly concentrate on the classical approach originally proposed by Press & Schechter [334]. The reason for this choice is twofold. First, the major part of alternative approaches to the mass function represents just modifications of this model. Second, despite its conceptual limitations, it is remarkably good in keeping the main features displayed by both observational data and cosmological N–body simulations.

In the framework of the Press & Schechter (PS) approach, let us assume the fluctuation field $\delta(\mathbf{x})$ to be Gaussian and introduce the smoothed field

$$\delta_\mathrm{R}(\mathbf{x}) = \int \delta(\mathbf{y}) \, W_\mathrm{R}(|\mathbf{x} - \mathbf{y}|) \, d\mathbf{y} \,. \tag{71}$$

Here $W_\mathrm{R}(|\mathbf{x}|)$ represents a suitable window function that suppresses the fluctuation modes at wavelengths $\lesssim R$. Accordingly, the variance $\sigma_\mathrm{R}^2$ of $\delta_\mathrm{R}(\mathbf{x})$ is given by eq.(55). If $V_\mathrm{R}$ is the volume associated to a window of size $R$, then the mass scale associated to the smoothing radius $R$ is $M = \bar\rho \, V_\mathrm{R}$, $\bar\rho$ being the mean matter density of the Universe. In the case of top–hat and Gaussian filters, it is

$$M = \frac{4}{3}\pi \, \bar\rho \, R^3 \qquad \text{and} \qquad M = (2\pi)^{3/2} \bar\rho \, R^3 \,, \tag{72}$$

respectively. Accordingly, the mass variance scales as $\sigma_{\rm M}^2 = (M_o/M)^\beta$, with $\beta = 1 + n/3$ for a power law spectrum $P(k) \propto k^n$. This indicates that for $n > -3$ the variance increases for small scale fluctuations, so that smaller structures go non–linear first and the clustering proceeds in a hierarchical way. Press & Schechter's idea for deriving the mass function was to identify today–observable structures of mass $M$ with overdensities of the primeval linear fluctuation field, that exceed a critical threshold $\delta_c$, once smoothed at a scale $R$. They suggested that an overdensity above $\delta_c$ will turn into an object of mass $M$ or greater, $M$ being related to the smoothing radius $R$ according to eq. (72). The critical density contrast $\delta_c$ is that required at the initial time, so as to give rise to an observable structure at the present epoch. If, for instance, we assume that structures become observable after recollapse, then linear theory for spherical collapse gives $\delta_c = 1.68$.

For a Gaussian field smoothed at the scale $R$, the probability of exceeding the threshold level, i.e. the mass fraction in objects with mass above $M$, reads

$$p(\delta_c, M) \;=\; \frac{1}{\sqrt{2\pi}\,\sigma_{\rm M}} \int_{\delta_c}^{\infty} \exp\left(-\frac{\delta^2}{2\sigma_{\rm M}^2}\right) d\delta \;=\; \frac{1}{2}\, {\rm erfc}\left(\frac{\delta_c}{\sqrt{2}\,\sigma_{\rm M}}\right). \tag{73}$$

From the above expression, we recognize a serious drawback of the PS approach. In fact, taking $dp(\delta_c, M) = -(\partial p(\delta_c, M)/\partial M) dM$ to be the fraction of the total mass in structures with mass between $M$ and $M + dM$, its integral over the whole mass spectrum is $\int_0^\infty dp(\delta_c, M) = \frac{1}{2}$ and it fails to account for half the mass in the Universe. The origin of this problem lies in the fact that eq.(73) does not actually provide the fraction of mass in structures greater than $M$. For this to be the case, the mass in regions where $\delta < \delta_c$ should be assigned to structures of a greater mass, when smoothing the field on a larger scale. Press & Schechter overcame this problem simply by adding a factor 2 in front of eq.(73) and interpreting it as due to the accretion of surrounding matter, according to the secondary infall paradigm (see, e.g., ref.[192]). Actually, a more rigorous derivation of the mass function have been proposed to account for the correct normalization [47], which however remarkably gives the same shape as the Press & Schechter approach.

Since eq.(73) can also be interpreted as the fraction of volume occupied by fluctuations that turn into structures of mass $> M$, the number density of objects with mass between $M$ and $M + dM$ is

$$n(M)\,dM \;=\; -\frac{2}{V_{\rm R}}\,\frac{\partial p(\delta_c, M)}{\partial M}\,dM\,. \tag{74}$$

Taking the top–hat filter and the power law shape for $P(k)$, eq.(74) gives

$$n(M)\,dM \;=\; \frac{\bar{\rho}}{\sqrt{\pi}}\left(1 + \frac{n}{3}\right)\left(\frac{M}{M^*}\right)^{\frac{1}{2} + \frac{n}{6}} \exp\left[-\left(\frac{M}{M^*}\right)^{1 + \frac{n}{3}}\right] \frac{dM}{M^2}\,. \tag{75}$$

The characteristic scale $M^* = (\delta_c/\sqrt{2})^{1/\beta}\,M_o$ corresponds to the scale above which the exponential tail starts to dominate the profile of the mass function. More complicated expressions for $n(M)$ are expected for more "physical" power spectra, like the CDM or HDM ones, which possess characteristic scales (see Section 5, below).

Other than the above mentioned problem, the PS approach also presents other conceptual difficulties, which make its qualitative agreement with observational data and N–body results even more surprising. For example, during gravitational collapse the profiles of the lumps probably do not maintain spherical symmetry, while asphericity is expected to be more important at small mass scales. In addition, other processes such as fragmentation or merging between perturbations at different scales could also play a relevant role, but it is not clear how they can be accounted for in the framework of the crude PS approach. Several alternative approaches have been proposed in recent years to work out the mass function in the framework of more realistic models of structure formation (see, e.g., refs.[256, 301] for comprehensive reviews on this subject).

Although the power law plus exponential tail displayed by observed luminosity functions seems to be a quite natural outcome of many theoretical models of mass function, it is however not clear how mass and luminosity are related in a given class of cosmic structures. Thus, apart from the difficulty to properly account for the non–linear gravitational dynamics, which originates the "true" $n(M)$, a further crucial problem is to understand the non–gravitational (hydrodynamical) processes, which determine the scale dependence of the $M/L$ ratio to be used when passing from luminosity to mass spectrum. In this context, it becomes very important to devise observational prescriptions to directly work out the mass of cosmic structures on the basis of their internal dynamics. For instance, Ashman et al. [14] determined the mass content of dark halos surrounding spiral galaxies at the optical disk radius. The resulting mass–luminosity relation, $M \propto L^{0.6}$ implies $n(M) \propto M^{-1.6}$ in the $10^{10}$-$10^{12}$ $M_\odot$ mass range. Thus, according to the PS prescription, an effective spectral index $n \sim -1$ is implied at such scales by the dark matter content of spiral galaxies.

Since galaxies represent virialized structures, it is quite easy to relate their observed internal dynamics to the respective mass content. The situation is however less clear when considering galaxy systems, that, at larger scales, have an uncertain degree of virialization. Some attempts to work out the mass function of galaxy groups have been pursued by Pisani et al. [321], with the result that discrepancies are found between mass functions obtained for groups selected with different criteria. This shows how crucial it is to find an objective way to identify physical galaxy systems out of extended and complete galaxy samples.

Bahcall & Cen [21] measured the mass function of clusters resorting to both velocity dispersion data of member galaxies and X–ray data. This allowed the authors to span a quite large mass range, $10^{12} \lesssim M \lesssim 10^{15} M_\odot$. The cumulative mass functions $n(>M)$ (i.e., the number density of clusters with mass $>M$) traced by the two sets of data are remarkably consistent and are well fitted by the expression

$$n(>M) \;=\; n^* \left(\frac{M}{M^*}\right)^{-1} \exp\left(-\frac{M}{M^*}\right), \tag{76}$$

with $n^* \simeq 4 \times 10^{-5} (h^{-1} \mathrm{\ Mpc})^{-3}$. A similar result has also been found by Biviano et al. [43], although on a narrower mass range. Bahcall & Cen also showed that their result is inconsistent with the standard CDM scenario in a flat Universe, while an open universe with $\Omega \simeq 0.2$ fares much better. This confirms once more the relevance of the cluster mass function as a useful constraint for theoretical models.

# 4 Using fractals to measure the Universe

In this section I introduce the concept of fractal structure and discuss its applications to the study of the LSS. After characterizing a scale–invariant structure through its fractal dimension, I show that different definitions of dimension can be given. In this context, the possibility to have scale–invariant structures with local scaling properties (multifractals) requires the introduction of an infinite set of dimensions, which corresponds to the infinite sequence of correlation functions. In order to measure this multifractal spectrum of dimensions, a list of algorithms has been introduced in the framework of the study of complex and chaotic systems. I describe these methods and show some tests of their reliability. This is a necessary step in order to assess the robustness of the results provided by the fractal analysis of galaxy and cluster distributions, where the limited amount of statistics and the presence of characteristic scales could affect the dimension estimates. I conclude the section with a review of the more important results obtained from the application of fractal analysis to the study of the LSS, while more results from other analyses will be presented in Section 6 and 7.

Quite recently, it appeared on this Journal an article by Coleman & Pietronero [99], who extensively applied fractal analysis methods to demonstrate that the Universe behaves like a self–similar structure up to arbitrarily large scales, without evidence of homogeneity. On the contrary, I will show in this and in following sections that statistical methods based on fractal concepts are very well suited to disprove this picture and that any self–similar clustering is confined at rather small scales.

## 4.1 Fractals and fractal dimensions

The concept of scale–invariance is of key importance in the characterization of many physical systems. For a long time, it has been recognized that scale–invariant behaviours are usually associated to the complexity displayed by a given structure, that renders completely inadequate the usual instruments based on differentiable geometry. A classical example is represented by the study of the Brownian motion, that at the beginning of the century has been also interpreted in terms of non–differentiable manifolds. Although the concept of non–differentiable geometry has been subsequently used in many physical and mathematical applications, the concept of "fractal object" has been explicitly introduced and formalized only quite recently by Mandelbrot [271]. A description in terms of fractals gives a good representation of a wide spectrum of phenomena, not only in physics, but also in biology, geology, economics, social sciences and so on. Particularly fruitful it was the application of fractal techniques to the study of chaotic dynamical systems. Many of these systems display completely unpredictable trajectories in the configuration space, while their position in the corresponding phase space shows a tendency to be located around a structure that is neither fixed, nor periodic, and is usually called *strange attractor*. This kind of structure can not be represented by means of a usual geometrical object;

instead it is a fractal. As an example, Figure 14a shows the attractor generated by the phase space of the Henon map [205]. Another important phenomenological application of fractal concepts is in the study of turbulence. Despite the extreme difficulty in solving exactly the flow dynamics in the regime of fully developed turbulence, their statistical characterization led to the discovery of relevant systematics. Starting from the pioneering work by Kolmogorov [244], it has been recognized that modelling the dissipation in turbulent flows by means of cascading processes with fluctuations in the energy transfer from large to small scales leads to a fractal description of such intermittent structures [37, 38].

Despite the great difference existing between the dissipative dynamics of fully developed turbulence and the non–dissipative gravitational dynamics, nevertheless several common aspects can be identified. First of all, both the Navier–Stokes equation of fluidodynamics and the BBGKY equations which describe the gravitational dynamics (see Section 5, below), do not contain intrinsic scales. Furthermore, numerical simulations of both non–linear gravity and turbulent flows are seen to generate small scale coherent structures arising from a large scale smooth background. On the ground of these similarities, we expect that the fractal description, so successful in characterizing the statistics of dissipative eddies in turbulent flows, can be usefully employed also to analyze the statistics of gravitational clustering.

### 4.1.1 What is a fractal ?

A rough definition of a fractal object can be given by referring to the scale–invariance displayed by these structures. In this sense, we say that a fractal is a geometrical structure which looks always the same (at least in a statistical sense), independently of the resolution at which it is observed. In Figure 14b I show an example of a fractal point distribution, that is generated by means of a cascading process, according to the prescription of the $\beta$–model of turbulence [157]. From this picture, it is apparent that each part of the distribution is an exact replica of the whole.

A more formal and correct definition of a fractal set, as given by Mandelbrot [271], is "a mathematical object whose fractal (Hausdorff) dimension $D_H$ is strictly larger than its topological dimension $D_T$". Thus, for a fractal point distribution in a $d$–dimensional ambient space it is $D_T = 0$ and $0 < D_H \leq d$. A fractal dimension $D_H = d$ characterizes a space–filling and homogeneous distribution. In order to rigorously define the Hausdorff dimension $D_H$ for a given set $\mathcal{A}$ embedded in a $d$–dimensional ambient space, let us consider for any $r > 0$ the set of all the possible coverings of $\mathcal{A}$ ($\Gamma_\mathcal{A}^r$), having diameters $r_i \leq r$. Then, for any value of $\beta > 0$, we define the $\beta$–dimensional outer measure over $\mathcal{A}$ as

$$H^\beta(\mathcal{A}) = \lim_{r \to 0} \inf_{\Gamma_\mathcal{A}^r} \sum_i r_i^\beta \,. \tag{77}$$

The above expression defines the Hausdorff dimension $D_H$ as the unique value of $\beta$ that renders finite $H^\beta(\mathcal{A})$, with it vanishing for $\beta > D_H$ and diverging for $\beta < D_H$.

Another characterization of a fractal set can be given in terms of the *capacity dimension*. Let us consider the number $N(r)$ of $d$–dimensional hypercubes, all having the same side $r$, that are needed to cover $\mathcal{A}$. In the limit $r \to 0$, we expect for a fractal structure

$$N(r) \sim r^{-D_C}, \tag{78}$$

$D_C$ being defined as the the capacity, or box–counting, dimension. It can be shown in general that $D_C \geq D_H$, while in most cases of practical interest the two definitions of dimension can be considered as equivalent. According to eq.(78), for a space–filling distribution we expect that $N(r)$ should decrease as $r^{-3}$, so that $D_C = 3$. In a similar way, for a filamentary structure it is $N(r) \sim r^{-1}$, while for a planar point distribution $N(r) \sim r^{-2}$, with resulting dimensions $D_C = 1$ and $D_C = 2$, respectively. In more general cases, non integer dimensions can also be expected.

Note that the two above definitions of dimension deal with the number of required coverings, with no regard to the number of points contained inside each of them. In this sense, such dimensions depend on the "shape" of the distribution, and provide a purely geometrical description, while no information is given about the clumpiness, as correlation functions do. In order to extend the description in terms of fractal dimensions, so to include the clustering properties of a distribution, we need to introduce a probability measure $d\mu$. Then, the coarse grained probability

$$p_i(r) = \int_{\Lambda_i} d\mu(\mathbf{x}) \tag{79}$$

gives the "mass" contained inside the hypercube $\Lambda_i$ of side $r$, with $i = 1, 2, \ldots, N(r)$. Accordingly, the set $P_r = \{p_i; i = 1, \ldots N(r)\}$ is the probability distribution over the $N(r)$ different states. The *information* content of the distribution [221] can be defined as

$$J(r, P_r) = \log_2 N(r) + \sum_{i=1}^{N(r)} p_i \log_2 p_i. \tag{80}$$

For a homogenous distribution, all the boxes are expected to be equally populated, that is, all the states are equally probable (maximum entropy configuration). Correspondingly, the quantity $J(r, P_r)$ vanishes, thus indicating the absence of any information carried by unclustered structures. Vice versa, the maximum information content is obtained when one single state has unity probability, while it is vanishing for all the other states (minimum entropy configuration). In this case, $J(r, P_r) = N(r)$, while in general $0 \leq J(r, P_r) \leq N(r)$. We define the Shannon information (or entropy),

$$I(r, P_r) = -\sum_{i=1}^{N(r)} p_i \log_2 p_i, \tag{81}$$

as the difference between the maximum information content and the actual information provided by the $P_r$ distribution. Therefore, the *information dimension*,

$$D_I = \lim_{r \to 0} \frac{I(r, P_r)}{\log_2(1/r)}, \tag{82}$$

is related to the rate of information loss as the resolution scale increases.

A further characterization of the scale–invariant properties of a fractal set is given in terms of the *correlation dimension*, originally introduced by Grassberger & Procaccia [185, 186]. For a given point $\mathbf{x}_i$ belonging to $\mathcal{A}$, let

$$C_i(r) \;=\; \frac{1}{N}\sum_{j=1}^{N}\Theta(r-|\mathbf{x}_i-\mathbf{x}_j|) \;=\; \frac{n_i(<r)}{N} \tag{83}$$

be the measure for the probability of finding $n_i(<r)$ out of the $N$ points of the set within a distance $r$ from $\mathbf{x}_i$. In eq.(83), $\Theta$ is the Heaviside step function. Then, we introduce the correlation integral

$$C(r) \;=\; \frac{1}{N}\lim_{N\to\infty} C_i(r) \tag{84}$$

whose scaling in the limit $r \to 0$ defines the correlation dimension, $D_\nu$, according to

$$C(r) \;\sim\; r^{D_\nu}. \tag{85}$$

Note that for a structure that behaves like a fractal at all the scales it is not possible to define an average density, since it turns out to depend on the dimension of the fractal itself. In fact, since eq.(85) gives the scaling of the number of neighbors, the density around the $i$–th point will scale as $r^{3-D_\nu}$, and, thus, unless $D_\nu = 3$, it decreases for increasing scales. Note that this kind of behaviour is not expected for the distribution of cosmic structures, which, on grounds of the Cosmological Principle, should reach homogeneity at sufficiently large scales. However, we can define fractal dimensions in a finite scale range, while taking homogeneity at large scales. In this case, following the definition of the 2–point correlation function given in §3.1, it is easy to see that it is related to the correlation integral of eq.(84) according to

$$C(r) \;=\; n\int_0^r d^3r'\,[1+\xi(r')] \;=\; \bar{N}\left[1+\left(\frac{r_c}{r}\right)^\gamma\right]. \tag{86}$$

Here, $\bar{N} = \frac{4}{3}\pi r^3 n$ is the number of neighbors within $r$ expected for a homogeneous distribution, while the clustering scale $r_c$ is related to the correlation length $r_o$ as $r_c = [3/(3-\gamma)]^{1/\gamma} r_o$. Thus, according to the definition (85) of correlation dimension, the observed power law shape of the 2–point correlation function implies that at $r \ll r_c$ the galaxy distribution behaves like a fractal with $D_\nu \simeq 1.2$, while assuming large scale homogeneity gives $D_\nu = 3$ at $r \gg r_c$.

As shown in Section 3, a complete statistical description of a given distribution requires the knowledge of correlations or moments of any order. In a similar way, we expect that a complete characterization of the scaling properties of a fractal set should require the introduction of a hierarchy of scaling indices, that generalize those already introduced and that account for the scaling of correlation functions of different orders. This will be realized in the following by introducing the concept of the multifractal spectrum of generalized dimensions.

### 4.1.2 Generalized dimensions

The various definitions of fractal dimension that I have introduced represent particular cases of a continuous sequence of scaling indices, known as multifractal spectrum of generalized dimensions [206, 37, 296] (see ref.[297] for a comprehensive review about multifractals). A first definition can be given in terms of the generalized Hausdorff dimensions, which represents the extension of the classical Hausdorff dimension of eq.(77). Let $p_i$ be the measure associated to a given set $\Lambda_i$, as defined by eq.(79), and let us introduce the partition function

$$\Gamma(q,\tau) = \begin{cases} \lim_{r \to 0} \inf_{\Gamma_{\mathcal{A}}^r} \sum_i \frac{p_i^q}{r_i^\tau} & \tau \leq 0 \ , \quad q \leq 1, \\ \lim_{r \to 0} \sup_{\Gamma_{\mathcal{A}}^r} \sum_i \frac{p_i^q}{r_i^\tau} & \tau \geq 0 \ , \quad q \geq 1. \end{cases} \qquad (87)$$

For each value of $q$, the respective $\tau(q)$ is defined as the unique value which makes $\Gamma(q,\tau)$ a finite constant. Then, the generalized Hausdorff dimensions are defined as

$$D(q) = \frac{\tau(q)}{(q-1)} \ , \quad D(1) = \lim_{q \to 1} D(q) \ . \qquad (88)$$

From this definition, it is easy to recognize that $D_H = D(0) = -\tau(0)$.

A further set of scaling indices is given by the Renyi dimensions [343]. Let us consider a covering of $\mathcal{A}$ formed by $N(r)$ cells of the same size $r$. Then, if $n_i$ is the number of points in the cell $i$, the probability $p_i = n_i(r)/\sum_j n_j$ is the measure associated to the $i$-th box. The Renyi dimensions are defined as

$$D_q = \frac{1}{q-1} \lim_{r \to 0} \frac{\log \sum_i p_i^q}{\log r} \qquad (89)$$

In this case, the capacity dimension corresponds to the $q = 0$ case, while the information dimension is recovered in the limit $q \to 1$. In general, it can be proved that $D(q) \leq D_q$, while in most cases of practical application the two definitions (88) and (89) of generalized dimensions can be considered as completely equivalent.

A slightly different definition is represented by the Minkowski–Bouligand dimensions. In this case, the covering of the fractal set is obtained by means of spheres of radius $r$, that are centered each at a point belonging to the fractal. If $n_i(< r)$ is the number of points within $r$ from the $i$-th point, the Minkowski–Bouligand dimensions are defined as

$$D'_q = \lim_{M \to \infty} \frac{1}{N^2} \lim_{r \to 0} \frac{1}{q-1} \frac{\log \sum_i n_i^{q-1}}{\log r} \ . \qquad (90)$$

and generalize the correlation dimension of eq.(85). It can be proved that the Renyi and Minkowski–Bouligand dimensions are completely equivalent [133].

An important class of fractals is represented by self–similar monofractals. These fractal sets are characterized by the fact that every part of the set represents an exact replica of the whole set (in a statistical sense), so that the scaling properties are the same around each point. For these fractals $D_q = D_H$ for any $q$, so that a single dimension gives a complete characterization of the whole set. More complex fractal sets are represented by the so-called multifractals. In this case, the entire spectrum of generalized fractal dimensions $D_q$ is required to describe the local character of the scaling properties. For a multifractal, it can be shown that the relation $D_q \leq D_{q'}$ when $q \geq q'$ is obeyed under general conditions. According to the definitions (87) of the $\Gamma$ partition function and (89) of Renyi dimensions, in the case $q \gg 0$ the summations are dominated by the densest regions in the set, while for $q \ll 0$ the least dense regions give the largest contribution. In this sense, for positive $q$'s the generalized dimensions provide information about the scaling properties of the distribution inside the regions of high density, as correlation functions do, while for $q \ll 0$ they account for the scaling inside the underdense regions, thus providing a comprehensive statistical description of the entire point distribution.

### 4.1.3 The spectrum of singularities

A further characterization of a fractal set can be given in terms of the so-called spectrum of singularities. For a given fractal structure, we can define the "local dimension" $\alpha$ through the scaling of the number of points $n_i(r)$ contained inside the $i$-th box. For a fractal structure, we expect that at small scales $r$ it is

$$n_i(r) \sim r^{\alpha_i}, \qquad (91)$$

where in general the scaling index $\alpha_i$ depends on the chosen box. Accordingly, we can group all the boxes that are characterized by a crowding index in the range $[\alpha, \alpha + d\alpha]$ into a subset $S(\alpha)$. Thus, in the scaling regime, the number of boxes $dN_\alpha(r)$ needed to cover $S(\alpha)$ behaves like

$$dN_\alpha(r) = d\rho(\alpha)\, r^{-f(\alpha)}, \qquad (92)$$

$f(\alpha)$ being defined as the Hausdorff dimension of $S(\alpha)$. In the above expression, the measure $d\rho(\alpha)$ represents the density of scaling indices in the interval $[\alpha, \alpha + d\alpha]$. In this description, a generic fractal set is interpreted as formed by interwoven sets, each having dimension $f(\alpha)$, and formed by the distribution of those singularities, whose scaling index is $\alpha$. Note that in a $d$–dimensional ambient space, those boxes which have $\alpha > d$ are not singularities (peaks) of the density field, but minima of the distribution.

In order to relate the description given in terms of the singularity spectrum to that based on the $D_q$ dimensions, note that the moments of the box counts appearing in eq.(89) can be written as integrals over $\alpha$, according to eq.(92):

$$\sum_{i=1}^{N(r)} p_i^q(r) \propto \int d\rho(\alpha)\, r^{\alpha q - f(\alpha)}. \qquad (93)$$

In the $r \to 0$ limit, the above integral can be computed with the usual saddle point technique,

so that the definition (89) of Renyi dimensions gives

$$D_q = \frac{1}{q-1} \min_\alpha [\alpha q - f(\alpha)]. \tag{94}$$

Thus the $D_q$ dimensions are obtainable from $f(\alpha)$, while the inversion of eq.(94) gives the $f(\alpha)$ spectrum of singularities from the multifractal dimensions. In the case of a monofractal structure, it is $D_q = D_o$ for any $q$ value, so that $f(\alpha)$ degenerates into a single point, whose coordinates are $(D_o, f(D_o) = D_o)$. In this sense, we can say that a monofractal structure has global scaling properties, since all the singularities have the same scaling index $\alpha = f(\alpha) = D_o$, while for a multifractal structure the scaling of the local density is different at different points.

According to eq.(94), for each $q$ value the corresponding $D_q$ is determined by the crowding index $\alpha(q)$, that satisfies the extremum conditions

$$q(\alpha) = \left.\frac{df(\alpha')}{d\alpha'}\right|_{\alpha'=\alpha} \quad ; \quad \tau_q = q\alpha - f(\alpha). \tag{95}$$

This shows that the pairs of variables $(q, \tau_q)$ and $(\alpha, f(\alpha))$ give equivalent descriptions of the scaling properties of a fractal set and are related by the Legendre transform (95). Furthermore, the decreasing trend of $D_q$ constrains $f(\alpha)$ to be a convex function ($f''(\alpha) < 0$). Since $S(\alpha)$ is a subset of the whole distribution, it follows that $N_\alpha(r) \leq N(r)$, so that $f(\alpha) \leq D_o$. The maximum allowed value of $f(\alpha)$ occurs in correspondence to $q = 0$, that is $f_{max} = D_o$, while the information dimension satisfies the relation $f(D_I) = D_I$. Inverting eq.(94), it is easy to see that the asymptotic values of $D_q$ for $q \gg 0$ and $q \ll 0$ are given by the minimum and maximum allowed $\alpha$ values:

$$\alpha_{min} = \lim_{q \to +\infty} D_q \quad ; \quad \alpha_{max} = \lim_{q \to -\infty} D_q. \tag{96}$$

This is a rather obvious consequence of the fact that, following the definition (91) of local dimension, its lowest and highest values dominate the scaling inside the most overdense and most underdense regions, respectively.

In Figure 15 I show the $D_q$ spectrum of generalized dimensions for the random $\beta$-model plotted in Figure 14$b$, along with the corresponding $f(\alpha)$ spectrum. The multifractality of the structure is apparent from the $D_q$ shape or, equivalently, from the spreading of the crowding index values over a quite large interval. The correspondence between $f_{max}$ and $D_o$ is clearly visible, while also the asymptotic relations (96) are well reproduced.

## 4.2 Methods of fractal analysis

All the definitions of fractal dimensions are given in the limit $r \to 0$, that can be reached only when an arbitrarily large number of points is allowed. However, in practical estimates of the scaling properties of a point set, one usually deals with a finite amount of data, so that only finite scales can be probed. The non–vanishing size and quite large mean separations of galaxies and galaxy clusters in observational data, as well as the presence of numerical smoothing in N–body simulations, put limits to the achievable small scale resolution. In addition, a homogeneous distribution at large scales, as expected in the cosmological context for the galaxy distribution, also limits the scaling range. It is therefore necessary to resort to approximate methods that provide information on the scaling properties over a limited range of scales. These methods clearly have to converge to the rigorous definitions when the resolution in the sample increases. In some cases, different "characteristic" or "effective" fractal dimensions may be properly associated with different scale ranges if a sufficiently extended "local" scaling behavior is observed, in the spirit of "intermediate asymptotics" [29, 335]. In this sense, from a physical point of view a fractal behavior cannot be separated by intermediate scaling properties.

The **box–counting** (BC) algorithm is one practical method for computing the spectrum of generalized dimensions. This method uses the classic definition of Renyi dimensions (see eq.[89]). In this approach, the statistics is described through the partition function

$$Z^B(q,r) = \sum_{i_b=1}^{N_b(r)} [p_{i_b}(r)]^q, \qquad (97)$$

where $N_b(r)$ is the number of boxes with side $r$, which are needed to completely cover the set, and $p_i(r) = n_i(r)/N$, $n_i(r)$ being the number of points in the $i$–th box and $N$ the total number of points. For a fractal set, one has $Z^B(q,r) \propto r^{\tau(q)}$ in the scaling range, with $D_q = \tau(q)/(q-1)$ providing an estimate of the generalized dimension of order $q$. The well known box–counting dimension is found for $q = 0$. By plotting $\log Z^B(q,r)$ versus $\log r$ one immediately realizes whether the partition function has a power law behavior. The generalized dimension $D_q$ is then obtained by least–square fitting $\log Z^B(q,r)$ versus $\log r$ or by evaluating the average value of the local logarithmic slope of $Z^B(q,r)$ in the region of power law behavior.

A second method to compute the dimension spectrum is based on the **correlation integral** (CI) method, proposed by Grassberger & Procaccia [185] and extended by Paladin & Vulpiani [296]. In this approach, which is based on the definition of Minkowski–Bouligand dimensions (see eq.[90]), one introduces the partition function

$$Z^C(q,r) = \frac{1}{N^2} \sum_{i=1}^{N} [n_i(<r)]^{q-1}. \qquad (98)$$

Here $n_i(<r)$ is the number of particles inside a sphere of radius $r$ centered at the $i$–th object and $N$ is the total number of points in the distribution. For a fractal distribution, the scaling relation $Z^C(q,r) \propto r^{\tau(q)}$ holds and provides another estimate of the $q$–th generalized dimension. The previously discussed correlation dimension is found for $q=2$.

A further method is based on the **density reconstruction** (DR) algorithm, that represents a sort of inversion of the CI method [184]; instead of measuring for each point the number of particles within a fixed distance, this method is based on the estimate of the minimum radius which includes a fixed number of points. The DR partition function is defined as

$$W(\tau,p) = \frac{1}{N}\sum_{i=1}^{N}[r_i(p)]^{-\tau} , \qquad (99)$$

where $r_i(p)$ is the radius of the smallest sphere centered on $i$ and containing $Np$ points, with $\frac{2}{N} \leq p \leq 1$. In the scaling regime, the DR partition function depends upon the probability $p$ according to $W(\tau,p) \propto p^{1-q}$. Note that in this case one obtains $q$, and consequently the dimension $D_\tau$, as a function of $\tau$.

Another useful algorithm is the **nearest–neighbor** (NN) method [15]. For each point $i$ of the set, let $_k\delta_i$ be the distance of the particle $i$ to its $k$–th neighbor. Then, the partition sum

$$_k\bar{\delta} = \left[\frac{1}{n}\sum_{i=1}^{n} {}_k\delta_i^\beta\right]^{\frac{1}{\beta}} , \qquad (100)$$

can be shown to scale as $_k\bar{\delta}^\beta \propto n^{-\frac{\beta}{h(\beta)}}$, where $n$ is the number of points in a randomly selected subsample of the whole distribution; the fixed point of $h(\beta)$, $h(D_C) = D_C$, is an estimate of the capacity dimension. Badii & Politi [16] have shown that eq.(100) can be used to estimate the Renyi generalized dimensions. Let us consider the partition function

$$G(k,n,\tau) = \frac{1}{n}\sum_{i=1}^{n}[_k\delta_i(n)]^{-\tau}. \qquad (101)$$

For a fractal distribution it is $G(k,n,\tau) \propto n^{q-1}$, independent of the neighbor order. Again, this methods provides $D_\tau$ as a function of $\tau$. Although the original formulation of the NN method was based on the identification of the nearest–neighbors, it is often very useful to consider higher–order neighbors, in order to eliminate small scale noise [16]. Neighbor orders $k = 3, 4$ are in general adequate, while taking even higher orders could miss the details of the small scale statistics.

A last method to compute the spectrum of generalized dimensions has recently been proposed by Martinez et al. [266] and by Van de Weygaert, Jones & Martinez [401]. This method is based on the calculation of the **minimal spanning tree** (MST) [292, 423] connecting the points of a subsample which has been randomly selected from the distribution. The MST is defined as the unique graph connecting all the points, with no closed loops and having minimal length. The construction of the MST graph proceeds as follows. Let us start with a randomly chosen point and connect it to its nearest neighbour. At this first step, the tree $T_1$ has only one branch of length $\lambda_1$. At the $k$–th step, we define the distance of the $i$–th point, still not belonging to the MST, from the $T_{k-1}$ tree as

$$\lambda_{i,T_{k-1}} = \min_{j\in T_{k-1}} \lambda_{ij} . \qquad (102)$$

For a distribution of $N$ points the MST is given by $T_{N-1}$ and contains the set of branch lengths $\{\lambda_i\}_{i=1}^{N-1}$. From its definition, it follows that the MST construction is unique and independent of the point which is chosen to start building the tree. There are several reasons why the MST is a useful tool in clustering analysis. First of all, it is completely determined only once the position of each single point is known, so that it conveys informations about correlations of any order. Moreover, when one branch is added to the tree, its position does not depend on that of the previously added branch, so that we can say that the MST construction is delocalized. For the above reasons, the MST is particularly suited to emphasize the main features of the global texture of a point distribution, such as its connectivity, filamentarity, etc. The MST has been applied in a cosmological context by Barrow, Bhavsar & Sonoda [31] who showed that it is efficient at discriminating between different kinds of models. Bhavsar & Ling [41] used the MST to study the filamentarity of the spatial galaxy distribution in the CfA redshift survey. Plionis et al. [327] performed a similar analysis of a redshift sample of Abell and ACO clusters, while in ref.[60] we realized a similar analysis by comparing the angular distributions of PBF clusters with simulated data sets.

The MST has been introduced in fractal analysis with the goal of providing a close estimate of the generalized Hausdorff dimensions. In fact, the construction of the MST is based on the search for a tree of minimal length. This is somewhat similar to the search for a minimal covering required by the definition of the Hausdorff dimension. For this reason the MST tries to estimate the generalized Hausdorff dimension $D(q)$, rather than the Renyi dimensions $D_q$ [401]. In this approach, the basic quantity is the partition function

$$S(m,\tau) \;=\; \frac{1}{m-1} \sum_{i=1}^{m-1} [\lambda_i(m)]^{-\tau}. \qquad (103)$$

For a fractal set it is $S(m,\tau) \propto m^{(q-1)}$. A fitting of this relation allows one to check the scaling properties of the sample and the dimensions $D_q$.

It is important to note that there is a crucial difference between the first two methods (BC and CI) and the remaining three (DR, NN and MST). In fact, the first two algorithms evaluate the partition function by a priori fixing the scale $r$. Therefore, the "effective" dimension $D_q(r)$ (as given by the local logarithmic slope of the partition function) is a function of the physical scale $r$. This fact allows to disentangle the contributions of different scaling regimes at different scales. The other methods, instead, evaluate the partition functions interms of the probability $p$ or of the number of points in random subsamples. All these quantities do not bear a one–to–one correspondence with the physical scale; for instance, in the DR method a given probability is associated with a broad distribution of scales. As a consequence, the behavior of the partition function at a given value of $p$, $n$ or $m$ mixes several contributions from different scale ranges. This may cause troubles in situations where different scaling regimes are present at different scales. In addition, the shape of the scale distribution is a function of $\tau$, being narrower for large positive values of $\tau$ and much broader for negative $\tau$'s. This dependence on $\tau$ leads to weighting the various scales in a different way at different values of $\tau$; a monofractal distribution with two scaling regimes at different scales may thus be spuriously viewed as a multifractal distribution when analyzed with these methods. This aspect is extremely relevant in the analysis of large scale clustering, where fractality of the galaxy distribution is detected at small scales, while homogeneity is expected to hold at large enough scales. For the above reasons, some differences should a priori be expected among the results provided by the various methods of analysis. The limited statistics normally encountered in the study of galaxy samples may be another source of problems. Analogously, the presence of boundary effects (related to the peculiar shapes of the galaxy surveys) may potentially affect the results.

In order to assess the reliability of the results provided by the different multifractal estimators when dealing with a finite number of data points, I show in the following the result of applying the different algorithms to fractal distributions with *a priori* known scaling properties. These tests are a necessary step in order to obtain reliable estimates of multifractal properties from galaxy data (see ref.[63], where this problem is addressed in more details).

In order to generate fractal structures with controlled dimensions, it is useful to resort to $\beta$–model and random $\beta$–model of turbulence [157, 158], which have recently been proposed also as simplified models of the large scale distribution of galaxies [98, 335, 336, 229]. Such models provide fractal point distributions through a cascading process, that, in the context of turbulence, mimics the energy transfer from large to small scales, where dissipation occurs. To implement the cascading process in three dimensions, let us start with a "parent" cube of side $L$, which breaks into $2^3$ "son" subcubes, having side $L/2$. Let $\beta$ be the fraction of the mass of the "parent" cube which is assigned to a given subcube. By repeating $k$ times this cascade iteration, we end up with $2^{3k}$ small cubes with side $l_k = L/2^k$. Accordingly, the mass contained inside a cube is

$$M_k \sim \prod_{j=1}^{k} \beta_j \,, \tag{104}$$

and depends on its fragmentation history $\{\beta_1, ..., \beta_k\}$. Thus, the $q$–th order moment for the mass distribution inside the $2^{3k}$ cubes reads

$$\langle M_k^q \rangle \sim \int \left\{ \prod_{j=1}^{k} d\beta_j \, \beta_j^q \right\} P(\beta_1, ..., \beta_k) \,, \tag{105}$$

where $P(\beta_1, ..., \beta_k)$ is the probability distribution for the mass redistribution after $k$ fragmentations. Assuming no correlation between different fragmentation iterations, then $P(\beta_1, ..., \beta_k) = \prod_{j=1}^{k} P(\beta_j)$. Since at each step the single cube is split into eight subcubes, $P(\beta)$ can be in general written as

$$P(\beta) = \sum_{i=1}^{8} c_i \, \delta_{\rm D}(\beta - f_i) \,, \tag{106}$$

being $\sum_{i=1}^{8} c_i = 1$. In the above equation, each $f_i$ represents the mass fraction assigned to a subcube and $\sum_{i=1}^{8} f_i = 1$, as required by mass conservation. Accordingly, the moment (105) evaluated at the scale $l_k = 2^{-3k}L$ is

$$\langle M_k^q \rangle \sim \left( \sum_{i=1}^{8} f_i^q \right)^k \propto l_k^{-\log_2(\sum_{i=1}^{8} f_i^q)} \qquad (107)$$

and the resulting spectrum of generalized dimensions reads

$$D_q = \frac{\log_2 \sum_{i=1}^{8} f_i^q}{1-q}. \qquad (108)$$

According to eq.(108), the number of non vanishing $f_i$'s determines the value of the Hausdorff dimension, while the asymptotic values $D_{-\infty}$ and $D_{+\infty}$ are fixed by the smallest and largest $f_i$, respectively. Once the final density field is obtained, its Monte Carlo sampling gives the required point distribution. A particularly simple case occurs when all the non vanishing $f_i$'s take the same value. In this case, eq.(108) gives a monofractal spectrum, with the dimension value uniquely fixed by the number of non vanishing $f_i$'s. A homogeneous space–filling distribution is obtained when the $f_i$'s are all equal, so that the mass is equally distributed between all the subcubes. Another interesting case occurs when the $f_i$ values change with the iteration step. The corresponding structure is not self similar, but has different scaling properties on different scale ranges, or a non–scaling behavior, depending upon the selected scale dependence of the $f_i$'s. This case has been discussed in detail in refs.[335, 336], along with its applications to describe the large scale galaxy clustering.

Due to its cosmological relevance, let us firstly consider a scale–dependent monofractal distribution, which is homogeneous ($D = 3$) at scales larger than an homogeneity scale $L_h$, and by $D = 1$ for scales smaller than $L_h$. This distribution is obtained by the cascading process previously discussed, by an appropriate choice of the $P(\beta)$ distribution in the two different scaling ranges. At scales below $L_h$, the probability of mass distribution is

$$P(\beta) = \frac{1}{4} \delta(\beta - 0.5) + \frac{3}{4} \delta(\beta) \qquad (109)$$

The homogeneity scale is chosen to be 1/4 of the size $L$ of the simulation box; there is a total of 30,000 points in the distribution. A comparison with the galaxy distribution is possible by requiring the point number density of the simulated distribution to be approximately equal to the observed average number density of bright galaxies, $n \simeq 0.01 (h^{-1} \text{ Mpc})^{-3}$. The density indicated above gives $L = 140\ h^{-1}$ Mpc and $L_h = 35\ h^{-1}$ Mpc in physical units. As an example of undersampling, a random subsample of the complete distribution containing 3000 points is also considered. Figure 16 shows the results of the multifractal analysis of the scale dependent monofractal distribution (solid circles) and of its random subsample (open triangles). For positive $q$'s, both CI and BC methods (first and second column, respectively) provide extremely reliable results for the complete distribution, indicating both the correct value of the dimension below $L_h$ ($D = 1$) and the transition to homogeneity above $L_h$. For $q = 0$, the BC method gives a correct estimate of the dimension, while CI provides a slight underestimate of the dimension. None of these methods is able to estimate at small scales the generalized dimensions for $q < 0$, since they are sensitive to the lack of statistics in the underdense regions. Note that CI gives an apparently stable (but incorrect!) estimate $D_q \simeq 0.5$ for $q = -2$. For the random subsample, neither CI nor BC provide reliable results at small scales. However, both methods still detect the transition to large scale homogeneity. In the case of the random subsample, the statistics is thus not sufficient to correctly sample the fractal behavior at scales smaller than $L_h$. We expect this effect to be even more apparent if we were considering a fractal point distribution with higher dimension. In fact, more and more points are needed to adequately sample less clustered structures. In the following of this section I will introduce a prescription which allows us to account for the Poissonian undersampling, so to recover the correct dimension estimate at small scales.

The results of the DR method are reported in the third column. For $\tau \geq -2$, this method provides a reliable estimate of the fractal dimension and of the transition to homogeneity in the case of the complete distribution. For the random subsample, the results are not correct and they provide spurious estimates of the fractal behavior. As already mentioned, a characteristic of this method is that it mixes different scales in the evaluation of the partition function at a given value of $p$. This mixing becomes more evident as the value of $\tau$ decreases; for $\tau \leq -4$ the results can hardly be interpreted, due to a strong mixing between the small scales (where $D = 1$) and the large scales (where $D = 3$). Care has thus to be taken when using this method for evaluating the negative $\tau$ dimensions on scale dependent fractal sets.

Columns 4 and 5 of Figure 16 report the results of the NN approach and of the MST method, respectively. The random subsample provides extremely scattered and unstable results when analyzed with these two methods. For positive $\tau$'s, the NN method gives somewhat reliable results for the complete distribution, with the caveat that scale mixing tends to fuzzy the true scale dependence of the fractal dimension for moderate values of $\tau$. The best scale separation is obtained here, as for the previous method, for large values of $\tau$. For $\tau < 0$ scale mixing becomes dramatic (even worse than for the DR method); for example, a small scale (large $n$) dimension $D_\tau \approx 2$ is evaluated for $\tau = -2$, and $D_\tau \approx 3$ for $\tau = -4$, suggesting (erroneously) the presence of a multifractal distribution. Thus, the use of this method can spuriously transform the presence of two scaling regimes at different scales into an apparent multifractality. The results provided by the MST are quite stable along the whole sequence of $\tau$ values. For the complete distribution, it always detects the correct values, $D = 1$ and $D = 3$, holding at small and large scales, respectively. However, because of scale mixing, only a smooth transition between these two values is detected, without any evidence of scale invariance over a finite interval. Again, for the smaller sample the limited statistics heavily affects the dimension estimate.

Other than detecting a change in the scaling regime, a reliable dimension estimator should also be able to follow the change of dimension along the $D_q$ spectrum for a multifractal structure. To address this further problem, I show the result of analysing a multifractal structure generated

according to

$$P(\beta) = \frac{1}{8}\delta(\beta - 0.6) + \frac{1}{4}\delta(\beta - 0.15) + \frac{1}{8}\delta(\beta - 0.1) + \frac{1}{2}\delta(\beta). \tag{110}$$

The resulting spectrum has $D_\infty = 0.8, D_0 = 2$ and $D_{-\infty} = 3$. The field produced by the random $\beta$–model has been sampled with a total of 50,000 Monte Carlo points. From the analysis of the scale dependent structure it follows that CI and BC methods are not adequate for $q < 0$, while NN and MST algorithms always suffer for limited statistics. In Figure 17 I plot the expected $D_q$ curve and that estimated by using the DR method. The result clearly shows that this algorithm is remarkably good to detect multifractality, at least for structures which do not have characteristic scales.

The general indication that can be drawn from the above considerations is that a great care must be payed when trying to work out the fractal structure of a point distribution, especially when the structure is not known *a priori* (as occurs in any physical situation), or the statistics is rather limited (as for the samples of galaxies and galaxy clusters). For these reasons, it is recommended to apply different methods, which are shown to be reliable in different regimes.

## 4.3 Fractal analysis of the galaxy distribution

As emphasized in Section 3, the application of several clustering measures indicates the presence of well defined scaling properties for the galaxy distribution. The detection of the power law shape for the 2–point correlation function, $\xi(r) \propto r^{-\gamma}$, for both galaxies and clusters, implies a small scale fractality of the clustering pattern, with correlation dimension $D_\nu = 3 - \gamma \simeq 1.2$ (see eq.[86]). The above picture is also supported by the hierarchical appearance of the large scale galaxy distribution, where objects of small size are nested inside larger structures. The resulting texture of the galaxy distribution shows the presence of big voids, filaments and huge galaxy concentrations (superclusters), whose sizes are comparable to those of the largest available redshift samples. Such a complexity led several authors to interpret the observed large scale clustering as a fractal process [140, 308, 271], having rigorously scale invariant statistical properties. This conclusion is clearly at variance with respect to previously presented results based on the analysis of correlation functions. In fact, while at small scales non–linear clustering gives $\xi(r) \gg 1$, at scales much larger than the correlation length it is $\xi(r) \ll 1$ and the distribution becomes essentially homogeneous.

However, based on the increase of the galaxy correlation length with the volume of the sample, as already detected by Einasto et al. [147], serious criticisms in the use of the $\xi(r)$ correlation function have been raised by Pietronero and coworkers [329, 100, 99]. Their main criticism was based on the fact that, while the definition (6) of $\xi(r)$ includes a normalization with respect to the mean object density $n$, the limited sizes of presently available galaxy redshift surveys does not allow to precisely fix the value of $n$. Nevertheless, in usual correlation analysis the observed galaxy distribution is normalized to a Poissonian distribution, so that the 2–point function is forced to vanish at large scales. In this way, large scale homogeneity is far from being verified, instead it is assumed. If the real distribution is such that the average density depends on the size of the sampled volume, as it should be for a fractal structure, the same distribution of objects appears to have different correlation lengths as the volume of the sample changes. In fact, for a fractal structure, the number density of neighbours within $r$ from a fixed object scales as

$$N(r) = B\, r^{D_\nu}, \qquad (111)$$

where $B$ is a constant and $D_\nu$ is the correlation dimension (see eq.[85]). As a consequence, if $R_s$ is the characteristic size of the sample, the average density inside a sphere of radius $R_s$,

$$n = \frac{N(R_s)}{V(R_s)} = \left(\frac{3}{4\pi}\right) B\, R_s^{3-D_\nu}, \qquad (112)$$

turns out to be a decreasing function of $R_s$, and the 2–point function is

$$\xi(r) = \frac{D_\nu}{3}\left(\frac{r}{R_S}\right)^{-(3-D_\nu)} - 1. \qquad (113)$$

Therefore, while the exponent of the power law, $\gamma = 3 - D_\nu$, is an intrinsic property of the distribution, the normalization of $\xi$ depends explicitly on $R_s$. Pietronero and collaborators drawn extreme conclusions from this argument and suggested that the LSS of the Universe can be explained as a self–similar fractal extending at least up to scales $\sim 200 h^{-1}$ Mpc, with no evidence of large scale homogeneity. In this picture, they interpreted the increase of the galaxy correlation length with the sample size, while the amplification of the cluster 2–point function is just due to the fact that clusters sample much larger scales than galaxies, with a subsequent increase of the correlation amplitude (see Figure 18). Consistently, instead of $\xi(r)$, these authors proposed as an alternative clustering measure the quantity

$$\Gamma(r) = n\left[\xi(r) + 1\right], \qquad (114)$$

that, by definition, is independent of the actual value of $n$ and does not rely on the assumption of large scale homogeneity. Making use of $\Gamma(r)$, Pietronero [329] analysed the CfA1 survey, dividing the whole sample into two volume limited subsamples. He found that the galaxy distribution behaves like a simple fractal, which extends at least up to the sample size, without any evidence that homogeneity is attained within the sample boundaries. Apart from the obvious problem that this kind of picture has to account for other striking observational facts, such as the high degree of homogeneity of the cosmic microwave background, several further evidences appeared, that the scale–invariant properties of the galaxies distribution were overestimated, if represented by means of a self similar fractal.

A classical criticism to a purely fractal description of galaxy clustering arises from the results about the amplitude of the 2–point angular function, $w(\vartheta)$. In fact for an angular sample having a depth $\mathcal{D}$, the Limber equation (36) predicts that $w(\vartheta) \propto (r_o/\mathcal{D})^\gamma \vartheta^{1-\gamma}$. Thus, if homogeneity holds at large scales (that is, $r_o = const$), then $w(\vartheta) \propto \mathcal{D}^{-\gamma}$. Vice versa, if fractality extends at arbitrarily large scales, then $r_o \propto \mathcal{D}$ and the angular correlation amplitude does not depend on the sample depth. In Figure 18$b$ we plot the amplitude of $w(\vartheta)$ as estimated for several angular samples, as a function of the depth. It is apparent that the data are by far much better represented by assuming large scale homogeneity than by modelling the clustering with a self–similar fractal at arbitrarily large scales.

This argument has been however severely criticised by the supporters of a fractal Universe. Coleman & Pietronero [99] claimed that angular data are much less reliable than redshift data to establish the existence of large scale homogeneity. Projection of a fractal structure onto a sphere could cause, differently from an orthogonal projection, a spurious homogenization of the distribution at large angular scales. It is however hard to expect that the effect of non–planar projection could play a significant role at angular scales of few degrees, where curvature corrections are negligible and where the projected galaxy correlation is shown to decline (see Figure 7). I will address in more detail the problem of preservation of scale–invariance after projection on a sphere during the presentation in Section 7 of the fractal analysis of projected cluster distributions.

Alternative interpretations of the observed scaling of $\xi(r)$ with the size of the sample volume have also been proposed, for example by invoking effects of luminosity segregation [53, 67, 267]. In fact, it is well known that the clustering strength is an increasing function of the absolute galaxy luminosity. Since galaxy samples are limited in apparent magnitude, enlarging the sampled volume one preferentially includes intrinsically more luminous objects, thus increasing the observed clustering as a spurious effect.

Martinez & Jones [265] divided the CfA1 sample into 10 volume limited subsamples, in order to carefully check the increase of the clustering length with the volume. They claimed that the observed increase of $r_o$ for the CfA1 galaxies is most probably due to the luminosity dependence of the clustering pattern. Moreover, the resulting power law shape of $\xi(r)$ implies self–similarity only at the small scales ($r \lesssim 5h^{-1}$ Mpc) of non–linear clustering, while homogeneity is attained at larger scales. More recently, Guzzo et al. [197] analysed the Perseus–Pisces redshift survey and claimed for the presence of two well defined scaling regimes: at scales $r \lesssim 4h^{-1}$ Mpc it behaves like a fractal with correlation dimension $D_\nu \simeq 1$, while it takes $D_\nu \simeq 2$ up to $r \simeq 30h^{-1}$ Mpc. While confirming previous indications by Dekel & Aarseth [120], the above result is also supported by further analyses realized on different samples [145, 80].

In order to investigate the whole spectrum of generalized dimensions, Martinez et al. [266] performed a multifractal analysis of the CfA1 sample. As a result, they found that the galaxy distribution is characterized by a non trivial scaling behaviour at the small scales of non–linear clustering, where self–similarity is detected. The resulting $D_q$ curve shows a remarkable multifractality, with Hausdorff dimension $D_o \simeq 2.2$ and correlation dimension $D_\nu \simeq 1.3$. At negative $q$ values it is $D_q \gtrsim 3$ thus indicating that points in the underdense regions are minima of the local density field, instead of singularities (see Figure 19).

Although these results indicate that the galaxy clustering can not be described by a self–similar fractal extending to arbitrarily large scales, nevertheless fractal analysis has been proved to be an extremely powerful statistical instrument to properly study the scaling properties of the large–scale structure of the Universe. In addition, a careful check of the scale of homogeneity for the galaxy distribution shows that it should be at least of the same order of the size of current redshift samples, and not much smaller, as required by a consistent analysis based on the $\xi(r)$ function.

In Sections 6 and 7 I will discuss the extensive application of methods of fractal analysis to answer two important questions. First, can the self–similar galaxy clustering at small scales be dynamically interpreted purely on the ground of non–linear gravitational evolution? Second, do clusters extend at larger scales the same self–similar clustering shown by galaxies, as the detected shapes of correlation functions seem to suggest?

## 4.4 Correlations and fractal dimensions

Through the application of both correlation and fractal analysis, it has been recognized that a scale invariant behaviour is always associated with the small scale clustering of galaxies. Here we address the problem of providing a unifying description of correlation and fractal properties. As already emphasized, the power law shape of the 2–point correlation function, $\xi(r) \propto r^{-\gamma}$, implies small scale fractality with a correlation dimension $D_\nu \simeq 1.2$. In a similar fashion, we expect the scaling of higher order correlations to be somehow connected with the $D_q$ spectrum of generalized dimensions.

After reviewing the basic statistical formalism to address this problem, I will show how correlation properties are connected to the multifractal structure of a point distribution.

### 4.4.1 The statistical formalism

As in the case of correlation functions, introduced through their generating functionals, moments of increasing orders can be defined in a similar fashion by successive differentiation of suitable generating functions. To see this, let us consider the cosmic matter density field as described by a random variable $\rho$, and let $p(\rho)$ be its probability density function (pdf). Then, the moment of order $q$ reads

$$m_q \equiv \langle \rho^q \rangle = \int d\rho\, p(\rho)\, \rho^q \,. \tag{115}$$

Following the expression (24) for the $q$–point joint probability of the density field, the corresponding moments are expressed in terms of correlation functions through the relation

$$m_q = \bar{\rho}^q \int_v d^3 x_1 \ldots \int_v d^3 x_q [1 + (\text{terms of order} < q) + \mu_q(\mathbf{x}_1, ..., \mathbf{x}_q)]\,. \tag{116}$$

In analogy with the $\mathcal{Z}[J]$ functional generator of correlation functions, introduced in §2.1, we define the moment generating function as the Laplace transform of pdf,

$$M(t) \equiv \int d\rho\, p(\rho) e^{t\rho} = \langle e^{t\rho} \rangle, \qquad (117)$$

in such a way that the moments $m_q$ are the coefficients of its McLaurin expansion

$$M(t) = \sum_{q=0}^{\infty} \frac{m_q}{q!} t^q \quad ; \quad m_q = \left.\frac{d^q M(t)}{dt^q}\right|_{t=0}. \qquad (118)$$

In a similar fashion, the cumulants or irreducible moments $k_q$ are defined through the generating function

$$K(t) \equiv \log M(t) = \sum_{q=0}^{\infty} \frac{k_q}{q!} t^q \quad ; \quad k_q \equiv \left.\frac{d^q K(t)}{dt^q}\right|_{t=0}, \qquad (119)$$

which is analogous to the $\mathcal{W}[J]$ generator of connected correlations. In fact, the cumulant turns out to be related to the connected functions according to

$$k_q = \bar{\rho}^q \int_v d^3 x_1 \ldots \int_v d^3 x_q\, \kappa_q(\mathbf{x}_1, ..., \mathbf{x}_q). \qquad (120)$$

Suitable relations between $k_q$ and $m_q$ can be found by successively differentiating eq.(119), which resembles the analogous relations between connected and disconnected correlation functions (see eq.[25]). According to eqs.(116) and (120), the scaling of correlation functions turns into a scaling of the respective moments. As an example, by taking for the connected function $\kappa_q$ the hierarchical expression of eq.(144), it follows from eq.(120) that the $q$-th order cumulant scales as $k_q \propto r^{(q-1)(3-\gamma)}$, being $\gamma$ as usual the logarithmic slope of the 2–point function.

Following eq.(117), it is possible to express the probability distribution in terms of the cumulant generating function as the inverse Laplace transform

$$p(\rho) = \frac{1}{2\pi i} \int_{-i\infty}^{+i\infty} dt\, e^{-t\rho} e^{K(t)}. \qquad (121)$$

In the case of Gaussian pdf,

$$p(\rho) = \frac{1}{\sqrt{2\pi\sigma^2}} \exp[-(\rho - \bar{\rho})^2/2\sigma^2], \qquad (122)$$

so that $K(t) = \bar{\rho}t + \frac{1}{2}\sigma^2 t^2$, and the statistics is completely specified by the average density and by the variance $\sigma^2$.

For a Poisson point distribution,

$$p(\rho) = \sum_{N=0}^{\infty} \frac{1}{N!} \bar{\rho}^N e^{-\bar{\rho}}\, \delta_{\mathrm{D}}(\rho - N), \qquad (123)$$

with the corresponding cumulant generating function

$$K(t) = \bar{\rho}(e^t - 1), \tag{124}$$

that does not contain any contribution of correlation terms, according to the expectation that a Poissonian process has vanishing correlation functions.

Although it is in general possible to introduce the moments for a given distribution, the convergence of the series, that defines the respective generating function, is not always guaranteed. The classic example is the *lognormal* density field [96], obtained by means of the exponential transformation

$$\chi(\mathbf{x}) = \exp[\rho(\mathbf{x})] \tag{125}$$

of the Gaussian–distributed field $\rho(\mathbf{x})$. Accordingly, the pdf of the $\chi(\mathbf{x})$ field reads

$$p(\chi) = \frac{1}{\sqrt{2\pi\sigma_\rho^2}} \exp\left[\frac{-(\log\chi - \bar{\rho})^2}{2\sigma_\rho^2}\right]\frac{1}{\chi} \tag{126}$$

with $\sigma_\rho^2$ the variance of the Gaussian field $\rho(\mathbf{x})$. The relevance of the lognormal distribution in cosmological context has been discussed in detail by Coles & Jones [96], who proposed it as a good approximation to describe the moderately non–linear gravitational evolution of a primordial Gaussian random field, as well as the correlation properties of the very high peaks of a Gaussian density distribution. According to eq.(126), the corresponding moment of order $q$ is

$$m_q = \exp\left(q\bar{\rho} + q^2\sigma_\rho^2/2\right). \tag{127}$$

As discussed in ref.[96], the divergence of the moment series is related to the fact that the lognormal distribution is not completely determined by the knowledge of its moments. Following the above expression for $m_q$, it is easy to verify that the lognormal pdf generates correlation functions of Kirkwood type (see eq.[27]). In fact, by substituting eq.(27) for the $q$–point correlations into eq.(116), we get

$$m_q = m_1^{\binom{q-2}{q}} m_2^{\binom{q}{2}}. \tag{128}$$

which is satisfied by the expression (127) for $m_q$.

If, instead of a continuous density field, we are dealing with a discrete point process, as it is for the galaxy distribution or for the particle distribution generated by N–body simulations, the local object count $N$ in a volume $v$ can be considered as given by a Poissonian sampling of the underlying continuous density field $\rho$. Then, according to the expression (124) for the moment generator of a Poissonian distribution, the discrete nature of the distribution is accounted for by changing the variable $t \to e^t - 1$ in the $M(t)$ function [308, 162]. That is,

$$M_{discr}(t) = M_{cont}(e^t - 1). \tag{129}$$

Evaluating the moments of counts by differentiating eq.(129) with respect to $t$, it turns out that new terms appear, which account for the discrete nature of the distribution. For the moments of first orders, one finds

$$\begin{aligned} \langle N \rangle &= \bar{N} \\ \langle N^2 \rangle &= m_1 + m_2 = \bar{N} + \bar{N}^2(1 + \bar{\xi}) \\ \langle N^3 \rangle &= m_1 + 3m_2 + m_3 = \bar{N} + 3\bar{N}^2(1 + \bar{\xi}) + \bar{N}^3(1 + 3\bar{\xi} + \bar{\zeta}), \end{aligned} \quad (130)$$

while more cumbersome expressions hold at higher orders. In eq.(130) $\bar{N} = nv$ is the average count,

$$\bar{\xi} = \frac{1}{v^2} \int_v d^3r_1 \, d^3r_2 \, \xi_{12} \quad (131)$$

the average 2–point function inside $v$, and $\bar{\zeta}$ the average 3–point function, which is defined by analogy with eq.(131). If $\bar{N} \ll 1$, then $\langle N^n \rangle \simeq \langle N \rangle$ and the moments are dominated by discreteness (shot–noise) effects. Vice versa, for $\bar{N} \gg 1$, the terms of lowest order in $\bar{N}$ become negligible and the continuous case is recovered.

Taking into account the effects of discreteness on the pdf expression, we get

$$p(\rho) = \int_{-\infty}^{+\infty} \frac{d\phi}{2\pi} e^{-i\phi\rho} e^{K(e^{i\phi}-1)}. \quad (132)$$

Since the variable $y = e^{i\phi}$ takes its value on the unit circle of the complex plane centered at the origin, $K$ turns out to be a periodic function. Accordingly, its Fourier transform, $p(\rho)$, is written as a series of Dirac $\delta$–functions as

$$p(\rho) = \sum_{N=-\infty}^{+\infty} \delta_{\mathrm{D}}(\rho - N) P_{\mathrm{N}}, \quad (133)$$

and the probability distribution vanishes except for integer $\rho$ values, as it should for a discrete density field. According to eq.(132), the coefficients $P_{\mathrm{N}}$ are

$$P_{\mathrm{N}} = \oint \frac{dy}{2\pi i} y^{-(N+1)} e^{K(y-1)}. \quad (134)$$

For analytical $K(t)$, the $P_{\mathrm{N}}$'s vanish for $N < 0$ and they acquire the meaning of probabilities of finding $N$ points inside the sampling volume. For $N = 0$, eq.(134) gives the void probability function,

$$P_0 = e^{K(-1)} = \exp\left[\sum_{n=0}^{\infty} \frac{(-1)^n}{n!} \int_v d^3x_1 \ldots \int_v d^3x_n \, \kappa_n(\mathbf{x}_1, \ldots, \mathbf{x}_n)\right]. \quad (135)$$

This shows that the void statistics contain information about correlations of any order, so that $P_o$ represents a suitable quantity to characterize the global properties of the galaxy distribution.

In a similar fashion, by further applying the Cauchy theorem to evaluate higher order residues from eq.(134), we get

$$P_{\text{N}} = \frac{1}{N!} \frac{d^N}{dy^N} e^{K(y-1)} \bigg|_{y=0} , \qquad (136)$$

and also the count–in–cell probabilities $P_{\text{N}}$ convey information about higher-order correlations. According to eq.(136),

$$K(y) = \sum_{N=0}^{\infty} \frac{P_{\text{N}}}{N!} (1-y)^N , \qquad (137)$$

so that, for a discrete distribution, the cumulant generating function of the background continuous field is the generator of the count probabilities $P_{\text{N}}$. The continuous limit is recovered for $\bar{N} \to \infty$ and the density variable is $\rho/\bar{\rho} = N/\bar{N}$. In this case, the pdf is obtainable from $P_{\text{N}}$ according to $\bar{N} P_{\text{N}} \to \bar{\rho} p(\rho)$.

Although the statistics based on the analysis of moments and count probabilities is strictly related to the correlation approach, nevertheless the relative feasibility of computing them from observational data and N–body simulations makes this method particularly suitable to get information about higher order clustering. Actually, the void probability analysis has been extensively applied to characterize the distribution of galaxies (see, e.g., refs.[165, 408]) and clusters [223, 82]. While the void probability function is a measure of the geometry, rather than of the clustering, of a point distribution, the count–in–cell probabilities $P_{\text{N}}$ weights in a different way regions having different densities. Detailed analyses of the $P_{\text{N}}$'s as a function of the scale have been realized on several galaxy redshift samples (see, e.g., refs.[11, 272, 68, 164]). The results of such investigations show remarkable scaling properties for the galaxy distribution, which confirm results based on the analysis of high–order correlation functions.

### 4.4.2 Relation to fractal dimensions

Let us consider a spatial point distribution containing a total number $N_t$ of particles, and suppose it is covered by a set of cubical boxes, all having the same volume $v = r^3$. If $V = L^3$ is the total volume occupied by the point distribution, then $B(r) = (L/r)^3$ is the total number of boxes. The box–counting partition function of order $q$ reads

$$Z(q,r) \equiv B(r) \sum_N \left(\frac{N}{N_t}\right)^q P_{\text{N}}(r) = B(r) \frac{\langle N^q \rangle_r}{N_t^q} . \qquad (138)$$

For $q = 0$, the partition function depends on the number of non–empty boxes and eq.(138) simplifies into

$$Z(q=0,r) = B(r)(1-P_0) . \qquad (139)$$

For integer $q \geq 2$, the scaling of the partition function can be related to the behaviour of the $q$–point correlation function, according to eqs.(130). Generalizing to moments of any positive

integer order, we get

$$\langle N^q \rangle = [\langle N \rangle + ... \text{discreteness terms of order } < q] \qquad (140)$$
$$+ n^q \int_v d^3r_1 ... \int_v d^3r_q \left[1 + (\text{disconnected terms of order } < q) + \kappa_q(\mathbf{r}_1, ..., \mathbf{r}_q)\right].$$

Here, all the discreteness terms become negligible in the $\bar{N} \gg 1$ case, while the scaling is dominated by Poissonian shot–noise for $\bar{N} \ll 0$. In this case, most of the cell counts will be one or zero, so that $\langle N^q \rangle \simeq \langle N \rangle \propto r^3$. As a consequence, the $Z(q,r)$ partition function becomes nearly independent of $r$ and $D_q \simeq 0$. This result can be interpreted by saying that, at the scales where discreteness effects dominate, we are not measuring the dimension of the fractal structure, but the dimension of each single point that, indeed, is just zero.

As usually done in the analysis of correlation functions, it is possible to recover from eq.(141) the expression of $\kappa_q$ by subtracting from the measured $\langle N^q \rangle$ all the lower–order moments generated by the discrete nature of the galaxy distribution (see, e.g., ref.[308]). In a similar fashion, it is also possible to apply the same prescription in fractal analysis, in order to recover the correct dimension when dealing with a sparse sampling [65]. In order to show how this method work, I plot in Figure 20 the result of its application to the same scale–dependent monofractal structure previously analyzed (see Figure 16). The considered distribution has 5000 points randomly selected from a larger sample of 128,000 points, which does not suffer for discreteness in the considered scale–range. We plot the moments of cell counts as a function of the cell size for $q = 2, 3$ and 4, along with the corresponding "continuous" moments, corrected according to eqs.(130). The reliability of this method to correct for sparse sampling noise is remarkable, on the shape of both $\langle N^q \rangle$ and $D_q(l)$, which recovers the correct value $D_q = 1$ in the scaling range. Note that, without any corrections no indications about the scaling of the distribution can be drawn from the local dimension, which only shows a transition from $D_q = 0$ at $L \ll L_h$ to $D_q = 3$ at $L > L_h$.

Although this method looks remarkably good, nevertheless it must be observed that it can be applied only to correct moments of integer order $q \geq 2$. For other $q$ values, according to eq.(138) the continuous limit can only be recovered by knowing *a priori* the behaviour of the $P_N$ probabilities in the $\bar{N} \gg 1$ regime. Therefore, this correction can be applied only to estimate the $q > 0$ tail of the dimension spectrum, while discreteness effects are expected to mostly pollute the statistics inside the underdense regions. Furthermore, the application of eq.(141) implies that we are recovering the scaling properties of a background structure, whose Poissonian sampling is the considered point distribution. For instance, if we select the high peaks of a continuous field, their distribution does not represent a Poissonian sampling. Thus, correcting according to the above prescription, we get scaling properties which in general do not coincide with those of the underlying continuous field.

According to eq.(138), the knowledge of the $P_N$ probabilities, and, thus, of the generating function, completely specifies the $D_q$ spectrum of generalized dimensions. Following the com-putations reported in Appendix A, it is possible to express the partition function in terms of the cumulant generating function as

$$Z(q,r) = -N_t^{-q} \frac{B}{N_c} \frac{\Gamma(q)}{2\pi i} \int_{(0,N_c)^+} dy \left[\log\left(1 - \frac{y}{N_c}\right)\right]^{-q} K'(-y/N_c) \, e^{K(-y/N_c)}. \tag{141}$$

(see also ref.[26]). Here we introduced the quantity

$$N_c = \bar{N}\,\bar{\xi} \propto r^{3-\gamma}, \tag{142}$$

which represents the average number of particles in excess of random within each box. In eq.(141), the integration contour $(0, N_c)^+$ on the complex plane runs counterclockwise around the real axis from the origin to $N_c$. Based on the above expression for the box–counting partition function, Balian & Schaeffer [26] worked out the fractal properties of their model, based on the scaling relation

$$\kappa_q(\mathbf{x}_1, ..., \mathbf{x}_q) = \lambda^{q-1} \kappa_q(\lambda \mathbf{x}_1, ..., \lambda \mathbf{x}_q) \tag{143}$$

for the connected correlation functions, which is implied by the hierarchical pattern of eq.(28). In a similar context, I investigated the fractal properties for a list of hierarchical distributions, showing how these generate different multifractal spectra [56].

The relevance of hierarchical models lies in the fact that both observational data and theoretical arguments converge to indicate that the connected part of the $q$–point correlation function is well represented by the hierarchical expression

$$\kappa_q(\mathbf{r}_1, ..., \mathbf{r}_q) = Q_q \sum_{a \neq b} \xi(|\mathbf{r}_a - \mathbf{r}_b|)^{q-1}, \tag{144}$$

(see also eq.[28]). The validity of eq.(144) has been tested against the galaxy distribution for $q = 3, 4$ [189, 166, 373] and is also confirmed by the outputs of cosmological N–body simulations [111, 276, 141, 395]. Even at very large scales, the hierarchical expression has been found to be consistent with data on the 3–point function of galaxy clusters [387, 226, 177, 61] (see Section 3). From the theoretical point of view, the hierarchical behaviour of correlation functions arises from a first–order perturbative approach to the evolution of density inhomogeneities [161], and is consistent with models of non–linear clustering, such as the solution of the BBGKY equations in the strongly non–linear regime [115, 160, 199] and the thermodynamical approach proposed by Saslaw and coworkers [358, 356].

In order to provide a phenomenological description of the galaxy clustering pattern, several authors introduced a variety of hierarchical probability distributions, that are able to generate the sequence (144) of $q$–point functions, with hierarchical coefficients depending on the model details (see, e.g., ref.[162]). Apart from the shape of the $q$–point functions, the assumption of hierarchical probability distribution also has precise implications on the behaviour of other observable quantities, such as the count–in–cell probabilities [84, 358, 162, 25], the void probability function [163, 148] and the mass function of cosmic structures [258, 92]. Detailed comparisons of such models with the observed galaxy distribution [163, 11, 357, 408] and with the outputs of cosmological N–body simulations [165, 356, 70] show that several aspects of the non–linear clustering can be well reproduced by any hierarchical model, while no strong preference for one particular model has been obtained as yet.

Without entering into the details of the computations, I outline in the following some aspects of the scale invariant properties associated to hierarchical correlations. Assuming eq.(144) for connected correlation functions gives for the cumulants

$$k_q = \bar{\xi}^{-1} q^{q-2} Q_q N_c^q ,\qquad(145)$$

so that the cumulant generating function can be written as

$$K(t) = \bar{\xi}^{-1} \sum_{q=1}^{\infty} \frac{q^{q-2} Q_q}{q!} (N_c t)^q .\qquad(146)$$

The above relation shows that the hierarchical generating functions depend upon the variable $N_c t$ and its shape is defined through the sequence of hierarchical coefficients $Q_q$. For integer $q \geq 2$, the shape of the multifractal spectrum implied at the non–linear clustering scale by any hierarchical model can be easily computed, without explicitly evaluating the integral in eq.(141). In fact, assuming negligible discreteness effects in eq.(141), we get

$$\langle N^q \rangle \simeq I_q q^{q-2} Q_q r^{3q-\gamma(q-1)}\qquad(147)$$

at the small scales of non–linear clustering ($\bar{\xi} \gg 1$), where the connected part of the $q$–point function gives the leading contribution to the $q$–th order moment. In the above expression, the quantity $I_q$ accounts for the integrals over the $\mathbf{r}_i$ variables, which appear in eq.(141), after making the rescaling $\mathbf{r}_i \to \mathbf{r}_i/r$. In this way, $I_q$ is a dimensionless quantity, as long as the 2–point function behaves as a power law up to the scale $r$, and represents only a geometrical factor. According to the scale dependence in eq.(147), we have for the partition function

$$Z(q,r) \propto r^{(q-1)(3-\gamma)} ,\qquad(148)$$

so that the corresponding $D_q$ spectrum reads

$$D_q = 3 - \gamma , \qquad \forall q \in \aleph , \qquad q \geq 2 .\qquad(149)$$

Thus, hierarchical correlation functions imply monofractality of the distribution inside the overdense regions, at least at the scale of non–linear clustering, independently of the detailed shape of the density distribution function. This also agrees with previous attempts to recognize the more adequate hierarchical model to reproduce the non–linear clustering (see, e.g., refs.[162, 163, 70]). Such results have shown that all these models are more or less equally efficient in accounting for the clustering inside the overdense structures. In fact, the shape of the $K(t)$ generating function in the non–linear ($\bar{\xi} \gg 1$) clustering regime is essentially specified by the positive order moments, that mostly weights the overdense regions. Consequently, an arbitrary variation of the distribution inside the "devoid" regions has only a small effect on the overall shape of $K(t)$. Vice versa, the negative $q$ tail of the $D_q$ spectrum turns out to be sensitive to the model details and, thus, can be usefully employed to discriminate between different hierarchical prescriptions. This point has been discussed in detail in the analysis of fractal properties of different hierarchical prescriptions (see ref.[56]). Therefore, the study of the multifractal dimension spectrum is a very promising approach, since the $D_q$ curve contains not only information about the more clustered regions in its positive $q$ part, but also accounts for the distribution inside the underdense regions for $q < 0$.

A further multifractal model, based on a lognormal distribution of cell counts, has been introduced by Paladin & Vulpiani [297] in the framework of time–series analysis and applied in cosmological context by Jones, Coles & Martinez [229]. In this approach, which approximates the $f(\alpha)$ spectrum of singularities to a parabolic function, the resulting spectrum of generalized dimensions is

$$D_q = (D_o - \alpha_o)q + D_o. \tag{150}$$

Here $\alpha_o > D_o$ is the value of the local dimension corresponding to the maximum $f(\alpha)$ value, that is $f(\alpha_o) = D_o$. Although Jones et al. [229] have shown that eq.(150) is quite successful in reproducing the $D_q$ spectrum for the CfA galaxy distribution for moderately low values of $|q|$, nevertheless it also has serious drawbacks. For instance, it gives negative generalized dimensions at positive $q$'s, which imply a very rapid decreasing of the density around strong singularities. In fact, a lognormal distribution is characterized by Kirkwood type correlation functions, as those provided by eq.(27) [96]. Therefore, neglecting the discreteness terms in eq.(141), we get

$$\langle N^q \rangle = n^q \int_v d^3r_1 \ldots \int_v d^3r_q \prod_{i>j}^{\binom{q}{2}} [1 + \xi(r_{ij})]. \tag{151}$$

At sufficiently small scales, the leading term in the integrand is $\xi^{q(q-1)/2} \propto r^{-\gamma q(q-1)/2}$. Thus, in the continuous limit, eq.(151) gives $\langle N^q \rangle \propto r^{3+(q-1)(3-\gamma q/2)}$ and the scaling of the partition function reads

$$Z(q,r) \propto r^{(q-1)(3-\gamma q/2)} \quad \Rightarrow \quad D_q = 3 - \frac{\gamma q}{2}. \tag{152}$$

The resulting dimension spectrum is a linearly decreasing function of $q$ and becomes negative for $q > 6/\gamma$. A comparison of this $D_q$ shape to that detected for the galaxy distribution (see Figure 19)shows that a linearly decreasing $D_q$ is acceptable only for $|q| \lesssim 2$. At large positive multifractal orders, the $D_q$ profile flattens, thus pointing toward a hierarchical pattern for the clustering, in agreement with the results based on correlation analysis.

As a concluding remark, I would like to stress that the reliability of studying the $D_q$ spectrum of generalized dimensions in order to characterize the global texture of the galaxy distribution and compare it to theoretical models. In fact, a single plot, namely the $D_q$ curve, contains a great amount of statistical information, ranging from the scaling properties of higher order correlation functions, implied by the $q > 0$ tail, to the scale dependence of the void probability function, which is strictly related to the definition of Hausdorff dimension, to the counts in the underpopulated cells, that specify the negative order dimensions and could be extremely useful in discriminating between different models. For these reasons, the fractal description of clustering surely represents a powerful tool in order to obtain precise hints about the nature of the non–linear gravitational dynamics and its scale–invariant properties.

# 5 The dynamics of structure formation

In the previous sections I described how the LSS of the Universe appears and how its statistical properties can be characterized. The cosmological relevance of providing an accurate clustering description resides in the fact that, in principle, we should be able to solve the dynamical evolution of the Universe back in time, in order to deduce the initial conditions on the basis of what is today observed. Other than referring to the large scale distribution of galaxies and galaxy clusters, a very efficient way to get information about the primordial Universe is represented by the study of the cosmic microwave background (CMB) anisotropies, which contain the imprint of the initial density inhomogeneities at the outset of the recombination epoch. However, although the continuous refinements of measurements and the recent exciting results obtained by COBE [376, 419], South Pole [168] and MAX [190] experiments have further restricted the set of possible initial conditions, no compelling evidences still exist in favour of one particular model. Data on the anisotropy of the CMB temperature indicate that the presently observed variety of structures should have evolved from extremely small fluctuations, with a characteristic amplitude $\delta \sim 10^{-5}$. If structure formation began after recombination, at a redshift $z \sim 1000$, a successful model for structure formation must be able to provide the observed global texture of the large scale structure, starting from this high degree of isotropy.

Although reconstructing initial conditions from the observed galaxy distribution is likely to be possible at sufficiently large scales, where linear gravitational dynamics still preserves memory of the initial conditions, it becomes extremely difficult when considering the small scales, where both non–linear gravity and dissipative hydrodynamical effects are responsible for the structure formation processes. However, even considering sufficiently large scales, problems arise due to incompleteness of galaxy samples, systematic errors in the estimate of the galaxy distances, or other observational biases. Going in the opposite direction, we can also choose the strategy of fixing initial conditions and let them dynamically evolve. A comparison of the results with the observed LSS should indicate whether the chosen model is viable or not. Cosmological N–body simulations are based on this approach.

In this section I describe the dynamics which determines the evolution of density fluctuations in the framework of the gravitational instability picture. The non–gravitational mechanism based on cosmic explosions [294, 217] is briefly described in §4.3. After writing the complete set of equations for the evolution of density perturbations in a Friedmann background, I discuss their solution in the linear regime. Although the non–linear dynamics can be adequately followed only by resorting to N–body simulations, some approximations haveeally been introduced in order to account for the quasi–linear or mildly non–linear gravity. Also mentioned are analytical approaches to account for partial aspects of the non–linear clustering (the treatment of non–linear clustering through N–body simulations is described in Section 6). Then, I discuss the problem of the initial conditions for LSS evolution, by describing the more important models for the primordial fluctuation spectrum. In this context, it will be emphasized the strict connection between the shape of the power spectrum and the matter content of the Universe. Motivated by the problems displayed by the standard dark matter models, I also discuss alternative scenarios, which violate the random–phase prescription provided by the canonical inflationary model. Finally, the problem of the galaxy formation out of the initial density perturbations is treated. Particular attention is devoted to "biased" models, in order to show how the history of formation of cosmic structures determines not only their inner morphology, but also their large scale clustering properties.

## 5.1 The evolution of density perturbations

Let us assume that the matter content of the Universe is described by a pressureless and self–gravitating Newtonian fluid. Then, the equations that describe its evolution (see, e.g., refs.[247, 308]) are the equation of continuity for mass conservation, Euler's equation for the description of the motion, and the Poisson's equation, that accounts for the Newtonian gravity. Let us choose a system in which $\mathbf{x}$ represents the comoving coordinate and $\mathbf{r} = a(t)\,\mathbf{x}$ the proper coordinate. If $\mathbf{v} = \dot{\mathbf{r}}$ is the physical velocity and $\mathbf{u} = a(t)\,\dot{\mathbf{x}}$ the peculiar velocity, then

$$\mathbf{v} = \dot{a}\mathbf{x} + \mathbf{u}\,, \qquad (153)$$

where the first term on the r.h.s. accounts for the Hubble flow. With such conventions, the dynamical equations for the evolution of density inhomogeneities in a Friedmann background read

$$\begin{aligned}
\frac{\partial \delta}{\partial t} + \nabla_x \cdot \mathbf{u} + \nabla_x \cdot (\delta \mathbf{u}) &= 0 \quad \text{(continuity)}; \\
\frac{\partial \mathbf{u}}{\partial t} + 2H\mathbf{u} + (\mathbf{u} \cdot \nabla_x)\mathbf{u} &= -\frac{\nabla_x \Phi}{a^2} \quad \text{(Euler)}; \\
\nabla_x^2 \Phi &= 4\pi G \bar{\rho} a^2 \delta \quad \text{(Poisson)}\,.
\end{aligned} \qquad (154)$$

Here $\nabla_x$ is the comoving gradient, $\delta(\mathbf{x})$ the density perturbation field, $\Phi(\mathbf{x})$ the potential fluctuations, and $H = \dot{a}/a$ the Hubble parameter. Newtonian theory is adequate to describe the evolution of density fluctuations on scales well inside the Hubble radius $\lambda_{\mathrm{H}} = ct$. However, on scales comparable or exceeding the horizon size, a general–relativistic treatment is required. In this regime, some complications occur due to the ambiguity arising in general relativity when fixing the gauge (i.e., the correspondence between background and physical space–time), in which fluctuation measurements are performed. In this case, a proper gauge–invariant description of fluctuation evolution is required and several prescriptions have been proposed by different authors (see ref.[285] for a recent review on this subject). At the present time, all the scales relevant for clustering studies are are well inside the horizon. However, since perturbation wavelengths scale with the redshift as $\lambda \propto (1+z)^{-1}$ and the horizon as $\lambda_{\mathrm{H}} \propto (1+z)^{-3/2}$, the typical scale of galaxy clustering, $\lambda \sim 10 h^{-1}$ Mpc, crosses the horizon at $z \sim 10^5$, roughly corresponding to the epoch of matter–radiation equality. Thus, a gauge invariant treatment is required when studying the formation of density fluctuations relevant to the present observed clustering. In the following I adopt a purely Newtonian treatment, which is adequate to follow the perturbation evolution after recombination, when structure formation starts.

### 5.1.1 The linear approximation

A particularly simple case occurs when dealing with very small inhomogeneities. This is the case either at very early epochs or at the present time, when considering sufficiently large scales, so that the variance of the galaxy number counts is much less than unity. In this regime, $\delta \ll 1$ and $ut/d_c \ll 1$, $t \sim 1/\sqrt{G\bar\rho}$ being the expansion time–scale and $d_c$ the characteristic coherence length of the fluctuation field. Accordingly, all the non–linear terms in eqs.(154) becomes negligible, so that

$$\frac{\partial^2 \delta}{\partial t^2} + 2H\,\frac{\partial \delta}{\partial t} \;=\; 4\pi G\bar\rho\delta\,. \qquad (155)$$

In the case of a flat Universe with $\Omega_o = 1$, no cosmological constant, and matter dominated expansion, the Friedmann equations (3) give $a(t) \propto t^{2/3}$, and eq.(155) has the general solution

$$\delta(\mathbf{x},t) \;=\; A(\mathbf{x})\,t^{2/3} + B(\mathbf{x})\,t^{-1}\,. \qquad (156)$$

The above expression essentially says that perturbation evolution is given by the superposition of a growing and a decaying mode, that becomes rapidly negligible as expansion goes on. It is interesting to note that $\delta(\mathbf{x})$ grows at the same rate as the scale factor $a(t)$. This is nothing but the consequence of the similarity between the gravitational collapse time–scale, $t_{dyn} \sim 1/\sqrt{G\rho}$, and the expansion time–scale, $t_{exp} \sim 1/\sqrt{G\bar\rho}$, that occurs in the $\delta \ll 1$ case. As a density fluctuation increases, it becomes $t_{dyn} \ll t_{exp}$ and $\delta(\mathbf{x},t)$ starts growing much faster than in the linear regime. This solution (156) is valid both for a $\Omega = 1$ Universe and for a non–flat Universe at sufficiently early times, when $1+z \gg |\Omega_o^{-1} - 1|$ and the spatial curvature can be neglected. If $\Omega < 1$, at later times the cosmic expansion rate increases, so that the fluctuation stops growing and their amplitudes are frozen. Thus, in order to allow observed structures to be formed , a not too low density parameter is required. Detailed calculations give the constraint $\Omega_o h^2 \gtrsim 0.006$, which is largely satisfied by all the estimates of the mean cosmic density, for any reasonable choice of the Hubble parameter. An interesting characteristics of the linear evolution is that the $\delta$ field grows with the same rate at all the points. Therefore, its statistics is left unchanged by linear gravity, apart from the increase of the correlation amplitude.

However, as the typical fluctuation amplitude approaches unity at a given scale, coupling terms in eqs.(154) start playing a role and the linear approximation breaks down. In this case, a rigorous treatment for inhomogeneity evolution has not yet been derived, although different prescriptions have been developed to get hints about non–linear aspects of gravitational clustering. As a direct approach, one can perturbatively expand eqs.(154) in terms of increasing order in $\delta$ and try to solve the system order by order. Attempts in this direction have been performed, at least at the lowest (second) perturbative order, and show some interesting features. While linear equations preserve initial Gaussianity, the inclusion of the lowest order non–linear terms gives rise to a non vanishing skewness of the fluctuation probability distribution [308], that is the signature of non–Gaussian statistics. The corresponding connected correlation functions are shown to follow at this order the hierarchical expression of eq.(28) [161]. This approximation is expected to hold in the $\delta \lesssim 1$ regime, while in the $\delta \gg 1$ regime the convergence of the perturbative expansion series is no longer guaranteed.

### 5.1.2 The Zel'dovich approximation

A simple and elegant approximation to describe the non–linear stage of gravitational evolution has been developed by Zel'dovich [425] (see the review by Shandarin & Zel'dovich [369], for an exhaustive description of the Zel'dovich approximation). In this approach, the initial matter distribution is considered to be homogeneous and collisionless. If the unperturbed (initial) Lagrangian coordinates of the particles are described by $\mathbf{q}$, then the Eulerian coordinates of the particles at the time $t$ are given by

$$\mathbf{r}(\mathbf{q}, t) = a(t) \left[ \mathbf{q} + b(t) \mathbf{s}(\mathbf{q}) \right]. \tag{157}$$

Here $a(t)$ is the cosmic expansion factor and $b(t)$ the growing rate of linear fluctuations. The velocity term $\mathbf{s}(\mathbf{q})$, which provides the particle displacement with respect to the initial (Lagrangian) position, is related to the potential $\Phi_o(\mathbf{q})$ originated by the initially linear fluctuations, according to

$$\mathbf{s}(\mathbf{q}) = \nabla \Phi_o(\mathbf{q}). \tag{158}$$

In order to better visualize the meaning of eq.(157), let us consider a pressureless and viscosity–free, homogeneous medium without any gravitational interaction. For this system, the Eulerian positions $\mathbf{x}$ of the particles at time $t$ are related to the Lagrangian positions $\mathbf{q}$ by the linear relation

$$\mathbf{x}(\mathbf{q}, t) = \mathbf{q} + \mathbf{v}(\mathbf{q}) t, \tag{159}$$

being $\mathbf{v}(\mathbf{q})$ the initial velocity. The above expression is essentially analogous to the Zel'dovich approximation (157), apart from the presence of the $a(t)$ term, which accounts for the background cosmic expansion, and of the $b(t)$ term, which accounts for the presence of gravity, providing a deceleration of particles along the trajectories (actually, $b(t) \propto t^{2/3}$ in a $\Omega = 1$ matter dominated Universe).

Since at $t > 0$ density inhomogeneities are created, mass conservation requires that $\rho(\mathbf{r}, t) \, d\mathbf{r} = \rho_o \, d\mathbf{q}$, so that the density field as a function of Lagrangian coordinates reads

$$\rho(\mathbf{q}, t) = \rho_o \left| \frac{\partial \mathbf{r}}{\partial \mathbf{q}} \right| = \frac{\bar{\rho}}{\left| \delta_{ij} + b(t) \frac{\partial s_i}{\partial q_j} \right|}. \tag{160}$$

Here the *deformation tensor* $\partial s_i/\partial q_j$ accounts for the gravitational evolution of the fluid, while $\bar{\rho} = (a_o/a)^3 \rho_o$ is the mean density at time $t$. At the linear stage, when $b(t)\,\mathbf{s}(\mathbf{q}) \ll 1$, eq.(160) can be approximated by

$$\rho(\mathbf{q},t) \simeq \bar{\rho}\left[1 - b(t)\nabla_q \cdot \mathbf{s}(\mathbf{q})\right], \qquad (161)$$

so that $\bar{\rho}\delta(\mathbf{x}) \simeq -b(t)\,\nabla_q \cdot \mathbf{s}(\mathbf{q})$ and we recover (the growing mode of) the linear solution.

More in general, since the expression (158) for $\mathbf{s}(\mathbf{q})$ makes the deformation tensor a real symmetric matrix, its eigenvectors define a set of three principal (orthogonal) axes. After diagonalization, eq.(160) can be written in terms of its eigenvalues $-\alpha(\mathbf{q})$, $-\beta(\mathbf{q})$ and $-\gamma(\mathbf{q})$, which give the contraction or expansion along the three principal axes:

$$\rho(\mathbf{q},t) = \frac{\bar{\rho}}{[1 - b(t)\,\alpha(\mathbf{q})]\,[1 - b(t)\,\beta(\mathbf{q})]\,[1 - b(t)\,\gamma(\mathbf{q})]}. \qquad (162)$$

If the eigenvalues are ordered in such a way that $\alpha(\mathbf{q}) \geq \beta(\mathbf{q}) \geq \gamma(\mathbf{q})$, then, as evolution increases $b(t)$, the first singularity in eq.(162) occurs in correspondence of the Lagrangian coordinate $\mathbf{q}_1$, where $\alpha$ attains its maximum positive value $\alpha_{max}$, at the time $t_1$ such that $b(t_1) = \alpha_{max}^{-1}$. This corresponds to the formation of a pancake (sheet–like structure) by contraction along one of the principal axes. For this reason, Zel'dovich [425] argued that pancakes are the first structures formed by gravitational clustering. Other structures like filaments and knots come from simultaneous contractions along two and three axes, respectively. Doroshkevich [130] evaluated the probability distribution for the three eigenvalues in the case of a Gaussian random field and concluded that simultaneous vanishing of more than one of them is less probable. Thus, in this scenario, pancakes are the dominant features arising from the first stages of non–linear gravitational clustering.

The Zel'dovich approximation predicts the first non–linear structure to arise in correspondence of the high peaks of the $\alpha(\mathbf{q})$ field and represents a significant step forward with respect to linear theory. For this reason, the Zel'dovich approach is very well suited in many studies of gravitational clustering. In particular, it is widely employed in the realization of cosmological N–body simulations, where its ability to better follow mildly non–linear clustering than the linear approximation permits one to fix much more accurately the initial conditions. Furthermore, it has been also used to dynamically reconstruct the potential and density field traced by large scale galaxy peculiar motions (see, e.g., ref.[119] and references therein). Laso, we have shown that it provides a reliable tool to realize large sets of simulations of cluster distribution, which closely reproduce results from "exact" N–body simulations [59].

However, within the Zel'dovich prescription, after a pancake forms in correspondence of crossing of particle orbits, such particles continue travelling along straight lines, according to eq.(157). Vice versa, in the framework of a realistic description of gravitational dynamics, we expect that the potential wells, that correspond to non–linear structures, should be able to retain particles and to accrete from surrounding regions. This is exactly what comes out by comparing simulations purely based on the Zel'dovich approximation to N–body experiments based on exactly solving the dynamics of eq.(154) (see Figure 21). It is clear that the failure of this approximation is expected to be more pronounced for spectra having a large amount of power at small scales, where non–linear structures rapidly develop and the typical time for orbit crossing is very small.

### 5.1.3 The adhesion approximation

In order to overcome the smearing of pancakes arising in the framework of the Zel'dovich approach, an alternative approximation has been proposed, that is based on the idea of gravitational sticking of particles, occurring after their orbits cross. The idea of sticking in a collisionless medium has been originally introduced in the cosmological context by Gurbatov & Saichev [193] and Gurbatov et al. [194], with the aim of avoiding some undesirable features of the Zel'dovich approximation. The fluctuation evolution with sticking is described by the Burgers equation [77, 78], that is well known in hydrodynamical studies of viscous fluids.

Starting from the approximate solution given by eq.(157), let us evaluate the velocity and the gravitational acceleration fields,

$$\mathbf{u} = \dot{\mathbf{x}} = a\dot{b}\mathbf{s}(\mathbf{q}) \quad ; \quad \dot{\mathbf{u}} = (\dot{a}\dot{b} + a\ddot{b})\mathbf{s}(\mathbf{q}) = \left(H + \frac{\ddot{b}}{\dot{b}}\right)\mathbf{u}. \tag{163}$$

Taking $b(t)$ as the new time variable and introducing the comoving density $\eta = a^3\rho$ (here we fix the scale factor to be unity at the present time), the first two of eqs.(154) becomes

$$\frac{\partial \eta}{\partial b} + \nabla_x \cdot (\eta \mathbf{s}) = 0$$
$$\frac{\partial \mathbf{s}}{\partial b} + (\mathbf{s} \cdot \nabla_x)\mathbf{s} = 0. \tag{164}$$

Consistently, the Zel'dovich approximation $\mathbf{x} = \mathbf{q} + b\mathbf{s}(\mathbf{q})$ is the solution of the second of the above equations, the first equation providing the continuity condition. A modification of the dynamics described by such equations to account for the effect of sticking is obtained by introducing a viscosity term, so that

$$\frac{\partial \mathbf{s}}{\partial b} + \nabla_x \cdot (\eta \mathbf{s}) = \nu^2 \nabla_x^2 \mathbf{s}. \tag{165}$$

The above equation is very well known in studies in fluidodynamics as the Burgers equation [77, 78]. For any non–vanishing $\nu$ value, no matter how small it is, it prevents the penetration of one particle stream into another and avoids the orbit crossing occurring in the Zel'dovich approximation. In the framework of Burgers dynamics, the forming structures are thin sheets, whose thickness depends on the viscosity parameter $\nu$. Since viscosity is relevant on a scale–length $\nu^{1/2}$ (below which dissipation occurs and velocity gradients are erased), we expect that this should also be the typical thickness of arising non–linear structures. For this reason, the Burgers dynamics in the $\nu \to 0$ limit provides structures of vanishing thickness. The resulting clustering is different from that provided by the Zel'dovich solution, which amounts to fix $\nu = 0$ in eq.(165). While the Burgers dynamics gives rise to extremely thin pancakes, which correspond to the formation of shocks, the Zel'dovich prescription generates more diffuse structures, since particles continue travelling after crossing of their orbits.

The application of the Burgers equation in the gravitational clustering study is usually called the *adhesion* approximation. Many attempts have been devoted in recent years to investigate the dynamics of LSS formation within this approach, by means of both analytical [195, 131] and numerical [412, 289] treatments. The relevance of the Burgers equation lies in the fact that, despite it is manifestly non–linear, it can be linearized in a straightforward way and its analytical solution can be explicitly evaluated. This permits one to characterize, at least partially, the statistics of the subsequent clustering [131]. In order to linearize eq.(165), let us consider the vectorial Hopf–Cole transformation for the velocity potential

$$s_i(\mathbf{q}, t) = -2\nu \frac{\partial}{\partial x_i} \log U(\mathbf{x}, b). \tag{166}$$

After substituting into the Burgers equation, we get

$$\frac{\partial U}{\partial b} = \nu \nabla_x^2 U, \tag{167}$$

which is the usual linear diffusion (heat) equation. Its solution for the velocity potential reads

$$\mathbf{s}(\mathbf{x}, b) = \frac{\int \frac{\mathbf{x} - \mathbf{q}}{b} \exp\left[-\frac{G(\mathbf{x}, \mathbf{q}, b)}{2\nu}\right] d\mathbf{q}}{\int \exp\left[-\frac{G(\mathbf{x}, \mathbf{q}, b)}{2\nu}\right] d\mathbf{q}}, \tag{168}$$

where

$$G(\mathbf{x}, \mathbf{q}, b) = \Phi_o(\mathbf{q}) + \frac{(\mathbf{x} - \mathbf{q})^2}{2b} \tag{169}$$

and $\Phi_o(\mathbf{q})$ is the potential of the initial field. Although a suitable geometrical method has been devised to study the structure of the potential velocity field given by eq.(168) (see, e.g., ref.[369]), a detailed characterization of the gravitational clustering described by Burgers dynamics has been made possible with the realization of numerical simulations based on the adhesion approximation (see, e.g., refs.[289, 412]). The reason for the great deal of attention to the adhesion approach lies essentially in the fact that the availability of the analytical solution for $\mathbf{s}(\mathbf{q})$ greatly reduces the computational costs with respect to usual N–body codes, while avoiding some of the limitations of the Zel'dovich approach. The trick of putting in by hand the viscous term in the r.h.s. of eq.(165) avoids the pancake smearing and simulates non–linear gravitational clustering as a sort of dissipative smoothing process. Although this could give a reasonable approximation at scales where not so much power is present in the fluctuation spectrum, it fails to describe the strong clustering regime, where "exact" N–body simulations show a greater variety of structures, that can not be generated only on the ground of a smoothing process.

Several other approximations have been proposed in order to go beyond the simple linear treatment (see, e.g., refs. [197, 76, 270, 17, 72]). A common characteristic of these approximations resides in the fact that, while they keep the essential features of quasi–linear clustering, they seems rather inadequate to account for the strongly non–linear gravitational dynamics at small scales. In the following, we discuss other approximations, that deal with the case of fully non–linear gravitational dynamics.

### 5.1.4 Self–similar clustering

The detection of the power law shape for the correlation functions led to the suggestion that gravitational clustering should proceed in a self–similar way. Although eqs.(154), which describe the evolution of density inhomogeneities, do not introduce characteristic scales, nevertheless these can be present in the Friedmann background or in the initial fluctuation spectrum. For this reason, self–similar clustering is based on the assumption that *a)* the Universe is spatially flat, with $\Omega_o = 1$, in order to have no characteristic time– or length–scales in the background metric, and *b)* the initial power spectrum is scale–free,

$$P(k) = Ak^n. \qquad (170)$$

In the above expression, the spectral index ranges in the interval $-1 < n < 1$, in order to allow for convergence of the peculiar velocity field at very small and very large scales. Actually, by introducing a small scale cutoff, smaller $n$ values are permitted. In any case, $n > -3$ is required in order to allow for gravitational clustering to proceed in a hierarchical way (see §3.2). Note, however, that cosmological spectra are expected to have characteristic scales (see §5.2, below). In this case, the argument of self–similar clustering still applies, at least if the effective spectral index takes a roughly constant value in a sufficiently large scale range.

In the linear regime, the variance at the comoving scale $x$ (see eq.[55]) for a matter-dominated expansion reads

$$\sigma_x^2(t) = x^{-(n+3)} t^{4/3}. \qquad (171)$$

If $x$ represents the comoving scale at which unity variance is attained, then it scales with time as $x \propto t^{-3(n+3)/4}$, and larger and larger scales go non–linear with time. Accordingly, the physical non–linearity scale is $r = a(t) x \propto x^{(5+n)/2}$. Moreover, the requirement for a non–linear lump of size $r$ to be stable is that its characteristic dynamical time–scale, $t_{dyn} = (G\rho)^{-1/2}$, should be equal to $t$. Thus, the average density inside non–linear lumps scales as $\rho \propto t^{-2} \propto x^{-(9+3n)/2}$ and the density–radius relation, $\rho \propto r^{-\gamma}$, has the logarithmic slope

$$\gamma = \frac{9 + 3n}{5 + n}. \qquad (172)$$

The above result was derived for the first time by Peebles [305] and expresses the power law shape of the 2–point function as a function of the primordial spectral index. According to this model, the value $\gamma = 1.8$, relevant to galaxy clustering, is produced by $n = 0$, which correspond to a white–noise initial spectrum. Starting from different scale–invariant initial conditions, Efstathiou et al. [141] tested the validity of the Peebles' scaling argument against evolved N–body simulations. They found that eq.(172) agrees quite well with their simulations only in the strongly non–linear, $\xi \gtrsim 100$, regime, while in the range $1 \lesssim \xi \lesssim 100$ the 2–point function declines more steeply than expected. This result leads to a quite paradoxical conclusion; since the observed $\xi(r)$ for galaxies is a unique power law up to scales ($\sim 10h^{-1}$ Mpc) greater than the correlation length, the assumption of an initial self–similar spectrum gives rise to characteristic scales in the evolved $\xi(r)$. Vice versa, the observed scale invariance of clustering forces one to conclude that the initial spectrum should have a scale dependence tuned in such a way that self–similarity is produced at the present.

It is however to be observed that the above scaling argument is based on assumptions, whose validity is far from being proved. First of all, linear theory is used to establish the time dependence of the size of lumps that reach the non–linear phase. This amounts to assuming that the dynamical range for the transition between linear and non–linear gravitational clustering is very narrow, so that linear theory can be safely applied to establish the initial conditions for non–linear evolution. Although this could be a reasonable approximation for $n \gtrsim 0$ spectra, which have strong small scale power and rapidly decline at large scales, it is expected not to be true for less steep spectra, which develop non–linear clustering over a broad scale range. In this case, some aspects of gravitational evolution are present, which are not accounted for in the above simplified picture. As an example, if significant power is present at large scales, the forming clumps start to accrete material from surrounding regions, so as to steepen the density profile of the resulting structures. On the other hand, the secondary accretion will not be very important if the characteristic amplitude of large scale fluctuations is very small. Hoffman & Shaham [211] evaluated the density profile of non–linear structures, by allowing initial clumps to accrete matter from the surrounding, according to the prescription of the secondary infall paradigm [176, 191]. In this way, they derived the logarithmic slope $\gamma$ of the density–radius relation to be

$$\gamma = \begin{cases} \frac{9+3n}{4+n} & -3 < n \leq -1, \\ 2 & -1 \leq n < 1. \end{cases} \quad (173)$$

A comparison of this expression with eq.(172) shows that there is no remarkable difference between the two expressions for $0 \lesssim n \lesssim 1$, in accordance with the expectation that secondary accretion is not very important in the case of steep spectra. Vice versa, for $n \lesssim 0$ eq.(173) systematically provides a larger slope for the density–radius relation, which significantly detaches from $\gamma \simeq 2$ only at $n \lesssim -1.5$. In this picture, the value $\gamma \sim 2$, expected on the ground of galaxy clustering and flatness of spiral rotation curves, is a much more natural outcome of non–linear gravitational clustering, than expected on the ground of the Peebles' scaling argument.

A similar conclusion has been also obtained by Saslaw [354], following a completely different approach to non–linear gravitational dynamics. In the framework of a thermodynamical approach, he suggested that $\gamma \sim 2$ is a sort of "attracting" solution for fully virialized structures, quite independently of initial conditions. It is remarkable to note that all these indications toward a unique density scaling relation produced by non–linear gravitational evolution confirm the earlier suggestion that Fournier d'Albe gave already in 1907 [155]. He observed that the requirement that the virial velocity $v = \sqrt{M/R}$ in a cluster of size $R$ does not diverge, but converges to a finite constant value, leads to the density–scale relation $\rho \propto R^{-\gamma}$ with $\gamma = 2$.

Although the above discussed descriptions of non–linear clustering evolution provide hints about the origin of the observed 2–point correlation function, nevertheless they say nothing about the scaling of higher–order correlations. A comprehensive treatment to include the whole hierarchy of $n$–point functions requires to solve the non–linear dynamics for the probability distribution function, which generates correlations of any order.

A classical approach in this direction is represented by the BBGKY (Bogoliubov–Born–Green–Kirkwood–Yvon) equations for a collisionless and self–gravitating fluid. Since a complete description of the analysis of the BBGKY equations is rather long, here we will present only a brief sketch, while comprehensive treatments for applications in the cosmological context can be found in refs.[308, 160, 199]. In order to describe the statistics of the fluid particles, let $f(\mathbf{x}, \mathbf{p}, t)$ be the phase–space distribution function, which depends on the comoving coordinate $\mathbf{x}$, on the moment $\mathbf{p} = ma^2\dot{\mathbf{x}}$ ($m$ being the mass of the single particle) and on time $t$. Accordingly, the fluid density reads

$$\rho(\mathbf{x}) = \frac{m}{a^3} \int d\mathbf{p}\, f(\mathbf{x}, \mathbf{p}, t), \qquad (174)$$

while the statistics of the system is described by the phase–space correlation functions

$$C^{(n)}_{1,\ldots,n} = \langle f_1 \ldots f_n \rangle. \qquad (175)$$

The evolution of the distribution function is described by Liouville's equation

$$\frac{\partial f}{\partial t} + \frac{p^\alpha}{ma^2} \frac{\partial f}{\partial x^\alpha} - m \frac{\partial \Phi}{\partial x_\alpha} \frac{\partial f}{\partial p^\alpha} = 0, \qquad (176)$$

where $\Phi$ is the gravitational potential related to the density field according to

$$\Phi(\mathbf{x}) = -G a^2 \int d\mathbf{x}\, \frac{\rho(\mathbf{x}) - \bar{\rho}}{|\mathbf{x}' - \mathbf{x}|}. \qquad (177)$$

From eq.(176), an infinite sequence of equations can be obtained by evaluating its moments, that is, by solving the equation for the time–evolution of the phase–space correlation functions. The sequence of all such equations represents the hierarchy of BBGKY equations, which are obtained through a sort of moment expansion of the Liouville's equation (176). A characteristic of this hierarchy is represented by the fact that the solution of the equation of order $n$ depends on the moment of order $n + 1$. This can be easily understood, by observing that the $\Phi$ term includes the integral of the distribution function, through eq.(174). Therefore, at the order $n$, this term turns into the integral of the $(n+1)$-th order correlation. Because of this peculiarity, suitable assumptions are needed to close the hierarchy and provide a consistent solution.

Several attempts have been devoted to solve the BBGKY equations, although all such solutions rely on approximations, which are expected to hold only in the strongly non–linear regime. Davis & Peebles [115] observed that, in the case of a flat Universe with scale–free initial spectrum, the BBGKY equations admit a self–similar solution, with spatial $n$–point correlation functions of the type

$$\kappa_n \propto r^{(n-1)\gamma}, \tag{178}$$

As usual, $\gamma$ is the slope of the 2–point function, so that eq. (178) is consistent with the hierarchical expression for the $n$–point function as provided by eq.(28). Both Fry [159, 160] and Hamilton [199] concluded that hierarchical correlations represent the solution of the BBGKY hierarchy in the strongly non–linear regime, although they found different sequences of hierarchical coefficients $Q_n$.

## 5.2 The spectrum of primordial fluctuations

In the description of the evolution of density inhomogeneities, the simplifying assumption that the initial power spectrum has the power law expression $P(k) \propto k^n$ is often done. Although the scale–free Zel'dovich spectrum ($n = 1$) is expected on the grounds of the classical inflationary scenario, nevertheless distortions of its shape should arise during the subsequent phases of cosmic expansion, and characteristic scales are imprinted on the form of $P(k)$ at the onset of the structure formation. Since the amount of such distortions and the scales at which they occur are strictly related to the nature of the fluctuations and to the matter content of the Universe, their knowledge becomes of crucial relevance in order to fix the initial conditions for the galaxy formation process.

Theoretical models for the determination of the power spectrum starts from the assumption of a primordial $P_{pr}(k)$ at a sufficiently high redshift, $z_{pr} \gg z_{eq}$ [$z_{eq} \simeq 4.2 \times 10^4 \, (\Omega h^2)$ is the redshift of the epoch of matter–radiation equality]. The usual choice $P_{pr}(k) = Ak$ corresponds to the Harrison–Zel'dovich spectrum. Due to the evolution of density perturbations, the slope of the power spectrum is left unchanged at wavelengths $\lambda \gg \lambda_{eq} \sim ct_{eq}$, that exceed the horizon size at $t_{eq}$. On the contrary, the shape of the spectrum at $\lambda \ll \lambda_{eq}$ crucially depends on the nature of the matter which dominates the expansion. In order to account for these effects, the post–recombination spectrum is usually written as

$$P(k) = T^2(k) P_{pr}(k), \tag{179}$$

where the *transmission factor* $T(k)$ conveys all the informations about the pre–recombination evolution and the nature of the matter content.

Before starting with the discussion about the different scenarios for the origin of the fluctuation spectrum, I mention an important distinction between two different kinds of primeval fluctuations, namely adiabatic (curvature) and isothermal (isocurvature) fluctuations. Adiabatic perturbations correspond to fluctuations in the energy density and both matter and radiation components are equally involved in such perturbations. The name adiabatic derives from the fact that the number density of any species relative to the entropy density is constant. On the contrary, isothermal fluctuations do not correspond to perturbations in the energy density, rather they are originated by variations of the local equation of state. Such fluctuations are called isothermal since temperature fluctuations are suppressed with respect to matter fluctuations by a factor $\sim \rho_m/\rho_r$, so that during the radiation–dominated expansion any $\delta T/T$ becomes negligible. Any difference between adiabatic and isothermal fluctuations makes sense only on super–horizon scales. Vice versa, for perturbations that are well inside the horizon, causal microphysical processes give rise to a local redistribution of the energy density, due to the presence of pressure gradients in isothermal fluctuations. As a consequence, the adiabatic condition is finally attained and any distinction is no longer important. However, at sufficiently early times, presently observable scales are outside the horizon and the above distinction becomes crucial. As far as the origin of adiabatic fluctuations is concerned, they can either be assumed to be imprinted in the initial conditions or can be generated by any mechanism which is able to push sub–horizon to super–horizon scales. This is the case for the inflationary expansion, that, indeed, is considered the classical mechanism to generate adiabatic fluctuations. On the other hand, isothermal perturbations require microphysical mechanisms to be generated, which cannot transport energy on super–horizon scales. Although suitable models have also been proposed to generate isothermal perturbations from inflation, special initial conditions are always required, while adiabatic perturbations comes out much more naturally [135, 243]. Stringent constraints about the nature of primordial fluctuations have been recently provided by COBE observations. A comparison of the temperature fluctuations predicted by a variety of isocurvature models [212] with the detected CMB temperature anisotropy seems to rule out all such models [419]. For these reasons, in the following I will only describe the evolution of adiabatic density perturbations.

### 5.2.1 The evolution of baryonic fluctuations

In order to follow the evolution of fluctuations of baryonic matter, let us consider the second of eqs.(154), with the inclusion of a pressure term due to the dissipative nature of the baryons:

$$\frac{\partial \mathbf{v}}{\partial t} + (\mathbf{v} \cdot \nabla)\mathbf{v} + \frac{1}{\bar{\rho}}\nabla p + \nabla \Phi \;=\; 0\,. \tag{180}$$

Accordingly, the linearized equation for the evolution of the $\delta$ field in Fourier space reads

$$\frac{\partial^2 \tilde{\delta}_{\mathbf{k}}}{\partial t^2} + 2H \frac{\partial \tilde{\delta}_{\mathbf{k}}}{\partial t} + \left( \frac{v_s^2 k^2}{a^2} - 4\pi G \bar{\rho} \right) \tilde{\delta}_{\mathbf{k}} = 0 \,. \tag{181}$$

Here

$$v_s = \left( \frac{\partial p}{\partial \rho} \right)^{1/2}_{adiabatic} \tag{182}$$

is the adiabatic sound speed in a medium with equation of state $p = p(\rho)$. According to eq.(181), the critical Jeans wavelength,

$$\lambda_J = v_s \left( \frac{\pi}{G \bar{\rho}} \right)^{1/2} \,, \tag{183}$$

discriminates between two different regimes for the perturbation evolution. For $\lambda > \lambda_J$ the pressure contribution can be neglected and the linear solution of eq.(156) is recovered. Vice versa, for $\lambda < \lambda_J$ the gravitational term becomes negligible, the perturbation is pressure–supported and the solution oscillates like an acoustic wave. Thus, while fluctuations on a scale greater than the Jeans length are not pressure–supported and are able to grow by gravity, at scales below $\lambda_J$ they behave like oscillating sound waves.

If $\rho_b$ is the average baryon density, we can define a baryon Jeans mass scale,

$$M_J = \frac{2}{3} \pi \rho_b \lambda_J^3 \,, \tag{184}$$

which is the mass of the smallest baryonic fluctuation that is able to grow. Before recombination, at $z_{rec} \simeq 10^3$, matter and radiation are tightly coupled by Thomson scattering. In this regime they behave like a single fluid with

$$v_s = \frac{c}{\sqrt{3}} \left( \frac{3}{4} \frac{\rho_m}{\rho_r} + 1 \right)^{-1/2} \,. \tag{185}$$

Since matter–radiation equality occurs at $z_{eq} = 4.2 \times 10^4 (\Omega h^2)$, the Jeans mass just before recombination is

$$M_J \simeq 10^{17} (\Omega_o h^2)^{-2} M_\odot \,, \tag{186}$$

of the same order of the mass of a supercluster. After recombination, however, photons are no longer coupled to matter, so that the equation of state rapidly changes and the baryonic component behaves like a monoatomic gas with

$$v_s = \left( \frac{5 k_B T}{3 m_p} \right)^{1/2} \,, \tag{187}$$

$m_p$ being the proton mass. At the recombination temperature $T_{rec} \simeq 3000\,K$, it corresponds to a Jeans mass

$$M_J \simeq 1.3 \times 10^6 (\Omega_o h^2)^{-1/2} M_\odot\,. \tag{188}$$

Thus, although before recombination the Jeans mass involves scales of superclusters, after matter and radiation decouple it drops by several orders of magnitude to the value of the mass of globular clusters, and fluctuations on small scales are able to start growing again.

It is worth comparing the Jeans mass before recombination with the baryon mass contained inside the Hubble radius $M_{H,b} = (4\pi/3)\rho_b(ct^3)$. According to eq.(185), it is

$$\frac{M_J}{M_H} \simeq 26 \left(\frac{3}{4}\frac{\rho_m}{\rho_r} + 1\right)^{-3/2}, \tag{189}$$

so that, until radiation dominates, the Jeans mass exceeds the mass inside the horizon and all the subhorizon fluctuations are constrained not to grow.

A further characteristic scale, which enters in the spectrum of baryon fluctuations, is due to the collisional damping occurring just before recombination. As recombination is approached, the coupling between radiation and baryons becomes no longer perfect and the photon mean free path starts increasing. Thus, photons can diffuse more easily from overdensities carrying with them matter, to which they are however still quite tightly coupled. The final effect is to damp fluctuations below the scale which corresponds to the distance travelled by a photon in an expansion time–scale. This is known as Silk damping [375] and an accurate evaluation of the smoothing mass scale in the post–recombination baryon spectrum [143] gives

$$M_D \simeq 2 \times 10^{12} (\Omega_o/\Omega_b)^{3/2} (\Omega_o h^2)^{-5/4} M_\odot\,. \tag{190}$$

The Silk damping increases by several orders of magnitude the mass–scale of the smallest fluctuation, which starts growing after recombination, smaller scale perturbations being heavily suppressed.

An accurate estimate of the transmission factor for the baryon fluctuation spectrum was made by Peebles [309], and the result is shown in Figure 22. The severe small scale suppression due to Silk damping is apparent. The oscillatory behaviour is due to the phases of the Jeans oscillation of the single mode at recombination. Although the simplicity of a purely baryonic model is rather attractive, nevertheless it suffers for a number of serious problems, which make it extremely unlikely. Even without referring to the difficulty of reconciling the predictions based on primordial nucleosynthesis, $\Omega_b \simeq 0.05$ [418], with both dynamical estimates of the mean cosmic density and the inflationary prejudice $\Omega_o = 1$, the baryonic spectrum gives too large fluctuations at the scale of $10$–$20 h^{-1}$ Mpc, with respect to what observed for the galaxy distribution. Even more, a purely baryonic model is ruled out since it predicts too high CMB temperature fluctuations with respect to current detections and upper limits [341, 376].

Peebles [313] suggested that agreement with the upper limits on the CMB anisotropy could be attained if we allow for reionization at a redshift substantially smaller than that of standard recombination ($z_{rec} \sim 1000$). In this case, a substantial reduction of the small scale anisotropies would be provided. In the framework of his "minimal" baryonic model, Peebles suggested an open purely baryonic Universe, in which an early formation of cosmic structures should provide the source for the ionizing radiation. Although this mechanism erases primary CMB temperature fluctuations, nevertheless secondary fluctuations are generated because of peculiar motions of electrons at the time at which the optical depth attains unity (see, e.g., refs.[407, 134]). Again, the resulting anisotropies turn out to severely constraint the purely baryonic model.

### 5.2.2 Non–baryonic models

A fundamental property of non–baryonic dark matter is that it is not coupled to radiation, at least at the epochs relevant for the origin of the primordial fluctuation spectrum. For this reason, no dissipative Silk damping occurs. However, a non–dissipative damping of fluctuations is present in any case, due to free-streaming of dark matter particles. In fact, until such particles are relativistic, they are able to freely cross the horizon within the Hubble time, thus washing out all the fluctuations below the horizon scale. This effect is no longer important when the temperature of the Universe drops below the mass of the DM particles and they become non relativistic. The size of the horizon at this epoch fixes the smallest scale of the fluctuations surviving free-streaming damping. Thus, a crucial parameter to establish the shape of the fluctuation spectrum in a DM model is the velocity distribution of the constituent particles.

The importance of following the evolution of the DM spectrum lies in the fact that it determines after recombination the spectrum of fluctuations of ordinary baryonic matter, so as to provide the seeds where dissipative processes occur and galaxy formation takes place. In fact, soon after recombination the Jeans mass for the baryonic component drops to a very small value, according to eq.(188). As a consequence, baryonic fluctuations starts growing again by gravitational instability, until their amplitude matches that of the non–baryonic DM perturbations.

#### The HDM spectrum

In the hot dark matter (HDM) model the mass contained inside the horizon when constituent particles become non relativistic is much larger than the typical mass of a galaxy ($\gg 10^{11} M_\odot$), so that particles with a low mass are required. A natural candidate for HDM constituent is the massive neutrino. If a neutrino species has non vanishing mass, then its contribution to the mean matter density is

$$\Omega_\nu \simeq \left(\frac{m_\nu}{100 eV}\right) h^{-2}, \qquad (191)$$

where $m_\nu$ is expressed in $eV$. Although in the classical version of the Standard Model for the electroweak interaction the neutrino is considered massless, nevertheless there exists no funda-mental reason which fixes $m_\nu = 0$. Vice versa, several theoretical models have been proposed to generate a non–vanishing neutrino mass [424, 420, 172]. The possibility of HDM models with massive neutrinos became very popular at the beginning of the 80s, after Lyubimov et al. [254] claimed the discovery of a non–vanishing mass, $m_{\nu_e} \simeq 30\,eV$, for the electron neutrino, which provides the closure density by taking $h = 0.55$ for the Hubble parameter. Although this result has been not confirmed by subsequent experiments, nevertheless the possibility of massive neutrinos received great attention due to its deep implications both in astrophysics and in elementary particle physics.

For a neutrino of mass $m_\nu$, the redshift at which it becomes non–relativistic is $z_\nu \simeq 6 \times 10^4\,(m_\nu/30\,eV)$, which corresponds to

$$M_{\nu,\mathrm{H}} \simeq 2 \times 10^{15} \left(\frac{m_\nu}{30\,eV}\right) M_\odot \qquad (192)$$

for the mass contained inside the horizon at that epoch, which is also the smallest mass scale surviving free–streaming. Numerical calculations carried out by Bond & Szalay [49] give the transmission function

$$T(k) = 10^{-(k/k_\nu)^{1.5}} \qquad ; \qquad k_\nu \simeq 0.4\,\Omega_o\,h^2\,\mathrm{Mpc}^{-2}\,, \qquad (193)$$

which suppresses all the fluctuation modes at wavelengths $\lambda < \lambda_\nu = 2\pi/k_\nu \simeq 40\,(m_\nu/30\,eV)^{-1}$ Mpc. In Figure 23 I show a log–log plot of $k^{3/2}|\delta(k)|$, which represents the typical mass fluctuation $(\delta\rho/\rho)_\lambda$ at the scale $\lambda = 2\pi/k$, versus the fluctuation wavelength. The deficit of power at scales below $\sim 10\,(\Omega_o h^2)^{-1}$ Mpc, due to neutrino free–streaming, is apparent. In the HDM scenario, the smallest fluctuations surviving recombination are roughly on the same scale as large galaxy clusters. Accordingly, structure formations proceeds in a "top–down" way; first large pancakes of mass $\sim 10^{15} M_\odot$ form, while galaxies originate later via fragmentation of structures at larger scales. Numerical simulations of structure formation in HDM dominated Universe have been done [416] and show the development of cellular structures, which are promisingly similar to those displayed by the redshift galaxy surveys. Big voids form on scales comparable to the characteristic scale $\lambda_\nu$, which are surrounded by galaxies and clusters forming at the intersection between three of such cells. Unfortunately, the agreement with the observed galaxy distribution is only apparent. In fact, since the characteristic size of the earliest forming structures is $\lambda_\nu$, the variance of the matter distribution at this scale should be around one. This is quite difficult to reconcile with the much smaller correlation length displayed by the galaxy distribution, $r_o \simeq 5 h^{-1}$ Mpc. In order to alleviate this problem, one can either assume that galaxy formation occurred very recently (at $z \lesssim 1$) so as to give no time for structures to become overclustered, or invoke some mechanism to reduce the galaxy clustering with respect to that of the underlying matter. It is however clear that, whatever way out we choose, additional problems arise. A too recent galaxy formation seems to be quite difficult to reconcile with the detection of high redshift ($z \gtrsim 3$) quasars. It is however not clear to what extent quasars can be considered progenitors of all the galaxies. If this were not the case, then the criticism of galaxy formation timing for HDM models would not be so stringent (see, e.g., refs.[274, 44]). Also requiring that galaxies are less clustered than matter is at variance with the expectation that dissipative galaxy formation should preferentially occur in the deep wells of the gravitational potential field, which would give an increase of their correlation with respect to the DM distribution. In addition, current upper limits on the CMB anisotropy at scales of some arcminutes are dangerously near to the HDM predictions. Hopefully, in the near future more precise observations will give the final word about the viability of the standard HDM model. It is however clear that, if a neutrino species were discovered to have a non–vanishing mass $m_\nu \simeq 30\,eV$, we are obliged to consider HDM as responsible for structure formation and try to overcome in some way all the above difficulties. To this purpose, variations of the standard HDM scenario have been proposed. As an example, Villumsen et al. [405] suggested that, if fluctuation evolution is driven by the presence of randomly distributed seeds (such as primeval black holes or non–topological solitons), then a considerable amount of small–scale power is added to the HDM spectrum and some of its undesirable features are avoided.

**The CDM spectrum**

The dark matter content of the Universe is said to be cold if particles become non–relativistic at sufficiently early times, so that the mass contained within the horizon at that time is much smaller than the typical galaxy mass. Thus, in the cold dark matter (CDM) scenario the free–streaming cut–off scale is too small to be of any cosmological relevance. The low velocity required for CDM particles can arise for two different reasons. Firstly, the particle mass is so large that they become non relativistic at a high temperature. This cannot be the case for massive neutrinos, for which a mass $\gtrsim 100\,eV$ give an exceedingly high contribution to the density parameter (see eq.[191]). However, supersymmetric theories provide a large variety of exotic CDM candidates, such as photinos, gravitinos, or other weakly interacting massive particles (WIMPs) with masses above $1\,GeV$. Secondly, there can be particles, like axions, that never were in thermal equilibrium, so to have a very low thermal velocity, despite their small mass ($\sim 10^{-5}\,eV$). See, e.g., ref.[149] for a review of CDM candidates.

Although neither free–streaming nor Silk damping introduce characteristic scales for CDM fluctuations, a distortion of the spectrum is however generated by the Meszaros effect [279], which suppresses the growth of CDM fluctuations matter which cross the horizon before non–relativistic matter start dominating. In order to see how this happens, let us consider the equation for the evolution of non–relativistic matter fluctuations in a relativistic background:

$$\frac{d^2\delta}{dt^2} + 2H\,\frac{d\delta}{dt} - 4\pi G \bar\rho_m \delta = 0\,. \qquad (194)$$

Here the relativistic background enters only in determining the cosmic expansion rate $a(t)$. By introducing the new time variable $\tau = \bar\rho_m/\bar\rho_r$ ($\tau \propto a$ since $\bar\rho_m \propto a^{-3}$ and $\bar\rho_r \propto a^{-4}$), eq.(194)

can be rewritten as

$$\frac{d^2\delta}{d\tau^2} + \frac{2+3\tau}{2\tau(1+\tau)}\frac{d\delta}{d\tau} - \frac{3}{2}\frac{\delta}{\tau(1+\tau)} = 0 \, . \tag{195}$$

The growing–mode solution of the above equation turns out to be $\delta \propto 1 + \frac{3}{2}\tau$. Thus, during radiation domination ($\tau \ll 1$), the fluctuation amplitude is frozen. Only when non–relativistic matter dominates ($\tau \gg 1$) the matter fluctuations start growing as $\delta \propto a$, as expected on the grounds of linear evolution. In order to evaluate the resulting distortion of the spectrum, let $\tilde{\delta}_k(t_i)$ be the amplitude of the fluctuation mode with wavenumber $k$ at some initial time $t_i$ before matter-radiation equality, and suppose that it crosses the horizon at $t_H$, also before equality. During this period the amplitude grows by a factor $t_H/t_i$. Since $t_H \propto k^{-2}$ during radiation domination, the amplitude of the perturbation after horizon crossing is frozen at the value $\tilde{\delta}_k(t_i) t_i^{-1} k^{-2}$ until matter starts dominating. Vice versa, fluctuations outside the horizon continue to grow according to linear theory so that no distortion of the spectrum occurs at such scales. The characteristic scale at which we expect a feature in the spectrum is that corresponding to the horizon size at $t_{eq}$,

$$\lambda_{eq} \sim 10\,(\Omega_o h^2)^{-1}\,\mathrm{Mpc}\, . \tag{196}$$

If $P(k) \propto k^n$ was the primordial spectrum, its shape after $t_{eq}$ will be preserved at scales larger than $\lambda_{eq}$, while for $\lambda \ll \lambda_{eq}$ the freezing of the fluctuation amplitude tilts the spectral index to the value $n - 4$. Precise computations of the processed spectrum in a CDM dominated Universe have been done by several groups (see, e.g., refs.[310, 48]). Bond & Efstathiou [48] evaluated the CDM transmission factor for adiabatic fluctuations

$$T(k) = [1 + (ak + (bk)^{1.5} + (ck)^2)^\nu]^{-1/\nu} \tag{197}$$

[$a = 6.4\,(\Omega_o h^2)^{-1}$ Mpc, $b = 3.0\,(\Omega_o h^2)^{-1}$ Mpc, $c = 1.7\,(\Omega_o h^2)^{-1}$ Mpc, $\nu = 1.13$], assuming the presence of three species of massless neutrinos and negligible contribution from the baryonic component ($\Omega_b \ll \Omega_{CDM}$). In Figure 23 the shape of the CDM power spectrum is plotted, by assuming a primordial Zel'dovich spectrum. According to eq.(197), as $k \to 0$, we have $T(k) \simeq 1$ and the primordial spectrum $P(k) \propto k$ is left unchanged. At small scales, $T(k) \propto k^{-2}$, so that $P(k) \propto k^{-3}$. In Figure 23, the presence of a characteristic scale at $\lambda_{eq}$ is apparent, although the bending of the spectrum is rather gradual.

As opposed to the HDM scenario, a considerable amount of small scale power is now present, so that the first fluctuations reaching non–linearity are at small scales. The resulting clustering proceeds in a "bottom–up" way, with structures of increasing size forming from the tidal interaction and the merging of smaller structures. It is clear that the possibility for small scale structures not to be disrupted as the hierarchical clustering goes on depends on the ability of the baryonic component to cool down and fully virialize before being incorporated within larger DM fluctuations. Taking into account dissipative effects allows one to identify the CDM fluctuations where galaxy formation takes place. Detailed investigations of galaxy formation in the CDM scenario (see, e.g., ref.[45]) have shown that the observed variety of galaxy morphology and the relative morphological segregation of the clustering can be nicely reproduced.

A series of detailed numerical simulations of structure formation in a CDM Universe have been originally realized by Davis, Efstathiou, Frenk & White (see, e.g., refs.[111, 417]). The N–body experiments show that the primordial CDM spectrum is able to account for many aspects of the observed galaxy clustering at small and intermediate scales. Once a suitable "biasing" prescription is assumed to identify galaxies in a purely dissipationless simulation, not only the correct correlation amplitudes are reproduced, but also the density profile of galaxy halos, small scale velocity dispersions, cluster richness and the mean number density of both galaxies and clusters (see refs.[112, 249] for recent reviews about CDM). Despite the remarkable merits of CDM in reproducing the observed clustering at scales $\lesssim 10\,h^{-2}$ Mpc, serious problems are encountered when dealing with larger scales. In fact, for fluctuation wavelengths above $\lambda_{eq}$, the shape of the CDM spectrum steepens toward the Zel'dovich profile, $P(k) \propto k$, with a reduction of the power at such scales. As a first consequence, the resulting 2–point correlation function goes negative already at $\sim 15\,(\Omega_o h^2)^{-1}$ Mpc. This is at variance with respect to the observed 2–point function for the cluster distribution, that shows no evidence of anticorrelation up to $\sim 50h^{-1}$ Mpc. Also the amplitude of the cluster correlation is too high with respect to that provided by CDM [417], even allowing for a substantial contamination of cluster clustering [136]. A further problem encountered by CDM is due to the high amplitude of large scale motions. For instance, Vittorio et al. [406] compared the predictions of CDM with the observed large scale ($\sim 50h^{-1}$ Mpc) galaxy motions and concluded that this model is ruled out to a high confidence level. This result has been further strengthen by the reconstruction of the density field from observed galaxy peculiar velocities, as proposed by Bertschinger et al. [40]. The CDM model ran into difficulties also in reproducing the large scale ($\gtrsim 20h^{-1}$ Mpc) galaxy clustering observed from recently compiled catalogues. Using the APM sample, Maddox et al. [263] have shown that the angular 2–point function declines at large scales much less steeply than predicted by CDM, thus revealing an excess of power at $\gtrsim 20h^{-1}$ Mpc (see Figure 7). A similar conclusion has also been reached by Efstathiou et al. [142] and Saunders et al. [359] from the moment analysis of the galaxy counts for the QDOT redshift sample, and By Loveday et al. [255] rom a similar analysis of the Stromlo–APM redshift survey.

It is clear that these problems could be alleviated by allowing for variations of the standard CDM model, which is based on the assumptions of $\Omega_o = 1$, Zel'dovich initial spectrum, Gaussian initial fluctuations, as provided by the classical inflationary paradigma, and galaxy clustering substantially enhanced with respect to that of the underlying dark matter. Saving the standard inflationary scenario, a first possibiliy to reconcile CDM with observations is allowing for galaxies to have a correlation amplitude similar to that of matter. If this is the case, then more time is required for gravitational clustering until it attains the observed level of correlation than previously thought. Therefore, the present Universe in N–body simulations should correspond to a dynamically more evolved configuration. One's hope is that more dynamical evolution should add more large scale power, so to alleviate the CDM problems. Attempts in this direction have been pursued by Couchman & Carlberg [105]. Using an admixture of numerical methods and analytical approximations, they estimated the projected galaxy correlation function from the three–dimensional N–body simulations and found excellent agreement with the APM data. Based on large N–body simulations, we recently realized a more direct approach to the problem of projected galaxy correlation [283]. We generated artificial Lick maps, by reproducing at best the observational setup. By comparing the resulting angular correlation with that of the APM sample it turns out that even allowing for more dynamical evolution does not add enough large scale power (see Figure 24).

A further possibility to add more large scale power could be lowering $\Omega_o$. According to eq.(196), this amounts to increase the horizon size at time of matter–radiation equality, so as to push to larger scales the bending of the CDM spectrum. In this case, the agreement with the inflationary requirement for a flat Universe can still be achieved by taking a non–vanishing cosmological constant term $\Lambda$. Efstathiou et al. [144] have shown that taking $\Omega_{CDM} = 0.2$ and $\Lambda = 0.8$ accounts for the excess power displayed at large scales by the APM survey. Another way to add large scale power is assuming a "tilted" post–inflationary spectrum, $P(k) \propto k^n$ with $n < 1$. Although it is in principle possible to achieve this kind of spectrum within acceptable inflationary scenario (see, e.g., ref.[257]) the spectral index value is quite strongly constrained by COBE data ($n > 0.6$, see ref.[419]) and the achieved improvement is rather marginal [249]. In the following, I will discuss the effect of abandoning the random–phase prescription, which ensures the Gaussian nature of primordial fluctuations.

**MDM models**

A further complication one can introduce in the previous DM models in order to alleviate their problems is to allow for a Universe dominated by comparable amounts of both hot and cold particles. The resulting mixed dark matter (MDM) model should be tuned in such a way to keep the "good" features of each one-component model; the CDM component gives at small scales the seeds for galaxy formation, while the HDM component provides large scale power. The resulting spectrum (see Figure 23) has a small scale behaviour which is similar to that provided by CDM, but with a much higher level of large scale power, thanks to the hot component. Depending on the composition of the mixture, a continuous sequence of spectra is obtainable. Although the MDM models have been originally introduced several years ago [394, 7, 212], they received a considerable attention best quite recently, after the COBE detection of the CMB temperature anisotropy. In fact, immediately after this measurement, it has been recognized that, with the large scale normalization provided by COBE, the only DM spectrum able to account for observational constraints at much smaller scales is the mixed one, with $\sim 30\%$ of HDM and $\sim 70\%$ of CDM [419]. This result significantly strengthened linear calculations (see, e.g., [397, 213]), which suggested cold+hot DM as a reliable model for LSS formation.

Recent non–linear calculations based on N–body simulations have been employed to study the clustering developed by MDM models. A complication in this kind of simulations with respect to the CDM ones resides in the presence of a hot particle component. In fact, their large thermal velocities require a very large number of such hot particles, in order to adequately sample the corresponding phase–space, thus increasing the computational cost. A first simplified calculation has been realized by Davis, Summers & Schlegel [117], who, however, limited the size of the simulation box to $14h^{-1}$ Mpc. More extended simulations have been performed by Klypin et al. [239], who considered simulation boxes up to 200 Mpc aside ($H_o = 50\,\mathrm{km\,s^{-1}\,Mpc^{-1}}$) and a MDM spectrum with a mixture of 60% CDM, 30% HDM and 10% of baryons. By applying several statistical tests, these authors found that such a hybrid model is remarkably efficient to account for several aspects of the observed large scale clustering. Simulations at very high resolution have been recently realized [241, 50], which allowed a careful analysis of the non–linear clustering developed by these hybrid models. It is however clear that, before drawing definitive conclusions about the reliability of this model, more and more tests must be realized, being this model still much less tested if compared to HDM and CDM. In this context, the computational cost of large MDM N–body simulations makes linear and quasi–linear analyses a useful tool to compare the available observational data to the large scale model prediction (see, e.g., refs.[386, 328]).

A fundamental problem, however, arises in any DM scenario, which includes both a cold and a hot components. In fact, this kind of model requires the presence of both a light particle, like a neutrino with $m_\nu \sim 10\,eV$ to give $\Omega_\nu \sim 0.3$, along with a heavy particle, with a mass of many orders of magnitude larger. Therefore, a fine tuning is required for the number densities of such particles, so to give comparable contributions to the global mass density. If the MDM picture were confirmed to furnish a reliable framework for the interpretation of large scale clustering, the presence of this fine tuning clearly calls for the presence of a dynamical mechanisms for the production of DM particles, which should be provided by a viable model of fundamental interactions.

As a general comment about the DM models that I discussed in this section, it is worth emphasizing that any comparison with observations is based on suitable assumptions to identify galaxies out of DM fluctuations, both in analytical calculations and N–body experiments. One may ask whether changing these prescriptions substantially modifies the model predictions. Indeed, several analyses show that this is often the case. Therefore, before drawing any extreme conclusion about the reliability of a model, it is appropriate to check whether any prediction strongly depends on the assumption of galaxy formation or is a robust outcome of the DM spectrum, quite independently of the relation between matter and galaxy distributions.

## 5.3 Do we need non–Gaussian perturbations ?

Naturalness arguments suggest that Gaussian statistics should characterize the density fluctuations at the early stages of their evolution. However, the random–phase prescription provided by classical inflation [28], joined with the simplest (HDM and CDM) dark matter models does not account for the complete body of observational constraints. The idea of abandoning initial Gaussianity surely enlarges the permitted parameter space. Although the increase of degrees of freedom could make completely arbitrary any choice of initial conditions, nevertheless observational constraints can hopefully significantly reduce the number of allowed models.

Still keeping the advantages of inflation, non–Gaussian initial conditions arise in several models of stochastic and power law inflation, or within models where the inflaton field driving the accelerated expansion is non–linearly coupled to an auxiliary scalar field (see, e.g., refs.[12, 293, 32]; see also ref.[351] and references therein). In recent years, non–Gaussian models based on the generation of global textures, cosmic strings and other topological defects during early phases of the cosmic expansion have become very popular and detailed investigations of the LSS they develop have been made.

Global textures are expected to be formed during a symmetry breaking phase–transition, which leads to the formation of topological defects [235]. Phase transitions and defects are naturally expected in any particle physics theory, whose fields transform according to some representation of a non–Abelian symmetry group [73]. Non-Gaussian fluctuations are produced through the formation of texture "knots", having size of the same order of the horizon at that epoch. Subsequently, such knots collapse down to a scale ($\sim 10^{-37}$ cm) corresponding to the energy of symmetry breaking ($\sim 10^{16}\,GeV$, as expected from GUTs), accreting the surrounding DM with a spherically symmetric pattern. The resulting matter distribution is characterized by the presence of isolated non–linear clumps, which act as seeds for the formation of cosmic structures. N–body simulations of a CDM Universe with textures [300] show that, because of the large amount of power added at small scales with respect to the standard scenario, high density peaks trace the matter distribution fairly well. Although several problems of large scale clustering are alleviated, the presence of massive clumps causes a too high velocity dispersion inside clusters.

Another kind of topological defect that arises from symmetry breaking is represented by cosmic strings ([427]; see also refs.[73, 243] and references therein). Different from texture formation, cosmic strings arise from the breaking of the Abelian $U(1)$ group. The resulting topological defect is one dimensional (a string, indeed). After the $U(1)$ symmetry breaks down, a string network arises, which is formed both by infinite strings and closed string loops. Once formed, this network evolves under the competing effects of string stretching, which is due to cosmic expansion and dominates on scales larger than the horizon, and of string tension, which dominates below the horizon size. If the evolution of the network were only described by string stretching, it would led to a catastrophe. In fact, during radiation domination the ordinary energy density in a comoving volume would decrease as $a^{-1}(t)$, while that associated to a string would increase as $a(t)$. As a consequence one should reach the undesirable result of a string–dominated Universe. Actually, the network evolution is not so trivial and a detailed numerical treatment is required [36]. The two essential aspects for the evolution of the network are: *a)* self–interaction of infinite strings, which produces finite loops, and *b)* loop oscillations, as they cross the horizon, and decay through the emission of gravitational waves. Both analytical arguments and numerical simulations suggest that at the time of matter–radiation equality there are few long strings crossing the horizon plus a distribution of string loops. The relevance of string loops is in that they are expected to be seeds for the formation of galaxy clusters. In fact, a string loop of radius $R$ produces a gravitational field that, at scales greater than $R$, is the same as that due to a point source of mass $M = \beta\mu R$, with $\beta$ a suitable numerical constant, which is related to the shape of the loop, and $\mu$ the string mass per unit length. For reasonable choices of the parameters, it turns out that, for a radius $R = 10^{-1}\lambda_{eq}$, it is $M \sim 10^{11} M_\odot$. Using linear theory for the growth of mass perturbations, we find that from $t_{eq}$ to the present time, such a loop accretes a mass of $\sim 10^{15} M_\odot$, which is the typical mass of Abell clusters. In this context, the correlation properties of the loop distribution should reproduce that of clusters. Turok [392] found that the 2–point function of string loops has a nearly power–law shape, with slope $\gamma \simeq 2$, surprisingly similar to that observed for clusters. It is however clear that a more detailed model for structure formation should also specify the dark matter content. The presence of loops as seeds for structure formation is expected to add power at small scales and, thus, could alleviate some drawbacks of HDM. Numerical simulations with cosmic strings and massive neutrinos [365] have shown that galaxy formation starts substantially earlier than in both standard CDM and HDM models, thus reconciling with observations of high–redshift QSOs. Due to the presence of accretion seeds, the final distribution contains isolated density peaks embedded in a smooth background.

The possibility of generating seeds of accretion for structure formation through a series of mechanisms, which also include primordial black holes and non–topological solitons, led Villumsen et al. [405] to study the general problem of seeded structure formation with N–body simulations. The general result is that galaxies again form at high redshifts and immediately after their formation they are strongly clustered. The availability in the near future of complete and extended QSO surveys at increasing redshifts will be a crucial test for these kind of models. An analytical treatment of the statistical properties of matter distribution with accretion seeds has also been recently provided by Scherrer & Bertschinger [364].

A further scenario, in which non–Gaussian fluctuations drive the formation of large scale clustering, is provided by the explosion model. Different from the gravitational instability picture, in the explosion scenario energy perturbations of non–gravitational origin drive material away from the seeds of the explosions, sweeping primordial gas into dense, expanding shells. As these shells cool, their fragmentation could give a further generation of objects which again explode, thus amplifying the process and giving rise to large scale structure formation. At the end, clusters of galaxies are expected to be placed at the intersection of three expanding bubbles, while galaxies are arranged on spherical shells [217, 294]. The explosion scenario became very popular after the compilation of redshift surveys suggested a "bubbly" geometry of the galaxy distribution, with nearly spherical voids surrounded by a sheet–like galaxy distribution (see ref.[13] for a description of large sca;e clustering in terms of bubbly geometry). A variety of physical mechanisms might generate such explosions, such as supermassive stars or supernovae from the earliest galaxies. However, it is at present not clear whether or not such energy sources can be sufficient to create the large voids that are observed. Holes on scales of some tens of Mpcs would require a fantastic amount of supernovae exploding coherently. Moreover, suitable initial conditions are anyway needed to generate primordial objects that act as seeds of explosions. Since both initial conditions for explosion generations and fragmentation processes for galaxy formation are poorly understood, only simplified versions of the explosion scenario have been investigated. For instance, Weinberg et al. [411] modelled explosions by using a random distribution of expanding shells with a power–law distribution of radii. After identifying clusters at the intersection of three shells, the statistics of their spatial distribution is quite well reproduced.

Rather than dealing with specific non–Gaussian models, which arise from theoretical prescriptions for primeval fluctuation origin, a further possibility is to analyze the development of clustering for a wide class of non–Gaussian primordial perturbations and check whether some of these are able to reproduce the observed large-scale texture of the galaxy distribution. Moscardini et al. [284] and Matarrese et al. [269] analysed CDM N–body simulations starting with both Gaussian and non–Gaussian initial conditions. They considered several non–Gaussian models, generated through local non–linear transformations of an underlying Gaussian field. The aim of their analysis was to show whether a CDM dominated Universe can be reconciled with the observed large scale clustering, once we take more general initial conditions than those provided by the random–phase prescription. As expected, remarkable differences with respect to the large scale clustering produced by Gaussian initial conditions are produced, which turns out to crucially depend on the sign of the initial skewness, $\langle \delta^3 \rangle$, of the density fluctuations. Positive skewness models, which have a predominance of concentrated overdense regions, rapidly develop extremely clumped structures with a resulting small coherence length. The resulting distribution suffers even more for all the problems of the standard Gaussian CDM model. Vice versa, for negative skewness models the dynamics of the clustering is dominated by the presence of expanding devoid regions, while the merging of the surrounding shells forms large scale filaments and knots. The resulting cellular structure resembles that arising in the explosion scenario, but with the fundamental difference that it is purely driven by gravitational instability, with large scale coherence produced by the initial phase correlations. The rich variety of structures produced at large scales goes in the right direction to reconcile the CDM model with large scale observational constraints. Weinberg & Cole [410] ran a variety of N–body simulations by considering a wider class of non–Gaussian models. Instead of checking whether the CDM model can be improved by adopting suitable initial conditions, these authors attempted to seek the features of large scale clustering which are to be ascribed to the presence of initial phase correlations. After applying a list of statistical tests, they concluded that, despite the remarkable variety of clustering realization obtainable with non–Gaussian initial conditions, the standard Gaussian model with $\Omega_o = 1$, biased galaxy formation and power spectrum $P(k) \propto k^{-1}$ at the scales relevant to galaxy clustering is the most efficient in reproducing the complete body of observational data. Since the required shape of $P(k)$ closely follows that of the CDM spectrum, at least at the scales probed by their simulations, these authors concluded that the problems displayed by the CDM model at scales $\gtrsim 10h^{-1}$ Mpc are not due to the standard random–phase assumption, rather they are more likely to be ascribed to the lack of large scale power.

This suggestion is however contradicted by results based on simulations of projected galaxy and cluster samples from non–Gaussian CDM models [97, 283, 60]. As an example, I plot in Figure 24 our results about the angular correlation function for simulated Lick maps, based on negative–skewness models [283], as compared with the APM data. It is apparent the high degree of large scale coherence generated by primordial phase-correlations, which significantly improve the performance of the CDM model. In Section 7 I will show the results about the scaling properties associated to simulated angular cluster samples, also based on non–Gaussian initial conditions.

As a final comment, it is worth observing the similarity between the large scale clustering pattern generated by skew–negative CDM models and Gaussian models with power spectra providing more large scale power. Indeed, in the second case, gravitational evolution generates phase correlations in the mildly non–linear regime in a manner which is coupled to the initial power spectrum. Therefore, it becomes difficult to distinguish the effects of large scale power from those of primordial phase correlation and accurate statistical measures should be devised to discriminate between Gaussian large scale power and coherence induced by non–Gaussian initial conditions.

## 5.4 Biased galaxy formation

A crucial step to test any theory about the formation and evolution of primordial inhomogeneities is to compare its predictions to the observed galaxy distribution. Differences between dark matter and galaxy distributions probably exist and their origin lies in the physical mechanisms and environmental effects occurring during the formation of cosmic structures. For this reason, it is of crucial relevance to understand which kind of processes are relevant to the formation of visible objects and whether they give rise to a segregation between luminous and dark matter. I will now discuss the main motivations which require substantial differences between the clustering of DM and of observable objects, also describing some mechanisms of galaxy formation, which could be naturally responsible for this "bias". Finally, I will point out analytical approaches to biasing, in order to show the relations between the statistics of the galaxy distribution and that of the matter density field.

### 5.4.1 Motivations

The original motivation, which led to the introduction of the concept of biasing in the distribution of cosmic structures, is the enhanced clustering displayed by rich clusters with respect to the galaxy distribution. As discussed in Section 3, both galaxies and galaxy clusters are characterized by a 2–point correlation function with the same power law shape, $\xi(r) \propto r^{-1.8}$, although holding at different scales and with a remarkably different amplitude. By comparing the value $r_{o,g} \simeq 5h^{-1}$ Mpc of the galaxy correlation length to that, $r_{o,c} \simeq 20h^{-1}$ Mpc, of rich clusters, it turns out that $\xi_c(r) \simeq 15\, \xi_g(r)$. As a consequence, the large scale distribution of matter in the Universe cannot be traced with the same efficiency both by galaxies and galaxy clusters. On the contrary, these results seem to suggest that neither galaxies nor clusters fairly trace the actual matter distribution. The large correlation amplitude for rich Abell clusters was the main reason that lead Kaiser [230] to introduce the concept of bias. According to this model, he postulated that rich Abell clusters arise only from those peaks of the background field, that exceed a limiting density threshold value, and consequently exhibit an enhanced clustering with respect to the underlying matter. Further supports in favour of a biased distribution of cosmic structures also come from the existing correlations between galaxy types and environment, and from luminosity and morphological dependence of galaxy clustering. On the ground of such results, it would be surprising if galaxy formation were not significantly affected by environmental effects, segregating somehow the luminous content of the Universe from the dark one.

In the framework of the standard CDM scenario, results of N–body simulations led to the conclusion that, if $\Omega_o = 1$, the large scale distribution of galaxies cannot be reproduced, unless the galaxy formation is biased (see ref.[111] and Section 6 below). In such simulations, the resulting 2–point correlation function steepens in time, so that, if mass traces the galaxy distribution, the stage of the simulation to be considered as the present time is reached when its logarithmic slope matches that ($\gamma = 1.8$) observed for galaxies. However, this evolutionary stage corresponds to a value of the clustering length $r_o \simeq 1\,(\Omega_o h^2)^{-1}$ Mpc, too small if compared with $r_{o,g} \simeq 5\, h^{-1}$ Mpc observed for galaxies, unless $\Omega_o h \lesssim 0.2$. Thus, assuming a flat CDM dominated Universe requires an enhanced clustering of galaxies with respect to the background, such that $\xi_g(r) = (5 - 20)\xi_m(r)$ (for $h = 0.5$ and $h = 1$). Also in the case of a HDM dominated Universe, some biasing in the galaxy distribution should be present. In this case, the lack of small scale power causes a high coherence length in the primeval spectrum. As a consequence, the requirement that the slope of the 2–point function is $\gamma = 1.8$ implies $r_o = 8\,(\Omega_o h^2)^{-1}$ Mpc for the neutrino correlation length. Here, the required bias is in the opposite sense because the resulting galaxies must be less clustered than the DM background (antibiasing).

Indications that the luminous matter should be segregated with respect to the dark matter also come form measurements of the mass–to–light ratio for cosmic structures of increasing size. Mass is in general estimated by using a relation of the type $M \simeq v^2 r / G$, where $v$ is some observed velocity involved in a structure of size $r$. While at the scales of individual galaxies ($\sim 10\, h^{-1}$ kpc) a typical value of the mass–to–light ratio for the stellar content is of the order $M/L \lesssim 10$ (in units of $M_\odot / L_\odot$), data on the virial analysis of groups and clusters of galaxies suggest $M/L \simeq 200$–$500\, h$ at scales $\sim 1 h^{-1}$ Mpc [45]. This indicates that luminous (baryonic) matter does not follow the DM distribution, instead it turns out to be preferentially segregated at small scales. By comparing the above $M/L$ value for clusters to that, $M/L \simeq 1600\, h$, required to close the Universe [139], we see that the contribution to the average density coming from the DM clustered at such scales gives $\Omega_o \simeq 0.2$–$0.3$. This value can be reconciled with $\Omega_o = 1$ only by allowing for the rising trend of $M/L$ to include even larger structures (superclusters), thus increasing the amount of biasing as structures at larger and larger scales are considered. The dynamical analysis of the amount of mass clustered at scales $\sim 1 h^{-1}$ Mpc is usually based on the application of the "cosmic virial theorem" [307, 308] to pairs of galaxies. This theorem, which expresses the condition for hydrostatic equilibrium of a self–gravitating system of collisionless particles, relates the quadratic velocity dispersion between pairs at a given separation $r$ to the corresponding 2–point correlation function $\xi(r)$, which measures the mass excess at the scale $r$ responsible for the galaxy motion. Accordingly,

$$v^2(r) \; \propto \; \Omega_o \, r^2 \, \xi(r) \,. \tag{198}$$

By comparing the measured $v^2(r)$ with the galaxy 2–point function according to this relation gives $\Omega_o \simeq 0.1$–$0.3$ [116, 35, 198]. If, however, galaxies cluster more than the underlying matter, the density contrast of the two distributions can be related as $\delta_g = b_g\, \delta_m$, $b_g > 1$ being the so–called galaxy biasing parameter. According to this linear biasing prescription, the galaxy 2–point function turns out to be amplified with respect to that of the background as

$$\xi_g(r) \; = \; b_g^2 \, \xi_m(r) \,. \tag{199}$$

Here $\xi_m(r)$ represents the matter 2–point correlation function. Since pairwise galaxy velocities are related to the mass excess, and not to the galaxy number excess, the matter correlation function $\xi_m(r)$ must be used in eq.(198). Then, the resulting value of $\Omega_o$ turns out to be amplified by a factor $b_g^2$ and agreement with a flat Universe is achieved for $b_g \simeq 2$–$3$.

Based on a similar approach, we realized a measurement of $\Omega_o$, by using the dynamical information coming from the observed rotation curves for a suitable sample of spiral galaxies [352]. We obtained the density excess associated with dark halos of spiral galaxies by considering, rather than the motion of a companion galaxy, the motion of test bodies rotating in the disk of spiral galaxies. This is done by means of a proper decomposition of the galaxy rotation curves into the contributions or dark and visible components (see, e.g., ref.[318] and §3.1.4). On scales of few tens of kpcs, the mass excess around galaxies, as evaluated through the virial estimate, turns out to scale with $r$ like the excess in number of spiral galaxies. On the assumption that the light traces the mass, the resulting density parameter lies in the range $\Omega_o = 0.2$–$0.4$. Also in this case, the requirement of a flat Universe gives $b_g \sim 2.5$ for the galaxy biasing factor. In this respect, it is interesting to observe that, according to results reported in §3.1.4, the galaxy–matter correlation length turns out to be $r_{o,gm} = (3.2 \pm 0.3)\,\Omega_o^{-1/\gamma} h^{-1}$ Mpc. If $\Omega_o = 1$, a galaxy biasing factor $b_g \simeq 2.5$ gives $r_{o,m} = b_g^{-1/\gamma} r_{o,gm} \simeq 2 h^{-1}$ Mpc, which agrees remarkably with the CDM correlation length, $r_{o,\text{CDM}} \simeq 1\, h^{-2}$ Mpc, coming from N–body simulations, once $h \simeq 0.5$ is taken for the Hubble parameter.

At the larger scales of some tens of Mpcs, measurements of $\Omega_o$ have been made by using all–sky redshift samples based on the IRAS point source catalogue, to check whether the acceleration of the Local Group converges within the sample depth. This kind of analysis provides $\Omega_o$ values not far from one, thus suggesting that IRAS galaxies are, at such scales, fairly good tracers of the matter distribution (see ref.[232] and references therein). This result can be interpreted by saying either that IRAS galaxies are substantially less clustered than optically selected galaxies or that the amount of biasing should depend on the scale, passing from $b_g \simeq 2$–$3$ at scales below a few Mpcs, where the clustering is non–linear, to $b_q \simeq 1$ at the larger scales of linearity. Although the above result is rather comfortable as far as the flatness of the Universe in concerned, it seems however at variance with respect to that of similar analysis based on the distribution of galaxy clusters. The dipole estimate for the spatial distribution of Abell and ACO clusters [325, 361] shows that the acceleration of the Local Group converges at larger distances ($\sim 150 h^{-1}$ Mpc) than indicated by IRAS galaxies ($\sim 40 h^{-1}$ Mpc), with a larger value of the dipole amplitude at the scale of convergence. The resulting estimate of the the density parameter gives the much lower value $0.05 \lesssim b_c^{-1.67} \Omega_o \lesssim 0.08$ (here $b_c$ is the biasing factor for galaxy clusters). Although this result is apparently in contradiction with previous estimates, nevertheless we should bear in mind that it is based on the distribution of rich galaxy clusters, which are structures even more biased than galaxies. By comparing the correlation amplitude for galaxies and clusters, we obtain that the biasing factor for clusters with respect to galaxies is $b_{gc} \simeq 3$. Thus, taking $b_g \simeq 2$, the global biasing factor for the cluster distribution with respect to matter is $b_c = b_{cg}\, b_g \simeq 6$, and gives an $\Omega_o$ value which is consistent with one. Note, however, that this result is based on the assumption that the relation found at $\sim 10 h^{-1}$ Mpc scales between the clustering amplitudes of galaxies and clusters can be extrapolated at the much larger scales of cluster dipole convergence. Whether or not this is a correct assumption surely requires further investigations.

### 5.4.2 Physical mechanisms for bias

It is clear that there does not exist a unique way to obtain an efficient segregation between luminous objects and DM distribution. Instead different bias mechanisms can arise in different cosmological scenarios, depending on the nature (cold or hot) of the dark matter, on the way of generating perturbations (gravitational instability or explosions) and so on (see refs.[118, 123] as reviews on biased galaxy formation). A fundamental component to determine the resulting amount of biasing is represented by the baryon distribution at the onset of galaxy formation. In particular, it could happen that the baryonic component is segregated from the non–baryonic DM, so that galaxy formation occurs only in certain regions. Alternatively, it is also possible that the large scale baryon distribution does trace the DM, but the efficiency with which baryons turn into luminous galaxies depends on other environmental effects, such as the local background density, or it may be the result of feedback from other galaxies. The effect may be destructive, suppressing galaxy formation, or constructive, enhancing galaxy formation in the neighborhood of other galaxies (e.g., explosions). An enhanced clustering of galaxies over the background matter can arise in a "bottom–up" scenario, if galaxies formed only from those peaks of the primordial density distribution, smoothed on the galactic scale $R$, that have amplitude at least $\nu$ times the r.m.s. value $\sigma_{\rm R}$. If the power spectrum has sufficient amplitude at small wavenumbers, high peaks occur with greater probability in the crests rather than near the minima of a large scale fluctuation mode, so they display an enhanced clustering (see Figure 25). In particular, correlation functions arising from N–body simulation in a CDM–dominated Universe reproduce the observed ones if galaxies are identified with initial peaks at $\nu \sim 2.5$. In this picture, the crucial point is however to understand what physical mechanisms provide a sharp cutoff in the efficiency of galaxy formation for density fluctuations $\delta_{\rm R} < \nu \sigma_{\rm R}$.

A mechanism that has been proposed to introduce a threshold effect in the galaxy formation process is the so–called *natural bias*. In order to describe this scenario, let us observe that for a galaxy to be visible at the present time, we must ask that the baryonic matter has been able to dissipate and turn into stars. In order for dissipation to occur, the redshift of collapse clearly needs to be sufficiently large that there is time for an object to cool between its formation at redshift $z_i$, when the density fluctuation attains the critical value $\delta = \delta_c$ (if this epoch is identified with the end of the recollapse of a spherical fluctuation, then $\delta_c \simeq 1.69$ in the linear model), and the present epoch. More massive objects take longer to cool. Then, the requirement for a fluctuation on a given mass–scale $M$ to have enough time from $t_i$ (corresponding to $z_i$) to cool down turns into the introduction of a mass–dependent threshold $\nu(M)\sigma_{\rm M} = \delta_c[1+z_{cool}(M)]$. This situation is described in the density–temperature plot of Figure 26. The cooling curve, above which $t_{cool} < t_{dyn}$ (here $t_{dyn} \sim 1/\sqrt{G\rho}$ is the gravitational free–fall time, while $t_{cool}$ is the cooling time), confines the region where the gas can contract and form stars [342, 45]. The plotted cooling curve is evaluated once a primordial gas composition is specified and under the assumption that the mean baryonic gas density is 10% of the total mass density. Each of the dotted diagonal lines indicates in the $n - T$ diagram the positions of all the structures having the same Jeans mass $M_J \simeq 100\, T^{3/2} n^{-1/2} M_\odot$. The almost vertical line $V_{crit}$, which has been introduced by Dekel & Silk [124], divides the permissible region for galaxy formation in two; a protogalaxy characterized by a virial velocity $> V_{crit}$ cannot expel a large fraction of its original gas content and form a normal galaxy. A protogalaxy with $V < V_{crit}$ can produce a supernova–driven wind, which would drive a substantial fraction of the protogalactic gas out, with a diffuse dwarf as final product. The dashed curves labeled by $\nu\sigma$ ($\nu = 1,3$) refer to fluctuations with $\delta M/M$ equal to $\nu$ times the r.m.s. value, for a CDM spectrum. The corresponding parallel dashed curves refer to the protogalactic gas clouds, after a contraction of a factor 10 inside isothermal halos, to densities that are comparable to the halo densities such that star formation is possible. The two vertical arrows indicate the largest galaxies that can form out of $1\sigma$ and $3\sigma$ peaks, respectively. Let us observe that the major part of galaxies arising from $1\sigma$ peaks have $V < V_{crit}$, so that they would turn into dwarf galaxies. Instead, the shaded area indicates the position of normal galaxies. It is also evident that most of them are originated from $2\sigma$ and $3\sigma$ peaks. According to such predictions, normal galaxies, arising from exceptionally high peaks, are expected to be much more correlated than the background fluctuations and lie preferentially in rich clusters. Vice versa, dwarf galaxies form from typical (i.e., $1\sigma$) peaks, consequently they are expected to be better tracers of the matter distribution. In a scenario of this kind, morphological segregation naturally arises once different galaxy types are identified with peaks of different height of the primordial fluctuation field.

### 5.4.3 Properties of the biased distribution

The possibility to devise a mechanism for interpreting the process of galaxy formation as a threshold effect on the initial density field allows us to relate the statistics of the matter distribution to that of the "biased" field. The simplest case occurs for Gaussian density fluctuations $\delta(\mathbf{x})$ (see ref.[8] for a detailed description of the properties of Gaussian random fields). In this case the statistics is completely specified by the 2–point correlation function $\xi(r)$. According to eq.(71), in order to identify "physical" structures of characteristic size $R$ out of $\delta(\mathbf{x})$, let us consider the smoothed fluctuations $\delta_{\mathrm{R}}(\mathbf{x})$, given by the convolution of $\delta(\mathbf{x})$ with a suitable window function, which suppresses fluctuation modes at wavelengths $< R$. Accordingly, the correlations of the smoothed field are related to those of $\delta(\mathbf{x})$ according to

$$\mu_{\mathrm{R},n}(\mathbf{x}_1,...,\mathbf{x}_n) \;=\; \int \left[\prod_{i=1}^{n} W_{\mathrm{R}}(|\mathbf{x}_i - \mathbf{y}_i|)\,d\mathbf{y}_i\right] \mu_n(\mathbf{y}_1,...,\mathbf{y}_n)\,, \qquad (200)$$

where $\mu_{\mathrm{R},n}$ is the "smoothed" correlation function.

Following Kaiser's original prescription [230], let us introduce the biased field

$$\rho_{\nu,\mathrm{R}}(\mathbf{x}) \;=\; \theta[\delta_{\mathrm{R}}(\mathbf{x}) - \nu\sigma_{\mathrm{R}}]\,, \qquad (201)$$

which assigns a unity probability that a fluctuation $\delta_{\mathrm{R}}(\mathbf{x}) > \nu\sigma_{\mathrm{R}}$ turns into an observable object, while structure formation is forbidden for those fluctuations lying below the critical threshold. Note that this implementation of the natural biasing substantially reproduces the already described Press & Schechter [334] approach to the mass function (see §3.4). In both cases, one refers to the initial Gaussian density field in order to identify those fluctuations which turn into today observable objects. In particular, eq.(73) for the fraction of mass in structure of mass larger than $M$ can be equivalently rewritten as

$$\langle \rho_{\nu,\mathrm{R}} \rangle = \int \mathcal{D}[\delta]\, \mathcal{P}[\delta]\, \rho_{\nu,\mathrm{R}}(\mathbf{x}) =$$
$$= \frac{1}{\sqrt{2\pi}\,\sigma_{\mathrm{R}}} \int_{\nu\sigma_{\mathrm{R}}}^{+\infty} \exp\left(-\frac{\delta^2}{2\sigma_{\mathrm{R}}^2}\right) d\delta = \frac{1}{2}\,\mathrm{erfc}\left(2^{-1/2}\nu\right), \qquad (202)$$

so as to represent the expectation value of the biased field, i.e. the fraction of volume above the threshold. It coincides with eq.(73) once we identify $\delta_c = \nu \sigma_{\mathrm{R}}$.

The statistics of the so-called "excursion set" identified by eq.(201) does not coincide with those of the density peaks above $\nu\sigma_{\mathrm{R}}$. However, the two distributions are expected to coincide when very high thresholds, $\nu \gg 1$, are considered. Bardeen et al. [27] gave a comprehensive description of the properties of peaks and excursion set for a random Gaussian field in relation to the natural biasing scheme.

An interesting property of the biased field $\rho_{\nu,\mathrm{R}}$ is that, due to the non-linear transformation of $\delta_{\mathrm{R}}$ provided by eq.(201), it turns out to have a non-Gaussian statistics. In this case, non-Gaussianity is not the consequence of the non-linear dynamical evolution of $\delta_{\mathrm{R}}(\mathbf{x})$, but has a statistical origin. The $n$-point correlation functions of $\rho_{\nu,\mathrm{R}}(\mathbf{x})$ can be evaluated in terms of the $n$-point joint probability

$$\langle \rho_{\nu,\mathrm{R}}(\mathbf{x}_1) \ldots \rho_{\nu,\mathrm{R}}(\mathbf{x}_n) \rangle = \int \mathcal{D}[\delta]\, \mathcal{P}[\delta]\, \rho_{\nu,\mathrm{R}}(\mathbf{x}_1) \ldots \rho_{\nu,\mathrm{R}}(\mathbf{x}_n). \qquad (203)$$

As for the 2-point function, an analytical expression can be given in the high threshold limit $\nu \gg 1$ [230], which reads

$$\xi_{\nu,\mathrm{R}}(r) \simeq \exp\left[\left(\frac{\nu}{\sigma_{\mathrm{R}}}\right)^2 \xi_{\mathrm{R}}(r)\right] - 1. \qquad (204)$$

Although the correlation amplitude turns out to be increased, the first zero crossing of the "biased" function occurs at the same scale as for the background function, so that peak selection does not introduce coherence at larger scales. In the weak correlation regime, $\xi_{\mathrm{R}}(r) \ll 1$, the expansion of the exponential term in eq.(204) gives

$$\xi_{\nu,\mathrm{R}}(r) \simeq \left(\frac{\nu}{\sigma_{\mathrm{R}}}\right)^2 \xi_{\mathrm{R}}(r), \qquad (205)$$

and the linear biasing prescription is recovered, with $b = \nu/\sigma_{\mathrm{R}}$. As the small scales of non-linearity are considered, then eq.(204) gives a peak correlation function which detaches from the power law shape expected for the matter correlation function. This is at variance with the detected 2-point function, which holds as a power law even in the $\xi > 1$ regime. Note, however, that at small scales the matter distribution is expected to have non-Gaussian statistics due to the effect of non-linear gravity, so that eq.(204) does not hold.

Still keeping $\nu \gg 1$ and $\xi_\mathrm{R}(r) \ll 1$, a closed expression can be given also for the connected $n$–point functions. Jensen & Szalay [222] have shown that the higher-order correlations for the biased field reproduce the Kirkwood expression of eq.(27). The corresponding connected 3–point function reads

$$\begin{aligned}\zeta_{\nu,\mathrm{R};123} &= \xi_{\nu,\mathrm{R}}(r_{12})\,\xi_{\nu,\mathrm{R}}(r_{23}) + \xi_{\nu,\mathrm{R}}(r_{12})\,\xi_{\nu,\mathrm{R}}(r_{13}) + \xi_{\nu,\mathrm{R}}(r_{13})\,\xi_{\nu,\mathrm{R}}(r_{23}) \\ &+ \xi_{\nu,\mathrm{R}}(r_{12})\,\xi_{\nu,\mathrm{R}}(r_{23})\,\xi_{\nu,\mathrm{R}}(r_{13})\,,\end{aligned} \qquad (206)$$

which, however, does not reproduce the hierarchical expression suggested by observational data (see Section 3). This could be well explained since in the weak correlation regime the cubic term in eq.(206) becomes negligible and it is difficult to detect any difference with respect to the hierarchical expression. On the other hand, the galaxy 3–point function is better determined at small scales, where both the weak clustering approximation and the assumption of Gaussian fluctuations break down.

The generalization of the above biasing scheme to the non–Gaussian case has also been pursued, for both the statistics of the excursion set [268, 187] and for the peak distribution [88]. In this case, however, analytical expressions for measurable quantities can be given only in a limited number of cases, thus making even more difficult any comparison with observations.

The above described approach, based on the natural biasing prescription, surely has several merits in keeping some essential features of the galaxy formation and to relate it to the observed clustering for different classes of cosmic structures. This procedure appears, however, to be too crude in several aspects: relating galaxy formation only to the height of primordial density peaks neglects other effects, such as merging within larger structures, tidal disruption, feedback from nearby forming objects, non–spherical geometry of the gravitational collapse, and so on. In this respect, selecting peaks in a yes/no fashion according to eq.(201) seems an exceedingly simplified representation of galaxy formation. Therefore, it becomes of crucial relevance asking whether or not this model is robust, that is whether or not a slight modification of the biasing prescription leaves nearly invariant the statistics of the biased field. Attempts in this direction have been pursued by several authors, both by considering specific expressions for $\rho_{\nu,\mathrm{R}}(\mathbf{x})$ [233, 27] or developing a general formalism to treat a wide class of biased fields for Gaussian [381] and non–Gaussian [36] background fluctuations. In this framework, the biased field can be expressed as

$$\rho_{\nu,\mathrm{R}}(\mathbf{x}) = f[\delta_\mathrm{R}(\mathbf{x}) - \nu\sigma_\mathrm{R}]\,, \qquad (207)$$

where $f(\alpha)$ describes a generic non–linear transformation of $\delta_\mathrm{R}(\mathbf{x})$ and generalizes eq.(201). By taking $0 \leq f \leq 1$, eq.(207) can be interpreted as the probability for a fluctuation to become an observable object. In order to check the robustness of the results provided by the $\theta$–threshold of eq.(201), we worked out some statistical properties for more general thresholds. In particular, we considered the effects of changing the criteria to select fluctuations on the mass function for groups and clusters of galaxies [57] and on the 2–point correlation function of rich clusters [54]. In both cases, we found that for suitable choices of the parameters determining the threshold profile, much better fits than those provided by the classical scheme are obtained. Far from meaning that such best–fit thresholds account for all the non–linear effects of structure formation, the results indicate that the criteria of fluctuation selection is a very critical issue; a marginal modification of the biasing scheme turns into important variations of the statistics of the biased field.

Although the generality of the non–linear transformation provided by eq.(207) gives rise to a wide spectrum of possible statistical properties, nevertheless these cannot be completely arbitrary. An example of this is the high–threshold ($\nu \gg 1$) limit for the correlations of $\rho_{\nu,\mathrm{R}}(\mathbf{x})$, which can be shown to have a universal behavior, independent of the form of the threshold function [57]. Furthermore, eq.(207) represents a local transformation of the underlying fluctuation field, so that it is expected not to introduce coherence at scales larger than the typical correlation length of $\delta(\mathbf{x})$. As a consequence, the lack of large scale power of the CDM spectrum cannot be alleviated by any local biasing prescription. A different situation occurs if the peak selection probability is somehow enhanced for the presence of other nearby selected peaks. Several aspects of this "cooperative" biasing prescription have been discussed by Bower et al. [71]. They showed that, despite the uncertainties concerning plausible astrophysical mechanisms, which were able to implement this scenario, the resulting increase of large scale coherence turns out to be really remarkable, even with a modest modulation of the peak selection. Once more, this suggests that, before drawing any conclusion about the large scale clustering provided by a given model, a clear understanding of the physical mechanisms of galaxy formation is required.

# 6 Non–linear clustering through N–body simulations

In order to study gravitational clustering in its non–linear phases, in this section I review some results about the scaling analysis of cosmological N–body simulations. After a description of some technical aspects concerning N–body codes, I present the results of analyses, based both on correlation and fractal approaches. In particular, the 3– and 4–point correlation functions for the background matter distribution and for biased subsets of the entire particle configurations, which are identified as the high peaks of the initial density field, will be discussed. As far as the fractal analysis is concerned, I show the result of applying the dimension estimator already introduced in Section 4, in order to investigate the scaling properties associated with non–linear gravitational dynamics. The presentation of the results about fractal properties is inspired to our analysis, which is presented in much more details in ref.[396].

## 6.1 Why use N–body simulations ?

A complete description of the processes of formation and evolution of cosmic structures is very complicated and far from having reached a satisfactory level of explanation. Formation of galaxies and galaxy clusters involves both gravitational dynamics and hydrodynamical processes. As discussed in the previous section, a full analytical treatment of non–linear gravity has been not yet formulated, while several approximations have been devised to account for partial dynamical aspects. The situation is less clear when trying to account for dissipative hydrodynamical effects. This really represents a serious limitation to our understanding of the large scale galaxy distribution, since the observational mapping of the Universe mostly passes through the detection of luminous structures, i.e. of the regions where dissipation plays a fundamental role. A complete description of the attempts devoted to account for the hydrodynamical effects involved in galaxy formation is beyond the scope of this article. I only mention that at present the most promising approach is probably represented by numerical N–body simulations which also include hydrodynamical and radiative effects. Although only preliminary steps have been made in this direction (see, e.g., refs.[89, 91]), nevertheless the emerging results are rather promising and it is to be expected that a further improvement of the "technology" will probably clarify in the next future our view of galaxy formation mechanisms. On the other hand, in the framework of the gravitational instability picture, any non–gravitational effect is expected to be relevant only at quite small scales, where the characteristic time–scale for gravitational collapse, $t_{dyn} \sim (G\rho)^{-1/2}$, becomes comparable to the cooling time–scale $t_{cool}$. The determination of $t_{cool}$ is surely less reliable than that of $t_{dyn}$, since it relies on the knowledge of cooling mechanisms, local chemical composition, etc. Nevertheless, it is reasonable to assume that at scales larger than that of a typical galaxy the dynamics is entirely determined by the non–dissipative gravitational interaction. At such scales, the description of the formation of the LSS is obtainable by solving the equations (154) for the evolution of the density inhomogeneities.

However, the difficulty of analytically following the gravitational evolution when such equations are not linearizable forces one to resort to numerical methods. In this context, N–body simulations furnish a fundamental contribution towards understanding the nature of gravitational dynamics. In fact, N–body codes describe the evolution of non–linear gravitational clustering by following particle trajectories under the action of the gravitational force. Initial conditions ( i.e., initial fluctuation and velocity fields) are fixed in a consistent way at a sufficiently early time, so that linear theory is a good approximation at all the relevant scales. Then, the final result of gravitational clustering is compared with the observational data, in order to assess the reliability of the initial condition model. It is clear that, since small scale virialized structures probably retain almost no memory of initial conditions, structures at larger scales (e.g., filaments, voids or galaxy aggregates) are much more useful in giving constraints about the nature of the primordial fluctuations.

A basic parameter which measures the capability of N–body codes to faithfully represent gravitational clustering is the width of their dynamical range for mass and length resolution. Mass resolution is fixed by the total number of particles employed. Since a given mass is assigned to each particle, we should require at the linear stage that fluctuations on a mass scale below that of a particle were negligible. The dynamical range for length resolution is fixed by the ratio of the size of the simulation box to the softening scale for the computation of the gravitational force. Very detailed tests are always required to measure the resolution of numerical codes, in order to be sure about the reliability of the subsequent clustering representation. Because of the limits imposed by computational costs and computer memory, different strategies can be adopted in order to compromise between numerical resource and extension of the dynamical range. Accordingly, four main categories of N–body simulations can be devised, which essentially differ in their prescriptions for evaluating the gravitational force between particles.

a) Direct integration of the force acting on each particle, due to the presence of all the other particles [2, 1]. Within this approach, the force softening scale is usually very small and the particle trajectories are calculated with great precision. However, the price to be payed for this accuracy is the high computational cost, which goes like $N^2$ ($N$ being the number of particles). Therefore, only a rather limited number of particles ($\lesssim 10^4$) is usually employed.

b) Tree–codes, which compromise between high force resolution and computational time. In this kind of codes, the particle are arranged in groups and subgroups (trees). Then, the force acting on a particle from a distant group is approximated by a single contribution from his center of mass. This approximation is better if the considered group is more distant, so that its internal structure is uninfluent. If, instead, the dimension of each group is comparable to the distance from the particle, then its substructure must be exploited, by further subdividing it into subgroups, until the required computational precision is attained (see, e.g., refs.[30, 207]).

**c)** Evaluation of the gravitational force by means of the "particle–mesh" (PM) method, in which Poisson's equation is solved by a mass assignment to a discrete grid [209]. This method is better suited when a large number of particles ($\gtrsim 10^5$) is required, although the small scale resolution is limited by the grid spacing. For this reason, PM codes fail to describe in detail the structure of small scale virialized clumps, although they are adequate to follow the evolution at intermediate scales.

**d)** Combination of the direct integration and of the PM method, in order to improve the force resolution of the PM code. The resulting "particle-particle–particle-mesh" (P$^3$M) code corrects the small scale force acting on each particle by summing over the contributions from neighbor particles [138, 209, 137]. This makes the resolution of the P$^3$M code considerably better than that of the PM scheme, especially when a strong clustering develops. In fact, in this regime the resolution of the PM code is in any case fixed by the mesh size, while that of the P$^3$M code essentially depends on the mean interparticle distance (much smaller than the mesh size inside clustered structures), and is only limited by the softening scale for the particle–particle force. In this section I will discuss results based on a P$^3$M code.

## 6.2  Numerical integrations

In the P$^3$M code the force acting on a particle is split between a long–range (PM) component and a short–range (PP) part (see the Hockney & Eastwood's book [209] and the Efstathiou et al. paper [137] for more technical details and applicative aspects). As a first step in the PM computation, the density field $\rho(\vec{x})$ is represented by a mass assignment to the grid point positions $\vec{y}_n$ according to

$$\rho(\vec{y}_n) \; = \; \frac{M^3}{N} \sum_{i=1}^{N} W(\vec{x}_i - \vec{y}_n)\,. \qquad (208)$$

Here $M$ is the total number of grid points, $N$ the particle number, $\vec{y}_n$ and $\vec{x}_i$ the vector positions of the $n$–th grid point and of the $i$–th particle, respectively. In the following, periodic boundary conditions are chosen, while simulation box side, gravitational constant and total mass are set to unity. The key quantity in eq.(208) is the mass assignment function $W$, whose shape should be chosen to be as smooth as possible in order to have a well behaved density field representation. In the code we used, a triangular–shaped cloud (TSC) interpolation scheme for mass assignment is adopted, in which the mass is assigned with suitable weights to the 27 nearest neighbor grid points around each particle (see ref.[209], §5.3).

With the density field of eq.(208), the Poisson's equation gives for the gravitational potential on the grid

$$\Phi(\vec{y}_n) \; = \; \frac{1}{M^3} \sum_{m=1}^{M^3} G\left(\vec{y}_n - \vec{y}_m\right) \rho(\vec{y}_m)\,, \qquad (209)$$

where $G$ is an approximation to the Green's function of the $\nabla^2$ operator. In the simulation, that I will describe, the mass distribution is described with $N = 32^3$ particles, while the density field is solved over $64^3$ grid points. After a suitable choice for an optimized Green's function, the potential $\Phi$ is solved by a fast Fourier transform (FFT) algorithm. Accordingly, the force at the grid points,

$$\vec{F}(\vec{y}_n) = -\frac{D_n \Phi}{N}, \tag{210}$$

is evaluated (here $D_n$ is the finite difference approximation to the gradient) and gives the force acting on the $i$–th particle,

$$\vec{F}(\vec{x}_i) = \sum_{n=1}^{M^3} W(\vec{x}_i - \vec{y}_n)\, F(\vec{y}_n)\,, \tag{211}$$

as its convolution with the mass assignment function.

As for the short–range (PP) force, it is evaluated by summing all the contributions from the neighbor particles within a distance $r_s$, while it is set to zero for interparticle separations $> r_s$. Within $r_s$ the force between two particles is represented by two attracting mass clouds, centered on particle positions, with linear density profile and radius $\eta/2$. For the present simulations, $r_s = 2.7/M$ and $\eta = 0.3/M$ for the smoothing scale of the interparticle force.

In a $\Omega = 1$ matter dominated Universe, the expansion factor scales as $a(t) \propto t^{2/3}$. We set $a(t_i) = 1$ at the initial time of the simulation. The simulation amounts to solve for each particle Newton's equation of motion, which in comoving coordinates reads

$$\frac{d\vec{v}_i}{dt} + 2H\vec{v}_i = \frac{\vec{F}_i}{a^3 m} \tag{212}$$

where $\vec{F}_i$ is the force acting on the $i$–th particle, which is evaluated according to the above prescriptions, and $\vec{v} = d\vec{x}_i/dt$. After introducing the new time variable $p = a^\alpha$, particle positions are displaced at each integration step according to the time–centered leapfrog scheme [137]. Particle positions at the $j$–th integration step are related to those at the $(j-1)$–th step and to the velocity at $j - 1/2$ according to

$$\vec{v}_{j-1/2} = \frac{\vec{x}_j - \vec{x}_{j-1}}{\Delta p}\,. \tag{213}$$

The integrations must be stopped when perturbations with wavelength of the order of the box length, $L$, are entering the non–linear regime. For physical spectra, like the CDM one, the comoving length of the simulation cube must be fixed ($L = 32.5h^{-1}$ Mpc at the present epoch, for the CDM simulations that I will discuss). On the contrary, scale fixing is completely arbitrary in the case of scale free spectra.

Initial conditions are set according to the Zel'dovich algorithm (see §5.1.2 and ref.[137]). Eulerian positions and velocities are given by

$$\begin{aligned} \vec{x} &= \vec{q} - b(t)\vec{\psi}(\vec{q}) \, ; \\ \vec{v} &= -\frac{db}{dt}\vec{\psi}(\vec{q}) \, . \end{aligned} \qquad (214)$$

Here $\vec{q}$ is the lattice (Lagrangian) coordinate of the particle, $b(t)$ the linear growth factor of the perturbation ($b \propto a$) and $\vec{\psi}(\vec{q})$ is the gradient of the gravitational potential

$$\Phi(\vec{q}) = \sum_{\vec{k}} \frac{\delta_{\vec{k}}}{k^2} e^{i(\vec{k}\cdot\vec{q}+\phi_{\vec{k}})} \, , \qquad (215)$$

which gives the initial particle displacement. Because of the finite discrete number of waves representing the continuous Fourier spectrum, the final result must be averaged over several integrations with equivalent statistical conditions. In eq.(215) the Fourier transform of the fluctuation field is related to the power spectrum $P_k$ according to $\delta_{\vec{k}} = \sqrt{-2\ln r_1}\, P_k$, while $\phi_{\vec{k}} = 2\pi r_2$ is the random phase, which ensures the Gaussianity of the initial fluctuation field, via the Central Limit Theorem. Here $r_1$ and $r_2$ are two random numbers between 0 and 1, so that different random sequences for $r_1$ and $r_2$ amount to generate different random realizations of statistically equivalent initial fluctuation fields. Especially for spectra having a considerable amount of large scale power, even starting with statistically equivalent intial conditions we end up with substantially different structures. For this reason, it is recommended to take several different random phase assignment for each model.

If $N$ particles are employed, wave modes along each axis having $\lambda < 2/N^{1/3}$ cannot be adequately sampled. Thus, initial perturbations at wavenumbers exceeding the Nyquist frequency $k_{Ny} = 2\pi\frac{N^{1/3}}{2}$ cannot be generated. Since for $k > k_{Ny}$ white noise dominates, the normalization of the initial spectrum has to be set to the white-noise level, $|\delta_k|^2 = 1/N$, at $k = k_{Ny}$. Hereafter we shall consider two scale-free integrations with spectral indexes $n = 1$ and $n = -2$ (SF+1 and SF-2 models, respectively), as well as a CDM initial spectrum. Scale-free spectra with $n = 0, -1$ represent only intermediate cases between the $n = -2$ and $n = 1$ cases [141]. Figure 27 reports some particle configurations at different evolutionary stages for the initial CDM spectrum. It is apparent that, despite initial conditions are nearly homogeneous, clustering rapidly develops and gives rise to a remarkable complexity of structures, with clumps and filaments extending over scales comparable to the box length.

In order to properly compare the outputs of evolving N-body simulations to the observed large scale clustering we need a suitable prescription to choose the dynamical epoch in the simulation to be identified with the present Universe. Different possibilities exist to define the present time, which relies on the comparison with the observed value of the variance $\sigma_R^2$ or of the $J_3$ integral for galaxies (see eq.[60]). A further approach, on which the simulations I consider are based, identifies the present time with the integration step at which the logarithmic slope of the 2–point correlation function matches the observational value $\gamma \simeq 1.8$, over a comparable scale–range. Once this normalization has been chosen, the amplitude of $\xi(r)$ is also fixed. However, with such a prescription, the DM correlation length at present time turns out to be much smaller than the canonical value $r_o \simeq 5h^{-1}$ Mpc, holding for galaxies. In order to overcome this problem, Davis et al. [111] suggested that, in the spirit of biased galaxy formation, the entire particle distribution in the simulations is not a faithful representation of the galaxy distribution. Instead, galaxies should be identified with those particles, which correspond to peaks of the initial fluctuations exceeding $\nu$ times the r.m.s. value of the smoothed density field. By taking $r_s = 2\pi/k_s \simeq 5 \times 10^{-3}$ (in units of the box length) for the smoothing length and $\nu \simeq 2$ peaks of an initial CDM spectrum, Davis et al. [111] found that the corresponding biased distribution has the correct clustering length when $\gamma = 1.8$. As an extra bonus, the resulting number density of selected particles reproduces quite well the observed galaxy number density, thus supporting the reliability of the CDM model to account for the observed clustering at intermediate scales.

It is clear that introducing biasing is not required by the analysis of scale–free spectra. In this case, when the slope of $\xi(r)$ takes the correct value, the correlation length $r_o$ can be matched to the observed one simply by suitably rescaling the box size. Nevertheless we prefer to apply biasing also for the SF+1 and SF–2 models, in order to check the effects of gravitational clustering on the statistics of peak distributions. In the following I will discuss the clustering properties of particle populations associated to $\nu = 0.5$ and 2, other than of the whole distribution, so to check the effect of progressive biasing.

## 6.3 Correlation properties

The 2–point function $\xi(r)$ for the CDM spectrum is shown in Figure 28, Only data about the present epoch $[a(t) = 2.5]$ are plotted, for the three different density thresholds. The effect of introducing a biasing in the distribution of particles, in the CDM case, is that of increasing the clustering. Indeed, in the presence of a spectrum with sufficient power at large scales, high peaks occur with enhanced probability in the crests rather than in the valleys of a large scale fluctuation mode, so that they exhibit an enhanced clustering.

Further informations about the nature of non–linear clustering are provided by higher–order correlation functions. Analyses of the 3–point correlation functions from N–body simulations have been discussed by several authors (see, e.g., refs.[111, 141]), and shows that it approaches the hierarchical expression as non–linear clustering takes place. In our analysis we evaluated correlations up to the fourth–order [395]; going a step forward in the correlation order is rather interesting, since finding the relations existing between different order correlations allows one to discriminate between different models of non–linear clustering. Actually the 4–point function represents the lowest correlation order to test this recurrence.

In order to evaluate higher–order correlations a very useful approach is represented by the moment method. This technique is based on the counting of the objects inside spherical shells centered on each object, rather than on the counting of $n$–plets of objects, as the direct counting approach does [372]. The major advantage of the moment method lies in its high computational speed with respect to the counting of multiplets. For this reason, it is particularly suitable for the investigation of higher order ($n > 3$) correlations (see ref.[395] for a detailed description of this method). Let $N_{kj}(\delta)$ the number of particles contained in a spherical shell from $r_k$ down to $r_{k-1} = r_k(1-\delta)$, centered on the $j$–th object. The $z$–th order moment, $\langle N^z \rangle_k \equiv N^{-1} \sum_{j=1}^N N_{kj}^z$, is then obtainable by averaging over all the $N$ objects taken as centers, while the *central moments*

$$s_k^{(z)} = \langle (N_{kj} - \langle N \rangle_k)^z \rangle_j \tag{216}$$

are related to the $(z+1)$–point correlation function. For $z = 2$ we have

$$s_k^{(2)} = \langle N \rangle_k + \left(\frac{N_k}{V(r_k)}\right)^2 \int_{r_k}^{r_{k+1}} dV_1 \, dV_2 \, [\zeta_{012} - \xi_{01}\xi_{02} + \xi_{12}], \tag{217}$$

while the case $z = 3$ reads

$$s_k^{(3)} = 3s_k^{(2)} - 2\langle N \rangle_k + \left(\frac{N_k}{V(r_k)}\right)^3 \int_{r_k}^{r_{k+1}} dV_1 \, dV_2 \, dV_3 \, [\eta_{0123} -$$
$$- 3\xi_{01}\zeta(r_2, r_3, r_{23}) + \zeta(r_{12}, r_{23}, r_{13}) + 2\xi_{01}\xi_{02} + \xi_{02}]. \tag{218}$$

Here $N_k$ represents the expected number of objects in the $k$–th shell for a uniform (random) distribution, and $V(r_k)$ is the volume of the $k$–th shell. In eqs.(217) and (218), $\zeta$ and $\eta$ represent as usual the connected parts of the 3– and 4–point correlation functions, respectively. In this analysis we assume for the 3–point function the model

$$\zeta_{012} = Q \left[\xi_{01}\xi_{02} + \xi_{01}\xi_{12} + \xi_{02}\xi_{12} + q \, \xi_{01}\xi_{02}\xi_{12}\right]. \tag{219}$$

For q=0, the above expression gives the hierarchical model of eq.(28). The Kirkwood superposition, arising from eq.(219) with $q = 1 = Q$, is predicted by analytical approaches to biased models, as shown in §5.4. As for the 4–point function, the hierarchical expression

$$\eta_{0123} = R_a[\xi_{01}\xi_{12}\xi_{23} + ... 12 \, terms] + R_b[\xi_{01}\xi_{02}\xi_{03} + ... 4 \, terms] \tag{220}$$

is assumed.

The results of the analysis for the SF+1 and CDM models are shown in Figures 29 and 30 for different dynamical epochs and biasing levels. The upper panels shows the scale dependence for the hierarchical coefficient $Q_h$, the central panels are for the Kirkwood coefficient $Q_k$ (actually $Q_k = 1$ for the Kirkwood model to hold), while the lower panels are for a suitable combination, $R_a + pR_b$ of the 4–point hierarchical coefficients (the precise value of $p$ depends on the shape of the 2–point function, but the value $p = 0.35$ is adequate in all the considered cases). In this kind of plot, the scales where such coefficients are nearly constant represent the range of validity of the respective models.

The Zel'dovich spectrum rapidly develops a strong small scale clustering, so that it is the first to attain the regime in which hierarchical correlations are expected to arise. In this case, no differences are detected between different biasing levels, in agreement with the expectation that threshold effects are unimportant when spectra with a small amount of large scale power are considered. The coefficient $Q_h$ rapidly takes a rather constant value over a wide range of scales, as non–linear evolution takes place. A similar behaviour is shown by the combinations $R_a + pR_b$ of the 4–point coefficients. As in the case of the 3–point function, the 4–point one turns out to approach the hierarchical expression as non–linear clustering develops. It is also interesting to note the behaviour of the coefficient $Q_k$ for the 3–point Kirkwood superposition. The very small value ($Q_k \simeq 10^{-3}$) it takes at $r \ll r_o$ indicates the non–validity of the Kirkwood model in the non–linear regime. Indeed, since $\xi \gg 1$, the presence of the $\xi^3$ term constrains $Q_k$ to be very small. In any case there are no indications supporting the validity of this model, for both the background and the biased particle distributions.

As for the CDM simulations, the presence of a relatively limited amount of small scale power causes a much slower development of non–linear clustering. As a consequence, gravitational dynamics is not able to generate hierarchical correlations with the same efficiency as in the SF+1 case. In fact, the hierarchical coefficients plotted in Figure 30 always display a remarkable scale dependence, which is only marginally weakened by leaving the simulation evolve or taking high peak distributions. This result agrees with more recent independent analyses of $N$–body simulations. Bouchet & Hernquist [69] analyzed moments of count–in–cells developed by initial CDM, HDM and white–noise initial spectra, while Lucchin et al. [259] realized a similar analysis assuming power–law initial spectra, having spectral index in the range $-3 \leq n \leq 1$. These analyses confirm that deviations from hierarchical correlations are anyway observed, which are more and more apparent as the relative amount of small scale power is decreased in the initial conditions. A possible explanation for this could be that gravitational clustering requires in some cases more time to relax to the hierarchical behaviour than is allowed by numerical limitations. This is particularly true for spectra having a considerable amount of large scale power. Indeed, in this case clustering simultaneously develops over a broad scale range, at large scales the the mildly non–linear regime is rapidly attained and simulations need to be stopped before small scale clustering has gone to a complete virialization. For these reasons, $N$–body simulations with a large dynamical range are needed to clarify whether hierarchical correlations really represent the asymptotic stage of non–linear clustering, as predicted by several theoretical models, or it is not a necessary consequence of gravitational dynamics.

In Figure 30 it is also plotted the 3–point hierarchical prediction by Fry [161], based on his second–order perturbative approach to fluctuation evolution. The corresponding $Q_h$ coefficient, plotted as open triangles, is normalized in such a way that $Q_h = 1$ should be observed for the Fry's model to be verified. Also in this case, $Q_h$ displays a steep scale-dependence, away from the prediction of perturbative theory. Only approaching the linearity scale the $Q_h$ value decreases to an adequate level, thus indicating that the dynamical range where the perturbative approach works should be rather narrow.

## 6.4 Fractality of the non–linear clustering

In §4.4 I have shown that statistical descriptions of clustering based based on correlation and fractal approaches are strictly connected. In particular, the hierarchical expression for correlation functions turns into a monofractal scaling inside the overdense regions. Although very useful, statistical informations based on high–order correlation functions suffer for two main limitations. First, correlation functions are sensitive only to the distribution inside the overdensities, while they say essentially nothing about the devoid regions. On the contrary, the $D_q$ spectrum of generalized dimensions gives a comprehensive description of the scaling properties of a distribution; positive $q$'s deal with overdensities, while negative $q$'s are for the underdensities. Second, extracting connected high–order correlations is a quite hard task, and statistical noise rapidly pollutes any signal as the correlation order is increased.

For these reasons, fractal analysis represents a valid alternative to correlation functions. Actually, it turns out to be extremely well suited to closely investigate the scaling properties associated to non–linear gravitational dynamics. Recently, several groups attempted to quantify this scaling by applying fractal analysis to cosmological N–body simulations [396, 104, 421], obtaining consistent results. In the following, I review the main results of our fractal analysis [396] and I refer to our original paper for more details about the implementation of the analysis.

In order to follow the formation and evolution of multifractal structures in the simulated universes, it is necessary to resort to the fractal dimension estimator described in Section 4. The relative partition functions for the three initial spectra (CDM, SF–2 and SF+1 models) are evaluated for each value of the expansion factor $a(t)$ and of the biasing level $\nu$. In general, we consider spatial scales above $\sim 10^{-2.5}$ (in units of the box length), since at smaller scales the results turn out to be affected by the finite numerical resolution.

The box–counting (BC) partition function $Z^B(q,r)$ is evaluated according to eq.(97) by dividing the simulation box in cells of varying size and counting the number of particles contained inside each of them. Cell size varies between $r = 1/640$ and $r = 1/3$ in units of the simulation box length, while $q$ values are taken between –8 and +8, with step $\Delta q = 0.25$.

For the correlation–integral (CI) partition function $Z^C(q,r)$ of eq.(98), we count the number of particles, $n_i(<r)$, contained inside a sphere of radius $r$, centered at the $i$-th particle. The values of $r$ vary between $1/500$ and $1/10$ of the simulation box–length. The values of $q$ for which $Z^C$ is evaluated are the same as those used for the box–counting method.

In the case of the density reconstruction (DR) method, in order to estimate the $W(\tau,p)$ partition function of eq.(99) the values of the probability $p$ are taken between $2/N$ and $0.1$. For each particle $i$ one has to find the radius $r_i$ of the smallest sphere centered on $i$ such that there are $Np \equiv m$ particles inside the sphere. Note, however, that the upper limit $p = 1$ cannot be defined for a periodic system, like the simulations we are considering, for which only scales

$\leq L/2$ are meaningful.

As far as the BC and CI methods are concerned, the scale range over which the fractal properties of the particle distributions are analyzed naturally follows from the choice of the cell size or of the sphere radius. In the case of the DR partition function, the evaluation of the local dimension is not performed in a range of physical scale, but in a range of probabilities $p$. Thus, in order to compare the scaling properties detected with such a method with those arising from the previous two methods, a suitable criterion is required to associate each value of $p$ to a characteristic scale. To this purpose, let us express the probability $p$ in terms of the 2–point correlation function $\xi(r)$, according to

$$p = \frac{\langle N(<r) \rangle}{N} = \frac{\bar{N}(<r)}{N}[1 + K_1\,\xi(r)]. \qquad (221)$$

Here $\langle N(<r)\rangle$ and $\bar{N}(<r)$ are the mean number of neighbors within a distance $r$ for the real data and for a random distribution, respectively. Moreover, if $\xi(r) = (r/r_o)^{-\gamma}$, then $K_1 = 3/(3-\gamma)$. Thus, solving eq. (221) in $r$ for each value of $p$ gives the relation between the probability range and the physical scale range. Note, however, that since different particle distributions have different 2–point functions, the interval of scales corresponding to a fixed interval of $p$ values is not uniquely determined for all simulations. As a further possibility, we can also take the value of $W(\tau = -1, p)$, which is, by definition, the average separation associated with a given probability. Similarly, $W(\tau = -2, p) - W^2(\tau = -1, p)$ provides an estimate of the variance of the distribution of the relative separations among the particles, so that its value is related to the amount of scale–mixing in the DR method; smaller values correspond to a better separation of scales.

The implementation of the nearest–neighbor (NN) method is based on the evaluation of the partition function $G(k, n, \tau)$, as defined by eq.(100), taking 16 randomly selected subsamples of the whole particle configuration. The poorest subsample always contains 500 points. The NN partition function is evaluated in correspondence of the same values of $\tau$ already selected for the DR function; the NN function has been computed up to the fourth neighbor order. The set of $\tau$ values for which the NN partition function $G$ has been evaluated is the same as that chosen for the DR method. In order to estimate the physical length scale associated to a given subsample containing $n$ particles, we evaluate the frequency distribution of the $_k\delta_i$ distances for each subsample and for each neighbor order $k$. Then, the typical scale associated with each choice of $n$ is given by the value of $_k\delta$ that corresponds to the peak of the frequency distribution. Note however that this procedure provides a rigorous scale identification only in the ideal case when the frequency distribution approaches a Dirac delta function; more generally, different scales contribute with different weights to the value of the partition function. Clearly, the scale resolution of this method is accurate only when the frequency distributions have a rather peaked shape.

For the implementation of the minimal spanning tree (MST) algorithm, we evaluate the partition function $S(m,\tau)$ of eq.(103) for 21 randomly selected subsamples. The selected values of $\tau$ are the same as for the previous two methods. Similarly to what has been done for the nearest neighbor method, the procedure which has been followed to estimate the range of physical scales sampled by the MST algorithm is based on finding the peak in the frequency distribution of the edge lengths inside the tree.

### 6.4.1 Analysis of the CDM simulations

Figure 31 shows the partition functions $Z^B(q,r)$ for the BC method. Each panel refers to a different epoch $a(t)$ and to a different value of $q$; time increases from left to right and the order $q$ of the moment increases from bottom to top. In each panel, the partition functions for the background ($\delta \geq -1$) and for two different biasing levels ($\nu \geq 0.5$ and $\nu \geq 2$, respectively) are plotted. Results for $q = 0,2,4,6$ are shown. At negative $q$'s, the partition function provides information on the distribution in the underdense regions. However, for $q < 0$ the partition function $Z^B$ is dominated at small scales by the contribution of boxes containing only one particle. Thus, $Z^B$ is quite independent of $r$ at scales smaller than the average separation among the particles in the underdense regions. As a result, the BC algorithm does not provide a good characterization of the distribution in the underdense regions, unless we have a very large number of particles. At positive $q$'s, the partition function $Z^B(q,r)$ shows an approximate power law behavior. The scaling regime becomes more and more evident as the clustering evolves. Note for example the existence of a plateau in $Z^B(q,r)$ at initial times, at scales up to about $10^{-2}L$, which is due to discreteness effects. This plateau is moved to smaller scales as the clustering piles up the particles in the fractal regime at small scales. The scaling behavior seems to be more evident at higher density thresholds and at higher $q$ values, suggesting that high–density regions attain a fractal distribution before the lower density ones. In a sense, the scaling properties at large $q$'s, even at the initial stages of the evolution, mimic those of the high peak distribution and anticipate the properties which will be attained by the distribution of the background particles at later evolutionary times. At large values of $q$, the differences between the background and the biased distributions become less pronounced, since only the largest peaks (present at all biasing levels) are weighted by the partition function.

In order to estimate the spectrum of generalized dimensions, a linear least–square fit of $\log Z^B(q,r)$ versus $\log r$ is required over the scale range where self–similarity is detected, that is below the linearity scale (i.e., the scale corresponding to $\xi(r) = 1$). The spectrum of generalized dimension is summarized in Figure 33 where I plot the $\tau_q$ curve obtained by means of all the employed methods in the case of the CDM spectrum. The open circles refer to the results of the BC method. From Figure 33 one sees that the generalized dimensions $D_q$ have a value of about one for positive $q$'s, with some evidence of a weak but systematic decrease for increasing values of $q$ and for increasing biasing levels. These results indicate that the generalized fractal dimensions (for positive $q$'s) decrease from an initial value $D_q \simeq 3$ to a value of order one as the non–linear gravitational clustering evolves. During the evolution, discreteness effects are also reduced as the particles approach each other and create the small scale fractal structure.

In order to analyse the $q < 0$ tail of the $D_q$ spectrum, the DR method is required (see §4.2). In Figure 32 I show the DR partition function $W(\tau, p)$ for different levels of bias, epochs $a(t)$ and $\tau = -6, -2, 2, 6$. According to eq.(221), the considered values of $p$ correspond approximately to the range of scales between $10^{-2}L$ and $10^{-1}L$. For $\tau \leq -2$ a good scaling regime is observed in the partition function $W(\tau, p)$. The generalized dimensions have a value exceeding three for $\tau \ll 0$, indicating that the scaling inside the underdense regions is characterized by the presence of minima of the density field. Note that such high values for the generalized dimensions are not generated by the fact that the DR partition function samples the large scales where homogeneity is present, since the chosen values for the probability $p$ correspond to physical scales that never exceed the non–linearity scale. On the other hand, for $\tau > 0$ the scaling behavior of the DR partition function is well established and it provides indications of a dimension $D_q \simeq 1$–1.5. The generalized dimensions tend to decrease to a value of about one as the biasing level increases. Note, in general, that some care has to be taken in the evaluation of the DR partition function, due to the existence of a numerical softening length at the lower limit of the scaling regime, which increases the dimension. Since the DR partition function tends to weight also scales smaller than those sampled by the previous methods, the slightly larger values of the dimensions corresponding to positive $q$'s found for the background could well be due to the effect of the softening length. For the highest level of biasing, however, a generalized dimension of about one is correctly reproduced; this is due to the fact that in this case there are fewer particles and a given level of probability corresponds to scales which are definitely larger than the softening length. The plateau observed for the biased distribution for $p > 10^{-1.5}$ and $\tau > 0$ is associated with the break in the small scale fractal regime and, correspondingly, with the transition toward large scale homogeneity.

Putting together the results of applying these and the other dimension estimator described in §4.2 one can draw the following picture, which is summarized in Figure 33. In this figure we report the value of $\tau(q) = (q-1)D_q$ versus $q$ are reported for the different methods of analysis. For $q > 0$, the three reliable methods are represented by the BC, CI and DR algorithms. The results of the analysis indicate that, due to non–linear gravitational evolution, the generalized dimensions for positive density fluctuations very rapidly approach to one. The dimension becomes slightly lower for higher values of $q$ and for the higher biasing levels, thus indicating a weak multifractality. The correlation dimension $D_2$ turns out to be between about one and 1.2 as expected from the classic correlation function studies. In general, during the gravitational evolution the dimensions of the distribution in positive density fluctuations become more and more close to one and to a monofractal (for positive $q$'s) behavior. Thus, the value of about one for the generalized fractal dimensions of the galaxy distribution seems to be built by the non–linear gravitational clustering.

For negative values of $q$, the reliable methods are represented by the DR partition function,

while some care must be used when applying the NN and MST algorithms. All these methods provide generalized dimensions exceeding three for extremely negative $\tau$'s, indicating that the particle distribution in the underdense regions contain minima of the density field, instead of singularities, as in the overdense regions. Among the generalized dimensions, a peculiar role is played by the capacity dimension $D_o$, owing to its independence of the statistics and to its pure geometrical meaning. For multifractal distributions, the capacity dimension $D_o$ is known to be a decreasing function of the threshold, its variation reflecting the trend of the generalized dimensions $D_q$ for the background. Thus, the dependence of the fractal properties on the biasing threshold should be significant in the case of $D_o$. However, the capacity dimension is quite difficult to compute in practice. In our case, both the BC and the CI methods do not have a clear scaling behavior for $q = 0$, due to strong discreteness effects. On the other hand, the DR, NN and MST methods indicate that $q = 0$ occurs for $\tau \simeq 3$, so that $D_o \simeq 3$ for the background. The capacity dimension slowly decreases to values between 2.5 and three as the gravitational evolution proceeds, since particles are moved into the more clustered regions and the underdense areas are less and less sampled (this effect would presumably disappear in the limit of a huge number of particles, that adequately sample the continuous matter field). For the biased distributions, the situation is more complex. In this case, all the methods provide values of $D_o$ at the final evolutionary stages between about 1.5 and 2.2. In Table 1 we show the values of $D_0$, obtained from the DR algorithm, for different levels of bias, epochs $a(t)$ and initial spectra. These values are rather interesting, being in agreement with the findings of Martínez & Jones [265], who determined a value $D_o \simeq 2$ for the 14.5 CfA sample. In fact the DR method gives a value of $D_0 \simeq 1.7$ for the peak distribution of the CDM model at the present epoch.

### 6.4.2 Dependence on the initial spectrum

The fractal analysis of the CDM model answered with a firm 'yes' to the question about the possibility to generate a self–similar (fractal) clustering through the development of non–linear gravitational dynamics. A further question one may ask is whether the resulting fractal structure is universal, i.e. independent of the initial spectrum, or it keeps memory of initial conditions. To answer this question, I discuss also the case of the two power law spectra, with spectral index $n = -2$ and $n = 1$. These spectra correspond to the two extreme cases of large power at large scales and at small scales, respectively.

For the case of an initial scale–free spectrum with spectral index $n = -2$ one expects a weaker evolution of the clustering behavior at the non–linear scales. In general, the scaling properties of the partition functions for the SF–2 model are quite similar to those found in the study of the CDM evolution, apart from some interesting details connected with the weaker clustering evolution. The case of the $n = 1$ spectrum is known to be of limited cosmological interest, as far as the formation of structures at small scales is concerned. However, as also shown by the correlation analysis, this spectrum represents the best model for studying scale–free spectra, since it gives to some extent a good characterization of the small scale distributions at very evolved stages. This is clearly due to the presence of a large amount of power at small scales.

Table 1: The generalized dimension $D_0$ as a function of $a(t)$ for different models and levels of bias. $D_0$ has been estimated using the DR method.

|  |  | $a(t)$ | $D_0$ |
|---|---|---|---|
| $\delta > -1$ | $CDM$ | 2.0 | $2.78 \pm 0.02$ |
|  |  | 4.5 | $2.68 \pm 0.02$ |
|  |  | 5.0 | $2.69 \pm 0.02$ |
|  | $n = -2$ | 4.2 | $2.73 \pm 0.02$ |
|  |  | 4.8 | $2.72 \pm 0.02$ |
|  |  | 5.5 | $2.72 \pm 0.02$ |
|  | $n = 1$ | 40.0 | $2.27 \pm 0.03$ |
|  |  | 50.0 | $2.24 \pm 0.03$ |
|  |  | 60.0 | $2.20 \pm 0.03$ |
| $\nu > 0.5$ | $CDM$ | 2.0 | $2.55 \pm 0.04$ |
|  |  | 4.5 | $2.42 \pm 0.03$ |
|  |  | 5.0 | $2.42 \pm 0.03$ |
|  | $n = -2$ | 4.2 | $2.50 \pm 0.02$ |
|  |  | 4.8 | $2.48 \pm 0.02$ |
|  |  | 5.5 | $2.48 \pm 0.02$ |
|  | $n = 1$ | 40.0 | $2.20 \pm 0.04$ |
|  |  | 50.0 | $2.18 \pm 0.04$ |
|  |  | 60.0 | $2.17 \pm 0.04$ |
| $\nu > 2.0$ | $CDM$ | 2.0 | $1.73 \pm 0.08$ |
|  |  | 4.5 | $1.50 \pm 0.06$ |
|  |  | 5.0 | $1.48 \pm 0.05$ |
|  | $n = -2$ | 4.2 | $1.56 \pm 0.05$ |
|  |  | 4.8 | $1.55 \pm 0.05$ |
|  |  | 5.5 | $1.54 \pm 0.05$ |
|  | $n = 1$ | 40.0 | $1.24 \pm 0.08$ |
|  |  | 50.0 | $1.19 \pm 0.07$ |

In Figure 34 I report the $D_\tau$ spectrum of dimensions for the three considered spectra at the last evolutionary stage. The dimensions have been estimated through the DR partition function, which is the only one that works fine both for positive and negative $\tau$'s, by a log–log linear regression in the scale range where self–similarity is detected. From left to right, results for CDM, SF–2 and SF+1 are plotted, while from bottom to top the background particle distribution and the two biasing levels, $\nu = 0.5$ and 2 are considered.

For the SF–2 model, at negative $\tau$'s the generalized dimensions are again above three. In this regime, no clear differences between the background and the biased distributions are visible. For $\tau > 0$ the dimensions for the background turn out to be larger than those for the CDM spectrum. In particular, $D_2 \simeq 1.2$ at the end of the evolution for the $\nu = 2$ biased distribution, while $D_2 \simeq 1.6$ for the background. The generalized dimensions of the background show a weak decrease at increasing values of $\tau$ and remain asymptotically larger than one ($D_8 \simeq 1.4$). For the biased distributions the generalized dimensions definitely tend to the value $D = 1$ at increasing $q$'s. Thus, while the largest density peaks have evolved and have reached the regime of non–linear gravitational clustering (as indicated by $D_\tau \simeq 1$ for $\tau > 0$), the background is still in a less evolved state. Again, the Hausdorff dimension plays an important role. The results for the SF–2 model show that $D_o \simeq 3$ for the background, while it becomes of order two for the biased distributions, analogous to what happens for the CDM spectrum. The fractal properties of this distribution, as well as the values $D_q \simeq 1$ for the biased distributions, seem to be well established, that is, they do not appear to be a quickly evolving transient regime.

Again, for the SF+1 model it is $D_\tau \gtrsim 3$ at negative $\tau$ values, for all the biasing levels. For $\tau > 0$ the generalized dimensions have now a value slightly lower than one for the highest biasing level, while the unity value is anyway recovered for lower biasing levels. In this case, the $D_o$ dimension turns out to be about two for both the background and the moderately biased distributions. However, differently from the other two models, we find $D_o \simeq 1$ for the biasing level $\nu = 2$. The rapid evolution typical of this spectrum has evidently induced a stronger dependence of $D_o$ on the biasing level. In the case of the $n = 1$ spectrum, essentially all particles have undergone non–linear gravitational clustering, and they no longer provide a good sampling of the background density field. By using a much larger number of particles the correct capacity dimension for the background, $D_o = 3$, should be recovered.

Apart from the details about the values of each single dimension, a rather interesting result concerns the overall shape of the $D_q$ spectrum at positive $q$'s. From Figure 34 it is apparent that, as models with stronger small scale clustering or more biased distributions are considered, the positive–$q$ tail of the dimension spectrum becomes flatter and flatter. According to what demonstrated in §4.4, this turns into correlation functions, which approach the hierarchical pattern, thus confirming the result based on correlation analysis. Further, this suggests that, although non–linear gravitational clustering tends *asymptotically* to erase memory of initial conditions by always generating $D_q \simeq 1$ for $q \gg 1$ (i.e., inside the most virialized structures), for $q \lesssim 4$ the initial spectrum affects the $D_q$ shape, up to the most evolved configurations that we considered.

### 6.4.3 Outlook

A general conclusion that has been reached from the fractal analysis is that a dimension of about one for large and positive positive $q$'s is in general produced by the non–linear gravitational evolution, quite independently of the shape of the initial spectrum. Some differences among the different spectra can however be detected. The clustering evolution is slower in the case of the CDM and $n = -2$ models. Consequently, the dimension of the particle distribution in the background remains slightly but systematically larger than one, approaching one only for the highest peaks. For these spectra a systematic decrease of the generalized dimensions with increasing biasing level is in general observed. The correlation dimension $D_2$ decreases from a value of about 1.3–1.4 for the background to a stability value of about one for the highest peaks. Vice versa, for $n = 1$ the evolution is faster and the small scale clustering is stronger; this is reflected in the lowering of the dimension slightly below one for positive $q$'s, without significant differences between different biasing levels. For example, the correlation dimension $D_2$ has a value of one for this case, with no dependence on the biasing level. The above results indicate that values of the dimensions between about 0.8 and 1.3 for $q > 1$ are naturally produced by non–linear gravitational clustering and by high peak selection. Thus, the outcome of non–linear clustering tends to be a monofractal with approximately equal generalized dimensions $D_q$ for positive $q$'s, especially at late evolutionary stages and for the highest biasing levels. In terms of correlation language, this is equivalent to say that higher–order correlation functions tend to be more hierarchical, with a two–point function which scales as $\xi(r) \sim r^{-\gamma}$, with $\gamma \sim 2$.

A further indication that the dimension $D \simeq 1$ is a favourite product of the non–linear gravitational evolution comes also from the observation that the density $\rho_v(r)$ in virialized isothermal galaxy halos has a dependence $\rho(r) \propto r^{-2}$, as derived from the rather flat profile of the rotation curves for $r$ larger than about 10 kpc (see, e.g., ref.[348]). This indicates that $D_2 \simeq 1$ on these scales (see also ref.[211]). We note that, while for the $n = +1$ spectrum our results are in agreement with the prediction of Peebles' scaling argument about the stability of non–linear clustering (see ref.[305] and §5.1.4), this is not true for the $n = -2$ spectrum. In this case, the above model predicts a fractal dimension $D_2 = 3 - \gamma = 2$ for the non–linear structures (see eq.[172]), remarkably different from the values that we found. Efstathiou et al. [141] have shown that the stability assumption is valid only when $\xi \gg 100$, on these scales the N–body results seem to give a quite good fit to the slope of the 2-point function according to Peebles' scaling argument. Numerical simulations of the kind employed both here and in ref[141], however, are limited in time by an epoch $a_{fin}$ when the fundamental mode of the cube approaches non–linearity. In the case $n = -2$, $a_{fin} \simeq 5$ and the range where one can test the predictions of models for non–linear clustering is very limited. On the other hand, the evaluation of $D_q$ with $q \gg 0$ selects regions of high density, i.e. where the clustering is more evolved. Furthermore, the fractal dimension for the biased distribution is smaller than for the background, and the analysis of other spectra shows that the biased particle distribution can be in a sense considered as a background population analysed at late times and, thus, at a more clustered stage. Thus, the conclusion that $D_q \simeq 1$ for the asymptotic stage of gravitationally evolved systems seems to be well supported. This result is consistent with the findings of Saslaw [354, 363], who argued that the value $\gamma = 2$ (i.e., $D_2 = 1$) is dynamically stable in the fully non–linear clustering regime. This is also in agreement with the Hoffman & Shaham [211] model for the density profile of virialized halos (see also §5.1.4). Based on the secondary infall paradigm, they found that nearly isothermal halos, having density profile $\rho(r) \propto r^{-\gamma}$ with $\gamma \simeq 2$, are a natural outcome of non–linear clustering, quite independently of the initial fluctuation spectrum (see eq.[173]). In this sense, the observed flatness of galaxy rotation curves, as well as the slope of the galaxy 2–point correlation function and the hierarchical behaviour of higher–order correlations, should be considered as products of the non–linear gravitational dynamics, while containing only limited information about the primordial fluctuation spectrum. Vice versa, memory of initial conditions should be better preserved at larger ($\gtrsim 10h^{-1}$ Mpc) scales, where clustering is still linear.

As for the capacity dimension, $D_o$, note that by definition it must be equal to three for the background particle distribution, as long as such particles represent a good sampling of the underlying density field. In fact, for a continuous density field no devoid regions can be found, so that the number of non–empty cells, $n_c(r)$, scales as $n_c(r) \propto r^{-3}$. This condition is well satisfied by the CDM and SF–2 models, while the SF+1 model shows $D_o \simeq 2$ for background particles. In fact, for this spectrum the strong small–scale clustering constraints most particles to be confined inside dense isolated clumps, thus leaving a small number of them to adequately sample the continuous density field. Also rather different for the different modes is the $D_o$ value as biasing is introduced. Selecting peaks creates devoid regions, so to lower the capacity dimension. For CDM and, even more, for SF–2, the peak distribution is rather bubbly, with devoid regions surrounded by "galaxies" (much as observed). The resulting geometry of the distribution is planar, and the corresponding Hausdorff dimension is $D_o \simeq 1.5$–$2$, as observed for the real galaxy distribution (see, e.g., ref.[265]). On the other hand, the geometry of the SF+1 peak distribution is dominated by virialized isothermal knots, so to give the lower value $D_o \simeq 1$.

The relevance of fractal analysis methods to investigate the scaling of non–linear gravitational clustering has been recently stressed by other authors. Colombi, Bouchet & Schaeffer [104] analysed the CDM models by using only the box–counting approach. They found that a lot of care must be payed in the interpretation of the results, when only this method of analysis is applied, because of the limited number of particles allowed in N–body simulations and of the finite cell sizes to sample the structure. Nevertheless, these authors recovered the results that at the small scales of non–linearity a fractal dimension $D \simeq 1$ is associated to the clustering inside the overdense regions. Yepes, Dominguez–Tenreiro & Couchman [421] applied fractal analysis methods with the aim of comparing the outputs of N–body simulations to the observed galaxy distribution of the CfA1 sample. They used the low–bias CDM simulations by Couchman & Carlberg [105] and extracted from them artificial galaxy samples. Despite the largely different galaxy identification procedure they adopted, their results about the stability of the $D = 1$ dimension to characterize the clustering of high density peaks agrees with the above discussed results. These authors also concluded that the fractal properties of the clustering in the simulation box are not significantly altered when selection effects for realistic samples are introduced, thus confirming the reliability of the multifractal analysis approach to characterize the large scale clustering of cosmic structures.

As a general conclusion, I would like to emphasize once more the relevance of demonstrating that fractal structures are naturally formed at the small scales where non–linear gravitational dynamics takes place. From one hand, this result implies that, in order to accept a purely fractal description of the galaxy distribution at arbitrary large scales, as suggested by Pietronero and coworkers [329, 100, 99], one is forced either to reject the quasi–homogeneity of the initial conditions or to imagine that the primeval field had a large scale coherence of non–gravitational origin. From the other hand, it represents one of the few examples of physical systems which develops scale invariant structures from dynamical equations, where fractality is not put in "by hand". A further example of this kind is provided by the Navier–Stokes equation, which is used in hydrodynamics to describe turbulent flows. However, the limited resolution achieved in simulations of fully developed turbulence from Navier–Stokes dynamics (see, e.g., ref.[374]) allows the detection of self–similar structures over a rather limited range of scales, narrower than that displayed by cosmological N–body simulations. Thus, apart from the well established cosmological relevance, N–body simulations of gravitational clustering have also a remarkable interest from the point of view of the study of dynamical systems.

# 7 Scaling in the cluster distribution

As I often mentioned in this article, the detection of well-defined scaling properties for the large-scale distribution of cosmic structures seems to suggest, at first glance, a fractal picture of the Universe. Due to the rather limited sizes of the volumes sampled by the available galaxy redshift surveys, no clear evidence has been found that the galaxy distribution reaches a high degree of homogeneity at the largest scales allowed by such samples. Instead, the texture of the galaxy distribution gives rise to structures, such as filaments and voids, that involve scales comparable to that of the whole sample (see Section 2). Even within this picture, evidence is emerging that the galaxy distribution has at least one characteristic scale ($\simeq 4h^{-1}$ Mpc), below which a scaling regime exists, with a correlation dimension $D_2 \simeq 1$ [265]. The analysis of the multifractal properties of cosmological $N$–body simulations, described in the previous section, shows that the gravitational instability picture with quasi–homogeneous initial conditions naturally gives rise to fractality of the matter as well as of the galaxy distribution, but only up to scales where the clustering is non–linear.

In order to search for scaling of the clustering pattern at scales larger than those allowed by the galaxy distribution, it seems appropriate to use the distribution of galaxy clusters. Although their amplified clustering has been explained in the framework of *biassed* models, it was also used as a further support to the idea that the large–scale structure is described by a self–similar fractal, extending at arbitrarily large scales [100]. In this picture, the amplification of clustering of rich galaxy systems is not an intrinsic properties, but only reflects the fact that cluster catalogues sample larger volumes with a consequent increase of the correlation strength (see §4.4). If this were the case, then we should expect that the galaxy correlation function maintains the same slope also at the scales where cluster clustering is detected, thus in disagreement with the above mentioned existence of different scaling regimes for the galaxy distribution.

In this section I review the results of fractal analysis for both angular and redshift cluster distributions. In turn, the scaling properties detected for observational data sets are compared to simulated cluster distributions, which are generated so to reproduce the same features (selection effects, observational biases, etc.) of the real samples. This allows to carefully investigate the constraints that the detected scaling for the cluster distribution imposes on the initial conditions for LSS formation.

## 7.1 The angular analysis

### 7.1.1 The samples

The angular scaling analysis that I present here is based on the Plionis, Barrow & Frenk (PBF) cluster sample [323], already shortly described in §2.3. Furthermore, I will also compare the results of such analysis with those coming from synthetic cluster samples, extracted from N–body simulations, after reproducing the observational set–up of real data.

The PBF cluster catalogues were identified, using an overdensity criterion, in the 'free–of–overlap' Lick galaxy counts. For a full description of the cluster selection procedure (cluster-finding algorithm, biases, etc.) I refer to the original paper (ref.[323]).

The cluster finding algorithm is based on identifying clusters as high peaks of the underlying galaxy cell counts, after smoothing on a suitable angular scale. If $n_{ij}$ is the unsmoothed galaxy count in the 10×10 arcmin cell, labelled by the indices $i,j$, the corresponding smoothed count is

$$n^*_{ij} = \sum_{l=i-1}^{i+1} \sum_{k=j-1}^{j+1} w_{lk}\, n_{lk}\,. \qquad (222)$$

The weights $w_{lk}$ are assigned so that $w_{ij} = 1/4$, $w_{i\pm 1,j} = w_{i,j\pm 1} = 1/8$, $w_{i\pm 1,j\pm 1} = 1/16$ and their sum is unity in order to preserve the total galaxy count. This procedure is roughly equivalent to smoothing the projected galaxy distribution with a Gaussian window on a 30 arcmin scale. After applying this smoothing procedure, the algorithm identify those cells whose smoothed count is larger than a fixed threshold value:

$$n^*_{ij} \geq \kappa \bar{n}\,, \qquad (223)$$

$\bar{n}$ being the average cell count, which is obviously preserved after smoothing. Connected cells whose galaxy count satisfies eq.(223) form a cluster. By choosing different values for the threshold parameter $\kappa$ in eq.(223), samples of clusters having different richness are constructed, the richer corresponding to higher $\kappa$ values. In ref.[61] four different thresholds have been considered, corresponding to $\kappa = 1.8, 2.5, 3$ and $3.6$ (C18, C25, C30 and C36 samples, respectively). Clusters identified at higher thresholds are also included in lower $\kappa$ samples. It is worth remembering that such a cluster identification algorithm is objective and, thus, does not introduce biases arising from a visual inspection of photographic plates, as it is believed to happen for the Abell [3] and Zwicky [428] samples. Results about correlation [324, 61] and fractal [64] analyses have shown a remarkable dependence on the clustering strength on the chosen threshold. In the following I will mainly concentrate on the results for the C36. To eliminate the gross effect of Galactic extinction, the analyzed catalogues contain clusters with $|b| \geq 40°$.

To generate the artificial samples, the same cluster–finding algorithm has been applied to simulations of the Lick map, obtained by suitably projecting N–body simulations. The description of the parent three–dimensional simulations and of the Lick map generation are described in details in refs.[97, 284]. Here I only remind that the simulations are based on a CDM spectrum, with both Gaussian and non–Gaussian initial fluctuations. As for the Gaussian case, two different epochs have been considered, corresponding to values of the biasing parameter $b = 1$ and $b = 1.5$. As for the less evolved configuration ($b = 1.5$), it turns out that it is not able to produce the observed amount of rich clusters and an adequate level of correlation amplitude. The more evolved cluster distribution ($b = 1$), instead, turns out to be more reliable. As for the non–Gaussian simulations, they are obtained by through local non–linear transformations of a Gaussian random field $w(\mathbf{x})$ (see, e.g., ref.[284] for more details about generation of non–Gaussian initial conditions in N–body simulations). The considered models are the chi–squared with one degree of freedom, based on the transformation $w(\mathbf{x}) \to w^2(\mathbf{x})$, and the lognormal one, $w(\mathbf{x}) \to \exp w(\mathbf{x})$. Each non–Gaussian model splits in two, according to the sign of the initial skewness of the primordial gravitational potential fluctuations. In positive–skewness models there is an excess probability of finding overdense regions, while in the negative–skewness ones there is more probability to find underdensities. Accordingly, the former generate through gravitational evolution dense and isolated clumps, while the latter give rise to a bubbly appearance of the large–scale clustering. Therefore, they are in principle better suited to supply the large–scale coherence, otherwise lacking in the Gaussian CDM case. In the following I will review results based on $b = 1$ Gaussian ($G_1$), negative–skewed chi–square ($\chi_n^2$) and negative–skewed lognormal ($LN_n$) models.

As an example of the resulting cluster distribution, Figure 35 shows the real as well as the simulated C25 samples in quasi equal–area coordinates

$$X = (b^{\mathrm{II}} - 90°) \sin l^{\mathrm{II}}$$
$$Y = (b^{\mathrm{II}} - 90°) \cos l^{\mathrm{II}}.$$

Due to the geometry of the simulated samples, only galactic latitudes $b^{\mathrm{II}} \geq 45°$ are covered. Already from a visual inspection, it is apparent the remarkable difference in the number of selected clusters and in their clustering pattern between different models. Note that the $LN_n$ model seems to produce too large features (voids, filaments and cluster condensations) that, even after projection to a $\sim 210h^{-1}$ Mpc depth, involve angular scales comparable to the width of the observational cone. The smaller number of clusters detected in the simulations could well be due to a marginal lack of resolution in the employed N–body simulations (see ref.[60] for a detailed discussion about this point).

### 7.1.2 Results

In the analysis of angular samples we should understand the effects of projection in order to correctly infer the scaling properties of the three–dimensional distribution. To this purpose, I will also discuss the case of an artificial cluster sample obtained by projecting a three–dimensional scale–invariant structure. This allows us to check to which extent a possible detection of scale invariance in a limited angular scale–range reflects the presence of a characteristic scale even in the spatial distribution or it could be just an effect of projection.

The only difference with respect to the fractal analyses presented in Section 6 is that all the introduced algorithms are now applied for distributions projected on a sphere. In the following I will discuss results based on the correlation integral (CI) and on the density reconstruction (DR) methods. As for the other methods, I recall that the box–counting one has been proven in §4.2 to give very similar answers to those of the CI estimator, while the minimal–spanning–tree and the nearest–neighbour methods heavily suffer for limited statistics and presence of characteristic scales. The CI method, which is based on the evaluation of moments for neighbour counts, is well suited for positive–order dimensions, while it suffers for discreteness effects when underdense regions are weighted at $q < 1$. Vice versa, the DR method has been shown to rapidly converge also for negative–order dimensions. Differently from the CI approach, which fixes the angular scale as the neighbour limiting distance, the DR method displays scale–mixing. To account for this, for each value of the probability $p$ in the estimate of the DR partition function (see eq.[99]), the frequency distribution for the radii $\vartheta_i(p)$ ($i = 1, \ldots, N$) of the disks containing $n = p/N$ objects is obtained. Then, the $\vartheta$ value that is associated to a given probability measure is that corresponding to the peak of the frequency distribution.

A delicate point when dealing with data sets encompassing a finite volume concerns the treatment of boundary effects. In usual correlation analysis, the presence of boundaries is usually accounted for by comparing the statistics of the actual distribution to that of a random point distribution having the same boundaries and selection effects as the real data. This kind of procedure amounts to assume that the object distribution becomes homogeneous well within the sample boundaries. However, the purpose of the fractal analysis is to *demonstrate*, rather than *assume*, that the large–scale structure of the Universe does not behave like a self–similar structure up to arbitrarily large scales. To this purpose, a suitable treatment of boundary effect is required, which does not spuriously introduce large–scale homogeneity as an artifact. For this reason, in the implementation of the CI algorithm, the partition function at a scale $\vartheta$ is evaluated by summing only over those clusters, whose distance from the sample boundary is $\geq \vartheta$. In a similar fashion, the partition function for the DR method is evaluated at the probability scale $p$ only by summing over those clusters whose distance from its $n$–th neighbour ($n = pN_{cl}$, with $N_{cl}$ the total number of clusters) is less than its distance from the boundary. However, this boundary correction gives more weight to the overdense regions that are near the border: a cluster lying in a dense region is likely to be included in the partition sum up to quite high $p$ values, even if it is close to the boundary. Nevertheless, a comparison of the results obtained in this way with those coming from the corrected CI method shows a remarkable agreement, thus supporting the reliability of the analysis method (see below).

As for the CI method, I show in Figure 36 the corresponding partition function, as defined by eq.(98). The disk radii range from $\vartheta = 0°\!.5$ up to $\vartheta = 20°$. The left panel refers to the multifractal order $q = 0$ (corresponding to the estimate of the Hausdorff dimension), while the right panel is for $q = 4$. Also plotted in the upper panels is the local dimension $D_q(\vartheta)$, obtained through a 3–point local linear regression on the partition function. The plotted error bars refer to the $1\sigma$ uncertainty in the local fit. Note that in the case $q = 0$ the local dimension exhibits an increasing trend going from $D_o \simeq 0$ at small scales to a value $D_o \simeq 2$ at angular scales $\vartheta \gtrsim 6°$. While the constant value $D_o \simeq 2$ indicates that the cluster distribution in the angular sample appears to be homogeneous at large $\vartheta$ values, the small number of available objects causes discreteness effects to appear at small scales. Indeed, for small radii many disks contain only one object and this number does not sensibly increase until a sufficiently large $\vartheta$ value is allowed. As a consequence, the partition function values changes only very slowly and the corresponding slope is quite small. On the contrary, the discreteness should disappear when larger $q$ values are considered. In fact, in this case the CI partition function weights only the very clustered regions, where even the smallest disks have non negligible probability to contain more than one object. In fact, for $q = 4$ the local dimension remarkably detaches from zero already at small angular scales. In this case, a range of $\vartheta$ values appears, over which the local dimension takes a nearly constant value $D_q \simeq 1$ (apart from small fluctuations due to the noise in the evaluation of the local slope), up to $\vartheta \sim 6°$. Again, at larger scales the dimension approaches the value $D_4 \simeq 2$, thus indicating that also the distribution of the objects in the overdense regions turns out to be homogeneous at such scales. The appearance of a $\vartheta$-range where the local dimension is quite flat indicates that at such scales the cluster distribution has well defined fractal properties. At larger scales, however, any self-similarity turns out to be broken and the local dimension starts increasing toward the homogeneity value. In principle one could expect this result to be spurious since, with the break of scale-invariance due to some systematic effects related, for instance, to angular projection of the three-dimensional structure. In the following I will describe the effects of projection and luminosity selection on a three-dimensional scale-invariant structure with *a priori* known fractal properties. This test confirms that no characteristic scale is introduced after projection in the relevant range of $\vartheta$ values.

Further doubts about the reality of the fractal behaviour at small angular scales could arise due to the limited number of objects available in the C36 sample. If this were the case, we would expect that different methods, which have different sensitivities to the limited statistics, give different answers. However, the application of the density reconstruction method to the C36 cluster sample confirms the scale invariant nature of the cluster distribution at small scales. This is shown in Figure 37, where the DR partition function is plotted for $\tau = -2$ (left panel) and for $\tau = 3$ (right panel), together with the respective local dimensions. For $\tau = -2$ the local dimension has a rather flat shape with a value $D_{\tau=-2} \simeq 2$. This suggests that the distribution in the underdense regions is essentially space-filling. On the contrary, for $\tau = 3$ the local dimension shows a rather flat shape for small probabilities (corresponding to small physical scales), with $D_{\tau=3} \simeq 1.2$, thus remarkably similar to that coming from the CI method. At larger $p$'s the dimension starts again increasing toward the homogeneity value $D_{\tau=3} \simeq 2$. A closer comparison with the results coming from the CI approach can be done once a range of angular scales is associated to the probability range, where the DR partition function is estimated. This is done by taking for each $p$ value the $\vartheta$ value corresponding to the peak of the radii frequency distribution. In Figure 38 this frequency distribution is shown for different probabilities. A change of the shape of the frequency distribution is apparent as larger and larger probabilities are considered. While the skewed distribution at small $p$'s indicates the presence of small–scale clustering, the roughly Gaussian profile centered at the mean disk radius, $\langle \vartheta \rangle$, for larger $p$'s is the signature of the large–scale homogeneity. In the box insert also plotted is the angular scale $\vartheta$ associated to each probability value.

Based on this plot, I show in Figure 39 the expanded view of the local dimensions of Figure 37, but as functions of the angular scale. It confirms the homogeneity of the cluster distribution in the underdense regions for $\tau = -2$, while the break of the scale–invariance at $\vartheta \sim 7°$ is even more apparent for $\tau = 3$, a result which is in agreement with that obtained from the correlation integral method. This further support the self–similar behaviour of the cluster distribution at small scales, followed by a transition toward homogeneity at larger scales.

It is however clear that projecting the a three–dimensional scale–invariant distribution onto a curved surface, like the sphere, could break the self–similarity at scales where the curvature becomes relevant. Therefore, one may ask whether the breaking of scale–invariance is a spurious effect of projection or it reflects the presence of a characteristic scale even in the spatial distribution.

In order to properly answer to the above question, it is appropriate to generate a three–dimensional scale–invariant structure, with controlled multifractal properties, and project it on a sphere In addition, after assigning luminosities to each point according to a suitable luminosity function, it is possible to generate magnitude limited angular samples. In this way we are able to account for the effects of luminosity selection on the scaling properties of the spatial structure. In order to generate the three–dimensional fractal structure, I resort to the multiplicative random $\beta$–model (see §4.2). The multifractal structure is the same as that analysed in §4.2, whose $D_q$ spectrum is plotted in Figure 15. The cascading is generated with 8 iteration steps and at the end of the process the number of non empty cells is roughly $6.5 \times 10^4$. A number of particles is assigned to each cell, that is proportional to the fraction of the total mass it contains, so that the total number of particles to $5 \times 10^4$. After generating the three–dimensional structure, a luminosity is given to each point according to the luminosity function

$$\phi(L) \propto L^\beta e^{-L/L_*} . \qquad (224)$$

($L_* = 0.8 \times 10^{13} L_\odot$ and $\beta = -2$, as given in ref.[18] for the luminosity function for Abell clusters and groups of galaxies). Although the shape of eq.(224) probably does not exactly fit the real luminosity function of the clusters we are dealing with, it is reasonable to expect that it represents a good approximation to seek the effects of luminosity selection on the scaling properties of a given spatial structure. Note also that, for a point distribution that is rigorously scale–invariant, the normalization of the luminosity function cannot be uniquely assigned since the mean object density depends on the sample size. Nevertheless, I *assume* that in this "fractal universe" the luminosity distribution is assigned according to eq.(224), while its normalization is fixed only after fixing the physical length–scale sampled by the point distribution. The angular sample is realized by projecting on the sphere having diameter equal to the box side all the points contained inside there and with apparent magnitude below a fixed limit. Because of the absence of any characteristic scale in the three–dimensional distribution the same angular sample can be obtained with a different choice for the size of the box and suitably rescaling the value of the limiting apparent magnitude. For instance, taking a box side of $64h^{-1}$ Mpc and $m = 8$ for the limiting magnitude, the number of selected points is $\sim 1800$. With this choice, the projected number density that is of the same order of that in the C36 sample. Therefore, the analysis of this synthetic sample enable us also to check to which extent the objects available in the observational data sets are adequate to safely realize a fractal analysis.

In Figure 40 I show the CI partition function evaluated for $q = 3$ and $q = 5$. It is remarkable to note that the local dimension shows no evidences of breaking of scale–invariance at large scales. Instead its behaviour indicates that self–similarity is essentially preserved after projection, at least in the relevant range of angular scales. This finding is of crucial relevance for the interpretation of the scaling properties for the distribution of C36 clusters. Indeed, it confirms that the breaking of scale–invariance at $\vartheta \sim 6°$ reflects the presence of a characteristic scale in the three–dimensional distribution and disproves the picture of a purely scale–invariant Universe.

A further important question concerns the origin of the scale–invariance for the cluster distribution. In fact, the detected self–similarity at scales of few tens of Mpcs cannot have the same origin as the scale–invariance detected from the analysis of N–body simulations, the latter being the natural product of non–linear gravitational dynamics. This is surely not the case for the former, since at the scales of cluster clustering, the background dynamics is expected to be still in the linear or quasi–linear regime. Thus, the question that arises is whether, at scales $\gtrsim 10h^{-1}$ Mpc, selecting the peaks of a moderately evolved Gaussian background can account for the observed fractality, that is the signature of a strongly non–Gaussian statistics, or we need something else.

To clarify this point, I discuss now the results of the scaling analysis for the simulated cluster samples. The results of this analysis for the CI method are shown in Figure 41. From left to right, the results for the $G_1$, $\chi_n^2$ and $LN_n$ models are plotted, while upper and lower figures are for $q = 0$ and $q = 4$, respectively. According to its definition, the amplitude of the partition function is not normalized to be the same for two distributions having the same scaling properties but a different number of points. For this reason, one is only interested to compare the slopes of $Z(q, r)$ for data and simulations and not their amplitudes.

For $q = 0$, the local dimension shows a smooth transition from $D_o \simeq 0$ at small scales, to $D_q \simeq 2$ at $\vartheta \sim 6°$. All the three models generate a homogeneous geometry of the distributions roughly at the same scale as observational data. At smaller scales, the best model is the $LN_n$ one, which correctly reproduces the partition function slope. The other two models produce a $D_o(\vartheta)$ which is slightly smaller than observed, although $\chi_n^2$ seems to be more successful than $G_1$, especially at small angles.

More interesting is the $q = 4$ case. Unlike the real data, the $G_1$ model does not succeed to generate any small–scale self–similarity. Instead, at $\vartheta \lesssim 2°$ the partition function rapidly flattens and the local dimension declines to zero. Differently from the $q = 0$ case, it is hard to expect that this is only an effect of discreteness for two main reasons. First, for $q > 1$ most weight is assigned to the overdense parts of the distribution, where the sampling should be good. Second, if the limited statistics were the reason, we should expect to find a similar behaviour even for the other two models, which produces a comparable number of C36 clusters. However, this is not the case for both the $\chi_n^2$ and $LN_n$ models. Instead, they better reproduces the small–scale flat shape of the local dimension, with a resulting fractal dimension $D_q \lesssim 1$ to characterize the clustering inside the overdense regions. In particular, note the remarkable good agreement between data and $LN_n$ model at $\vartheta \lesssim 6°$, although at larger scales it generates too much clustering and, consequently, a too small dimension.

In Figure 42 I plot the $D_\tau$ and $f(\alpha)$ spectra for the two non–Gaussian models and for real data by using the $W(\tau, p)$ partition function. No similar plot are produced for the $G_1$ model, since it does not generate any fractality in the cluster distribution. Both the $\chi_n^2$ and the $LN_n$ models produce slightly smaller dimensions than observed. While this difference is not significant for $\tau < 0$, it is for $\tau > 0$. This is also reflected by the values taken by the local dimension $\alpha$; note that the $\alpha_{min}$ value is always smaller for the non–Gaussian simulations than for the real data, thus indicating the presence of stronger singularities.

However, it is worth stressing that the major aim of this analysis is not to find the non–Gaussian model, which best fits observational data. Rather, the relevance of these results is that the scaling properties for the cluster distribution represents a useful constraint to test models about the nature and the origin of primordial fluctuations.

## 7.2 The spatial analysis

### 7.2.1 The samples

The scaling analysis of the spatial cluster distribution is based on redshift subsets of the Abell and ACO cluster catalogs [3, 5]. The Abell catalog is the PGH redshift sample already mentioned in §2.3. It has geometrical boundaries $|b^{II}| \geq 30°$, $\delta \leq -27°$ and $z \leq 0.1$, with a total area of 4.8sr. These constraints result in a total number of 206 clusters distributed in the Northern (NGC) and Southern (SGC) Galactic Cap. For the ACO catalog the selection criteria are $m_{10} \leq 16.4$, $b^{II} \leq -20°$ and $\delta \geq -17°$. This subsample has an area of 1.8 sr and 103 clusters. In this subset, 19 clusters have redshift estimated form the $m_{10} - z$ relation (see, e.g., ref.[325]), otherwise redshift are directly taken from the original catalog or complemented with those measured by a number of authors. All the clusters with richness $R \geq 0$ are included in the samples, since no significant differences are expected between the clustering properties of the $R = 0$ and $R \geq 1$ cluster samples out to $z \approx 0.2$ [332, 327].

The selection function are both in galactic latitude and redshifts. The selection function in galactic latitude is

$$P(b^{II}) = 10^{\alpha(1-csc(|b^{II}|))}, \qquad (225)$$

with $\alpha = 0.3$ (0.2) for the Abell (ACO) subsample, while the redshift selection function reads

$$P(z) = \begin{cases} 1, & z < z_c \\ A\exp^{-z/z_0}, & z \geq z_c \end{cases}. \qquad (226)$$

Here $z_c$ is the completeness redshift, that is the maximum redshift at which the cluster distribution follows that of a uniform one. Following ref.[327], it is $z_c = 0.081, 0.063$ and $0.066$ for NGC, SGC and ACO, respectively.

As for the generation of the simulated cluster samples, they are based on the application of the Zel'dovich approximation (see §5.1), which, at the scales where cluster clustering develops, should give a fair representation of the gravitational dynamics. A detailed description of these simulations can be found in ref.[327]. Here I just give the basic sketch of them. A given number of points $N_p$ is randomly distributed in a cube of size 640 $h^{-1}$ Mpc with $N_g^3 = 64^3$ grid points. The points are then displaced from their positions according to the Zel'dovich algorithm. The density fluctuation spectrum $\delta_{\vec{k}}$ is chosen to have a Gaussian distribution with random phases and power spectrum

$$P(k) = \langle \delta_{\vec{k}}^2 \rangle = A k^n \exp\left(-|\mathbf{k}|^2/\Lambda^2\right) \Theta(|\mathbf{k}|). \qquad (227)$$

Following Postman et al. [333], $\lambda^{-1} = 0.1 h^{-1}$ Mpc, $\Theta(|\mathbf{k}|) = 0$ for $|\mathbf{k}| > (80 h^{-1}$ Mpc$)^{-1}$ and $\Theta(|\mathbf{k}|) = 1$ otherwise. To each particle is tagged a $\nu$ value, such that $\nu = \delta_{\vec{g}}/\sigma$, being $\delta_{\vec{g}}$ the value at the nearest grid point of the fluctuation of the linear density field smoothed with a Gaussian window of size $R = 10 h^{-1}$ Mpc, and $\sigma$ the corresponding r.m.s. fluctuation value. The number of points in the simulations, $N_p$, must be chosen so that, after having applied to the simulated samples the survey boundary limits and all the selection functions of the real data, we end up with the same number of clusters as in the real samples. Taking $N_p = 73000$ results to about $\sim 12000$ points being associated with peaks with $\delta > \nu\sigma$ with $\nu = 1.3$. The parameters of the simulations, namely $A$, $n$ and $\nu$, are chosen in such a way to reproduce the observed cluster 2–point correlation function with $r_o \simeq 20 h^{-1}$ Mpc and $\gamma \simeq 1.8$ in the appropriate scale range.

From the simulated cluster distributions, artificial redshift samples are built by taking volume elements of the simulation box having the same boundaries and the same selection functions as the real samples. The analysis of simulated cluster samples that I will review is based on 50 different initial phase assignment for statistically equivalent initial conditions. The final results are the average over such a large number of synthetic NGC, SGC and ACO redshift samples. Plionis et al. [327] observed that such simulations, which have initial conditions tuned so to produce the correct $\xi(r)$, gives as an extra bonus also a 3–point correlation function which is consistent with the observed one.

### 7.2.2 Results

Here I present results based on the correlation–integral (CI) approach, while a more comprehensive treatment also including the density–reconstruction approach is given in ref.[63]. The major problem in the analysis of the spatial cluster distribution arise due to the finite volume of the samples and on the presence of selection functions both in redshift and in galactic latitude. In ref.[63] we also discussed the technical details of how such effects can be accounted for.

As for the border corrections, different procedures can be devised, each having advantages and drawbacks. A first possibility, which makes no assumptions about the distribution outside the sample volume, consists in discarding from the partition sum of eq.(98) at the scale $r$ all the centers whose distance from the nearest boundary is less than $r$. However, this kind of procedure severely limits the statistics and, moreover, the analyzed sample is statistically different at different scales. This represents a serious problem in our case, since the geometry of the boundaries and the limited number of clusters do not allow to consider scales larger than $\sim 70 h^{-1}$ Mpc. A further possibility is to consider for each center $i$ the fraction $f_i(r)$ of the sphere of radius $r$ centered on it, which falls within the boundaries. In this way, if $\tilde{n}_i(r)$ is the counted number of neighbors, the corrected one is $n_i(r) = \tilde{n}_i(r)/f_i(r)$. Actually, this procedure also allows to account for other systematics of the cluster sample, such as the dependence of the local cluster density on redshift and galactic latitude, as provided by the selection functions of eqs.(225) and (226). Accordingly, the corrected local count is

$$n_i(r) = \frac{1}{f_i(r)} \sum_{j=1}^{N} \frac{\theta(|\mathbf{x}_i - \mathbf{x}_j| - r)}{P_j(b^{\text{II}}) P_j(z)}, \qquad (228)$$

where $P_j(b^{\text{II}})$ and $P_j(z)$ are the values of the galactic latitude and redshift selection functions at the position of the $j$-th cluster. In order to implement this correction, a Montecarlo sampling with rejection is realized, so to measure the corrected count $n_i(r)$. Note, however, that at distances much larger than the completeness redshift, $P(z)$ rapidly declines, with a subsequent increase of the noise in the correction procedure. For this reason, I will present only results based on clusters within the distance $d = 200 h^{-1}$ Mpc, where redshift selection is not dramatic for all the samples.

A potential problem with this procedure is that it relies to some extent on the assumption of large–scale homogeneity. Therefore, one expects that a breaking of scale–invariance could be induced on an otherwise self–similar structure and spurious homogeneity can be detected. To check this properly, I also analyze a simulated cluster sample drawn from a fractal structure with controlled dimensionality, which also includes the same selection functions, boundaries and number of objects as the real one. As a first test of this kind, I plot in Figure 43 the resulting local dimension for a simulated Abell sample extracted from a monofractal structure with $D = 1$ below the homogeneity scale $L_h = 40 h^{-1}$ Mpc and $D = 3$ at larger scales. It is evident that both the correct scaling and the presence of the characteristic scale are correctly detected.

The lower panel refers to the same analysis done on a purely fractal cluster sample, without large–scale homogeneity. From this plot, it appears that the correct scaling is again measured, apart from oscillations of the local dimension due to the lacunarity of the point distribution, without any large–scale homogeneity spuriously induced by the analysis method. Therefore, these tests allows us to conclude that the adopted procedure to correct for border effects does not significantly "homogenize" the distribution, at least at the scales we are interested in.

The results of the scaling analysis for real and Zel'dovich–simulated cluster samples are plotted in Figure 44. For Abell clusters (upper panel) there is no evidence for the existence on an extended scaling regime. Instead, the local dimension exhibit a smooth increasing trend until homogeneity is reached at about $100h^{-1}$ Mpc. A rather different result is obtained for the ACO sample (central panel), which shows a well–defined fractal scaling up to $\sim 35h^{-1}$ Mpc, with a characteristic dimension $D_q \gtrsim 1$, only slightly decreasing with the multifractal order. At larger scales self–similarity is broken and $D_q(r)$ starts increasing, much like in the upper panel of Figure 44. In this regime both Abell and ACO samples attain homogeneity at about the same scale, an evidence which is against the picture of a purely fractal Universe. It is interesting to note that, if $\xi(r) = (r_o/r)^\gamma$ is the cluster 2–point function, then the characteristic scale at which the power–law shape starts dominating the scaling of neighbour counts is $r_c = [3/(3-\gamma)]^{1/\gamma} r_o$ (see eq.[86]). For $r_o = 20h^{-1}$ Mpc and $\gamma = 1.8$, it is $r_c \simeq 33h^{-1}$ Mpc, thus very close to the scale at which fractal scaling breaks. In the lower panel I plot the results for the Zel'dovich cluster simulations. It is evident that no scaling occurs and that such simulations reproduce much better the behaviour of the Abell sample than that of the ACO one.

Therefore, the results of this analysis indicate that a substantial difference exist between the scaling behaviours of Abell and ACO clusters, the latter developing a self–similar clustering with $D \gtrsim 1$ up to $\sim 35h^{-1}$ Mpc. The value of the fractal dimension is consistent with that detected from the angular analysis previously discussed, although there is some difference between the corresponding characteristic scales, being $r_c \simeq 20h^{-1}$ Mpc for the C36 PBF sample. Apart from the different nature of the clusters selected in the two samples, this discrepancy in the $r_c$ values is likely to be ascribed to the effects of projection, which mix different scales in the line–of–sight direction. As a consequence, also results at small angular scales contains contamination from the clustering at large physical scales, so to apparently reduce the scale at which the fractal behaviour breaks down.

As for the difference between the clustering of Abell and ACO samples, it goes in the same direction as suggested by previous comparison between simulated and observed cluster distributions. Plionis et al. [327] found some significant differences between the clustering patterns of Abell and ACO samples, the latter being systematically at variance with respect to the clustering provided by the evolution of an initially Gaussian fluctuation spectrum. A similar finding has also been found from the analysis of the topological genus (see ref.[326]); a systematic "meatball" shift for ACO clusters is detected, which is not observed neither in the Abell sample nor in the same set of simulations. The conservative assumption about the different statistics of the two samples is that for ACO the higher sensitivity of the emulsion plates used to construct the catalogue implies that it traces the cluster distribution in a better way than Abell. Otherwise, one should be willing to accept the idea of a real difference in the statistical properties of clusters in different regions of the sky.

Hence, the main conclusion that one consistently draws from the scaling analysis of both angular and redshift cluster samples is that they shows a well defined fractal behaviour at scales of some tens of Mpcs, with a characteristic dimension $D \simeq 1-1.4$, followed by a transition toward large–scale homogeneity. From the one hand this result disproves the view of a purely fractal Universe. From the other hand it puts non–trivial constraints on initial condition models; for instance, I showed that Gaussian initial fluctuations do not reproduce the observed scaling, even with a spectrum tuned so to reproduce low–order clustering measures, such as the 2– and 3– point cluster correlation functions. It is however clear that, before going to extreme conclusions about this point, one needs exact N–body simulations with physical initial fluctuation spectra and including a reliable prescription to select clusters from them as close as possible to the observational procedure.

# 8  Summary

In this Article I reviewed the problem of characterizing the scale–invariant properties of the large–scale distribution of galaxies and galaxy clusters and explaining their origin in the framework of the gravitational instability picture. In this context, one of the most astonishing results is represented by the quite simple statistical descriptions of the LSS, in spite of the apparent complexity of the galaxy distribution, as emerging from the most extended surveys today available (see Section 2): far from being uniformly distributed, galaxies are arranged to form larger structures, like clusters, superclusters and filaments, while leaving almost devoid large patches of the Universe. On the other hand, statistical measures of the galaxy distribution (see Section 3) show that this non–trivial clustering pattern can be accounted for on the basis of rather simple scaling properties.

The classical example is represented by the 2–point correlation function, which remarkably scales like a power–law, $\xi(r) = (r_o/r)^\gamma$, over a rather large dynamical range for both galaxies and galaxy clusters, with almost the same index, $\gamma \simeq 1.8$, although with different correlation lengths, $r_o$, reflecting their different positions along the hierarchical sequence of cosmic structures. Even going to higher–order correlation measures, this similarity between the clustering of galaxies and clusters seems to be preserved: for both of them $N$–point correlation functions are simply related to $\xi(r)$ according to the hierarchical expression of eq.(28), while also the the cluster void probability function lies on the larger-scale extension of that relevant to galaxies (see, e.g., ref.[82]).

Based on this sort of self–similarity of the large–scale clustering, Pietronero and collaborators suggested that the Universe behaves like a pure fractal, extending at least up to the today largest sampled scales, thus rejecting any evidence of large–scale homogeneity (see ref.[99] in this same Journal for a detailed review about the motivations for a fractal Universe). The same authors suggested that the large size of inhomogeneities present in galaxy redshift samples is not to be interpreted in the light of a non "fairness" of these samples. Instead, they are nothing but the natural consequence of fractality: as more and more extended surveys are available, more and more extended structures appear, without any upper cutoff in their extension. In this picture, any measures of the 2–point correlation function, as defined by eq.(6) loose any significance, since it assumes the possibility of defining an average density. In fact, the criticism raised by the fractal Universe supporters to the usage of $\xi(r)$ is that the standard galaxy correlation length, $r_o \simeq 5h^{-1}$ Mpc, has no meaning for a distribution, like that of galaxies, which has filaments and voids extending at least up to scales of $\sim 50h^{-1}$ Mpc. Instead, they claim that $r_o$ does not measure any statistical property intrinsic to the LSS, instead it only depends on the sample size (see eq.[113]). On this ground, they also explain the detected increase of the galaxy correlation length with the sample size [147] as well as the enhanced correlation amplitude of rich clusters with respect to galaxies.

Even more important, the hypothesis of a fractal LSS would have a dramatic impact on the basis of the "standard" cosmology: a Universe of this kind were not homogeneous *by definition*, thus conflicting with the Cosmological Principle, which assumes that homogeneity must be reached well within the size of the present horizon ($\sim 3000h^{-1}$ Mpc). If we were going to accept the idea of a fractal Universe, an alternative explanation is required for the fundamental measurements of the CMB, whose temperature and degree of isotropy are both of the same order of that expected from general arguments in the framework of the "standard" cosmology.

On the other hand, by adopting this standard picture, there are no reasons to expect a priori any similarity between the clustering of galaxies and of clusters, which develop on different scale ranges. As pointed out in Section 5, the history of the expanding Universe is expected to imprint characteristic scales on the shape of the post–recombination spectrum of density fluctuations, even starting from a scale–invariant post–inflationary spectrum. Furthermore, according to the gravitational instability picture, the dynamics underlying the galaxy clustering at scales of few Mpcs is highly non–linear. Instead, nearly linear gravity is expected to hold at the scales of some tens of Mpcs, where the cluster clustering develops and initial conditions are better preserved.

To clarify these points, I extensively reviewed in this Article results about the study of LSS clustering based on fractal analysis methods. Far from assuming that the Universe behaves like a fractal structure at all the scales, such statistical methods probably represent the most efficient way to prove or disprove the paradigma of self–similar clustering. From a more technical point of view, the advantage of using the fractal approach resides in the detailed information provided about the existence and the extension of scale–invariance in the distribution of cosmic structures. The introduction of the concept of the multifractal spectrum of generalized dimensions (see Section 4) allows one to collect in a compact way informations not only about the statistics inside the overclustered regions, as correlation functions do, but also on the scaling of the void probability function and on the distribution in the underdense parts of the distribution. This is a welcome property, since the nature of the distribution inside underdensities turns out to be much more model discriminant than the statistics of the overdense regions (see §4.4). Despite the fact that fractal methods represent powerful tools for clustering analysis, nevertheless they are based on different approximations to the formal definition of fractal dimension. Furthermore, their introduction to the study of complex dynamical systems and deterministic chaos has been motivated by their reliability when applied to well sampled fractal structures, while the presence of characteristic scales in the galaxy distribution and the limited sizes of data sets could seriously affect their performance. For these reasons, it is always necessary to carefully test the reliability of different fractal dimension estimators on point distributions with a priori known statistical properties (see §4.2).

A first question to be addressed is therefore whether the detected power–law shapes for the galaxy correlation functions implies a fractal scaling over a limited scale–range. If yes, it remains to establish which is the dynamical origin of this scaling and whether it tells us something about initial conditions for structure formation. A detailed answer to this question has been given in Section 6 from the fractal analysis of P$^3$M N–body simulations. The result of this analysis shows that a scale–invariant fractal behaviour is always associated to the non–linear gravitational clustering for the distribution of matter as well as of peaks of the initial density field, which are associated to galaxies according to the biasing prescription [230, 27]. This is a robust outcome of the analysis, which comes out independently of initial conditions (see also ref.[396]). As small-scale self–similarity is established, it progressively extends to larger scales, as the typical clustering length increases. In this picture, the correlation length $r_o$ acquires a precise dynamical meaning, since it represents the transition scale linear and non–linear gravitational dynamics, which generates a self–similar clustering at $r \lesssim r_o$. A further remarkable result is that a fractal dimension near to one always characterizes the clustering inside the overdense regions, while dependence on the initial conditions only determines details of the $D_q$ dimension spectrum. This means that both the isothermal density profile of galaxy halos, $\rho(r) \propto r^{-2}$, indicated by the rather flat profile of spiral rotation curves, and the shape of the galaxy 2–point function, $\xi(r) \propto r^{-1.8}$, are quite natural products of the non–linear gravitational dynamics. The flat shape of the $D_q$ curve for $q > 0$ also indicates that monofractal statistics characterizes the overdense structures, in agreement with the theoretical expectation of hierarchical correlation functions (see eq.[28] and §4.4).

From the point of view of the study of complex dynamical systems, the non–linear gravitational clustering represents one of the few known examples of dynamically generated fractal structure. In this context, it would be of interest to verify whether the detected fractality is the signature of an underlying chaotic dynamics. If this were the case, then a suitable analysis of N–body simulations should indicate an exponential divergence of the phase space trajectories of the systems during the evolution, which is the signature of the unpredictable (chaotic) character of the dynamics. The evaluation of the corresponding Lyapunov exponents (see, e.g., ref.[297]) should quantify the degree of chaotic behaviour of non–linear gravity. Although "exact" N–body simulations represent a necessary ingredient for this kind of investigation, analytical approaches could also be usefully employed, for instance by resorting to approximate dynamical models accounting for the essential features of non-linear gravity.

Therefore, although fractality is dynamically generated at the small scales ($\lesssim 5h^{-1}$ Mpc) of non–linear clustering, at larger scales the linear dynamics preserves the the nearly homogeneous initial conditions. For this reason, the results coming from N–body simulations are perfectly consistent with the picture of a fractal galaxy distribution confined to the small scales of few Mpcs.

A different situation should however be expected as the larger scales ($\gtrsim 10h^{-1}$ Mpc) traced by the cluster distribution are considered, where the gravitational dynamics is still in the linear regime. At such scales, there are no special reasons to expect a fractal scaling for clusters. Nevertheless, the fractal analysis of the angular as well as redshift cluster samples (see Section 7) shows that the distribution of rich galaxy systems displays a remarkable scale–invariant behaviour, with dimension $D_q \simeq$ 1–1.4 characterizing the clustering inside the overdensities ($q > 1$). Also in this case, the scale–invariance does not extend over an arbitrarily large scale range. Instead it is confined to scales $\lesssim 35h^{-1}$ Mpc, after which it breaks down. This result should not be very surprising, since it is perfectly consistent with the shape and the amplitude of the cluster 2–point correlation function; the slope $\gamma \simeq 1.8$ is associated to the correlation dimension $D_\nu = 3 - \gamma \simeq 1.2$, while the correlation length $r_o \simeq 20h^{-1}$ Mpc is associated to the characteristic scale $r_c = [3/(3-\gamma)]^{1/\gamma} \simeq 33h^{-1}$ Mpc (see eq.[86]). What is knew is that in the fractal analysis the effects of large–scale homogenization are under control. Therefore, the observed breaking of fractal scaling is hardly questionable and points against the picture of a fractal Universe.

A different important question concerns the origin of the scaling in the cluster distribution. In the case of galaxies the non–linearity of the clustering is *dynamically* originated, while for clusters it has a *statistical* origin; high peak selection on the underlying density field amplify the clustering. From the one hand, this represents a fundamental difference between the nature of galaxy and cluster clustering and makes even more astonishing the similarity of their statistics. From the other hand it leads one to ask which kind of constraints this poses of the nature of the underlying density fluctuations and on the mechanism for cluster formation. Since fractal scaling is the signature of a non–Gaussian (spatially intermittent) distribution, the arising question is whether it can be accounted for by a peak–selection procedure on a Gaussian background (see §5.4) or it requires something more. To answer this question, I presented in Section 7 the same fractal analysis realized on simulations of both angular and spatial cluster samples.

As for the angular analysis, artificial cluster samples are based on projecting three-dimensional CDM N–body simulations based on both Gaussian and skewed initial fluctuations. Once the projected galaxy samples have been created, clusters are selected from them by following as close as possible the observational set up (see ref.[60]). As a remarkable result, I showed that the standard CDM Gaussian fluctuations are not able to reproduce the scaling observed in real data sets. On the contrary, the primordial large–scale coherence associated to negative–skewed models generates a fractal behaviour in the resulting cluster distribution. Far from meaning that this strongly points in favour of non–Gaussian primordial fluctuations, it surely indicates that the scaling of the cluster distribution represents a non trivial constraints for initial condition models.

A similar result comes also from the analysis of simulated redshift samples. In this case, simulations are based on the Zel'dovich approximation, which should give a reliable representation of gravitational dynamics at the relevant scales, joined with the peak selection, according to the biasing suggestion. Gaussian initial fluctuations are chosen to have a spectrum, so to reproduce the 2– and 3–point cluster correlation functions. Despite this fact, the fractal analysis shows that no scaling is produced even in a limited scale–range. This confirms that either the nature of the initial conditions or the cluster identification procedure are not correct. In this respect, it is clear that accurate simulations of large–scale clustering involving a dynamical range as large as possible are required in order to understand the consequence of a self–similar cluster distribution at scales of some tens of Mpcs on current models for the primordial fluctuations generation and structure formation.

As a concluding remark, I would like to point out that, despite the great amount of statistical information provided by the scaling analyses reviewed in this Article, nevertheless it is difficult to expect that they provide an exhaustive representation of the observed large–scale structure in the Universe. Several other methods, such as topology analysis, count–in–cell statistics and void probability functions, though connected to the methods we used, could be better suited to reveal certain aspects of clustering (see Section 3). In addition, the study of the large–scale peculiar motions should provide a more and more precise representation of the cosmic matter distribution as a larger amount of data will be available in the near future. Furthermore, the increasing sensitivity of measurements of CMB temperature anisotropies at various angular scales will lead in few years to a much more precise understanding of the initial conditions for structure formation. Extended galaxy redshift surveys are now going on, while even more ambitious projects are expected. The possibility of mapping in detail the three dimensional galaxy distribution is surely a unique opportunity to test dynamical models of structure formation. For these reasons, I believe that the study of the LSS of the Universe will represent in the near future an even more exciting field of investigation and the development and refinement of methods of analysis will greatly contribute to provide a much more clear statistical and dynamical picture.

**Acknowledgements.** This article has been extracted from my Ph.D. Thesis discussed at SISSA in Trieste. For this reason, I would like to thank my supervisor Prof. G.F.R.Ellis and Prof. D.W.Sciama for their continuous advice, encouragement and support during the preparation of my Thesis. In this article I also presented several results obtained in collaboration with several people. In alphabetic order they are: Prof. S.A.Bonometto, Dr. P. Coles, Dr. Y.P.Jing, Prof. F. Lucchin, Dr. V.J. Martínez, Prof. S. Matarrese, Dr. A. Messina, Dr. L. Moscardini, Dr. G. Murante, Dr. M.A. Pérez, Dr. M.Persic, Dr. M.Plionis, Dr. A.Provenzale, Dr. P.Salucci, Dr. R.Valdarnini.

# A. Partition function from moment generator

In this Appendix I show how the box-counting partition function, which characterizes the fractal properties of a given distribution, is related to the corresponding cumulant generating function (see §4.3). In particular, I derive the expression (141) for $Z(q,r)$ starting from its definition

$$Z(q,r) \;=\; B(r) \sum_N \left(\frac{N}{N_t}\right)^q P_{\rm N}(r) \tag{A1}$$

(see eq.[138]). The scheme of the following calculations is similar to that described by Balian & Schaeffer [26] to work out the multifractal spectrum of their scale-invariant model.

From the integral representation of the $\Gamma$-function (see, e.g., ref.[6]), it is

$$N^{q-1} \;=\; -\frac{\Gamma(q)}{2\pi i} \int_{(0,+\infty)^+} dz\, e^{-Nz} (-z)^{-q}\,. \tag{A2}$$

Inserting this expression into eq.(A1) and taking also into account the relation (134) between the $P_{\rm N}$ count probabilities and the cumulant generating function, we get

$$Z(q,r) \;=\; -N_t^{-q}\, B(r)\, \frac{\Gamma(q)}{(2\pi i)^2} \int_{(0,+\infty)^+} dz \oint dt \sum_{N=1}^{\infty} \frac{N\, e^{-Nz}}{t^{N+1}} (-z)^{-q}\, e^{K(t-1)}\,. \tag{A3}$$

After summing analytically the series appearing in the previous equation,

$$\sum_{N=1}^{\infty} \frac{N\, e^{-Nz}}{t^{N+1}} \;=\; \frac{e^{-z}}{(t-e^{-z})^2}\,,$$

the integral in eq.(A3) can be easily evaluated by applying the Cauchy theorem:

$$Z(q,r) \;=\; -N_t^{-q}\, B(r)\, \frac{\Gamma(q)}{2\pi i} \int_{(0,+\infty)^+} dz\, (-z)^{-q}\, e^{-z}\, K'(e^{-z}-1) \exp[K(e^{-z}-1)]\,. \tag{A4}$$

Since any hierarchical generating function turns out to depend upon the variable $N_c t$, it is convenient to change the integration variable in eq.(B4) according to

$$y \;=\; N_c\,(1-e^{-z})\,, \tag{A5}$$

so as to render adimensional the argument of the generating function. Accordingly, the $Z$ partition function takes the expression

$$Z(q,r) \;=\; -N_t^{-q}\, \frac{B}{N_c}\, \frac{\Gamma(q)}{2\pi i} \int_{(0,N_c)^+} dy \left[\log\left(1-\frac{y}{N_c}\right)\right]^{-q} K'(-y/N_c)\, e^{K(-y/N_c)}\,, \tag{A6}$$

which exactly coincides with eq.(141).

# References

[1] Aarseth, S.J., 1984, in Methods of computational Physics, eds. J.U.Brackhill, B.J.Cohen (Academic, New York)

[2] Aarseth, S.J., Gott, J.R., Turner, E.L., 1979, ApJ, 236, 43

[3] Abell, G.O., 1958, ApJS, 3, 211

[4] Abell, G.O., 1961, AJ, 66, 607

[5] Abell, G.O., Corwin, H.G., Olowin, R.P., 1989, ApJS, 70, 1

[6] Abramowitz, M, Stegun, I.A., 1972, Handbbok of Mathematical Functions, (10th ed.; New York: Dover)

[7] Achilli, S., Occhionero, F., Scaramella, R., 1985, ApJ, 299, 577

[8] Adler, R.J., 1981, The Geometry of Random Fields (John Wiley: New York)

[9] Albrecht, A., Stebbins, A., 1992, Phys. Rev. Lett., 68, 2121

[10] Albrecht, A., Turok, N., 1989, Phys. Rev., D 40, 973

[11] Alimi, J.-M., Blanchard, A., Schaeffer, R., 1990, ApJ, 349, L5

[12] Allen, T.J., Grinstein, B., Wise, M.B., 1987, Phys. Lett. B, 197, 66

[13] Amendola, L., Borgani, S., 1994, MNRAS, 266, 191

[14] Ashman, K.M., Salucci, P., Persic, M., 1993, MNRAS, 260, 610

[15] Badii, R., Politi,A., 1984, Phys. Rev. Lett., 52, 1661

[16] Badii, R., Politi,A., 1985, J. Stat. Phys., 40, 725

[17] Bagla, J.S., Padmanabhan, T., 1994, MNRAS, 266, 227

[18] Bahcall, N.A., 1979, ApJ, 232, 689

[19] Bahcall, N.A., 1988, ARAA, 26, 631

[20] Bahcall, N.A., Burgett, W.S., 1986, ApJ, 300, L35

[21] Bahcall, N., Cen, R., 1993, ApJ, 398, L81

[22] Bahcall, N.A., Soneira, R.M., 1983, ApJ, 270, 20

[23] Bahcall, N.A., West, M.J., 1992, ApJ, 392, 419

[24] Balian, R., Schaeffer, R., 1988, ApJ, 335, L43

[25] Balian, R., Schaeffer, R., 1989, A&A, 220, 1

[26] Balian, R., Schaeffer, R., 1989, A&A, 226, 373

[27] Bardeen, J.M., Bond, J.R., Kaiser, N., Szalay, A.S. 1986, ApJ, 304, 15

[28] Bardeen, J.M., Steinhardt, P.J., Turner, M.S., 1983, Phys. Rev. D, 28, 679

[29] Barenblatt, G., 1980, Similarity, Self-Similarity and Intermediate Asymptotics (Plenum Press:New York)

[30] Barnes, J., Hut, P., 1986, Nature, 324, 826

[31] Barrow, J.D., Bhavsar, S.P. Sonoda, D.H., 1984, MNRAS, 210, 19p

[32] Barrow, J.P., Coles, P., 1990, MNRAS, 244, 188

[33] Batuski, D.J., Bahcall, N.A., Olowin, R.P., Burns, J.O., 1989, ApJ, 341, 599

[34] Baumgart, D.J., Fry, J.N., 1991, ApJ, 375, 25

[35] Bean, A.J., Efstathiou, G.P., Ellis, R.S., Peterson, B.A., Shanks, T., 1983, MNRAS, 205, 605

[36] Bennet, D.P., Bouchet, F.E., 1989, Phys. Rev. Lett., 60, 257

[37] Benzi, R., Paladin, G., Parisi, G., Vulpiani, A. 1984, J. Phys. A, 17, 3521

[38] Benzi, R., Biferale, L., Paladin, G., Vulpiani, A., Vergassola, M., 1991, Phys. Rew. Lett., 67, 2299

[39] Bertschinger, E., Dekel, A., 1989, ApJ, 336, L5

[40] Bertschinger, E., Dekel, A., Faber, S.M., Dressler, A., Burstein, D., 1990, ApJ, 364, 370

[41] Bhavsar S.P., Ling E.N., 1988, ApJ, 331, L63

[42] Birsdall, C.K., Langdon, A.B., 1985, Plasma Physics via Computer Simulation (McGraw-Hill)

[43] Biviano, A., Girardi, M., Giuricin, G., Mardirossian, F., Mezzetti, M., 1993, 411, L13

[44] Blanchard, A., Buchert, T., Klaffl, R., 1993, A&A, 267, 1

[45] Blumenthal, G.R., Faber, S.M., Primack, J.R., Rees, M.J., 1984, Nature, 311, 517

[46] Bogart, R.S., Wagoner, R.V., 1973, ApJ, 181, 609

[47] Bond, J.R., Cole, S., Efstathiou, G., Kaiser, N., 1990, ApJ, 379, 440

[48] Bond, J.R., Efstathiou, G., 1984, ApJ, 285, L45

[49] Bond, J.R., Szalay, A.S., 1983, ApJ, 274, 443

[50] Bonometto, S.A., Borgani, S., Ghigna, S., Klypin, A.A., Primack, J.R., 1993, MNRAS, submitted

[51] Bonometto, S.A., Borgani, S., Persic, M., Salucci, P., 1990, ApJ, 356, 350

[52] Bonometto, S.A., Lucchin, F., Matarrese, S., 1987, ApJ, 323, 19

[53] Bonometto, S.A., Scaramella, R., 1988, SISSA preprint, 92/87/A

[54] Borgani, S., 1990, A&A, 240, 223

[55] Borgani, S., 1990, Master Thesis, International School for Advanced Studies, Trieste, Academic Year 1989/90

[56] Borgani, S., 1993, MNRAS, 260, 537

[57] Borgani, S., Bonometto, S.A., 1990, ApJ, 338, 398

[58] Borgani, S., Bonometto, S.A., Persic, M., Salucci, P., 1991, ApJ, 374, 20

[59] Borgani, S., Coles, P., Moscardini, L., 1943, MNRAS, submitted

[60] Borgani, S., Coles, P., Moscardini, L., Plionis, M. 1994, MNRAS, 266, 524

[61] Borgani, S., Jing, Y.P., Plionis, M., 1992, ApJ, 395, 339

[62] Borgani, S., Martínez, V., Pérez, M.A., Valdarnini, R., 1994, ApJ, in press

[63] Borgani, S., Murante, G., Provenzale, A., Valdarnini, R., 1993, Phys. Rev. E, 47, 3879

[64] Borgani, S., Plionis, M., Valdarnini, R., 1993, ApJ, 404, 21

[65] Borgani, S., Murante, G., 1994, Phys. Rev. E, in press

[66] Börner, G., Mo, H.J., 1990, A&A, 227, 324

[67] Börner, G., Mo, H.J., Zhou, Y., 1989, A&A, 221, 191

[68] Bouchet, F.R., Strauss, M.A., Davis, M., Fisher, K.B. Yahil, A., Huchra, J.P., 1993, ApJ, in press

[69] Bouchet, F.R., Hernquist, L., 1992, ApJ, 400, 25

[70] Bouchet, F.R., Schaeffer, R., Davis, M., 1991, ApJ, 383, 19

[71] Bower, R.G., Coles P., Frenk C.S., White, S.D.M. 1993, ApJ, 405, 403

[72] Brainerd, T.G., Scherrer, R.J., Villumsen, J.V., 1993, ApJ, 418, 570

[73] Brandenberger, R. Physics of the Sarly Universe; Proceedings of the $36^{th}$ SUSSP, ed. J.A.Peacock, A.F.Heavens and A.T.Davis (Edinburgh: SUSSP publications)

[74] Broadhurst, T.J., Ellis, R.S., Koo, D.C., Szalay, A.S., 1990, Nature, 343, 726

[75] Brown, E.M. Groth, E.J., 1989, ApJ, 338, 605

[76] Buchert, T., 1992, MNRAS, 254, 729

[77] Burgers, J.M., 1948, Adv. Appl. Mech., 1, 171

[78] Burgers, J.M., 1974, The Nonlinear Diffusion Equation (Dordrecht:Reidel)

[79] Burstein, D., 1990, Rep. Progr. Phys., 53, 421

[80] Calzetti, D., Giavalisco, M., Meiksin, A., 1992, 398, 429

[81] Cappi, A., Maurogordato, S., 1992, A&A, 259, 423

[82] Cappi, A., Maurogordato, S., Lachieze-Rey, M. 1991, A&A, 243, 28

[83] Carruthers, P., 1991, ApJ, 380, 24

[84] Carruthers, P., Minh Duong-Van 1983, Phys. Lett., 131B, 116

[85] Carruthers, P., Sarcevic 1989, Phys. Rev. Lett., 63, 1562

[86] Carruthers, P., Shih, C.C., 1983, Phys. Lett., 127B, 242

[87] Castagnoli, C., Provenzale, A., 1991, A&A, 246, 634

[88] Catelan, P., Lucchin, F., Matarrese, S., 1988, Phys. Rev. Lett., 61, 267

[89] Cen, R.Y., 1992, ApJS, 78, 341

[90] Cen, R.Y., Gnedin, R.Y., Kofman, L.A., Ostriker, J. P., 1992, ApJ, 399, L11

[91] Cen, R.Y., Ostriker, J.P., 1992, ApJ, 393, 22

[92] Colafrancesco, S., Lucchin, F., Matarrese, S., 1989, ApJ, 345, 3

[93] Coles, P., 1988, MNRAS, 234, 509

[94] Coles, P., 1989, MNRAS, 238, 319

[95] Coles, P., 1990, Nature, 346, 446

[96] Coles, P., Jones, B.J.T., 1991, MNRAS, 248, 1

[97] Coles, P., Moscardini, L., Plionis, M., Lucchin, F., Matarrese, S., Messina, A., 1993, MNRAS, 260, 572

[98] Coles, P., Plionis, M., 1991, MNRAS, 250, 75

[99] Coleman, P. H., Pietronero, L., 1992, Phys. Rep., 213, 311

[100] Coleman, P. H., Pietronero, L., Sanders, R. H., 1988, A&A, 200, L32

[101] Coleman, P. H., Saslaw, W.C., 1990, ApJ, 353, 354

[102] Collins, C.A., Heydon-Dumbleton, N.H. MacGillivary, H.T., 1989, MNRAS, 236, 7p

[103] Collins, C.A., Nichol, R.C., Lumsden, S. L., 1992, MNRAS, 254, 295

[104] Colombi S., Bouchet F.R., Schaeffer R., 1992, A&A, 263, 1

[105] Couchman, H.M.P., Carlberg, R.G., 1992, ApJ, 389, 453

[106] Crane, P., Saslaw, W.C., 1986, ApJ, 301, 1

[107] da Costa, L.N., et al., 1988, ApJ, 327, 544

[108] da Costa, L.N., et al., 1989, AJ, 97, 315

[109] Dalton, G.B., Efstathiou, G., Maddox, S.J., Sutherland, W.J., 1992, ApJ, 390, L1

[110] Davies, A.G., Coles, P., 1993, MNRAS, 262, 591

[111] Davis, M., Efstathiou, G., Frenk, C., White, S.D.M. 1985, ApJ, 292, 371

[112] Davis, M., Efstathiou, G., Frenk, C., White, S.D.M. 1992, Nature, 356, 489

[113] Davis, M., Geller, M.J., 1976, ApJ, 208, 13

[114] Davis, M., Huchra, J., 1982, ApJ, 254, 437

[115] Davis, M., Peebles, P.J.E., 1977, ApJS, 35, 425

[116] Davis, M., Peebles, P.J.E., 1983, ApJ, 267, 465

[117] Davis, M., Summers, F.J., Schlegel, M., 1992, Nature, 359, 393

[118] Dekel, A., 1987, in Observational Cosmology, ed. A.Hewitt, G.Burbidge, and L.Z.Fang, (Dordrecht: Reidel) p.145

[119] Dekel, A., 1991, in Thw Distribution of Matter in the Universe, Proceeding of the 2nd DAEC Meeting, ed.G.A.Mamon and D.Gerbal

[120] Dekel, A., Aarseth, S.J., 1984, ApJ, 283, 1

[121] Dekel, A., Bertshinger, E., Yahil, A., Strauss, M.A., Davis, M., Huchra, J., 1993, ApJ, 412, 1

[122] Dekel, A., Blumenthal, G.R., Primack, J.R., Olivier, S., 1989, ApJ, 338, L5

[123] Dekel, A., Rees, M.J., 1987, Nature, 326, 455

[124] Dekel, A., Silk, J., 1986, ApJ, 303, 39

[125] de Lapparent, V., Geller, M.J., Huchra, J.P., 1986, ApJ, 302, L1

[126] de Lapparent, V., Geller, M.J., Huchra, J.P., 1988, ApJ, 332, 44

[127] de Lapparent, V., Geller, M.J., Huchra, J.P., 1989, ApJ, 343, 1

[128] de Lapparent, V., Geller, M.J., Huchra, J.P., 1991, ApJ, 369, 273

[129] de Lapparent, V., Kurtz, M.J., Geller, M.J., 1986, ApJ, 304, 585

[130] Doroshkevich, A.G., 1970, Astrophisica, 6, 320

[131] Doroshkevich, A.G., Kotok, T.V., 1990, MNRAS 246, 10

[132] Dressler, A., 1980, ApJ, 236, 351

[133] Dubuc, B., Quiniou, J.F., Roques-Carmes, C., Tricot, C., Zucker, S.W., 1989, Phys. Rew., A39, 1500

[134] Efstathiou, G., 1988, Large-Scale Motions in the Universe, a Vatican Study Week, eds. V.C.Rubin, and G.V.Coyne (Cittá del Vaticano: Pontificia Academia Scientiarum)

[135] Efstathiou, G., 1990, Physics of the early Universe; Proceedings of the $36^{th}$ SUSSP, ed. J.A.Peacock, A.F.Heavens and A.T.Davis (Edinburgh: SUSSP publications)

[136] Efstathiou, G., Dalton, G.B., Sutherland, W.J., Maddox, S., 1992, 257, 125

[137] Efstathiou, G., Davis, M., Frenk, C., White, S.D.M. 1985, ApJS, 57, 241

[138] Eftathiou, G., Eastwood, J.W., 1981, MNRAS, 194, 503

[139] Efstathiou, G., Ellis, R.S., Peterson, B.A., 1988, MNRAS, 232, 431

[140] Efstathiou, G, Fall, S.M., Hogan, G., 1979, MNRAS, 189, 203

[141] Efstathiou, G., Frenk, C., White, S.D.M., Davis, M. 1988, MNRAS, 235, 715

[142] Efstathiou, G., Kaiser, N., Saunders, W., Lawrence, A., Rowan-Robinson, M., Ellis, R.S., Frenk, C.S., 1990, MNRAS, 247, 10p

[143] Efstathiou, G., Silk, J., 1983, Fund. Cosmic Phys., 9, 1

[144] Efstathiou, G., Sutherland, W.J., Maddox, S.J., 1990, Nature, 348, 705

[145] Einasto, M., 1991, MNRAS, 252, 261

[146] Einasto, J., Gramann, M., Saar, E., Tago, E., 1993, MNRAS, 260, 705

[147] Einasto, J., Klypin, A.A., Saar, E., Shandarin, S.F. 1984, MNRAS, 206, 529

[148] Elizalde, E., Gatzanaga, E., 1992, MNRAS, 254, 247

[149] Ellis, J., 1990, Physics of the early Universe; Proceedings of the $36^{th}$ SUSSP, ed. J.A.Peacock, A.F.Heavens and A.T.Davis (Edinburgh: SUSSP publications)

[150] Eyles, C.J., et al., 1991, ApJ, 376, 23

[151] Faber, S.M., Jackson, R.E., 1976, ApJ, 204, 668

[152] Fall, S.M., Tremaine, S., 1977, ApJ, 216, 682

[153] Felten, J.E., 1985, Comm. Ap. Sp. Sci., 11, 53

[154] Field, G.B., Saslaw, W.C., 1971, ApJ, 176, 199

[155] Fournier d'Albe, E.E., 1907, Two New Worlds (London: Longmans Green)

[156] Freeman, K.C., 1970, ApJ, 160, 811

[157] Frisch, U., Sulem, P., Nelkin, M., 1978, J. Fl. Mech., 87, 719

[158] Frisch, U., Parisi, G., 1985, in Turbulence and Predictability in Geophysical Fluid Dynamics and Climatology, eds. R.Benzi, G.Parisi A.Sutera (North-Holland)

[159] Fry, J.N., 1982, ApJ, 262, 425

[160] Fry, J.N., 1984, ApJ, 277, L5

[161] Fry, J.N., 1984, ApJ, 279, 499

[162] Fry, J.N., 1985, ApJ, 289, 10

[163] Fry, J.N., 1986, ApJ, 306, 358

[164] Fry, J.N., Gatzañaga, E., 1993, ApJ, 413, 447

[165] Fry, J.N., Giovanelli, R., Haynes, M.P., Melott, A., Scherrer, R.J., 1989, ApJ, 340, 11

[166] Fry, J.N., Peebles, P.J.E., 1978, ApJ, 221, 19

[167] Fukugita, M., Hogan, C.J., Peebles, P.J.E., 1993, preprint IASSNS-AST 93/12

[168] Gaier, T., Schuster, J., Gundersen, J.O., Koch, T., Meinhold, P.R., Seiffert, M., Lubin, P.M., 1992, ApJ, 398, L1

[169] Geller, M.J., de Lapparent, V. Kurtz, M.J., 1984, ApJ, 287, L55

[170] Geller, M.J., Huchra, J.P., 1988, Large-Scale Motions in the Universe, ed. V.C. Rubin G.V. Coyne (Princeton: Princeton University Press), p.3

[171] Geller, M.J., Huchra, J.P., 1989, Science, 246, 897

[172] Gell'man, M., Ramond, P., Slansky, S., 1980, Supergravity, eds. D.Z.Freedman and P.Van Nieuwenhuizen (North Hollan, New York)

[173] Giavalisco, M., Mancinelli, B., Mancinelli, P., Yahil, A., 1993, ApJ, 411, 9

[174] Giovanelli, R., Haynes, M.P., 1988, in Large-Scale Structure of the Universe, IAU Symposium No. 130, ed. J. Auduze, et al. (Dordrecht)

[175] Giovanelli, R., Haynes, M.P., Chincarini, G.L. 1986, ApJ, 300, 77

[176] Gott, III, J.R., 1975, 201, 296

[177] Gott III, J.R., Gao, B., Park, C., 1991, ApJ, 383, 90

[178] Gott III, J.R., Mao, S., Park, C., Lahav, O. 1992, ApJ, 385, 26

[179] Gott III, J.R., Melott, A.L., Dickinson, M., 1986, ApJ, 306, 341

[180] Gott III, J.R., Park, C., Juszkiezicz, R., Bies, W.E., Bennet, D.P., Bouchet, F.R., Stebbins, A., 1990, ApJ, 352, 1

[181] Gott III, J.R., Turner, E.L., 1977, ApJ, 216, 357

[182] Gott III, J.R., Turner, E.L., 1979, ApJ, 232, L79

[183] Gott III, J.R., et al., 1989, ApJ, 340, 625

[184] Grassberger, P., Badii, R., Politi, A., 1988, J. Stat. Phys., 51, 135

[185] Grassberger, P., Procaccia, I., 1983, Phys. Rew. Lett., 50, 346

[186] Grassberger, P., Procaccia, I., 1983, Phys. Rew., A 28, 2591

[187] Grinstein, B., Wise, M.B., 1986, ApJ, 310, 19

[188] Groth, E.J., Peebles, P.J.E., 1976, A&A, 53, 131

[189] Groth, E.J., Peebles, P.J.E., 1977, ApJ, 217, 385

[190] Gundersen, et al., 1993, ApJ, 413, L1

[191] Gunn, J.E., 1977, ApJ, 218, 592

[192] Gunn, J.E., Gott III, J.R., 1972, ApJ, 176, 1

[193] Gurbatov, S.N., Saichev, A.I., 1984, Izv. Vyssh. Uchebn. Zaved. Radiofiz., 27, 456

[194] Gurbatov, S.N., Saichev, A.I., Shandarin, S.F., 1985, Sov. Phys. Dokl., 30, 921

[195] Gurbatov, S.N., Saichev, A.I., Shandarin, S.F., 1989, MNRAS, 236, 385

[196] Guth, A., 1981, Phys. Rev., D 23, 347

[197] Guzzo, L., Iovino, A., Chincarini, G., Giovanelli, R., Haynes, M., 1991, ApJ, 382, L5

[198] Hale-Sutton, D., Fong, R., Metcalfe, N., Shanks, T. 1989, MNRAS, 237, 569

[199] Hamilton, A.J.S., 1988, ApJ, 332, 67

[200] Hamilton, A.J.S., Gott III, J.R., Weinberg, D., 1986, ApJ, 309, 1

[201] Harrison, E.R., 1970, Phys. Rev., D 1, 2726

[202] Hauser, M.G., Peebles, P.J.E., 1973 , ApJ, 185, 757

[203] Haynes, M.P., Giovannelli, R., 1986, ApJ, 306, L55

[204] Hawking, S.W., Ellis, G.F.R., 1980, in The Large Scale Structure of Space–Time, (Cambridge: Cambridge University Press)

[205] Henon, M., 1976, Comm. Math. Phys., 81, 229

[206] Hentschel, H.G.E., Procaccia, I., 1983, Physica, D8, 435

[207] Hernquist, L., 1987, ApJS, 64, 715

[208] Heydon-Dumbleton, N.H., Collins, C.A., MacGillavray, H.T., 1989, MNRAS, 238, 379

[209] Hockney, R.W., Eastwood, J.W., 1981, Computer Simulations using Particles, (Mc Graw-Hill)

[210] Hoessel, J.G, Gunn, J.E., Thuan, T.X., 1980, ApJ, 241, 486

[211] Hoffman, Y., Shaham, J., 1985, ApJ, 297, 16

[212] Holtzman, J.A., 1989, ApJS, 71, 1

[213] Holtzman, J.A., Primack, J.R., 1993, ApJ, 405, 428

[214] Huchra, J.P., Davis, M., Latham, D., Tonry, J., 1983, ApJS, 52, 89

[215] Huchra, J.P., Geller, M.J., 1982, ApJ, 257, 423

[216] Huchra, J.P., Geller, M.J., de Lapparent, V., Corwin, H.G.Jr., 1990, ApJS, 72, 433

[217] Ikeuchi, S., 1981, Publ. Astr. Soc. Japan, 33, 211

[218] Ikeuchi, S., Turner, E.L., 1991, MNRAS 250, 519

[219] Inagaki, S., Itoh, M, Saslaw, W.C., 1992, ApJ, 386, 9

[220] Itoh, M, Inagaki, S., Saslaw, W.C., 1988, ApJ, 331, 45

[221] Jaynes, E.T., 1957, Phys. Rew., 106, 620

[222] Jensen, L.G., Szalay, A.S., 1986, ApJ, 305, L5

[223] Jing, Y.P., 1990, A&A, 233, 309

[224] Jing, Y.P., Mo, H.J., Börner, G., 1991, A&A, 252, 449

[225] Jing, Y.P., Plionis, M., Valdarnini, R., 1992, ApJ, 389, 499

[226] Jing, Y.P., Valdarnini, R., 1991, A&A, 250, 1

[227] Jing, Y.P., Valdarnini, R., 1993, ApJ, 406, 6

[228] Jing, Y.P. Zhang, J.L., 1989, ApJ, 342, 639

[229] Jones, B.J.T., Coles, P., Martínez, V.J., 1992, MNRAS, 259, 146

[230] Kaiser, N., 1984, ApJ, 284, L9

[231] Kaiser, N., 1987, MNRAS, 227, 1

[232] Kaiser, N., 1991, Proceeding of the Texas/ESO-CERN Conference, Brighton, December 1990

[233] Kaiser, N., Davis, M., 1985, ApJ, 297, 365

[234] Kaiser, N., Peacok, J.A., 1991, ApJ, 379, 482

[235] Kibble, T.W.B., 1976, J. Phys., A 9, 1387

[236] Kiefer, J. Wolfowitz, J., 1956, Ann.Math.Stat., 27, 887

[237] Kirkwood, J.C., 1935, J. Chem. Phys., 3, 300

[238] Kirshner, R.P., Oemler, A., Jr., Schechter, P.L., Shechtman, S.A., 1981, ApJ, 248, L57

[239] Klypin, A.A., Holtzman, J., Primack, J., Regös, E. 1993, ApJ, 416, 1

[240] Klypin, A.A., Kopilov, A.I., 1983, Sov. Astr. Lett., 9, 41

[241] Klypin, A.A., Nolthenius, R., Primack, J.R., 1993, in preparation

[242] Kofman, L.A., Linde, A.D., 1935, Nucl. Phys., B282, 555

[243] Kolb, E.W., Turner, M.S., 1990, The Early Universe (Addison-Wesley Publ. Company)

[244] Kolmogorov, A.N., 1941, Dokl. Acad. Nauk SSSR, 30, 299

[245] Lahav, O., Edge, A.C., Fabian, A.C., Putney, A. 1989, MNRAS, 238, 881

[246] Lake, G., Tremaine, S., 1980, ApJ, 238, L13

[247] Landau, L.D., Lifshitz, E.M., Fluid Mechanics (Pergamon Press)

[248] Lauberts, A., 1982, The ESO/Uppsala Survey of the ESO(B) Atlas (München:European Southern Observatory)

[249] Liddle, A.R., Lyth, D.H., 1993, Phys. Rep., 231, 1

[250] Lilje, P.B., Efstathiou, G., 1988, MNRAS, 231, 635

[251] Limber, D.N., 1953, ApJ, 117, 134

[252] Linde, A., 1982, Phys. Lett., B 108, 389

[253] Linde, A., 1984, Rep. Prog. Phys., 47, 925

[254] Lyubimov, V.A., et al., 1980, Phys. Lett., B94, 266

[255] Loveday, J., Peterson, P.A., Efstathiou, G., Maddox, S.J., 1992, ApJ, 390, 338

[256] Lucchin, F., 1989, in Morphological Cosmology, Proceeding of the $11^{th}$ Crakow Cosmological School, Springer-Verlag, Ed.: P.Flin and H.W.Duerbeck

[257] Lucchin, F., Matarrese, S., 1985, Phys. Rev., D 32, 1316

[258] Lucchin, F., Matarrese, S., 1988, ApJ, 330, 535

[259] Lucchin, F., Matarrese, S., Melott, A., Moscardini, L., 1994, ApJ, 422, 430

[260] Lucchin, F., Matarrese, S., Vittorio, N., 1986, ApJ, 330, L1

[261] Lumsden, S.L., Nichol, R.C., Collins, C.A., Guzzo, L., 1992, MNRAS, 258, 1

[262] Lynden-Bell, D., et al., 1988, 326, 19

[263] Maddox, S.J., Efstathiou, G., Sutherland, W.J., Loveday, J., 1990, MNRAS, 242, 43p

[264] Martínez, V.J., 1990, Vistas in Astronomy, 33, 337

[265] Martínez, V.J., Jones, B.J.T., 1990, MNRAS, 242, 517

[266] Martínez, V.J., Jones, B.J.T., Dominguez-Tenreiro, R., van de Weygaert, R., 1990, ApJ, 357, 50

[267] Martínez, V.J., Portilla, M., Jones, B.J.T., Paredes, S., 1993, A&A, in press

[268] Matarrese, S., Lucchin, F., Bonometto, S.A., 1986, ApJ, 310, L21

[269] Matarrese, S., Lucchin, F., Messina, A., Moscardini, L., 1991, MNRAS, 253, 35

[270] Matarrese, S., Lucchin, F., Moscardini, L., Saez, D. 1992, 259, 437

[271] Mandelbrot, B.B., 1982, The Fractal Geometry of the Nature (Freeman)

[272] Maurogordato, S., Schaeffer, R., da Costa, L.N., 1992, ApJ, 390, 17

[273] Meiksin, A., Szapudi, I., Szalay, A.S., 1992, ApJ, 394, 87

[274] Melott, A.L., 1983, ApJ,, 264, 59

[275] Melott, A.L., 1990, Phys. Rep., 193, 1

[276] Melott, A.L., Fry, J.N., 1986, ApJ, 305, 1

[277] Melott, A.L., Weinberg, D., Gott III, J.R., 1988, ApJ, 328, 50

[278] Messina, A., Moscardini, L., Lucchin, F., Matarrese, S., 1991, MNRAS, 245, 244

[279] Meszaros, P., 1975, A&A, 38, 5

[280] Mo, H.J., 1991, Ph.D. Thesis, Ludwig-Maximilian-Univesität München, Dept. of Physics

[281] Mo, H.J., Jing, Y.P. Börner, G., 1992, ApJ, 392, 452

[282] Moore, B., et al., 1992, MNRAS, 256, 477

[283] Moscardini, L., Borgani, S., Coles, P., Lucchin, F., Matarrese, S., Messina, A., Plionis, M., 1993, ApJ, 413, L55

[284] Moscardini, L., Matarrese, S., Lucchin, F., Messina, A., 1991, MNRAS, 248, 424

[285] Mukhanov, V.F., Feldman, H.A., Brandenberger, R.H. 1992, Phys. Rep., 215, 203

[286] Nash, C., Sen, S., 1983, Topology and Geometry for Physicists (London: Academic Press)

[287] Nichol, R.C., Collins, C.A., Guzzo, L., and Lumsden, S.L., 1992, MNRAS, 255, 21p

[288] Nilson, P., 1973, Uppsala General Catalogue of Galaxies, Nova Acta, Reg. Soc. Sci. Upsaliensis, Ser. V: A, Vol. 1

[289] Nusser, A., Dekel, A., 1990, ApJ, 362, 14

[290] Oemler, A., 1974, ApJ, 194, 1

[291] Olivier, S., Blumenthal, G.R., Dekel, A., Primack, J.R., Stanhill, D., 1990, ApJ, 356, 1

[292] Ore, O., 1962, Amer. Math. Soc. Colloq. Publ., 38

[293] Ortolan, A., Lucchin, F., Matarrese, S., 1989, Phys. Rev., D 40, 290

[294] Ostriker, J.P., Cowie, L., 1981, ApJ, 243, L127

[295] Padmanabhan, T., 1993, Structure Formation in the Universe (Cambridge: Cambridge University Press)

[296] Paladin, G., Vulpiani, A., 1984, Lett. Nuovo Cimento, 41, 82

[297] Paladin, G., Vulpiani, A., 1987, Phys. Rep., 156, 147

[298] Park, C., Gott III, J.R., da Costa, L.N., 1992, ApJ, 392, L51

[299] Park, C., Gott III, J.R., Melott, A.L., Karachentsev, I.D., 1992, ApJ, 387, 1

[300] Park, C., Spergel, D.N., Turok, N., 1991, ApJ, 372, L53

[301] Peakock, J.A., 1990, in Particle Astrophysics: The Early Universe and Cosmic Structures, eds. J.M.Alimi et al. (Editions Frontieres)

[302] Peakok, J.A., 1991, MNRAS, 253, 1p

[303] Peakok, J.A., Nicholson, D., 1991, MNRAS, 253, 307

[304] Peakok, J.A., West, M.J., 1992, MNRAS, 259, 494

[305] Peebles, P.J.E., 1974, ApJ, 189, L51

[306] Peebles, P.J.E., 1975, ApJ, 196, 647

[307] Peebles, P.J.E., 1976, Ap. Sp. Sc., 45, 3

[308] Peebles, P.J.E., 1980, The Large Scale Structure of the Universe (Princeton: Princeton University Press)

[309] Peebles, P.J.E., 1981, ApJ, 248, 885

[310] Peebles, P.J.E., 1982, ApJ, 258, 415

[311] Peebles, P.J.E., 1984, ApJ, 284, 439

[312] Peebles, P.J.E., 1986, Nature, 321, 27

[313] Peebles, P.J.E., 1987, ApJ, 315, L73

[314] Peebles, P.J.E., 1993, Concepts of Physical Cosmology (Princeton: Princeton University Press)

[315] Peebles, P.J.E., Hauser, M.G., 1974, ApJS, 28, 19

[316] Peebles, P.J.E., Groth, E.J., 1975, ApJ, 196, 1

[317] Persic, M., Salucci, P., 1988, MNRAS, 234, 131

[318] Persic, M., Salucci, P., 1990, MNRAS, 245, 577

[319] Persic, M., Salucci, P., 1990, ApJ, 355, 44

[320] Pietronero, L., 1987, Physica, 144A, 257

[321] Pisani, A., Giuricin, G., Mardirossian, F., Mezzetti, M., 1992, ApJ, 389, 68

[322] Plionis, M., 1988, MNRAS, 234, 401

[323] Plionis, M., Barrow, J.D., Frenk, C., 1991, MNRAS, 249, 662

[324] Plionis, M., Borgani, S., 1991, MNRAS, 254, 306

[325] Plionis, M., Valdarnini, R., 1991, MNRAS, 249, 46

[326] Plionis, M., Valdarnini, R., Coles, P., 1992, MNRAS, 258, 114

[327] Plionis, M., Valdarnini, R., Jing, Y.P., 1992, ApJ, 398, 12

[328] Pogosyan, D.Y., Starobinsky, A.A., 1993, preprint

[329] Pokorski, S., Gauge Field Theories, (Cambridge: Cambridge University Press)

[330] Politzer, D., Wise, M., 1984, ApJ, 285, L1

[331] Postman, M., Geller, M.J., Huchra, J.P., 1986, AJ, 91, 1267

[332] Postman, M., Huchra, J.P., Geller, M.J., 1992, ApJ, 384, 407

[333] Postman, M., Spergel, D.N., Sutin, B., Juszkiewicz, R., 1989, ApJ, 346, 588

[334] Press, W.H., Schechter, P., 1974, ApJ, 187, 425

[335] Provenzale, A., 1992, in: Applying Fractal in Astronomy, A. Heck and J. Perdang Eds. (Springer: Berlin)

[336] Provenzale, A., Galeotti, P., Murante, G., Villone B., 1992, ApJ, 401, 455

[337] Raychaudhury, S., 1989, Nature, 342, 251

[338] Ramella, M., Geller, M.J., Huchra, J.P., 1989, ApJ, 344, 57

[339] Ramella, M., Geller, M.J., Huchra, J.P., 1990, ApJ, 353, 51

[340] Ramond, P., 1981, Field Theory: a Modern Primer (New York: Wiley)

[341] Readhead, A.C.S., Lawrence, C.R., Myers, S.T., Sargent, W.L.W., Hardebeck, H.E., Moffet, A.T., 1989, ApJ, 346, 566

[342] Rees, M.J., Ostriker, J.P., 1977, MNRAS, 179, 541

[343] Renyi, A., 1970, Probability Theory (North Holland: Amsterdam)

[344] Rice, S.A., Grey, P., 1965, The Statistical Mechanics of Simple Liquids (Princeton: Princeton University Press)

[345] Riordan, J., 1958, An Introduction to Combinatorial Analysis, (New York:Wiley)

[346] Rood, H.J., 1976, ApJ, 207, 16

[347] Rubin, V.C., Burstein, D., Ford, W.K., Thonnard, N., 1985, ApJ, 289, 81

[348] Rubin, V.C., Ford, W.K., Tonnard, N. 1980, ApJ, 238, 471

[349] Rudnicky, K., Dworak, T.Z., Flin, P., Baranovski, B., Sendrakowski, A., 1973, Acta Cosmologica, 1, 7

[350] Salopek, D.S., 1992, Phys. Rev., D 45, 1139

[351] Salopek, D.S., Bond, J.R., Bardeen, J.M., 1989, Phys. Rev., D 40, 6

[352] Salucci, P., Persic, M., Borgani, S., 1993, ApJ, 405, 459

[353] Sandage, A., Saha, A., Tammann, G.A., Panagia, A., Macchetto, D., 1992, ApJ, 401, L7

[354] Saslaw, W.C., 1980, ApJ, 235, 299

[355] Saslaw, W.C., 1985, Gravitational Physics of Stellar and Galactic Systems, Cambridge Univ. Press

[356] Saslaw, W.C., Chitre, S.M., Itoh, M., Inagaki, S., 1990 ApJ, 365, 419

[357] Saslaw, W.C., Crane, P., 1991, ApJ, 380, 315

[358] Saslaw, W.C., Hamilton, A.J.S., 1984, ApJ, 276, 13

[359] Saunders, W., Frenk, C., Rowan-Robinson, M., Efstathiou, G., Lawrence, A., Kaiser, N., Ellis, R., Crawford, J., Xia, X.-Y., Parry, I., 1991, Nature, 349, 32

[360] Scaramella, R., Baiesi-Pillastrini, G., Chincarini, G., Vettolani, G., Zamorani, G., 1989, Nature, 338, 562

[361] Scaramella, R., Vettolani, G., Zamorani, G. 1991, ApJ, 376, L1

[362] Schechter, P., 1976, ApJ, 203, 297

[363] Schectman, S., ApJS, 57, 77

[364] Scherrer, R.J., Bertschinger, E., 1991, ApJ, 381, 349

[365] Scherrer, R.J., Melott, A.L., Bertschinger, E., 1989, Phys. Rev. Lett., 62, 379

[366] Seldner, M., Peebles, P.J.E., 1977, ApJ, 215, 703

[367] Seldner, M., Siebers, B., Groth, E.J. Peebles, J.E. 1977, AJ, 84, 249

[368] Shane, C.D., Wirtanen, C.A., 1967, Publ. Lick Obs., vol. 22

[369] Shandarin, S.F., Zel'dovich, Ya.B., 1989, Rev. Mod. Phys., 61, 185

[370] Shapley, H., 1930, Harward Obs. Bull., 874, 9

[371] Shapley, H., Ames, A., 1932, Harward Annals, 88, Part II

[372] Sharp, N.A., 1979, A&A, 74, 308

[373] Sharp, N., Bonometto, S.A., Lucchin, F., 1984, A&A, 130, 79

[374] She, Z.-S., Jackson, E., Orszag, S.A., 1990, Nature, 344, 226

[375] Silk, J., 1968, ApJ, 151, 459

[376] Smoot, G.F., et al., 1992, ApJ, 396, L1

[377] Strauss, M.A., Davis, M., Yahil, A., Huchra, J.P. 1990, ApJ, 361, 49

[378] Strauss, M.A., Huchra, J.P., Davis, M., Yahil, A., Fisher, K.B., Tonry, J., 1992, ApJS, 83, 29

[379] Struble, M.F., Rood, H.J., 1987, ApJS, 63, 543

[380] Struble, M.F. Rood, H.J., 1991, ApJ, 374, 395

[381] Sutherland, W., 1988, MNRAS, 234, 159

[382] Sutherland, W., Efsthatiou, G., 1991, MNRAS, 258, 159

[383] Szalay, A.S., 1988, ApJ, 333, 21

[384] Szalay, A.S., Schramm, D.M., 1985, Nature, 314, 718

[385] Szapudi, I., Szalay, A.S., Boschan, P., 1992, 390, 350

[386] Taylor, A.N., Rowan-Robinson, M., 1992, Nature, 359, 396

[387] Tóth, G., Hollósi, J., Szalay, A.S., 1989, ApJ, 344, 75

[388] Totsuji, H., Kihara, T., 1969, Publ. Astron. Soc. Japan, 25, 287

[389] Tully, R.B., 1986, ApJ, 303, 25

[390] Tully, R.B., Fisher, J.R., 1977, A&A, 54, 661

[391] Tully, R.B., Scaramella, R., Vettolani, G., Zamorani, G., 1992, ApJ, 388, 9

[392] Turok, N., 1985, Phys. Rev. Lett., 55, 1801

[393] Turok, N., 1986, Cosmology, Astronomy and Fundamental Physics, ed. G. Setti and L. Van Hove (ESO, Garching bei München), p. 175

[394] Valdarnini, R., Bonometto, S.A., 1985, A&A, 146, 235

[395] Valdarnini, R., Borgani, S., 1991, MNRAS, 251, 575

[396] Valdarnini, R., Borgani, S., Provenzale, A., 1992, ApJ, 394, 422

[397] van Dalen, A., Schaefer, R.K., 1992, ApJ, 398, 33

[398] van den Berg, S., 1992, Science, 258, 421

[399] van der Kruit, P.C., 1987, A&A, 173, 59

[400] van de Weygaert, R., 1991, MNRAS, 249 159

[401] van de Weygaert, R., Jones, B.J.T., Martínez, V.J., 1992, Phys. Lett. A, 169, 145

[402] de Vaucouleurs, G., 1981, in Tenth Texas Symposium on Relativistic Astrophysics (eds. Ramaty, R. Jones, F.C.) Ann. N.Y. Acad. Sci. 375, 90

[403] de Vaucoleurs, G., de Vaucoleurs, G., 1964, Reference Catalogue of Bright Galaxies, (Austin: University of Texas Press)

[404] Vilenkin, A., 1985, Phys. Rep., 121, 265

[405] Villumsen, J.V., Scherrer, R.J., Bertschinger, E. 1991, ApJ, 367, 37

[406] Vittorio, N., Juszkiewicz, R., Davis, M., 1986, Nature, 323, 132

[407] Vishniac, E.T., 1987, ApJ, 322, 597

[408] Vogeley, M.S., Geller, M.J., Huchra, J.P., 1991, ApJ, 382, 44

[409] Vogeley, M.S., Park, C., Geller, M.J., Huchra, J.P. 1992, ApJ, 391, L5

[410] Weinberg, D.H., Cole, S., 1992, MNRAS, 259, 652

[411] Weinberg, D.H., Ostriker, J.P., Dekel, A., 1989, ApJ, 336, 9

[412] Weinberg, D.H., Gunn, J.E., 1990, MNRAS, 247, 260

[413] Weinberg, S., 1972, Gravitation and Cosmology, (New York: Wiley)

[414] West, M.J., van den Bergh, S., 1991, ApJ, 373, 1

[415] White, S.D.M., 1979, MNRAS, 186, 145

[416] White, S.D.M., Frenk, C.S., Davis, M., 1983, ApJ, 287, 1

[417] White, S.D.M., Frenk, C.S., Davis, M. Efstathiou, G., 1987, ApJ, 313, 505

[418] Walker, T.P., et al., 1991, 376, 51

[419] Wright, E.L., et al., 1992, ApJ, 396, L13

[420] Yanagida, T., 1979, KEK Lecture Notes, unpublished

[421] Yepes G., Domínguez–Tenreiro R., Couchman H.M.P. 1992, ApJ, 401, 40

[422] Yoshioka, S., Ikeuchi, S., 1989, ApJ, 341, 16

[423] Zahn, C.T., 1971, IEEE Trans. Comp., C20, 68

[424] Zee, A., 1980, Phys. Lett., B 93, 389

[425] Zeldovich, Y.B., 1970, A&A, 5, 84

[426] Zeldovich, Y.B., 1972, MNRAS, 160, 1

[427] Zeldovich, Y.B., 1980, MNRAS, 192, 663

[428] Zwicky, F., Herzog, E., Karpowicz, M., Koval, C.T. 1961-1968, Catalogue of Galaxies and Clusters of Galaxies (Pasadena, Calif.: California Institute of Technology)

# Figure Captions

**Figure 1.** Position of galaxies in the Zwicky catalogue, with $m \leq 15.5$ in the northern galactic cap (panel $a$) and in the southern galactic cap (panel $b$). The coordinates are cartesian.

**Figure 2.** Equal–area projection of the APM galaxy distribution, centred on the southern galactic pole. Nearly two million galaxies are reported, with apparent magnitude in J–band $-17 \leq m \leq -20.5$ (after [263])

**Figure 3.** ($a$) Cone diagram for galaxies in the region of the Perseus–Pisces chain. ($b$) Cone diagram for a complete sample of galaxies with $m \leq 15.5$ in the declination range $26°\!.5 \leq \delta \leq 32°\!.5$. ($c$) Cone diagram for a complete sample covering the declination range $26°\!.5 \leq \delta \leq 44°\!.5$. Note the "Great Wall" that runs across the survey. ($d$) Cone diagram for a nearly complete sample covering the declination range $8°\!.5 \leq \delta \leq 14°\!.5$ (after [171])

**Figure 4.** A 360° view that shows a relation between the "Great Wall" and the Perseus–Pisces chain. The slice covers the declination region $20° \leq \delta < 40°$ and contains all the 6112 galaxies with detected redshift $cz \leq 15,000\,km\,s^{-1}$. The regions that appear to be almost devoid of galaxies are obscured by the galactic plane (after [171])

**Figure 5.** Sky distribution of galaxies of the QDOT survey. Also shown are the areas not included in the survey owing to incomplete satellite coverage, source confusion or redshift incompleteness (after [359])

**Figure 6.** All–sky distribution in supergalactic coordinates of the 4073 Abell clusters contained in the Abell and ACO samples. The symbol size has been scaled by distance class: the $D = 0$ clusters are represented by large open circles, while $D = 7$ clusters corresponds to small dots (after [5])

**Figure 7.** Left panel: estimate of the galaxy 2–point angular correlation function from the APM galaxy survey (after [263]). Closed circles are for the APM data, while open ones are for the Lick map, scaled to the APM depth. The dotted and solid lines correspond to the $w(\vartheta)$ predicted by CDM models with $h = 0.5$ and $h = 0.4$. Right panel: estimate of the galaxy 2–point spatial correlation function from the CfA1 redshift survey; crosses on the right represent the quantity $1 + \xi(r)$, while the dashed line is a power–law with exponent $\gamma = 1.8$ (after [116])

**Figure 8.** The mean halo matter density at the optical disk radius, $\bar{\rho}_{h,opt}$ as a function of the optical disk radius, $R_{opt}$, for the PS90 sample of 58 Sb–Sc spiral galaxies with extended rotation curves.

**Figure 9.** Upper panel: the dependence of the correlation function on the richness of the system. A range of the mean richness is shown for each cluster point. The solid line indicate an approximate dependence on richness. Lower panel: the dependence of the correlation function on the mean separation of objects. The solid line represents the $d^{1.8}$ dependence (after [20]).

**Figure 10.** The angular 2–point correlation functions for the northern (left panel) and southern (right panel) clusters of the PBF samples. Filled squares, empty squares and crosses represent data for the C36, C25 and C18 samples, respectively. Dashed lines represent the best–fit power laws with slope $\gamma = 2$.

**Figure 11.** The power spectrum in dimensionless form, as the variance per $\ln k$: $\Delta^2 = d\sigma^2/d\ln k$. The solid line is the best–fit power spectrum to APM data. Filled Points are for radio galaxies [303], with $P(k)$ reduced by a factor 3; open circles are are for IRAS [142, 359] and crosses are for CfA [34]. After [302].

**Figure 12.** Isodensity contours about the mean value in a Gaussian random field (from [179]). Left panel: regions above the mean density. Right panel: regions below the median density. It is apparent how the two regions are one the complement of the other.

**Figure 13.** The genus curves for Abell clusters and CfA galaxies, plotted together with the best fit random–phase approximation of eq.(65). Filled circles are the mean values between bootstrap resamplings and errorbars are the bootstrap one. Open squares are from raw data. It is apparent the shift toward meatball topology for the CfA sample (after [183]).

**Figure 14.** ($a$) The phase space structure of the Henon attractor [205], which is generated by the iterative map $x'_1 = 1 - 1.4x_1^2 + x_2$, $x'_2 = 0.3x_1$; 15,000 point are plotted. ($b$) Two–dimensional projection of a three–dimensional multifractal structure generated by the random $\beta$–model (see text). The rich variety of structures is apparent. Note also the presence of big voids, which also survive after projection. Also for the random $\beta$–model 15,000 points are plotted.

**Figure 15.** ($a$) The multifractal dimension spectrum, $D_q$, for the random $\beta$–model of Figure 14b and ($b$) the corresponding $f(\alpha)$ spectrum of singularities, related to $D_q$ according to eq.(95).

**Figure 16.** The local dimension as estimated from the slope of the partition functions for the scale–dependent structure; the local slopes have been obtained as a running least–square fit over three adjacent values of the partition function. Filled circles refer to the complete distribution and open triangles refer to the 3000 points random subsample. Column 1 reports the results of the CI method, column 2 refers to the BC method, column 3 to the DR method, column 4 to the NN method and column 5 to the MST. In columns 1 and 2, the five panels refer to $q = -2, 0, 2, 4$ and 6 from bottom to top. In column 3, the five panels refer to $\tau = -4, -2, 0.1, 4, 6$ from bottom to top. In columns 4 and 5, the panels refer to $\tau = -6, -4, -2, 0.1, 2$ again from bottom to top.

**Figure 17.** Spectrum of generalized dimensions $D_\tau$ versus $\tau$ for the multifractal structure defined by eq.(110). The solid line indicates the theoretical values of the dimension while filled circles indicate the dimension estimates obtained by the DR method.

**Figure 18.** Left panel: The characteristic scales for the galaxy and cluster distribution are plotted as a function of the sample size $R_s$. Open circles correspond to the smallest scale where $\xi(r)$ vanishes, while filled circles are the correlation length $r_o$ defined as the scale at which $\xi(r)$ takes unity value (after [100]). Right panel: the amplitude of the angular 2–point function is plotted for different samples as a function of the sample depth. Straight lines correspond to a Universe with large scale homogeneity and to a purely fractal Universe (after [26]).

**Figure 19.** The multifractal dimension spectrum for a volume–limited subsample of the CfA1 galaxy distribution (after [275]).

**Figure 20.** The moments of count–in–cells for the scale–dependent monofractal distribution (see text). From left to right I plot the $\langle N^q \rangle$ for $q = 2, 3, 4$ (filled dots), along with the corresponding moments corrected for Poissonian shot–noise according to eq.(141) (open dots). Also plotted are the corresponding local dimensions, evaluated from a 5 point log–log linear regression on the moment values.

**Figure 21.** Comparison between an exact N–body simulation and the Zel'dovich approximation in a $\Omega_o = 1$ Universe for a CDM spectrum evolved until the present time. The same initial phase assignment is taken for both particle configurations (after [270]).

**Figure 22.** Plot of $B^2(k) = k^4 T^2(k)$ for adiabatic baryonic perturbations. Each curve is labeled by the corresponding value of $\Omega h^2$ and the parameter $b$ dividing $k$ takes the value 0.065, 0.100, 0.179, 0.392 for curves from bottom to top. The spectrum amplitude is arbitrary (from [309]).

**Figure 23.** The post–recombination spectrum for adiabatic perturbations in both HDM,CDM and hydrid scenarios. Primordial Zel'dovich spectrum is always assumed. The plotted quantity is $k^{3/2}|\delta(k)|$ and represents the typical density fluctuation at the scale $\lambda$, $(\delta\rho/\rho)_\lambda$. All the spectra are normalized to have unity variance within a sharp–edged sphere of radius $8h^{-1}$ Mpc. Each cure is labelled by the fraction of the HDM component.

**Figure 24.** The angular two–point correlation function, $w(\vartheta)$, for a more evolved Gaussian CDM model, corresponding to the bias parameter $b = 1$ (left panel), and for non–Gaussian CDM models with initial negative skewness (see ref.[283] for more details). Dashed lines are for the simulated Lick maps, while filled dots are for the APM correlation [263].

**Figure 25.** Enhancement of the clustering of peaks higher than a fixed threshold with respect to the background peaks for a one–dimensional fluctuation field.

**Figure 26.** Gas number density vs. virial velocity; the formation of dwarfs vs. "normal" galaxies in CDM halos, and the origin of biased galaxy formation (after [124]).

**Figure 27.** Projections of the particle distributions in the evolving CDM N–body simulation. The box–side is normalized to have length $32.5\,h^{-2}$ Mpc at the present epoch. Different panels

correspond to different evolutionary stages, starting from the initial condition to the most evolved configuration.

**Figure 28.** The two–point correlation function, $\xi(r)$, versus radial distance $r$ for a CDM model at $a(t) = 2.5$. The $\xi(r)$ for $\delta > -1$ corresponds to the whole particle population, while those for $\nu > 0.5$ and $\nu > 2.0$ are calculated from a subset of particles chosen according to their initial positions (see text). Open diamonds refer to the results reported in ref.[111].

**Figure 29.** The coefficient $Q_h$ (hierarchical), $Q_k$ (Kirkwood) and $R_a + pR_b$ are plotted for different expansion factors $a(t)$ and biassing levels $\nu$ in the case of the SF+1 model. Error bars represent the scatter within the ensemble; for the sake of clarity we plot them only for $a(t) = 20$. At later $a(t)$ the bars are not very different from the potted ones, expecially at the non–linear scales where the stability of the clustering is expected. The arrow marks the linearity scale, where $\xi = 1$.

**Figure 30.** The same as Figure 29, but for the CDM initial spectrum. Open triangles refer to Fry's perturbative model [160].

**Figure 31.** The partition function $Z^B(q,r)$ versus the box size $r$ for the box–counting method, as calculated from eq.(97), for the CDM model at various epochs. Different symbols refer to different levels of bias. Here $L$ is the box length. The BC partition function is plotted for $q = 0, 2, 4, 6$.

**Figure 32.** The partition function $W(\tau,p)$ versus the probability $p$, as defined by eq.(99). Here we show $W(\tau,p)$ only for $\tau = -6, -2, 2, 6$. The expansion factors $a(t)$ and the biassing levels are the same as in the preceding figures. The length scale associated with a given $p$ is defined through eq.(221).

**Figure 33.** The $\tau(q)$ spectrum for the CDM model at $a(t) = 3.39, 4.51, 4.95$ and for different levels of bias. Here, open circles, open squares and open triangles refer to BC, CI and DR methods. Filled triangles and squares for the NN and MST methods, respectively. For the BC and CI algorithms, $\tau(q)$ at a given $q$ is defined as the slope of the least square fit of the corresponding partition function. For the other methods one obtains $q$ as a function of $\tau$. We accept a scaling regime, and plot $\tau(q)$, only when the linear regression coefficient of the fit is $R > 0.98$. The straight lines correspond to monofractal structures with dimensions $D = 1, 2, 3$.

**Figure 34.** The generalized dimensions $D_\tau$ obtained by using the DR partition function are plotted here at the final time of the integrations for the three different spectra: CDM (a), $n = -2$ (b) and $n = +1$ (c). The generalized dimensions are defined as $D_\tau = \tau/(q_\tau - 1)$.

**Figure 35.** The distribution of C25 clusters in the quasi equal area coordinates, for both real and simulated samples.

**Figure 36.** The correlation–integral partition function $Z^C(q,\vartheta)$ for the C36 sample is plotted

as a function of the angular scale for $q = 0$ (left panel) and $q = 4$ (right panel). Also plotted are the respective local dimensions, $D_q(\vartheta)$.

**Figure 37.** The density–reconstruction partition function, $W(\tau,p)$, for the C36 sample is plotted as a function of the probability measure $p$ for both $\tau = -2$ (left panel) and $\tau = 3$ (right panel).

**Figure 38.** The frequency distribution $F_W(\vartheta)$ of disk radii containing a given number $n = p/N$ of C36 clusters (here $N$ is the total number of objects and $p$ is the probability measure) is plotted for different $p$ values. The solid line is for $p = 2.6 \times 10^{-3}$, the dotted line is for $p = 1.6 \times 10^{-2}$ and the dashed line is for $p = 0.2$. The box insert gives the angular scale associated to each value of $p$, that is estimated as the $\vartheta$ value corresponding to the peak of the corresponding frequency distribution.

**Figure 39.** Expanded view of the local dimensions reported in Figure 37, once an angular scale is associated to each $p$ value. It is apparent that for $\tau = 3$ a well definite scaling develops up to $\vartheta \simeq 7°$, with a subsequent breaking. For $\tau = -2$ we see the homogeneity of the distribution of C36 clusters in the underdense regions.

**Figure 40.** The correlation–integral partition function for the simulated angular sample, obtained after projecting a three–dimensional multifractal structure. It is apparent that no significant features in the scale–invariant structure are introduced by projection and luminosity selection effects.

**Figure 41.** The correlation–integral partition function, $Z(q,\vartheta)$, for both real and simulated C36 clusters for $q = 0$ (upper panels) and $q = 4$ (lower panels). From left to right I report results for the $G_1$, $\chi_n^2$ and $LN_n$ models. Also plotted is the local dimension, $D_q(\vartheta)$. Filled circles are for PBF clusters, as also plotted in Figure 36, while open squares are for simulations.

**Figure 42.** The multifractal dimension spectrum, $D_\tau$, and the singularity spectrum, $f(\alpha)$, for the $\chi_n^2$ (left panels) and the $LN_n$ (right panels) clusters (open circles), as compared to the PBF sample (filled circles). The dimension values are obtained from the slope of the $W(\tau,p)$ partition function in the $p$ range of values where a good scaling is observed for both real and simulated data. Error bars are $1\sigma$ standard deviations for the log–log linear regression on the partition function.

**Figure 43.** The correlation–integral local dimension for simulated cluster samples extracted from fractal point distribution, which have been generated with the $\beta$–model. Only results for multifractal orders $q = 3$ and 5 are presented. The upper panel is for a scale–dependent fractal structure with $D = 1$ below $L_h = 40h^{-1}$ Mpc and homogeneous above this scale. The lower panel is for a monofractal $D = 1$ distribution, without large–scale homogeneity.

**Figure 44.** The same as in Figure 43, but for real Abell and ACO samples (upper and central panel, respectively), as well as for the Zel'dovich–simulated Abell sample (lower panel).